\pdfoutput=1

\documentclass[12pt,reqno,preprint]{article}
\usepackage{jheppub}
\usepackage{amsmath,amsthm,amssymb,amsfonts,mathrsfs}
\usepackage[normalem]{ulem}
\usepackage{hyperref}
\hypersetup{
    pdftex,
    pdfstartview=FitH,
    bookmarksnumbered=true,
    pagebackref,
    colorlinks,
    breaklinks,
    urlcolor=Maroon,
    linkcolor=Maroon,
    citecolor=Bittersweet
}
\usepackage[dvipsnames]{xcolor}
\usepackage{tikz}
\usetikzlibrary{calc,matrix,decorations.pathmorphing,decorations.markings,arrows,positioning,intersections,mindmap,backgrounds}
\usepackage{booktabs}
\usepackage[yyyymmdd,hhmmss]{datetime}

\newcommand{\fixed}{}
\newcommand{\verified}{}
\newcommand{\measure}[2]{{\langle{#1}\mathrm{d}^{#2}{#1}\rangle}}
\newcommand{\form}[3]{{\langle{#1}\mathrm{d}^{#3}{#2}\rangle}}

\newcommand{\flag}[1]{}

\begin{document}

\title{One-Loop Integrals from\\Spherical Projections of Planes and Quadrics}

\author{Nima Arkani-Hamed and Ellis Ye Yuan}
\affiliation{School of Natural Sciences, Institute for Advanced Study,\\ Einstein Drive, Princeton, NJ 08540, USA}
\emailAdd{arkani@ias.edu, yyuan@ias.edu}

\abstract{We initiate a systematic study of one-loop integrals by investigating
the connection between their singularity structures and geometric
configurations in the projective space associated to their Feynman
parametrization. We analyze these integrals by two  recursive
methods, which leads to two independent algebraic algorithms that
determine the symbols of any one-loop integrals in arbitrary spacetime
dimensions. The discontinuities of Feynman diagrams are shown to arise
from taking certain ``spherical contour'' residues in  Feynman
parameter space, which is geometrically interpreted as a projection of
the quadric surface (associated to the Symanzik polynomial at one
loop) through faces of the integration region (which is a simplex).
This geometry also leads to a manifestly Lorentz-invariant
understanding for perturbative unitarity at one loop.}

\maketitle


\newpage

\section{Introduction and Summary}\label{sec:intro}

Recent years have seen the surprising emergence of novel mathematical 
structures, in combinatorics and geometry, underlying the physics of 
scattering amplitudes. At loop level, much of this  progress has been made 
at the level of the ``integrand'' of amplitudes \cite{Arkani-Hamed:2013jha,Arkani-Hamed:2013kca,Arkani-Hamed:2017vfh}, with comparatively less 
understanding about the final ``integrated'' amplitudes themselves (see, e.g., \cite{Dennen:2015bet,Dennen:2016mdk,Prlina:2017azl,Prlina:2017tvx,Abreu:2014cla,Abreu:2015zaa,Abreu:2017enx,Abreu:2017mtm} for some more recent explorations). 
One reason for this is that the way in which locality, causality and unitarity 
are encoded in scattering is not yet properly understood even in perturbation
theory at the level of the amplitude itself. This physics is supposed to be encoded in the analytic structure of scattering amplitudes, 
especially in their intricate structure of branch cuts and 
discontinuities. In the simplest cases, the integrated amplitudes are polylogarithmic functions
of kinematical variables, and the fantastic control of this intricate branch cut structure 
offered by ``symbology'' \cite{1996alg.geom..1021G,Goncharov.A.B.:2009tja,Goncharov:2010jf,Duhr:2011zq} has led to a great deal of progress in understanding this structure, 
but while many important consistency conditions and patterns have been 
uncovered, especially for 
${\cal N}=4$ SYM amplitudes, there is still no first-principle 
understanding of what the symbol for general amplitudes should be. 

Indeed, there are still many simple, qualitative questions about integrated amplitudes even 
at one-loop. Consider some general one-loop integral in $D$ dimensions. It is well-known that the integrals 
are polylogarithms of weight at most $D/2$ (for $D$ even).  Why are they polylogarithms at all? After all, loop integrals are 
non-compact integrals with poles quadratic in loop momenta, whereas polylogs are integrals 
over simplices of forms with linear poles. Why is the transcendentality $1/2$ of the number of loop integration 
variables? And why are the obvious discontinuities of the amplitudes associated with cutting {\it two} propagators at a time? For instance for the box integral in four dimensions, the first discontinuity is the unitarity cut, which puts two propagators on shell, while the second discontinuity is given by the leading singularity of the integrand, which puts four propagators on shell. Related to this, how can we understand unitarity even in Lorentz-invariant way, even at 1-loop? The textbook understanding of the cutting rules picks out a special time direction and makes use of the largest time equation, but obviously the final relation between discontinuities and cuts is completely Lorentz invariant, and it would be nice to have a manifestly Lorentz-invariant understanding of it.  Finally, and of course most practically, how can we compute loop integrals without actually doing any integrals? It is by now very clear that loop amplitudes are associated with deep algebraic and geometric
structures, most extensively seen at low enough loop orders (and for special enough amplitudes) that are polylogarithmic. It is clear that the
symbols of these polylogs are answering geometric questions, but which questions precisely, and how can we use the algebra and geometry to
write down the answer? 

In this paper we will undertake a systematic analysis of one-loop integrals in Feynman parameter space, with a goal of giving a conceptual answer to these qualitative questions. We will find a geometric characterization of the integrals, via ``spherical projections'' of quadrics and planes naturally arising in the problem. In this introduction we highlight some of the novel points in our presentation, before embarking on a more detailed exposition in the main body of the paper. 

We will understand the symbol of the (polylogarithmic) functions that appear from two perspectives; the ``differentiation" and the ``discontinuity" points of view. Our understanding of the discontinuities in particular involves taking a novel kind of residue of rational functions, which we associate with a ``spherical contour". This will lead to a completely algebraic operation for extracting the symbols, including any rational prefactors, without any need for first carrying out integral reduction to a canonical basis. 

Polylogarithms are  characterized by ``simplification" under differentiation, and taking discontinuities across branch cuts. A (pure) polylogarithmic function of weight $n$ has the property that its derivative simplifies to a sum over weight $(n-1)$ polylogs as
\begin{equation}\label{eq:difaction}
\mathrm{d} F_n = \sum \mathrm{d}\log(g_{\alpha})\, F_{\alpha, n-1}.
\end{equation}
We can express the $\mathrm{d}$ of all the lower weight polylogs in a similar way, and the entire content of this chain of derivatives is captured by the symbol: 
\begin{equation}
\mathcal{S}F_n=\sum_{\alpha_i} g_{\alpha_1 \alpha_2 \ldots \alpha_n}\otimes g_{\alpha_1\alpha_2\ldots\alpha_{n-1}}\otimes\cdots\otimes g_{\alpha_1\alpha_2}\otimes g_{\alpha_1}.
\end{equation}
By \eqref{eq:difaction} obviously the symbols admit of algebraic properties
\begin{equation}
\cdots\otimes(ab)\otimes=\cdots\otimes a\otimes\cdots+\cdots\otimes b\otimes\cdots,\quad
\cdots\otimes (\text{constant})\otimes\cdots=0.
\end{equation}
The length of each term in $\mathcal{S}F_n$ is identical to its weight, since the differentiation terminates after acting $n$ times. At the level of the symbols, the differentiation $\mathrm{d}$ acts from the right to the left: chopping off a last symbol entry at each time, and the remaining part associated to it is the symbol of the corresponding lower-weight polylog
\begin{equation}
\mathrm{d}:\;\mathcal{S}F_n\equiv\sum_{\alpha}\mathcal{S}F_{\alpha,n-1}\otimes g_{\alpha}\longmapsto\sum\mathrm{d}\log(g_\alpha)\,\mathcal{S}F_{\alpha,n-1}.
\end{equation}
The operation of taking discontinuities, on the other hand, acts in the opposite direction: the zero or pole of a first entry represents a branch point, and the discontinuity around that branch point  itself has a symbol given by dropping off that entry.
\begin{equation}
\mathrm{Disc}_{\alpha'}:\;\mathcal{S}F_n\equiv\sum g_{\beta'}\otimes\mathcal{S}F_{\beta',n-1}\longmapsto \mathcal{S}F_{\alpha',n-1},
\end{equation}
where $\alpha',\beta'$ instead label specific branch cuts.

The ``differentiation" and ``discontinuity" understanding of polylogarithms is familiar from the standard presentation of polylogs as integrals of forms with linear poles  over simplices, as already encountered in the simplest example of the logarithm, which we can express as
\begin{equation}
I = \int_c^d \mathrm{d}z \left(\frac{1}{z-a} - \frac{1}{z-b} \right) = {\rm log} \left(\frac{(d-a)(c-b)}{(d-b)(c-a)}\right).
\end{equation}
For the integral to be well-defined we imagine that the poles $a,b$ are not inside the interval $(c,d)$ on which the integral is defined. 
The ``differentiation" method notes that this (one-dimensional) integral simplifies to a one-lower (zero-dimensional) integral upon taking a derivative with respect to the ``external data" $a,b,c,d$. This is esepcially obvious for the derivatives wrst the boundary points $c,d$. For derivatives wrst $a,b$, we also observe that the derivative of the {\it integrand} wrst external data is itself a total derivative: $\partial_a \frac{1}{z-a} = - \partial_z \frac{1}{z-a}$, and thus $\partial_a I$ also localizes to the boundaries 
of the integration region. The integral expression also makes it clear that $I$ has branch cuts and tells us how to compute the discontinuity across the cut. Suppose we carry $a$ through a closed loop, such that along the way $a$ does path through the integration region $(c,d)$, To keep the integral well-defined we have to deform the integration contour. By the time $a$ returns to its original location, the contour differs from the original one, by a small circle enclosing the pole at $z=a$. This shows that $I$ has a discontinuity, and this discontinuity is computed by the residue of the integrand. Note that the discontinuity itself is a ``simpler" function--just the constant ``$2 \pi i$". 

This integral representation of the logarithm is naturally generalized to ``Aomoto Polylogs", which are a function of two simplices $\overline{\Delta}$ and $\underline{\Delta}$ in an $n$-dimensional projective space \cite{aomoto1982,Goncharov.A.B.:2009tja}. The integrand is a form with simple poles on the boundaries of the simplex $\underline{\Delta}$, and the integral is performed over the interior of the simplex $\overline{\Delta}$. We will find it illuminating to describe the most general object of this type, where the simplices above are replaced by general convex polytopes $\overline{P}$ and $\underline{P}$: 
\begin{equation}
\Lambda(\overline{P},\underline{P}) = \int_{\overline{P}} \Omega_{\underline{P}}.
\end{equation}
Here $\Omega_{\underline{P}}$ is the canonical form with logarithmic singularities on the boundaries of $\underline{P}$, and the integration contour is $\overline{P}$. It is useful to specify $\overline{P}$ by giving it's faces $H_i$, i.e. if the co-ordinates on the projective space are $X^I$, the interior of $\overline{P}$ is specified by the equations $H_{i I} X^I \geq 0$. It is similarly useful to specify the $\underline{P}$ by giving the vertices of the polytope $Z^I_a$. 

The differentiation method naturally leads to a beautiful geometric expression for the symbol of $\Lambda(\overline{P},\underline{P})$. 
We define an admissible sequence $\{ Z_{a_1}, Z_{a_2}, \cdots, Z_{a_n} \}$ by the following property: for each $\alpha$, $Z_{a_\alpha}$ must be the pre-image of a vertex of the polytope obtained by projecting $\underline{P}$ through the vertices $Z_{a_1}, \cdots, Z_{a_{\alpha -1}}$. Similarly, we define an admissible sequence of facets of $\overline{P}$, $\{H_{i_1}, H_{i_2}, \cdots, H_{i_n}\}$ by the property that $H_{i_\alpha}$ is a face of the polytope restricted sequentially to the boundaries $H_{i_1},\cdots, H_{i_{\alpha-1}}$. Note that any allowed sequence $\{Z_{a_1}, \cdots,Z_{a_n}\}$ can be thought of projecting the entire polytope down to one dimension. A one-dimensional polytope is just an interval,
so all of $\{Z_{a_1}, \cdots, Z_{a_n}\}$ are projected to one boundary. Thus, all the other vertices of the original polytope have either been projected down to the same boundary or are all
on the same side of this boundary. In other words, for all the vertices $Z_b \, \notin\,\{Z_{a_1}, \cdots, Z_{a_{n-1}}\}$ for which
$\langle Z_{a_1} \cdots Z_{a_{n-1}}Z_b \rangle \neq 0$, all the signs of $\langle  Z_{a_1} \cdots Z_{a_{n-1}}Z_b \rangle$ are the same. We call this sign $s[Z_{a_1} \cdots Z_{a_{n-1}}Z_b]$. 
There is a similiar sign associated with the admissible sequence of hyperplanes, $s[H_{i_1} \cdots H_{i_n}]$. With these definitions, the symbol is given by 
\begin{equation}\label{eq:polytopesymbols}
\begin{split}
\mathcal{S}\Lambda(\overline{P},\underline{P})=&\sum s[Z_{a_1}\cdots Z_{a_n}]\,s[H_{i_1}\cdots H_{i_n}]\times\\
&\times (Z_{a_1}\cdots Z_{a_n},H_{i_1}\cdots H_{i_n})\otimes\cdots\otimes(Z_{a_1}Z_{a_2},H_{i_1}H_{i_2})\otimes(Z_{a_1},H_{i_1}).
\end{split}
\end{equation}
Here the summation is over all the admissible sequences $\{Z_{a_1},\ldots,Z_{a_n}\}$ and $\{H_{i_1},\ldots,H_{i_n}\}$, and the pairing, e.g., $(Z_1Z_2,H_1H_2)=(Z_1Z_2)^{[I_1I_2]}(H_1H_2)_{[I_1I_2]}$, where the group of indices on both sides are antisymmetrized. 
It is easy to intuitively understand how this form emerges from the ``differentiation" method. For instance, clearly when differentiating with respect to the planes $H_i$ of $\overline{P}$, the integral localizes to the corresponding boundary. Similarly, as we will see the derivative of the form $\Omega$ with respect to one of the vertices $Z_a$ of $\underline{P}$ is related to an $X$ derivative of the one-lower canonical form of the $(n-1)$ dimensional polytope obtained by projecting through $Z_a$. 

Quite beautifully, we can see explicitly that exactly the same symbol, read from the left to the right, captures the discontinuities of $\Lambda(\overline{P},\underline{P})$, with the role of faces and vertices of the polytopes interchanged. As with the example of the logarithm, the discontinuities are associated with the {\it residues} of the integrand. Now a residue of $\Omega(X,\underline{P})$ is associated with a {\it face} $G$  of $\underline{P}$, and as we move the external data through a closed cycle, there is a discontinuity when this hyperplane ``slices off" one of the vertices $V$ of the integration region polytope $\overline{P}$. Thus the branch point (and so a first symbol entry) is given by $(G  \cdot V)$, and the discontinuity itself is evaluated by computing the integral of the (one-lower-dimensional residual form taking the residue of $\Omega$ on $X \cdot G \to 0$) over the projection of the integration region polytope $\overline{P}$ through the vertex $V$. But this is exactly reproduced by reading the expression we have given for the symbol forwards! 
This is due to a simple geometric fact about the sequences $\{ Z_{a_1}, \cdots, Z_{a_n} \}$. An equivalent characterization of this sequence is that for all $\alpha$, 
$\{ Z_{a_{\alpha+1}},\cdots, Z_{\alpha_n}\}$ is a face of the polytope, contained in the face $\{Z_{a_{\alpha}} Z_{a_{\alpha + 1}} \cdots Z_{a_n}\}$. It is this simple geometric duality between ``projection through points" and ``inclusion in faces" that underlies the dual, ``backwards-forwards" interpretation of the symbol as keeping track of ``derivatives" and ``discontinuities". 

Note that, given the expression for the ``standard" Aomoto polylogs associated with a pair of simplices $(\overline{\Delta}, \underline{\Delta})$, we could of course obtain the formula for general polytopes $(\overline{P}, \underline{P})$ simply by triangulating each of $\overline{P}, \underline{P}$ as a sum over simplices. But such an expression would have spurious contributions to the symbol that only cancel in the sum. The expression we have given is instead completely intrinsic and invariant, and is independent of any triangulations\footnote{An application of this understanding of ``polytope" polylogarithms to the computation of cosmological correlators associated with ``cosmological polytopes" has been given in \cite{Arkani-Hamed:2017fdk}}.

The above example will be the template with which we will attempt to understand one-loop integrals. Working in Feynman parameter space, one-loop integrals have the general form of an integral over a simplex in some $n$-dimensional projective space, where the integral has quadratic poles in the coordinate $X$, so that we will be interested in general integrals of the form 
\begin{equation}
\int_\Delta \frac{\measure{X}{n}\, T_{I_1 \cdots I_{2p - n - 1}} X^{I_1} \cdots X^{I_{2p - n - 1}}}{(Q_{IJ} X^I X^J)^{p}},
\end{equation}
where $\langle\cdots\rangle$ denotes the contraction by the Levi-Civita symbol. It will be convenient to think of the simplex as given in a canonical form with facets given by basis vectors, i.e. a point $X^I$ is in the simplex if $X=[x_0:x_1:\cdots:x_n]$ with $x_i/x_0 \geq 0$ ($\forall i=1,\ldots,n$). 

For completely generic one-loop integrals in $D$ dimensions, the tensor $T$ and the quadric $Q$ are not completely generic, but our methods will apply to all the interesting degenerate cases as well. For clarity of presentation, we begin be discussing the special case with integrals over odd-dimensional spaces with trivial numerator, which we can naturally normalize as 
\begin{equation}
e_{2n}= \int_\Delta \frac{\measure{X}{2n-1}\, \sqrt{-\det Q}}{(QXX)^n}
\end{equation}
With this normalization this integral will turn out to be a pure polylog of weight $n$. Of course this integral has been computed in antiquity, but we will describe it in a way that will be useful for the most general case. The differentiation method is quite straightforward. Once again we find that derivatives of the integrand wrst the external data $Q$ is itself a total derivative, and we localize to a one-lower dimensional boundaries of the original simplex. But the new integrand on the boundaries turns out to be a total derivative by itself, and so we further localize to co-dimension 2 boundaries of the original simplex, which is guaranteed to be an integral of the same type but with $n \to (n-1)$, and  with a new simplex and new quadric. The final expression for the symbol takes an interesting form: 
\begin{equation}
\mathcal{S}e_{2n}=\mathcal{C}\sum_{\rho}\underline{\rho_{2n-1}\rho_{2n}}\otimes\cdots\otimes\underline{\rho_3\rho_4}\otimes\underline{\rho_1\rho_2},
\end{equation}
where $\mathcal{C}$ is merely some overall constant, and the summation $\rho$ is over all ordered partition of the $2n$ labels into $n$ symmetric pairs. Let us explain the notation for a specific partition $\rho$. To compute the entry $\underline{\rho_{2k-1}\rho_{2k}}$ for some $k$, we first truncate the quadric $Q$ into $Q_{\widehat{\{\rho_1,\ldots,\rho_{2k-2}\}}}$ by deleting its rows and columns labeled by $\{\rho_1,\ldots,\rho_{2k-2}\}$, and then pick out the $2\times2$ submatrix of its inverse that are labeled by $\{\rho_{2k-1},\rho_{2k}\}$. This submatrix specifies a quadric in a one-dimensional projective space and naturally induces a log whose argument become the corresponding entry $\underline{\rho_{2k-1}\rho_{2k}}$. Explicitly we have that
\begin{equation}
\underline{\rho_{2k-1}\rho_{2k}}=r\left((Q_{\widehat{\{\rho_1,\ldots,\rho_{2k-2}\}}})^{-1}_{\{\rho_{2k-1}\rho_{2k}\}}\right),
\end{equation}
where the function $r(\mathbf{M})$ of a $2\times2$ matrix $\mathbf{M}$ is defined to be
\begin{equation}\label{intro:ratioofroots}
r(\mathbf{M})=\frac{\mathbf{M}_{12}-\sqrt{(\mathbf{M}_{12})^2-\mathbf{M}_{11}\mathbf{M}_{22}}}{\mathbf{M}_{12}+\sqrt{(\mathbf{M}_{12})^2-\mathbf{M}_{11}\mathbf{M}_{22}}}.
\end{equation}
We name this as the \textit{ratio of roots} of $\mathbf{M}$, as it is obviously the ratio of the two roots of the corresponding 1d quadric.

Let us now discuss the ``discontinuity" understanding of the same integral. Here we have a basic obstruction: the integrand $\measure{X}{n}/(XX Q)^n$ clearly does not have any ordinary residues at all! Fortunately, there a different notion of ``residue" is relevant. In order to motivate it, consider the 1-form with a double-pole in one variable, $\mathrm{d}z/z^2$. Clearly, any integral of this form over a  closed contour gives zero. A fancier way of saying this is to think of this as a 1-form with a double-pole on ${\mathbf P}^1$, $\langle \lambda\mathrm{d} \lambda \rangle/\langle \lambda a \rangle^2$. Any ``integral" of this form would have to give us a function of weight $-2$ in $a$, but no such invariant exists since $\langle a a \rangle = 0$. But let's now consider a 2-form with a double pole in two variables, $-\mathrm{d}y \mathrm{d}w/(y-w)^2$. Putting $y=-1/z$, the form becomes $\mathrm{d}z \mathrm{d}w/(z w + 1)^2$. 
But now there is a natural ``spherical" contour on which we can integrate this form to get something non-zero: 
\begin{equation}
\int_{w = \bar{z}} \frac{\mathrm{d}z \mathrm{d}w}{(w z + 1)^2} = 2 \pi \int_0^\infty \frac{\mathrm{d}r\, r}{(r^2 + 1)^2} = \pi
\end{equation}
We can again say this in a fancier way, thinking in terms of a form on ${\bf P}^1 \times {\bf P}^1$. This time, there is a natural ``residue" that we can algebraically associate with the form, with the correct projective weights: 
\begin{equation}
\int_{{\rm spherical \,  contour}} \frac{\langle \lambda \mathrm{d} \lambda\rangle [\tilde \lambda \mathrm{d} \tilde \lambda]}{\langle\lambda| Q |\tilde \lambda]^2} = \pi \times \frac{1}{\det Q}
\end{equation}
Indeed, just as ordinary residues can be defined simply as an algebraic operation on the form, similarly spherical contours can be defined on any form on ${\bf P}^1 \times {\bf P}^1$. Either by computing the integral on the same contour with $w = \bar{z}$, or by same algebraic logic as the above, we can define the spherical contour for a general form as
\begin{equation}
\int_{{\rm spherical \,  contour}}\frac{\langle \lambda \mathrm{d} \lambda\rangle [\tilde \lambda \mathrm{d} \tilde \lambda] \lambda_{\alpha_1} \cdots \lambda_{\alpha_m} \tilde \lambda_{\dot{\alpha}_1} \cdots \tilde \lambda_{\dot{\alpha}_m}}{\langle \lambda|Q|\tilde \lambda]^{2 + m}} \propto\pi\times\frac{ Q^{-1}_{\alpha_1 \dot{\alpha}_1} \cdots Q^{-1}_{\alpha_m \dot{\alpha}_m}}{\det Q}
\end{equation}
Now, given any quadric $Q$ in any number of variables, with poles given by powers of $(XQX)$, for generic $Q$ we can always take a spherical residue in pairs of variables. We  pick a pair of the variables $x_i, x_j$. By a linear transformation we can always put the quadric in the form $XQX = a x^{\prime}_i x^{\prime}_j + b$, and then we can take the ${\bf P}^1 \times {\bf P}^1$ spherical residue on this pair of variables. We can say this more invariantly by defining a natural ${\bf P}^2$ notion of spherical residue.  Consider a general quadric in two variables $x_i$ of the form $x_i x_j q_{ij} + x_i c_i + d$. Let's group these variables into $X^I = (1, x_1, x_2)$, we also have a natural plane at infinity $L_I = (1,0,0)$. Then, the spherical residue is defined as 
\begin{equation}
\int_{X \cdot L>0} \frac{\measure{X}{2} X^I}{(XQX)^2} =\pi \times \frac{(Q^{-1} L)^I }{\sqrt{\epsilon^{I_1I_2I_3}\epsilon^{J_1J_2J_3}Q_{I_1J_1}Q_{I_2J_3} L_{I_3} L_{J_3}}}
\end{equation}
and we can obtain the most general spherical contour integral with extra numerator factors by differentiating with respect to $q$. Note that the spherical contour is well-defined so long as the quadric is non-degenerate. 

Armed with the notion of spherical contours and residues, we can now discuss the ``discontinuity" method for computing the symbol of general integrals with powers of quadric poles, as usual reading the symbol in the opposite direction as the ``differentiation" method. This method will allow us to compute the symbol even for ``non-pure" integrals.  We again pick a pair of Feynman parameters $x_i,x_j$, and first consider setting all other parameters to zero. This is (projectively) a 1-dimensional space $V_{ij}$. Now, the choice of $(ij)$ is associated with a natural block decomposition of the full quadric $Q$ as 
\begin{equation}
Q=\left(\begin{matrix}Q_{\{i,j\}}&P_{\{i,j\}}^{\rm T}\\ P_{\{i,j\}}&Q_{\widehat{\{i,j\}}}\end{matrix}\right),
\end{equation}
Said geometrically, $Q_{\{i,j\}}$ is the restriction of the quadric to the projective line $V_{ij}$.  This quadric associates to a log, whose argument is a first entry of the symbol. Explicitly, it is again computed by the ratio of roots
\begin{equation}
r(Q_{\{i,j\}}^{-1}),
\end{equation}
following the same formula \eqref{intro:ratioofroots}. Moving on to the next entry, we encounter a new quadric $Q^{(ij)}$ resulting from the spherical contour integral discussed above. Geometrically, this is understood to be the original quadric $Q$, {\it projected through} the line $V_{(ij)}$. Algebraically, we have that 
\begin{equation}
Q^{(ij)}= Q_{\widehat{\{i,j\}}}-P_{\{i,j\}}(Q_{\{i,j\}})^{-1}P_{\{i,j\}}^{\rm T}.
\end{equation} 
We then begin with this new quadric, choose another pair $(i',j')$, and following the same prescription the second entry following the one worked out above is
\begin{equation}
r[(Q^{(ij)}_{\{i',j'\}})^{-1}].
\end{equation}
We then continue taking spherical contours in any way we can, recording the symbol entries as we go. The process ultimately terminates when we can no longer take any spherical contours, when the quadric is degenerate to a certain degree. There may still be many integration variables left at this point, but it is guaranteed that the final integral yields just a rational expression, and this rational is exactly the coefficient of the symbol term obtained in this way. For pure integrals, this way of determining  the symbol coincides precisely  with the ``differentiation" picture read in the opposite direction, for a similar geometric reason as seen for Aomoto polylogs.

Let us illustrate how the spherical contour computation of the symbol works with a simple example familiar from ${\cal N}=4$ SYM computations:  the finite one-loop hexagon integral with ``mixed numerator". 
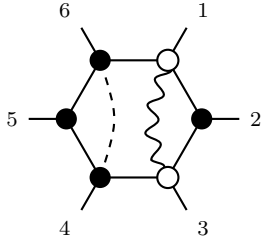
\begin{center}
\begin{tikzpicture}[decoration=snake]
\coordinate [label=60:{\scriptsize $1$}] (e1) at (60:1.4cm);
\coordinate [label=0:{\scriptsize $2$}] (e2) at (0:1.4cm);
\coordinate [label=-60:{\scriptsize $3$}] (e3) at (-60:1.4cm);
\coordinate [label=-120:{\scriptsize $4$}] (e4) at (-120:1.4cm);
\coordinate [label=180:{\scriptsize $5$}] (e5) at (180:1.4cm);
\coordinate [label=120:{\scriptsize $6$}] (e6) at (120:1.4cm);
\coordinate (v1) at (60:.9cm);
\coordinate (v2) at (0:.9cm);
\coordinate (v3) at (-60:.9cm);
\coordinate (v4) at (-120:.9cm);
\coordinate (v5) at (180:.9cm);
\coordinate (v6) at (120:.9cm);
\draw [black,thick] (v1) -- (v2) -- (v3) -- (v4) -- (v5) -- (v6) -- cycle;
\draw [black,thick] (v1) -- (e1);
\draw [black,thick] (v2) -- (e2);
\draw [black,thick] (v3) -- (e3);
\draw [black,thick] (v4) -- (e4);
\draw [black,thick] (v5) -- (e5);
\draw [black,thick] (v6) -- (e6);
\draw [black,thick,decorate] (v1) .. controls (.2cm,0) .. (v3);
\draw [black,thick,dashed] (v4) .. controls (-.2cm,0) .. (v6);
\path [draw=black,fill=white,thick] (v1) circle(4pt);
\fill [black] (v2) circle(4pt);
\path [draw=black,fill=white,thick] (v3) circle(4pt);
\fill [black] (v4) circle(4pt);
\fill [black] (v5) circle(4pt);
\fill [black] (v6) circle(4pt);
\end{tikzpicture}
\end{center}
It is conveniently expressed as an integral over lines $(AB)$ in momentum-twistor space, and we can easily go to Feynman parameter space: 
\begin{equation}
\begin{split}
I&=\underset{AB}{\int}\frac{\langle AB(612)\cap(234)\rangle\langle AB46\rangle}{\langle AB12\rangle\langle AB23\rangle\langle AB34\rangle\langle AB45\rangle\langle AB56\rangle\langle AB61\rangle}\\
&=\frac{\langle6124\rangle\langle2346\rangle}{\langle6134\rangle\langle1245\rangle\langle2356\rangle}\int_0^\infty\frac{\langle X\mathrm{d}X^5\rangle\,T[X^2]}{(XQX)^4},
\end{split}
\end{equation}
where the quadric
\begin{equation}
Q=\frac{1}{2}\left(\begin{matrix}0&0&1&1&1&0\\0&0&0&u_1&1&1\\1&0&0&0&u_2&1\\1&u_1&0&0&0&u_3\\1&1&u_2&0&0&0\\0&1&1&u_3&0&0\end{matrix}\right)
\end{equation}
only depends on the crossratios
\begin{equation}
u_1=\frac{\langle1234\rangle\langle4561\rangle}{\langle1245\rangle\langle3461\rangle},\quad
u_2=\frac{\langle2345\rangle\langle5612\rangle}{\langle2356\rangle\langle4512\rangle},\quad
u_3=\frac{\langle3456\rangle\langle6123\rangle}{\langle3461\rangle\langle5623\rangle}.
\end{equation}
And the numerator is a degree-2 polynomial
\begin{equation}
\begin{split}
T[X^2]=&XQX\!-\!6\left(\!\frac{u_2}{\langle2346\rangle} x_5-\frac{\langle1256\rangle}{\langle6124\rangle\langle2356\rangle} x_6\!\right)\!\!\left(\!-\frac{\langle6124\rangle\langle2345\rangle}{\langle1245\rangle\,u_2} x_2+\langle2346\rangle x_3\!\right).
\end{split}
\end{equation}
This integral is known to have uniform transcendental weight 2 but is not pure.

Let's now work out one of its symbol terms together with the corresponding rational coefficient using spherical contours. To be specific, let us project through the $(46)$ direction. The corresponding first symbol entry is determined by the $2\times2$ submatrix $Q_{\{4,6\}}=\left(\begin{matrix}0& u_3\\u_3&0\end{matrix}\right)$ is $u_3^2\otimes\cdots$ (in the computation the quadric under consideration is a bit special, such that the ratio of roots is not directly applicable, and one needs to refer to the general formula \eqref{eq:discentryformula}). To obtain the integral associated to the projection, we perform an affine transformation to block diagonalize $Q$ such that entries labeled by $\{4,6\}$ are isolated fro the rest, i.e.,
\begin{equation}
x_4\mapsto\frac{\sqrt{u_3}x_4-x_2-x_3}{u_3},\quad
x_6\mapsto\frac{\sqrt{u_3}x_6-x_1-u_1x_2}{u_3},
\end{equation}
and then assign $x_4=re^{i\theta}$, $x_6=re^{-i\theta}$ and integrate over $r\in[0,\infty)$, $\theta\in[0,2\pi]$. The resulting projected quadric reads
\begin{equation}
-\frac{1}{u_3}x_1x_2-\frac{u_1}{u_3}x_2^2+\left(1-\frac{1}{u_3}\right)x_1x_3-\frac{u_1}{u_3}x_2x_3+x_1x_5+x_2x_5+u_2x_3x_5+x_4x_6.
\end{equation}
Next we do another projection through the $(12)$ direction. The analysis is the same as what we did just now. The first two entries are thus determined as $u_3^2\otimes u_1u_3\otimes\cdots$. As a result of these two projections we obtain the integral
\begin{equation}
\begin{split}
\frac{\langle6124\rangle\langle2346\rangle}{8\langle6134\rangle\langle1245\rangle\langle2356\rangle}\int_0^\infty\!\!\frac{(u_1x_3-x_5+u_1x_5+u_5x_5)(u_5^{-1}u_2x_3-x_3+u_3x_3+u_3x_5)}{(u_1(u_3-u_1)x_3^2+(1-u_1-u_2-u_3+2u_1u_3)x_3x_5+u_3(u_1-1)x_5^2)^2},
\end{split}
\end{equation}
with $u_5=\frac{\langle1246\rangle\langle2345\rangle}{\langle2346\rangle\langle1245\rangle}$. This remaining one-dimensional integral turns out to be a rational function $\frac{1}{4}\frac{\langle1346\rangle}{\langle1345\rangle\langle1356\rangle}$. This indicates that the symbol term terminates at length two, and the integral above computes exactly the rational coefficient in front. Hence from this sequence of projections we read out the term
\begin{equation}
\frac{1}{2}\frac{\langle1346\rangle}{\langle1345\rangle\langle1356\rangle}\,u_3\otimes(u_1u_3).
\end{equation}
This can be decomposed into two terms $u_3\otimes u_1$ and $u_3\otimes u_3$. The former indeed appears in the symbols for this diagram with the expected coefficient, the latter cancel away with some other terms from other choices of projections.

The picture of spherical contours also give us a transparent connection with cuts of the loop integrand. Indeed, taking a spherical contour on a pair of Feynman parameters $x_i,x_j$ corresponds precisely to cutting the corresponding propagators in the loop integrand! In other words, the new Feynman parameter integrand obtained by taking the $(i,j)$ spherical contour, exactly corresponds to the Feynman parametrization of the cut integral putting the $(i,j)$ propagators on shell. 

We have now seen two integral representations for polylogarithims: the apparently more ``obviously" polylogarithmic Aomoto form, with discontinuities computed by ordinary residues, and the form with integrals over forms with quadric poles, associated with the apparently more exotic spherical contours. But in fact quite beautifully, the ``quadric" form is in fact more fundamental, and reproduce the Aomoto polylogs as a special case! This is seen in the by-now familiar expression of the canonical form $\Omega_{\underline{P}}$ with logarithmic singularities on the polytope $\underline{P}$, as a volume integral over the dual polytope $\widetilde{\underline{P}}$. This gives us a more fundamental expression for the Aomoto polylog
\begin{equation}
\Lambda(\overline{P},\underline{P}) = \int_{X \subset \overline{P}, Y \subset \widetilde{\underline{P}}} \frac{\measure{X}{n}\,\measure{Y}{n}}{(X \cdot Y)^{1+n}}
\end{equation}
Note that this is exactly of the general form of an integral over a $(2n)$ dimensional projective space, with a pole given by a power of the quadric $(X \cdot Y)$! Thus the story of spherical projections through quadrics and planes also gives an understanding for the ``standard" presentation of polylogarithms as well. 

We hope that this kind of geometric characterization and algebraic determination of one-loop integrals will extend in a natural way beyond one-loop, the exploration of which we leave to future work. But let us conclude this high-level summary of some of the main results of the present paper with an invitation to the subsequent sections. 
The main contents of the paper divide into two relatively independent parts. The first part (Section \ref{sec:dif1} to \ref{sec:feynman2}) investigates the differentiation method, while the second part (Section \ref{sec:dis1} to \ref{sec:unitarity}) introduces the discontinuity method.

In Section \ref{sec:dif1}, we inspect two elementary classes of quadric integrals, which we denote as $E_n$ and $E_{n,1}$. With these integrals we present the general features in using the differentiation method to determine symbols. In particular, as mentioned above for quadric integrals every differentiation reduces the (effective) number of integrations by 2, and each specific symbol entry is shown to be identical to the ratio of two roots of some quadratic polynomial, for which we derive the general formula. Technical details of some derivations and proofs in this section are collected in Appendix \ref{app:derivation} and \ref{app:cancelation}.

In Section \ref{sec:feynman1}, we apply the technique in Section \ref{sec:dif1} to scalar Feynman diagrams with $n<d+2$. Explicit examples include massive and massless diagrams in 4d and 6d. Among these, the complete result for 6d scalar hexagon is listed in Appendix \ref{app:hexagon}.

In Section \ref{sec:dif2}, we generalize the above method to arbitrary quadric integrals, including situations when the Feynman parametrization involves non-trivial tensor structure, as well as when the quadric is degenerate. General structural patterns of the quadric integrals are commented at the end of this section.

In Section \ref{sec:feynman2}, we analyze more general Feynman diagrams at one loop following the discussion in Section \ref{sec:dif2}. To illustrate the technical details we provide two examples of finite hexagons in 4d, one with and the other without unit leading singularities, as well as example of parity-odd diagrams, and diagrams that give rise to a summation of transcendental functions with different weights.

In Section \ref{sec:dis1}, we switch to the discontinuity analysis. We motivate and define the basic tool in extracting discontinuities of quadric integrals: what we referred to above as spherical contours. Correspondingly we provide a new prescription for determining the symbols of elementary quadric integrals, where the discontinuities play the essential role. We prove the equivalence between this discontinuity algorithm and that from the differentiation as discussed in Section \ref{sec:dif1}. At the end we also present a duality among $E_n$ integrals, with a detailed proof contained in Appendix \ref{app:duality}.

In Section \ref{sec:dis2}, we generalize the analysis of spherical contours to arbitrary quadric integrals. Some features in this analysis are further illustrated by applying the method explicitly to the same set of examples as in Section \ref{sec:feynman2}.

In Section \ref{sec:contour}, we explain in detail the geometric interpretation of the spherical contours, as projections. We also discuss about a new notion of $S^1$ contours, which leads to a fibration of the spherical contours, and also makes properties of integrals in odd spacetime dimensions transparent. We then generalize these discussions to arbitrary projections and explain their connections to contours and discontinuities of the quadric integrals. A general proof for the integrability of the resulting symbols is provided at the end.

In Section \ref{sec:unitarity}, we explain the equivalence between the representation of discontinuities from the spherical contours and the Feynman parametrization of unitarity cut of diagrams, which leads to a proof for unitarity of Feynman diagrams at one loop that is manifestly Lorentz-invariant. Feynman  parametrization for more general cuts is discussed in Appendix \ref{app:sec:morecuts}.

In both the differentiation and the discontinuity method situations with degenerate quadric call for extra care. While we collect some results in Section \ref{sec:dif1}, \ref{sec:dif2} and \ref{sec:contour}, and detailed discussion on this is provided in Appendix \ref{app:degenerateQ}.

Finally in Appendix \ref{app:aomoto} we present a detailed account on the Aomoto polylogarithms using both the differentiation and the discontinuity analysis, together with their geometric interpretation, and explain their connection to the quadric integrals. 

A brief summary of some notational conventions used in this paper is collected in Appendix \ref{app:convention} for easy loop-up.

\section{Differentiation I: Basic Quadric Integrals}\label{sec:dif1}

\flag{No further revisions.}

\subsection{The basic quadric integrals}

We begin with two basic families of integrals, indexed by an integer $n$
\begin{align}
\fixed
E_n\equiv E_{n,0}&\equiv\int_{\Delta}\frac{\measure{X}{n-1}}{(XQX)^{\frac{n}{2}}},\\
\fixed
E_{n,1}&\equiv\int_{\Delta}\frac{\measure{X}{n-1}\,(LX)}{(XQX)^{\frac{n+1}{2}}},
\end{align}
where
\begin{equation}
\measure{X}{n-1}=\frac{\epsilon_{I_1\ldots I_n}}{(n-1)!}\,X^{I_1}\mathrm{d}X^{I_2}\wedge\cdots\wedge\mathrm{d}X^{I_n},\quad\epsilon_{1\ldots n}=1,
\end{equation}
and $\Delta$ denotes a simplex contour in $\mathbb{CP}^{n-1}$, specified by its vertices $\{V\}$ or boundary hyperplanes $\{H\}$. Whenever there is no ambiguity we omit explicit tensor contractions but write according to the sequence in which the indices are contracted, e.g., $XQX\equiv X^IQ_{IJ}X^J$ and $LX\equiv L_IX^I$. More about the conventions used in this paper are summarized in Appendix \ref{app:convention}.

By definition these are projective integrals in $\mathbb{CP}^{n-1}$, invariant under an arbitrary $\mathrm{PGL}(n)$ action on the coordinates $X$. This implies that whatever function arising from such an integral has to be a function \emph{only} of the invariants constructed out of $\{\Delta,Q,L\}$, e.g., $V_iQV_j$. 

In practice one can always transform into a frame where the simplex is canonical, i.e., the $i^{\rm th}$ boundary is defined by $x_i=0$, so that $Q_{ij}\equiv V_iQV_j$. We call this frame the canonical frame, and throughout the paper we will assume to always work in it, unless otherwise stated. This frame is also what one naturally lands on from a straightforward Feynman parametrization.

For physical applications, $d$-gons and $(d+1)$-gons in $d$ dimensions are of the type $E_d$ and $E_{d+1,1}$, respectively. But to generalize the discussion we consider arbitrary $\{Q,L\}$, which do not necessarily originate from an actual Feynman integral.

More generally, we refer the name ``quadric integral'' to any projective integral whose integrand singularity only comes from a single quadric and whose contour is a simplex, i.e., integrals of the form
\begin{equation}
E_{n,k}=\int_{\Delta}\frac{\measure{X}{n-1}T[X^k]}{(XQX)^{\frac{n+k}{2}}},
\end{equation}
where $T[X^k]\equiv T_{I_1\ldots I_k}X^{I_1}\cdots X^{I_k}$ denotes some symmetric tensor of rank $k$.

The family of $E_n$ integrals stand out as essential objects in our discussion. On the one hand, they are ``minimal'': as we will show later, an arbitrary quadric integral is a linear combination of $E_n$ integrals. Also every quadric $Q$ uniquely defines an $E_n$ integral. On the other hand, they are ``maximal'': the symbol of an arbitrary quadric integral is in a sense always embedded inside the symbols for the $E_n$ integral defined by the quadric $Q$ therein. Hence it is crucial to understand in great detail the structure of these objects. 

For the time being we assume that the quadric is non-degenerate, i.e.,
\begin{equation}
Q\neq0.
\end{equation}

\subsection{$E_{n,1}$}

Any $E_{n,1}$ integral involves an obviously trivial integration, because its integrand is always exact
\begin{equation}\label{eq:En1integrand}
\fixed
\frac{\measure{X}{n-1}\,(LX)}{(XQX)^{\frac{n+1}{2}}}
=\frac{1}{n-1}\,\mathrm{d}_X\!\!\left[\frac{\form{(Q^{-1}L)X}{X}{n-2}}{(XQX)^{\frac{n-1}{2}}}\right],
\end{equation}
which can be verified using the Schouten identity
\begin{equation}
\fixed
\mathrm{d}X^I\form{AX}{X}{n-2}+A^I\measure{X}{n-1}-X^I \form{A}{X}{n-1}=0,\quad \forall A^I.
\end{equation}
Thus we are able to localize the integral on each of the codim-1 boundaries $\{H_i\}$. 

Since we work in the canonical frame, each boundary $H_i$ is defined by $x_i=0$; the localization onto $H_i$ is merely to set to zero the $i^{\rm th}$ component of $X$ (denote it as $X_{(i)}$), and correspondingly we obtain a new integral in one lower dimensions, with the new quadric $Q_{(i)}$ obtained by deleting entries labeled by $i$. Explicitly, we have
\begin{equation}
\fixed
E_{n,1}=\sum_{i=1}^{n}\frac{H_iQ^{-1}L}{n-1}\int_{\Delta_{(i)}}\frac{\measure{X_{(i)}}{n-2}}{(X_{(i)}Q_{(i)}X_{(i)})^{\frac{n-1}{2}}}.
\end{equation}
In the above $(H_iQ^{-1}L)$ merely extracts the $i^{\rm th}$ component of the vector $Q^{-1}L$. $\Delta_{(i)}$ denotes the facet on the hyperplane $H_i$, which is again canonical in $\mathbb{CP}^{n-2}$.  This confirms that each $E_{n,1}$ integral is merely a linear combination of $E_{n-1}$'s.

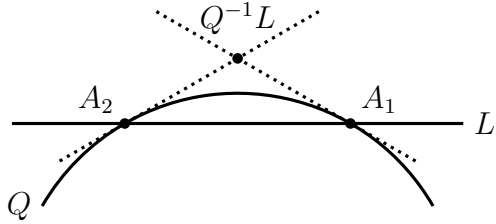
\begin{figure}[h]
\begin{center}
\begin{tikzpicture}
\draw [black,very thick] (30:3) arc [start angle=30,end angle=150,radius=3];
\draw [black,very thick] ($(60:3)+(1.5,0)$) -- ($(120:3)+(-1.5,0)$);
\coordinate [label=80:$A_1$] (p1) at (60:3);
\coordinate (p2) at ($(60:3)+(-30:1)$);
\coordinate [label=100:$A_2$] (p3) at (120:3);
\coordinate (p4) at ($(120:3)+(210:1)$);
\draw [black,very thick,dotted] (p2) -- ($(p2)!4!(p1)$);
\draw [black,very thick,dotted] (p4) -- ($(p4)!4!(p3)$);
\fill [black] (intersection of p2--p1 and p4--p3) circle(2pt);
\fill [black] (p1) circle(2pt);
\fill [black] (p3) circle(2pt);
\node [anchor=east] at (150:3) {$Q$};
\node [anchor=west] at ($(60:3)+(1.5,0)$) {$L$};
\node at (0,4) {$Q^{-1}L$};
\end{tikzpicture}
\caption{Geometric interpretation of $Q^{-1}L$.}
\label{fig:QL}
\end{center}
\end{figure}

Before we move on, note that the decomposition obtained above is not unique, since one can always add some exact $(n-2)$-form inside $\mathrm{d}_X$ in \eqref{eq:En1integrand}. However, it is special because the coefficients in the decomposition constitute a vector $Q^{-1}L$, which is actually the intersection point of all hyperplanes tangent to $Q$ at points in the intersection of $Q$ and $L$, because
\begin{equation}
\fixed
AQ(Q^{-1}L-A)=0,\quad\forall A\text{ s.t.~}AQA=LA=0.
\end{equation}
This simplest case in 2d is illustrated in Figure \ref{fig:QL}.

\subsection{$E_n$}

To make the analysis clean, we properly normalize $E_n$ integrals into
\begin{equation}
\fixed
e_n=\sqrt{-q}\,E_n,\quad q\equiv\det Q.
\end{equation}

We choose to study the structure of $e_n$ using the differentiation method. In order to quickly observe the expected structure, it is convenient to differentiate wrst the boundaries $H_i$. Note that there always exists a frame where $Q$ is purely an identity, so that the information about the integral is fully pushed to the definition of the integration region. Hence it suffices to check the differentiation wrst $H_i$. Explicitly we have
\begin{equation}
\begin{split}\fixed
\mathrm{d}_{H_i}e_n&=\int_{-\infty}^{+\infty}\frac{\sqrt{-q}\,\measure{X}{n-1}}{(XQX)^{\frac{n}{2}}}(\mathrm{d}H_iX)\delta(H_iX)\prod_{k\neq i}\theta(H_kX)\\
&=-\int_{\Delta_{(i)}}\frac{\sqrt{-q}\,\measure{X_{(i)}}{n-2}\,((\mathrm{d}H_i)_{(i)}X_{(i)})}{(X_{(i)}Q_{(i)}X_{(i)})^{\frac{n}{2}}},
\end{split}
\end{equation}
where $\mathrm{d}_{H_i}\equiv\mathrm{d}H_{iI}\partial_{H_{iI}}$ (no summation on $i$). This explicitly becomes an $E_{n-1,1}$ integral on the codim-1 boundary $\Delta_{(i)}$. Following the previous discussion it can be further localized onto the set of codim-2 boundaries bordering $\Delta_{(i)}$, giving rise to
\begin{equation}
\fixed
\mathrm{d}_{H_i}e_n=\sum_{j\neq i}\frac{-1}{n-2}\sqrt{\frac{q}{q_{(ij)}}}\,(\mathrm{d}H_i)_{(i)}Q_{(i)}^{-1}(H_j)_{(i)}\int_{\Delta_{(ij)}}\frac{\sqrt{-q_{(ij)}}\,\measure{X_{(ij)}}{n-3}}{(X_{(ij)}Q_{(ij)}X_{(ij)})^{\frac{n-2}{2}}}.
\end{equation}
So differentiation turns the original $e_n$ into a linear summation of $e_{n-2}$'s. It can further be verified that each coefficient above is actually a dlog. However, let us work this out by differentiating wrst $Q$. (If we consider generalizing the integral to polynomials of higher degrees in the denominator, then the data cannot be fully passed to the contour by a $\mathrm{PGL}(n)$, so that the differentiation wrst the curve is not just a necessity by actually a must.)

Applying $\mathrm{d}_Q\equiv \mathrm{d}Q_{IJ}\partial_{Q_{IJ}}$, although not obvious, we observe that the integrand turns into an exact form
\begin{equation}
\begin{split}\fixed
\mathrm{d}_Qe_n&=-\frac{\sqrt{-q}}{2}\int_\Delta\mathrm{d}_X\!\!\left[\frac{\form{(Q^{-1}\mathrm{d}QX)X}{X}{n-2}}{(XQX)^{\frac{n}{2}}}\right]\\
&=-\frac{\sqrt{-q}}{2}\sum_{i=1}^{n}\int_{\Delta_{(i)}}\frac{\measure{X_{(i)}}{n-2}\,(H_iQ^{-1}\mathrm{d}QP(i)X_{(i)})}{(X_{(i)}Q_{(i)}X_{(i)})^{\frac{n}{2}}},
\end{split}
\end{equation}
where $P(i)_{\;I'}^{I}\equiv \delta_{\;I'}^{I}$ is inserted just to account for the mismatch between the range of indices $I=1,\ldots,n$ and $I'=1,\ldots,\hat{i},\ldots,n$ contracted on the two sides.

Now in the second line each term is explicitly an $E_{n-1,1}$ integral, so that we can further localize onto the set of codim-2 boundaries (again let us assume $Q_{(i)}$ is non-degenerate for any $i$)
\begin{equation}\label{dO}
\fixed
\mathrm{d}_Qe_n=\sum_{1\leq i<j\leq n}\frac{C_{(ij)}}{2(n-2)}\,\int_{\Delta_{(ij)}}\frac{\sqrt{-q_{(ij)}}\,\measure{X_{(ij)}}{n-3}}{(X_{(ij)}Q_{(ij)}X_{(ij)})^{\frac{n-2}{2}}},
\end{equation}
with all the subscripts $(ij)$ symmetric in its labels. Correspondingly the explicit expression of the coefficient is
\begin{equation}\label{Cijgeneric}
\fixed
C_{(ij)}=-\frac{\sqrt{q}}{\sqrt{q_{(ij)}}}\left(H_iQ^{-1}\mathrm{d}QP(i)Q_{(i)}^{-1}P^{\rm T}(i)H_j+(i\leftrightarrow j)\right).
\end{equation}
Note that each term associated to a codim-2 boundary $\Delta_{(ij)}$ (residing on the intersection of hyperplanes $H_i\cap H_j$) receives two contributions via localizations, one via the codim-1 boundary $\Delta_{(i)}$ and the other via $\Delta_{(j)}$.

Despite of the complicated appearance, the coefficient $C_{(ij)}$ turns out to be exactly a dlog associated to a $2\times2$ symmetric matrix
\begin{equation}\label{eq:genericCij}
\fixed
C_{(ij)}=\sqrt{-1}\,\mathrm{d}_Q\log\left(r(\mathbf{H}_{\{i,j\}})\right),\quad \mathbf{H}_{ij}=H_iQ^{-1}H_j,
\end{equation}
where the subscript $\{i,j\}$ indicates the entries to be extracted in the submatrix (for both rows and columns). Here the function $r(\mathbf{M})$ of a $2\times2$ symmetric matrix $\mathbf{M}$ is defined as
\begin{equation}\label{ratioofroots}
\fixed
r(\mathbf{M})=\frac{\mathbf{M}_{12}-\sqrt{(\mathbf{M}_{12})^2-\mathbf{M}_{11}\mathbf{M}_{22}}}{\mathbf{M}_{12}+\sqrt{(\mathbf{M}_{12})^2-\mathbf{M}_{11}\mathbf{M}_{22}}}.
\end{equation}
Obviously $r(\mathbf{M})$ is the ratio of the two roots of the quadratic polynomial $(\substack{x\\1})^{\rm T}\mathbf{M}(\substack{x\\1})$. The detailed derivation for \eqref{eq:genericCij} is provided in Appendix \ref{app:genericCijderivation}. While we have an extra constant factor $\frac{\sqrt{-1}}{2(n-2)}$ in front of the dlog, it is universal to all the codim-2 boundaries, thus not essential. 

The remaining integral in each term of \eqref{dO} is again of the type $e$ but with its index reduced by two. This immediately reveals a recursive structure for each $e$ integral: we repeat the differentiation for the induced integral on each codim-2 boundary, and then further on each codim-4 boundary, etc. In each step the index of the remaining $e$ integral drops by two, and it is guaranteed that a pure dlog coefficient is produced in front of each term in the localization (apart from an overall constant factor).

In order that the integral has a well-defined transcendental weight, this iteration has to properly terminate at a rational function. Since the ratio of roots is always algebraically related to $Q$, the integral clearly stays inside the regime of iterated integrals / multiple polylogarithms.

If $n$ is even, the recursion goes all the way down to $e_2$ integrals, which in our convention explicitly have the form
\begin{equation}\label{eq:e2result}
\verified{}
e_2=-\frac{1}{2}\log\,r(Q^{-1}).
\end{equation}
Of course a further differentiation is trivially a $\mathrm{d}_Q\text{log}$. So in this case $e_n$ has weight $\frac{n}{2}$. If $n$ is odd, the recursion terminates at $e_1$ integrals, which are by definition trivially $\sqrt{-1}$, and so $e_n$ has weight $\frac{n-1}{2}$. Hence any $e_n$ integral is a pure function of weight $\lfloor \frac{n}{2}\rfloor$.

Furthermore, the recursion of differentiation straightforwardly determines the entire symbol for a given $e_n$ integral, and organize it in the form
\begin{equation}
\fixed
\mathcal{S}e_n=\frac{-(\sqrt{-1})^{\frac{n-2}{2}}}{2^{\frac{n}{2}}(n-2)!!}\sum_{\rho}\underline{\rho_1\rho_2}\otimes\underline{\rho_3\rho_4}\otimes\cdots\otimes\underline{\rho_{n-1}\rho_{n}},\quad \text{even }n,
\end{equation}
where the summation $\rho$ runs over all ordered partitions of labels $\{1,2,\ldots,n\}$ into $\frac{n}{2}$ pairs, and the notation $\underline{\rho_{2k-1}\rho_{2k}}\equiv\underline{\rho_{2k}\rho_{2k-1}}$ is defined to be symmetric (see below). We also have
\begin{equation}
\fixed
\mathcal{S}e_n=\frac{(\sqrt{-1})^{\frac{n+1}{2}}}{2^{\frac{n-1}{2}}(n-2)!!}\sum_{\rho}\underline{\rho_2\rho_3}\otimes\underline{\rho_3\rho_4}\otimes\cdots\otimes\underline{\rho_{n-1}\rho_{n}},\quad \text{odd }n,
\end{equation}
where $\rho$ runs over all partitions into one label followed by ordered sequence of symmetric pairs. Hence in general this is a summation of $n!/2^{\lfloor \frac{n}{2}\rfloor}$ symbol terms.

The above summation structure inherits the stratification of the simplex contour $\Delta$ into its facets, as is obvious from the contour for the induced integral associated with each of the dlog coefficient. Specifically, let us refer the notation $\Delta_{(i_1\ldots i_k)}$ to the codim-$k$ face of $\Delta$ that lies in the intersection of the $k$ hyperplanes $\{H_{i_1},\ldots,H_{i_k}\}$ (thus symmetric in the labels in the subscript; equivalently speaking this is the face excluding vertices $\{V_{i_1},\ldots,V_{i_k}\}$), then each term in $\mathcal{S}e_n$ is uniquely associated to one incidence relation
\begin{equation}
\fixed
\underline{\rho_1\rho_2}\otimes\underline{\rho_3\rho_4}\otimes\cdots\otimes\underline{\rho_{n-1}\rho_{n}}\;\Longleftrightarrow\; 
\underset{\scriptsize\text{codim-}n}{\Delta_{(\rho_1\ldots\rho_{n})}}\subset \underset{\scriptsize\text{codim-}(n-2)}{\Delta_{(\rho_3\ldots\rho_{n})}}\subset\cdots\subset \underset{\scriptsize\text{codim-}2}{\Delta_{(\rho_{n-1}\rho_{n})}},
\end{equation}
in the case of even $n$, and similarly for odd $n$. Correspondingly,  the entry $\underline{\rho_{2k-1}\rho_{2k}}$ is computed by deleting the rows and columns in $Q$ labeled by $\{\rho_{2k+1},\ldots,\rho_{n}\}$, calculating the inverse of the remaining matrix, and then extracting the submatrix labeled by $\{\rho_{2k-1},\rho_{2k}\}$, i.e.,
\begin{equation}
\fixed
\underline{\rho_{2k-1}\rho_{2k}}=r\left((Q_{\{\widehat{\rho_{2k+1}},\widehat{\rho_{2k+2}},\ldots,\widehat{\rho_{n}}\}})^{-1}_{\{\rho_{2k-1},\rho_{2k}\}}\right),
\end{equation}
(we use hat to denote entries that are deleted).

We see that the ratio of roots is a universal recurring algebraic structure in the symbols of the $E$ integrals. The only content that differs between symbol entries is the specific $2\times2$ symmetric matrix induced from $Q$.  Due to the projective nature of this quantity, the ratio of roots $r(\mathbf{M})$ remains invariant under an $\mathbb{R}_+\times\mathbb{R}\times\mathbb{Z}_2$ action on $\mathbf{M}$. Explicitly we have
\begin{equation}
r(\mathbf{M})=r(\mathbf{S}^{\rm T}\mathbf{M}\mathbf{S})=r(\mathbf{T}^{\rm T}\mathbf{M}\mathbf{T}),\qquad
\mathbf{S}=\left(\begin{matrix}s_1&0\\0&s_2\end{matrix}\right),\quad
\mathbf{T}=\left(\begin{matrix}0&1\\1&0\end{matrix}\right),\quad s_1s_2>0.
\end{equation}
In practice this allows us to simplify $\mathbf{M}$ before computing the ratio $r(\mathbf{M})$.

The discussion so far already equips us with the necessary tools for understanding a large class of Feynman integrals: arbitrary $d$-gons and $(d+1)$-gons in $d$ dimensions, with arbitrary massive loop propagators. Readers who are eager to get a taste of actual applications can directly jump to Section \ref{subsec:massivebox}.

Before ending this subsection, we would like to point out here that the above result for $d$-gons in $d$ dimensions with $d$ even has been worked out in previous literature \cite{Spradlin:2011wp}, by applying the mixed tate motives introduced in \cite{1996alg.geom..1021G}. In the above we provided a straightforward derivation for it, where the basic method is equally applicable to general one-loop integrals, as will be developed later on.

\subsection{Divergence in the localization}

So far in the discussion we have been assuming generic $Q$ and actually implicitly also generic $\mathbf{M}$ for every symbol entry. In actual computations $r(\mathbf{M})$ may produce singular symbol entries even when $Q$ itself is non-degenerate, and so this generic rule does not straightforwardly apply.

When this happens, of course one can always regularize it by slightly deforming $Q$ with some infinitesimal parameter $\varepsilon$, and then after determining the entire symbols take the limit $\varepsilon\to0$. In the limit, what one explicitly observe is that some of the symbol entries approximates to
\begin{equation}
\fixed
\varepsilon^{-1}(w+O(\varepsilon))\quad\text{or}\quad\varepsilon w+O(\varepsilon^2),
\end{equation} 
where $w$ is some expression that remains finite. This indicates that the induced integrals on the corresponding boundaries from the localization procedure actually diverge.

However, even though an individual induced integral may diverge, it does not necessarily mean that the original integral diverges. To understand why this is possible. Let us consider some $e_n$ integral with one of the integrand vertex, say $Z_i$ sitting on the quadric, $Z_iQZ_i=0$. This is a situation when the singularity of the integrand start to touch the integration contour, which potentially may generates a divergence. In the vicinity of this corner, controlled by a scale $\delta$, we see that the measure $\langle X\mathrm{d}X^{n-1}\rangle$ scales as the volume $\delta^{n-1}$ while the denominator $(XQX)^{n/2}\sim\delta^{n/2}$. So the integral can remain finite as long as we have $n>2$. However, from the previous discussions we see that each of the induced integral corresponds to slicing the geometric configuration by some boundary hyperplanes. So once we localize onto the dim-1 boundaries, correspondingly $n=2$, we observe a log divergence from the $Z_i$ corner for those induced integrals.

Such geometric configuration occurs frequently in Feynman integrals. As we see from the Feynman parametrization, whenever some loop propagator $i$ is massless, the Symanzik polynomial is free of the $x_i^2$ terms, leading to the situation discussed above.

If the original integral actually stay finite, one should explicitly observe that the leading divergence piece in the $\varepsilon\to0$ limit cancel away, as a result of algebraic properties of the symbols, and the remaining terms recover the correct symbols for the integral. However, it is good to gain a better understanding of whether any issue occurs with the localization procedure in this case, and further how to determine the result without the need of a regularization. It suffices to focus only on the last symbol entries, correspondingly a specific matrix $\mathbf{H}_{\{i,j\}}$, since the other entries always comes as the last entries of the induced $e$ integrals on some faces of $\Delta$.

$r(\mathbf{H}_{\{i,j\}})$ becomes singular whenever $\mathbf{H}_{ii}\equiv H_iQ^{-1}H_i$ or $\mathbf{H}_{jj}\equiv H_jQ^{-1}H_j$ or both of them vanish. Geometrically
\begin{equation}
HQ^{-1}H=0\quad\Longleftrightarrow\quad
H\text{ is tangent to }Q.
\end{equation}
To understand this, recall that the point $Q^{-1}H$ is a common point for all hyperplanes that are tangent to $Q$ at some point $A$ belonging to the intersection of $Q$ and $H$. When this special point itself lies on the quadric, i.e., $(HQ^{-1})Q(Q^{-1}H)\equiv HQ^{-1}H=0$, $H$ is obviously tangent to $Q$ at $Q^{-1}H$.

When $H_iQ^{-1}H_i=0$ for some $i$, it turns out that in localizing $e_n$ onto its codim-1 boundaries we obtain an $E_{n-1,1}$ integral with a degenerate quadric on the boundary $H_i$, because
\begin{equation}
\fixed
\det Q_{(i)}\equiv(\det Q)\,(\mathbf{H}_{ii})=0.
\end{equation}
This forbids us to directly apply a further localization for the $E_{n-1,1}$ induced on $\Delta_{(i)}$ in the way as described before.

Let us first deviate a little bit to discuss a generic $E_{n-1,1}$ with degenerate $Q$ in the following subsection, and then return to this problem.

\subsection{Case of degenerate $Q$}\label{sec:ecaseofdegenerateQ}

For simplicity, in the case of a degenerate $Q$ we only focus only the $E_{n-1,1}$ integrals where corank$Q=1$ in this section, and postpone a general analysis to Appendix \ref{app:degenerateQ}.

Given that corank$Q=1$ can be treated as a special limit of a generic $Q$, it is expected that an $E_{n-1,1}$ integral should again directly localize onto codim-1 boundaries of $\Delta$. Regardless of the specific properties of $Q$, for an arbitrary vector $W$ the following identity always holds
\begin{equation}
\fixed
\mathrm{d}_X\!\!\left[\frac{\form{WX}{X}{n-3}}{(XQX)^{\frac{n-2}{2}}}\right]=(n-2)\frac{(WQX)\,\measure{X}{n-2}}{(XQX)^{\frac{n}{2}}}.
\end{equation}
If we are able to identify $WQ\equiv L$, then the integrand of $E_{n-1,1}$ is explicitly exact. Differing from the discussions before, here we are not able to simply inverse the relation since $\det Q=0$. Correspondingly the collection of hyperplanes tangent to $Q$ at the intersection points of $Q$ and $L$ exactly intersect at the unique null point $N$ of $Q$, and so we cannot simply identify $W$ as this point either (since that means $(WQ)^I\equiv0$).

If fact it is not hard to see the above ansatz does not apply to arbitrary $L$. The reason is $WQN\equiv 0$, but $LN\neq0$ for a generic $L$. When this does not hold, we necessarily have to seek for more involved ansatz. But for our purpose in the previous subsection it suffices to assume the condition
\begin{equation}\label{LNcondition}
\fixed
LN=0
\end{equation}
on $L$, in other words, the hyperplane $L$ contains the null space of $Q$. 

To solve the ansatz, note that given some reference covector $R$ we can construct a non-vanishing invariant $\mathring{q}$ of $Q$ together with its associated modified inverse $\mathring{Q}^{-1}$ (this inverse depends on the choice of $R$)
\begin{equation}
\fixed
\mathring{q}=\frac{1}{(n-1)!}\parbox{3cm}{
\tikz{
\node [anchor=center] at (0,0) {$Q$};
\node [anchor=center] at (.6,0) {$Q$};
\node [anchor=center] at (1.2,0) {$\cdots$};
\node [anchor=center] at (1.8,0) {$Q$};
\node [anchor=center] at (2.4,.2) {$R$};
\node [anchor=center] at (2.4,-.2) {$R$};
\draw [black,thick] (0,.6) -- (2.4,.6);
\draw [black,thick] (0,.6) -- (0,.25);
\draw [black,thick] (.6,.6) -- (.6,.25);
\draw [black,thick] (1.8,.6) -- (1.8,.25);
\draw [black,thick] (2.4,.6) -- (2.4,.45);
\draw [black,thick] (0,-.6) -- (2.4,-.6);
\draw [black,thick] (0,-.6) -- (0,-.25);
\draw [black,thick] (.6,-.6) -- (.6,-.25);
\draw [black,thick] (1.8,-.6) -- (1.8,-.25);
\draw [black,thick] (2.4,-.6) -- (2.4,-.45);
}},\qquad
\mathring{Q}^{-1}=\frac{\mathring{q}^{-1}}{(n-2)!}\;\parbox{2.7cm}{
\tikz{
\node [anchor=center] at (.6,0) {$Q$};
\node [anchor=center] at (1.2,0) {$\cdots$};
\node [anchor=center] at (1.8,0) {$Q$};
\node [anchor=center] at (2.4,.2) {$R$};
\node [anchor=center] at (2.4,-.2) {$R$};
\draw [black,thick] (0,.6) -- (2.4,.6);
\draw [black,thick] (0,.6) -- (0,.25);
\draw [black,thick] (.6,.6) -- (.6,.25);
\draw [black,thick] (1.8,.6) -- (1.8,.25);
\draw [black,thick] (2.4,.6) -- (2.4,.45);
\draw [black,thick] (0,-.6) -- (2.4,-.6);
\draw [black,thick] (0,-.6) -- (0,-.25);
\draw [black,thick] (.6,-.6) -- (.6,-.25);
\draw [black,thick] (1.8,-.6) -- (1.8,-.25);
\draw [black,thick] (2.4,-.6) -- (2.4,-.45);
}}.
\end{equation}
Here to simplify notation we use a half-ladder graph to denote contractions with each Levi-Civita symbol; details are explained in Appendix \ref{app:convention}. In order that $\mathring{q}\neq0$ and $\mathring{Q}^{-1}$ well-defined, we only require that $RN\neq0$, because
\begin{equation}\label{LQQR0}
\fixed
N\propto\parbox{3cm}{\tikz{
\node [anchor=center] at (0,0) {$Q$};
\node [anchor=center] at (.6,0) {$Q$};
\node [anchor=center] at (1.2,0) {$\cdots$};
\node [anchor=center] at (1.8,0) {$Q$};
\node [anchor=center] at (2.4,-.2) {$R$};
\draw [black,thick] (0,.6) -- (2.4,.6);
\draw [black,thick] (0,.6) -- (0,.25);
\draw [black,thick] (.6,.6) -- (.6,.25);
\draw [black,thick] (1.8,.6) -- (1.8,.25);
\draw [black,thick] (2.4,.6) -- (2.4,.45);
\draw [black,thick] (0,-.6) -- (2.4,-.6);
\draw [black,thick] (0,-.6) -- (0,-.25);
\draw [black,thick] (.6,-.6) -- (.6,-.25);
\draw [black,thick] (1.8,-.6) -- (1.8,-.25);
\draw [black,thick] (2.4,-.6) -- (2.4,-.45);
}}.
\end{equation}
Under the condition \eqref{LNcondition} we then have the identification
\begin{equation}
\verified{}
W=\mathring{Q}^{-1}L.
\end{equation}
This leads to the decomposition of $E_{n-1,1}$ into
\begin{equation}\label{EdecompositionQcorank1}
\verified{}
E_{n-1,1}=\sum_{i=1}^{n-1}\frac{1}{(n-2)}\frac{H_i\mathring{Q}^{-1}L}{\sqrt{-q_{(i)}}}\int_{\Delta_{(i)}}\frac{\sqrt{-q_{(i)}}\,\measure{X_{(i)}}{n-3}}{(X_{(i)}Q_{(i)}X_{(i)})^{\frac{n-2}{2}}}.
\end{equation}
Note that in general $\det Q_{(i)}\neq0$. Also by construction $W\equiv \mathring{Q}^{-1}L$ is independent of the choice of $R$, but in each term of the above decomposition the coefficient is in general $R$ dependent.

\subsection{The tangent configurations}\label{sec:tangentconfig}

Let us now go back to the situation when $Q$ itself is non-degenerate, but without loss of generality assume the tangency condition $H_iQ^{-1}H_i=0$ for some $i$. 

In such singular configuration when we deform the quadric in doing differentiation we naturally desire to maintain the tangency, and so the actual conditions under consideration are
\begin{equation}\label{eq:actualtangency}
\fixed
H_iQ^{-1}H_i=0,\quad\text{and}\quad\mathrm{d}_Q(H_iQ^{-1}H_i)\propto\parbox{3.1cm}{
\tikz{
\node [anchor=center] at (0,.2) {$H_i$};
\node [anchor=center] at (0,-.2) {$H_i$};
\node [anchor=center] at (.6,0) {$Q$};
\node [anchor=center] at (2*.6,0) {$\cdots$};
\node [anchor=center] at (3*.6,0) {$Q$};
\node [anchor=center] at (4*.6,0) {$\delta Q$};
\draw [thick,black] (0,.6) -- (4*.6,.6);
\draw [thick,black] (0,.6) -- (0,.45);
\draw [thick,black] (.6,.6) -- (.6,.25);
\draw [thick,black] (3*.6,.6) -- (3*.6,.25);
\draw [thick,black] (4*.6,.6) -- (4*.6,.25);
\draw [thick,black] (0,-.6) -- (4*.6,-.6);
\draw [thick,black] (0,-.6) -- (0,-.45);
\draw [thick,black] (.6,-.6) -- (.6,-.25);
\draw [thick,black] (3*.6,-.6) -- (3*.6,-.25);
\draw [thick,black] (4*.6,-.6) -- (4*.6,-.25);
}}=0.
\end{equation}

He we encounter an $E_{n-1,1}$ integral on $\Delta_{(i)}$ where corank$Q_{(i)}=1$. Before applying the technique in the previous subsection we need to verify the condition \eqref{LNcondition}. In this specific case the LHS of \eqref{LNcondition}, i.e., $H_iQ^{-1}\mathrm{d}QP(i)N_{(i)}$, is proportional to
\begin{equation}
\fixed
\parbox{6cm}{\tikz{
\node [anchor=center] at (0,0) {$Q$};
\node [anchor=center] at (.6,0) {$\cdots$};
\node [anchor=center] at (2*.6,0) {$Q$};
\node [anchor=center] at (3*.6,.2) {$H_i$};
\node [anchor=center] at (3*.6,-.2) {$H_i$};
\node [anchor=center] at (4*.6,0.2) {$R$};
\node [anchor=center] at (5*.6,0) {$\mathrm{d}Q$};
\node [anchor=center] at (6*.6,-.2) {$H_i$};
\node [anchor=center] at (7*.6,0) {$Q$};
\node [anchor=center] at (8*.6,0) {$\cdots$};
\node [anchor=center] at (9*.6,0) {$Q$};
\draw[black,thick] (0,.6) -- (4*.6,.6);
\draw[black,thick] (0,.6) -- (0,.25);
\draw[black,thick] (2*.6,.6) -- (2*.6,.25);
\draw[black,thick] (3*.6,.6) -- (3*.6,.45);
\draw[black,thick] (4*.6,.6) -- (4*.6,.45);
\draw[black,thick] (0,-.6) -- (5*.6,-.6);
\draw[black,thick] (0,-.6) -- (0,-.25);
\draw[black,thick] (2*.6,-.6) -- (2*.6,-.25);
\draw[black,thick] (3*.6,-.6) -- (3*.6,-.45);
\draw[red,thick] (5*.6,-.6) -- (5*.6,-.25);
\draw[black,thick] (5*.6,.6) -- (9*.6,.6);
\draw[black,thick] (5*.6,.6) -- (5*.6,.25);
\draw[black,thick] (7*.6,.6) -- (7*.6,.25);
\draw[black,thick] (9*.6,.6) -- (9*.6,.25);
\draw[black,thick] (6*.6,-.6) -- (9*.6,-.6);
\draw[red,thick] (6*.6,-.6) -- (6*.6,-.45);
\draw[red,thick] (7*.6,-.6) -- (7*.6,-.25);
\draw[red,thick] (9*.6,-.6) -- (9*.6,-.25);
}}=(n-3)
\parbox{6cm}{\tikz{
\node [anchor=center] at (0,0) {$Q$};
\node [anchor=center] at (.6,0) {$\cdots$};
\node [anchor=center] at (2*.6,0) {$Q$};
\node [anchor=center] at (3*.6,.2) {$H_i$};
\node [anchor=center] at (3*.6,-.2) {$H_i$};
\node [anchor=center] at (4*.6,0.2) {$R$};
\node [anchor=center] at (5*.6,0) {$Q$};
\node [anchor=center] at (6*.6,-.2) {$H_i$};
\node [anchor=center] at (7*.6,0) {$Q$};
\node [anchor=center] at (8*.6,0) {$\cdots$};
\node [anchor=center] at (9*.6,0) {$Q$};
\node [anchor=center] at (10*.6,0) {$\mathrm{d}Q$};
\draw[black,thick] (0,.6) -- (4*.6,.6);
\draw[red,thick] (0,.6) -- (0,.25);
\draw[red,thick] (2*.6,.6) -- (2*.6,.25);
\draw[red,thick] (3*.6,.6) -- (3*.6,.45);
\draw[red,thick] (4*.6,.6) -- (4*.6,.45);
\draw[black,thick] (0,-.6) -- (5*.6,-.6);
\draw[black,thick] (0,-.6) -- (0,-.25);
\draw[black,thick] (2*.6,-.6) -- (2*.6,-.25);
\draw[black,thick] (3*.6,-.6) -- (3*.6,-.45);
\draw[black,thick] (5*.6,-.6) -- (5*.6,-.25);
\draw[black,thick] (5*.6,.6) -- (10*.6,.6);
\draw[red,thick] (5*.6,.6) -- (5*.6,.25);
\draw[black,thick] (7*.6,.6) -- (7*.6,.25);
\draw[black,thick] (9*.6,.6) -- (9*.6,.25);
\draw[black,thick] (10*.6,.6) -- (10*.6,.25);
\draw[black,thick] (6*.6,-.6) -- (10*.6,-.6);
\draw[black,thick] (6*.6,-.6) -- (6*.6,-.45);
\draw[black,thick] (7*.6,-.6) -- (7*.6,-.25);
\draw[black,thick] (9*.6,-.6) -- (9*.6,-.25);
\draw[black,thick] (10*.6,-.6) -- (10*.6,-.25);
}}
\end{equation}
where the expression on RHS is obtained by a Schouten identity (on the red legs shown in LHS, following \eqref{app:eq:schouten}). Applying a further Schouten identity (on the red legs on RHS) decomposes it into two terms, each of which is proportional to each of the tangency conditions in \eqref{eq:actualtangency}, and thus the condition \eqref{LNcondition} is satisfied.

The localization onto the codim-2 boundaries is then straightforward following \eqref{EdecompositionQcorank1}. We do not bother to write out the resulting $e_{n-2}$ integral in this case, which is the same as what we obtained before. However, its coefficient changes to
\begin{equation}\label{coefH1H2dQ}
\fixed
C_{ij}=-\sqrt{\frac{q}{q_{(ij)}}}
H_iQ^{-1}\mathrm{d}QP(i)\mathring{Q}_{(i)}^{-1}P(i)^{\rm T}H_j.
\end{equation}
Note here we do not put the parentheses in the subscript as the above $C_{ij}$ accounts only for the contribution from localization to $\Delta_{(ij)}$ via $\Delta_{(i)}$. After simplification, this coefficient dramatically reduces to
\begin{equation}\label{tangentCij}
\fixed
C_{ij}=\sqrt{-1}\,\text{sign}(\mathbf{H}_{ij})\left[
\mathrm{d}_Q\log(q \mathbf{H}_{ij})
-\mathrm{d}_Q\log\left(\parbox{2.5cm}{\tikz{
\node [anchor=center] at (0*.6,0) {$Q$};
\node [anchor=center] at (1*.6,0) {$\ldots$};
\node [anchor=center] at (2*.6,0) {$Q$};
\node [anchor=center] at (3*.6,.2) {$H_i$};
\node [anchor=center] at (3*.6,-.2) {$R$};
\draw [black,thick] (0,.6) -- (3*.6,.6);
\draw [black,thick] (0,.6) -- +(0,-.35);
\draw [black,thick] (2*.6,.6) -- +(0,-.35);
\draw [black,thick] (3*.6,.6) -- +(0,-.15);
\draw [black,thick] (0,-.6) -- (3*.6,-.6);
\draw [black,thick] (0,-.6) -- +(0,.35);
\draw [black,thick] (2*.6,-.6) -- +(0,.35);
\draw [black,thick] (3*.6,-.6) -- +(0,.15);
}}\right)\right],
\end{equation}
which is already a pure dlog by itself. However, the second term looks worrisome as it explicitly depends on $R$, which we originally introduced as an arbitrary reference.

It turns out that the second term above is redundant. It is eliminated after we sum up all the codim-2 contributions bordering the same codim-1 boundary. This is achieved by noticing that
\begin{equation}\label{spuriousterms}
\fixed
\sum_{\Delta_{(ik)}\subset\Delta_{(i)}}\text{sign}(\mathbf{H}_{ik})\int_{\Delta_{(ik)}}\frac{\sqrt{-q_{(ik)}}\,\measure{X_{(ik)}}{n-3}}{(X_{(ik)}Q_{(ik)}X_{(ik)})^{\frac{n-2}{2}}}=0.
\end{equation}
The proof for \eqref{spuriousterms} is provided in Appendix \ref{app:cancelation}. The only point to be emphasized here is that this identity always holds, regardless of whether any of the $H_k$ ($k\neq i$) is tangent to $Q$ or not. As a result, we have
\begin{equation}
\fixed
C_{ij}=
\sqrt{-1}\,\text{sign}(\mathbf{H}_{ij})\,\mathrm{d}_Q\log(q \mathbf{H}_{ij}).
\end{equation}

At this point, one might worry whether the other contribution to the same integral on the codim-2 boundary $\Delta_{(ij)}$ (for some $j$) could cause trouble, because in the case when none of the $H$'s are tangent to $Q$ it is only the sum of the two contributions that turns into a dlog. It turns out that
\begin{equation}
\fixed
C_{ji}=\sqrt{-1}\,\text{sign}(\mathbf{H}_{ij})\begin{cases}\mathrm{d}_Q\log(q\,\mathbf{H}_{ij})-\mathrm{d}_Q\log q_{(j)},&\mathbf{H}_{jj}\neq0,\\\mathrm{d}_Q\log(q\,\mathbf{H}_{ij}),&\mathbf{H}_{jj}=0.\end{cases}
\end{equation}
Details are provided in Appendix \ref{app:tangentCijderivation}.

Collecting all the situations we conclude that
\begin{equation}
\fixed
C_{(ij)}=\sqrt{-1}\begin{cases}\mathrm{d}_Q\log(r(\mathbf{H}_{\{i,j\}})),&\mathbf{H}_{ii}\neq0,\mathbf{H}_{jj}\neq0,\\{\displaystyle \mathrm{d}_Q\log\left(\frac{(q\,\mathbf{H}_{ij})^2}{q_{(j)}}\right)^{\text{sign}(\mathbf{H}_{ij})}},&\mathbf{H}_{ii}=0,\mathbf{H}_{jj}\neq0,\\{\displaystyle \mathrm{d}_Q\log\left(\frac{(q\,\mathbf{H}_{ij})^2}{q_{(i)}}\right)^{\text{sign}(\mathbf{H}_{ij})}},&\mathbf{H}_{ii}\neq0,\mathbf{H}_{jj}=0,\\\mathrm{d}_Q\log(q\,\mathbf{H}_{ij})^{2\,\text{sign}(\mathbf{H}_{ij})},&\mathbf{H}_{ii}=\mathbf{H}_{jj}=0.\end{cases}
\end{equation}

\subsection{Summary of the general results}

The above result is for the last symbol entries. Most generally, we may encounter similar issue with other symbol entries. Without loss of generality, let us focus on some $\underline{\rho_{2k-1}\rho_{2k}}$. Again working in the canonical frame, the corresponding quadric is then $Q_{\{\rho_1,\rho_2,\ldots,\rho_{2k}\}}$. Let us abbreviate
\begin{equation}
\fixed
h_{11}\equiv(Q_{\{\rho_1,\rho_2,\ldots,\rho_{2k}\}})^{-1}_{\rho_{2k-1}\rho_{2k-1}},\quad
h_{22}\equiv(Q_{\{\rho_1,\rho_2,\ldots,\rho_{2k}\}})^{-1}_{\rho_{2k}\rho_{2k}},\quad
h_{12}\equiv(Q_{\{\rho_1,\rho_2,\ldots,\rho_{2k}\}})^{-1}_{\rho_{2k-1}\rho_{2k}}.
\end{equation}
The tangency condition is then encoded in $h_{11}$ and $h_{22}$. We have
\begin{equation}\label{eq:symbolentryfromdif}
\fixed
\underline{\rho_{2k-1}\rho_{2k}}=\begin{cases}
r((Q_{\{\rho_1,\ldots,\rho_{2k}\}})^{-1}_{\{\rho_{2k-1},\rho_{2k}\}}),&h_{11}\neq0,h_{22}\neq0,\\
{\displaystyle \left(\frac{(q_{(\rho_{2k+1}\ldots\rho_n)}h_{12})^2}{q_{(\rho_{2k}\rho_{2k+1}\ldots\rho_n)}}\right)^{\text{sign}(h_{12})}},&h_{11}=0,h_{22}\neq0,\\
{\displaystyle \left(\frac{(q_{(\rho_{2k+1}\ldots\rho_n)}h_{12})^2}{q_{(\rho_{2k-1}\rho_{2k+1}\ldots\rho_n)}}\right)^{\text{sign}(h_{12})}},&h_{11}\neq0,h_{22}=0,\\
(q_{(\rho_{2k+1}\ldots\rho_n)}h_{12})^{2\,\text{sign}(h_{12})},&h_{11}=h_{22}=0.
\end{cases}
\end{equation}

If we choose to work instead in a generic frame, we just need to first construct $\mathbf{H}_{ij}=H_{iI}(Q^{-1})^{IJ}H_{jJ}$, and then use the matrix $\mathbf{H}$ in the above formula.

\subsection{Integrability of the symbols}

In general an arbitrary combination of abstract tensor expressions made of $\otimes$ may not be the symbol for some actual function. Consider an expression of depth $r$
\begin{equation}
\fixed
\sum_{i}w_{i,1}\otimes w_{i,2}\otimes\cdots\otimes w_{i,r},
\end{equation}
where $i$ labels the symbol term. In order that it is an actual symbol, it has to meet the integrability condition
\begin{equation}
\fixed
\sum_{l=1}^{r-1}\sum_{i}(\mathrm{d}\log(w_{i,l})\wedge\mathrm{d}\log(w_{i,{l+1}}))\,w_{i,1}\otimes\cdots\otimes\widehat{w_{i,l}}\otimes\widehat{w_{i,{l+1}}}\otimes\cdots\otimes w_{i,r}=0.
\end{equation}

It is not obvious that the expressions for the symbols that we determined automatically satisfy this condition, and so we need to check it. Here we just focus on a generic $e_n$, because the singular cases can always be treated as a certain limit of a generic one and also a general quadric integral always admits an expansion on them (as we will show in later discussions). 

One-loop Feynman integrals are actually very special functions, in the sense that a finer condition for their symbols is satisfied, as follows
\begin{equation}\label{seedintegrability}
\fixed
\sum_{S_4}\mathrm{d}\log(\underline{\rho_{2k-1}\rho_{2k}})\wedge\mathrm{d}\log(\underline{\rho_{2k+1}\rho_{2k+2}})=0,\qquad\forall k,
\end{equation}
where the summation is over any permutation of the labels in a fixed cardinality-4 subset $\{\rho_{2k-1},\rho_{2k},\rho_{2k+1},\rho_{2k+2}\}$ of $\{1,2,\ldots,n\}$, given that $\{\rho_{2k+1},\ldots,\rho_{n}\}$ is also fixed. We postpone the proof to Section \ref{sec:symbolentrygeometry}.

\section{Scalar Diagrams: $n<d+2$}\label{sec:feynman1}

\flag{No further revisions.}

\begin{figure}[ht]
\begin{center}
\begin{tikzpicture}
\draw [black,thick] (60:1cm) -- (0:1cm) -- (-60:1cm) -- (-120:1cm) -- (180:1cm) -- (120:1cm);
\draw [black,thick] (60:1cm) -- +(60:.4cm);
\draw [black,thick] (0:1cm) -- +(0:.4cm);
\draw [black,thick] (-60:1cm) -- +(-60:.4cm);
\draw [black,thick] (-120:1cm) -- +(-120:.4cm);
\draw [black,thick] (180:1cm) -- +(180:.4cm);
\draw [black,thick] (120:1cm) -- +(120:.4cm);
\draw [black,thick,dashed] (60:1cm) -- (120:1cm);
\node [anchor=center] at (-120:1.8cm) {\scriptsize $1$};
\node [anchor=center] at (180:1.8cm) {\scriptsize $2$};
\node [anchor=center] at (120:1.8cm) {\scriptsize $3$};
\node [anchor=center] at (60:1.8cm) {\scriptsize $n-2$};
\node [anchor=center] at (0:1.8cm) {\scriptsize $n-1$};
\node [anchor=center] at (-60:1.8cm) {\scriptsize $n$};
\node [anchor=center] at (-90:1.4cm) {$y_1$};
\node [anchor=center] at (-150:1.4cm) {$y_2$};
\node [anchor=center] at (150:1.4cm) {$y_3$};
\node [anchor=center] at (-30:1.4cm) {$y_n$};
\node [anchor=center] at (30:1.4cm) {$y_{n-1}$};
\node [anchor=center] at (0,0) {$y$};
\end{tikzpicture}
\end{center}
\vspace{-1.5em}\caption{Scalar $n$-gon.}
\label{fig:scalarngon}
\end{figure}
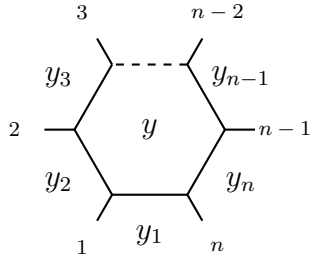

In this section we apply the results in the previous section to the scalar integrals in $d$ dimensions with $n$ external lines. Without loss of generality we assume a planar ordering of the external line as shown in Figure \ref{fig:scalarngon}. 
Here we assume that the $i^{\rm th}$ propagator can have mass $m_i$. It is convenient to introduce the region variables $y_i^\mu$ ($i=1,2,\ldots,n$) such that the momenta $k_i^\mu=y_i^\mu-y_{i+1}^\mu$. In deriving Feynman parametrization we put the kinematics data in the Euclidean region.

We further use the embedding formalism by mapping arbitrary $y^\mu$ to the projective light cone in an ambient space of two higher dimensions
\begin{equation}\label{embeddingmap}
\fixed{}
y^\mu\longmapsto Y^M=(1,y^2,y^\mu),
\end{equation}
where we use the light-cone coordinates, i.e., the metric $\eta_{+-}=\eta_{-+}=-\frac{1}{2}$, $\eta_{\mu\mu}=1$ ($\forall \mu$) while all other entries vanish. Then we have
\begin{equation}
\fixed{}
(y_1-y_2)^2=-2Y_1\cdot Y_2.
\end{equation} 
With these we set the convention of the scalar integrals to
\begin{equation}
I^\phi_{n,d}=\int\frac{\mathrm{d}^dy}{\prod_{i=1}^n((y-y_i)^2+m_i^2)}=\int[\mathrm{d}^dY]\frac{(-2\,Y\cdot Y_\infty)^{n-d}}{\prod_{i=1}^n(-2\,Y\cdot \mathcal{Y}_i)},
\end{equation}
where in the last expression we use the notation $\int[\mathrm{d}^dY]\equiv\int\frac{\mathrm{d}^{d+2}Y\,\delta(Y^2)}{\text{vol.}\mathrm{GL}(1)}$ ($\mathrm{GL}(1)$ acts as an overall rescaling of the $Y$ coordinates), and we use the notation
\begin{equation}
\mathcal{Y}_i^M=(1,y_i^2-m_i^2,y_i^\mu),
\end{equation}
for the ``mass-deformed'' vector (outside of the light cone), to distinguish it from the one obtained from the map \eqref{embeddingmap}. Obviously $\mathcal{Y}_i\rightarrow Y_i$ in the limit $m_i\rightarrow0$. Furthermore
\begin{equation}
Y_\infty^M=(0,1,0,\ldots,0),
\end{equation}
denotes a point at ``infinity''. The factor $(Y\cdot Y_\infty)^{n-d}$ is necessary in order that the last expression is genuinely an integral over the projective light cone.

Introducing the Feynman parameters we obtain
\begin{equation}
\begin{split}
\verified{}
I^\phi_{n,d}&=\int_0^\infty\measure{X}{n-1}\int[\mathrm{d}^dY]\frac{(Y\cdot Y_\infty)^{n-d}}{(-2)^d(Y\cdot W)^n}\\
&=\frac{(-1)^n\Gamma(d)}{2^d\Gamma(n)}\left(Y_\infty^M\frac{\partial}{\partial W^M}\right)^{n-d}\int_0^\infty\measure{X}{n-1}\int[\mathrm{d}^dY]\frac{1}{(Y\cdot W)^d},
\end{split}
\end{equation}
where $W=\sum_{a=1}^nx_a\mathcal{Y}_a$. By the constraint of conformal weight we directly observe that up to an overall constant (not relevant for the discussion) $\int\mathrm{d}^dY\frac{1}{(Y\cdot W)^d}=(W\cdot W)^{-d/2}$. As a consequence
\begin{equation}
I^\phi_{n,d}=\int_0^\infty\frac{\langle X\mathrm{d}X^{n-1}\rangle\,(Y_\infty\cdot W)^{n-d}}{(W\cdot W)^{n-\frac{d}{2}}}\equiv\int_0^\infty\frac{\measure{X}{n-1}\,(LX)^{n-d}}{(XQX)^{n-\frac{d}{2}}},
\end{equation}
where $L=[1:1:\ldots:1]$ and
\begin{equation}
Q_{IJ}=\mathcal{Y}_I\cdot\mathcal{Y}_J,\quad I,J=1,2,\ldots,n.
\end{equation}

The above result has an immediate important consequence. As we see the entries of $Q$ as a matrix is always a Lorentz product of two $\mathcal{Y}$'s in the ambient space, with the labels for the $\mathcal{Y}$'s those corresponding to the row and the column respectively. In $(d+2)$-dimensional spacetime we cannot have more than $d+2$ independent vectors. This indicates that
\begin{equation}
\mathrm{rank}Q=\begin{cases}n&n\leq d+2,\\ d+2&n>d+2.\end{cases}
\end{equation}
Especially, $\det Q=0$ whenever $n>d+2$.

\subsection{Case of $n<d+2$}

If $n<d$, we have extra linear factors in the denominator. This can be brought to $E$ integrals by lifting to higher dimensions (as will be discussed in Section \ref{sec:linearfactorsindenominator}).

In the case $n=d$, $I_{d,d}$ is exactly an $E_{d}$ integral, which has transcendental weight $\lfloor\frac{d}{2}\rfloor$. Up to an overall coefficient it is a pure function.

In the case $n=d+1$, $I_{d+1,d}$ is an $E_{d+1,1}$ integral, which also has the uniform transcendental weight $\lfloor\frac{d}{2}\rfloor$, but in general is not pure.

\subsection{E.g.~1: massive scalar box in 4d}\label{subsec:massivebox}

For simplicity we assume the external legs to be massless, but all the four loop propagators have mass $m$. This integral depends on two ratios
\begin{equation}
u=\frac{4m^2}{-s},\qquad v=\frac{4m^2}{-t}.
\end{equation}
The Feynman parametrization of this integral is of the $E_4$ type, and so can be made pure by a proper normalization factor. Explicitly we have
\begin{equation}
I_{\text{e.g.}1}=\frac{1}{4}\int_0^\infty\frac{\sqrt{st(st-4m^2(s+t))}\,\measure{X}{3}}{(-sx_1x_3-tx_2x_4+m^2(x_1+x_2+x_3+x_4)^2)^2}.
\end{equation}
The symbol of this integral is \cite{Davydychev:1993ut,Caron-Huot:2014lda}
\begin{equation}\label{eq:eg1knownresult}
\mathcal{S}I_{\text{e.g.}1}=\frac{\sqrt{1+u}-1}{\sqrt{1+u}+1}\otimes\frac{\sqrt{1+u+v}-\sqrt{1+u}}{\sqrt{1+u+v}+\sqrt{1+u}}+(u\leftrightarrow v).
\end{equation}

Since the integral only depends on $u$ and $v$ we can first rescale all the Feynman parameters by $m^{-1}$ to bring the quadric to
\begin{equation}
Q=\left(\begin{matrix}1&1&1+\frac{2}{u}&1\\1&1&1&1+\frac{2}{v}\\1+\frac{2}{u}&1&1&1\\1&1+\frac{2}{v}&1&1\end{matrix}\right).
\end{equation}
The contour is the canonical one. So in determining the last symbol entries the corresponding matrix $\mathbf{H}=Q^{-1}$. 

Let us work in the region where both $u$ and $v$ are positive. None of the diagonal entries of $\mathbf{H}$ vanishes, and so the ratio of roots applies to all the last symbol entries. Given that $\det Q>0$, applying the symmetries of the ratio of roots as discussed before, we can simplify the submatrices of $\mathbf{H}$. Specifically we have
\begin{equation}
\begin{split}
\mathbf{H}_{\{1,2\}}&\sim\mathbf{H}_{\{3,4\}}\sim
\mathbf{H}_{\{1,4\}}\sim\mathbf{H}_{\{2,3\}}\sim\left(\begin{matrix}-1&-1\\-1&-1\end{matrix}\right),\quad\\
\mathbf{H}_{\{1,3\}}&\sim\left(\begin{matrix}-u&2+u+2v\\2+u+2v&-u\end{matrix}\right),\quad
\mathbf{H}_{\{2,4\}}\sim\left(\begin{matrix}-v&2+2u+v\\2+2u+v&-v\end{matrix}\right).
\end{split}
\end{equation}
The matrices in the first line only lead to ratios of roots that are constants, which we can directly through away. So the structure of the symbols simplifies to
\begin{equation}
\mathcal{S}I_{\text{e.g.}1}=\underline{24}\otimes\underline{13}+\underline{13}\otimes\underline{24},
\end{equation}
where in the last entries
\begin{equation}
\underline{13}=\left(\frac{\sqrt{1+u+v}-\sqrt{1+v}}{\sqrt{1+u+v}+\sqrt{1+v}}\right)^2,\quad
\underline{24}=\left(\frac{\sqrt{1+u+v}-\sqrt{1+u}}{\sqrt{1+u+v}+\sqrt{1+u}}\right)^2,
\end{equation}
as can be worked out from the matrices $\mathbf{H}_{\{1,3\}}$ and $\mathbf{H}_{\{2,4\}}$. Then for the first entry $\underline{24}$ we take
\begin{equation}
Q_{\{2,4\}}=\left(\begin{matrix}1&1+\frac{2}{v}\\1+\frac{2}{v}&1\end{matrix}\right),
\end{equation}
and determine the ratio of roost from $(Q_{\{2,4\}})^{-1}$, which is
\begin{equation}
\underline{24}=\left(\frac{\sqrt{1+v}-1}{\sqrt{1+v}+1}\right)^{2}.
\end{equation}
Obviously the $\underline{13}$ in the first entry shares the same expression but with $v$ replaced by $u$. Recall the algebraic properties obeys by generic symbols, all the square exponent can be brought in front, resulting in an overall factor $4$. Hence we recover the well-known result \eqref{eq:eg1knownresult}.

Note here the region of kinematics in which we did the computation is chosen just for a matter of convenience. In principle one can work in other regions and they all lead to the same result.

\subsection{E.g.~2: massless scalar hexagon in 6d}\label{sec:hexagon6d}

As was discussed before when a loop propagator $i$ is massless the decomposition of the symbols resulted from the generic differentiation method suffers from divergence, essentially because the corresponding Feynman parameter $x_i$ enters only linearly in the quadratic Symanzik polynomial.

Here we illustrate this by the massless scalar hexagon in 6d. We assume here that all the external legs are massless as well. After normalizing to a pure function, it depends only on three variables, which are (Here we adopt the convention in \cite{ArkaniHamed:2010gh}, for the convenience of comparison in later examples.)
\begin{equation}
u_1=\frac{s_{23}s_{56}}{s_{234}s_{123}},\quad
u_2=\frac{s_{34}s_{61}}{s_{345}s_{234}},\quad
u_3=\frac{s_{45}s_{12}}{s_{123}s_{345}}.
\end{equation}
So as in the previous case we can first rescale the parameters to get
\begin{equation}
I_{\text{e.g.}2}=\frac{1}{8}\int_0^\infty\frac{\sqrt{D}\,\measure{X}{5}}{(x_1x_3+x_1x_4+u_1x_2x_4+x_1x_5+x_2x_5+u_2x_3x_5+x_2x_6+x_3x_6+u_3x_4x_6)^3},
\end{equation}
with $D\equiv(-1+u_1+u_2+u_3)^2-4u_1u_2u_3$. For notational convenience we also define
\begin{align}
x_\pm&=\frac{-1+u_1+u_2+u_3\pm\sqrt{D}}{2u_1u_2u_3}.
\end{align}

When working out the symbol $\mathcal{S}I_{\text{e.g.}2}$, it is good to assume a certain range of the variables, in order to unambiguously determine the relative signs between different symbol terms. Here for convenience we assume $0<u_i<1$ ($i=1,2,3$) and $D<0$.

For the last symbol entries, none of the diagonals of $\mathbf{H}$ is zero, and so we apply the generic algorithm, computing the ratio of roots, finding that
\begin{align}
\underline{14}=\underline{25}=\underline{36}&=\frac{x_+}{x_-},\\
\underline{13}=\underline{46}&=\frac{1-u_3x_-}{1-u_3x_+},\\
\underline{15}=\underline{24}&=\frac{1-u_1x_-}{1-u_1x_+},\\
\underline{26}=\underline{35}&=\frac{1-u_2x_-}{1-u_2x_+},\\
\underline{12}=\underline{45}&=\frac{1-u_2u_3x_+}{1-u_2u_3x_-}\equiv\frac{x_+(1-u_1x_-)}{x_-(1-u_1x_+)},\\
\underline{16}=\underline{34}&=\frac{1-u_1u_2x_+}{1-u_1u_2x_-}\equiv\frac{x_+(1-u_3x_-)}{x_-(1-u_3x_+)},\\
\underline{23}=\underline{56}&=\frac{1-u_1u_3x_+}{1-u_1u_3x_-}\equiv\frac{x_+(1-u_2x_-)}{x_-(1-u_2x_+)}.
\end{align}

Let us then directly look at the first entries. There the diagonals are always zero, and so $\underline{ij}=(\det Q_{\{i,j\}}(Q_{\{i,j\}}^{-1})_{i,j})^2$. The only non-tivial symbol entries are
\begin{equation}
\underline{24}=u_1^2,\quad
\underline{35}=u_2^2,\quad
\underline{46}=u_3^2.
\end{equation}
Here a direct computation yields an extra factor $1/4$, which we through away since they makes no contribution to the symbols. Consequently $\mathcal{S}I_{\text{e.g.}2}$ reduces to a summation of $18$ terms, starting with the above three first entries.

Let us just focus on details of the terms starting with $\underline{24}\otimes\cdots$, which are
\begin{equation}
\underline{24}\otimes\left(\underline{13}\otimes\underline{56}+\underline{15}\otimes\underline{36}+\underline{16}\otimes\underline{35}+\underline{35}\otimes\underline{16}+\underline{36}\otimes\underline{15}+\underline{56}\otimes\underline{13}\right).
\end{equation}
To study the second entries, we compute the relavent matrices
\begin{equation}
\begin{split}
(q_{(56)}Q_{(56)}^{-1})_{\{1,3\}}&=\left(\begin{matrix}0&\frac{u_1^2}{8}\\\frac{u_1^2}{8}&0\end{matrix}\right),\quad
(q_{(36)}Q_{(36)}^{-1})_{\{1,5\}}=\left(\begin{matrix}0&\frac{u_1(u_1-1)}{8}\\\frac{u_1(u_1-1)}{8}&0\end{matrix}\right),\\
(q_{(35)}Q_{(35)}^{-1})_{\{1,6\}}&=\left(\begin{matrix}\frac{u_1u_3}{4}&\frac{u_1}{-8}\\\frac{u_1}{-8}&0\end{matrix}\right),\quad
(q_{(16)}Q_{(16)}^{-1})_{\{3,5\}}=\left(\begin{matrix}0&\frac{u_1^2u_2}{8}\\\frac{u_1^2u_2}{8}&0\end{matrix}\right),\\
(q_{(15)}Q_{(15)}^{-1})_{\{3,6\}}&=\left(\begin{matrix}\frac{u_1u_3}{4}&\frac{u_1^2}{8}\\\frac{u_1^2}{8}&0\end{matrix}\right),\quad
(q_{(13)}Q_{(13)}^{-1})_{\{5,6\}}=\left(\begin{matrix}\frac{u_1u_3}{4}&\frac{u_1u_3}{-8}\\\frac{u_1u_3}{-8}&0\end{matrix}\right).
\end{split}
\end{equation}
The submatrices associated with second entries $\underline{13}$, $\underline{15}$ and $\underline{35}$ have both diagonal entries vanishing, and so the symbol entries are square of their off-diagonal elements
\begin{equation}
\underline{13}=u_1^4,\qquad
\underline{15}=\frac{1}{u_1^2(1-u_1)^2},\qquad
\underline{35}=u_1^4u_2^2,
\end{equation}
where again we drop the trivial factors. The expression for $\underline{15}$ is inversed since the element $(Q_{(36)}^{-1})_{\{1,5\}}$ is negative. For the other second entries, their associated submatrices involve one non-vanishing diagonal, and so applying \eqref{eq:symbolentryfromdif}
\begin{equation}
\underline{16}=\frac{q_{(135)}}{q_{(35)}^2(Q_{(35)}^{-1})_{1,6}^2}=\frac{u_3}{u_1},\quad
\underline{36}=\frac{q_{(15)}^2(Q_{(15)}^{-1})_{3,6}^2}{q_{(135)}}=\frac{u_1^3}{u_3},\quad
\underline{56}=\frac{q_{(135)}}{q_{(13)}^2(Q_{(13)}^{-1})_{5,6}^2}=\frac{1}{u_1u_3}.
\end{equation}
Using the identity
\begin{equation}\label{appuidentity}
\prod_{i=1}^3\frac{1-u_ix_-}{1-u_ix_+}=\frac{x_-^2}{x_+^2},
\end{equation}
the above six terms thus simplifies to
\begin{equation}
2\left(u_1\otimes(1-u_1)\otimes\frac{x_-}{x_+}+u_1\otimes u_2\otimes\frac{x_+(1-u_3x_-)}{x_-(1-u_3x_+)}+u_1\otimes u_3\otimes\frac{x_+(1-u_2x_-)}{x_-(1-u_2x_+)}\right).
\end{equation}
Note in particular that at first sight there should be the factor $u_1$ sitting in the second entry, but one can explicitly check those contributions sum up to
\begin{equation}
u_1\otimes u_1\otimes\frac{x_+^6\prod_i(1-u_ix_-)^3}{x_-^6\prod_i(1-u_ix_+)^3},
\end{equation}
and so by the identity \eqref{appuidentity} these contributions completely drop away.

The terms starting with $\underline{35}$ and with $\underline{46}$ works similarly, and it can be checked that the corresponding results are related to the above one by cyclic permutations. They sum up to the correct $\mathcal{S}I_{\text{e.g.}2}$ as has be computed in previous literature, e.g., \cite{Dixon:2011ng}.

For later reference, we list in Appendix \ref{app:hexagon} the expression for each of the 18 contributing symbol terms.

\subsection{E.g.~3: pentagon in 4d}

Next let us look at a pentagon in 4d to illustrate the $E_{n,1}$ integrals. This time we assume that the loop propagators are all massless while the external legs can be massive. Its Feynman parametrization is an $E_{5,1}$, with its $Q$ and $L$ as
\begin{equation}
Q_{IJ}=(x_I-x_J)^2\equiv x_{IJ}^2,\qquad L=[1:1:\ldots:1].
\end{equation}
Its integrand is exact, and so the integrals can be localized onto the five codim-1 boundaries of the canonical simplex in $\mathbb{CP}^4$. It is easy to observe that upon the boundary $\Delta_{(i)}$ the reduced integral becomes exactly the $e_4$ integral coming from the box obtained by deleting the $i^{\rm th}$ loop propagator from the pentagon. And so the localization leads to a box expansion.

Here we are not going to work out the entire result but just focus on the coefficient in front of each $e_4$ integral. From previous discussions the coefficient for the boundary $\Delta_{(i)}$ is
\begin{equation}
\frac{H_iQ^{-1}L}{\sqrt{q_{(i)}}}=\frac{\sum_{j=1}^5(Q^{-1})_{ij}}{\sqrt{\det Q_{\{\hat{i}\}}}}.
\end{equation}

\section{Differentiation II: Integrals with Tensor Numerators}\label{sec:dif2}

\flag{No further revisions.}

\subsection{Integrals with a tensor numerator}

We now extend our previous analysis to a general quadric integral, with arbitrary tensor numerator
\begin{equation}
\verified{}
E_{n,k}=\int_\Delta\frac{\measure{X}{n-1}\,T[X^k]}{(XQX)^{\frac{n+k}{2}}}.
\end{equation}
Since $T$ is always contracted with $k$ $X$'s, without loss of generality we assume it is completely symmetric. Still, $Q$ is non-degenerate in the present discussion.

From the experience with $e_n$ integrals we see that the integral representation we start with in general contains more integrals than the number that is necessary to lift the transcendental weight to the actual value. The extra integrals are trivial ones, in the sense that they should always be cleanly localized via Stokes' theorem.

This phenomenon is expected to occur for the more general integrals with tensor. Especially, as pointed out before, case $E_{n,1}$ with a rank-1 tensor is always fully localized, and so for higher rank tensors it is good to check to what extent this is still true.

First we can observe the following relation
\begin{equation}\label{eq:nondegeneratetensordecomposition}
\begin{split}
\fixed
\frac{\measure{X}{n-1}\,T[X^{k}]}{(XQX)^{\frac{n+k}{2}}}=&\frac{1}{n+k-2}\mathrm{d}_X\!\!\left[\frac{\form{(Q^{-1}T)[X^{k-1}]X}{X}{n-2}}{(XQX)^{\frac{n+k-2}{2}}}\right]\\
&+\frac{k-1}{n+k-2}\frac{\measure{X}{n-1}\,(\text{tr}_{Q}T)[X^{k-2}]}{(XQX)^{\frac{n+k-2}{2}}}.
\end{split}
\end{equation}
In the above $(Q^{-1}T)$ is to contract one index of $Q^{-1}$ with one index of $T$ to make up a rank-$(1,k-1)$ tensor, still completely symmetric in its lower indices; and
\begin{equation}\label{eq:troperator}
\fixed
(\text{tr}_QT)_{I_3\ldots I_k}:=(Q^{-1})^{I_1I_2}T_{I_1\ldots I_k},
\end{equation}
which can be regarded as taking trace of $T$ wrst to $Q^{-1}$. The RHS of the above identity consists of a total derivative and an additional piece of the same type as the LHS but with a lower-rank tensor numerator. Moreover, the total derivative piece becomes the integrand for $E_{n-1,k-1}$ after localized onto the codim-1 boundaries. This indicates that schematically
\begin{equation}
\fixed
E_{n,k}\sim E_{n-1,k-1}+E_{n,k-2},
\end{equation}
(``$\sim$'' refers to ``equal to a linear combination of''.)

Apart from $\text{tr}_{Q}$, let us introduce some more notations for operations on a completely symmetric tensor. First we define
\begin{equation}
\fixed
(\mathfrak{c}_Q^{(i)}T)_{I_2\ldots I_k}\equiv(Q^{-1})^{iI_1}T_{I_1I_2\ldots I_k},\quad I_2,\ldots,I_k\neq i.
\end{equation}
Note that here while the dummy index $I_1$ runs over the full label set, we defined the operation $\mathfrak{c}_Q^{(i)}$ to also restrict every other index to run in a smaller range excluding $i$. Using $\text{tr}_Q$ and $\mathfrak{c}_Q^{(i)}$ we further define two composite operations
\begin{align}
\fixed
\label{eq:ccoperator}\mathfrak{cc}_Q^{(ij)}T&\equiv\mathfrak{c}_{Q_{(i)}}^{(j)}\mathfrak{c}_Q^{(i)}T+\mathfrak{c}_{Q_{(j)}}^{(i)}\mathfrak{c}_Q^{(j)}T,\\
\fixed
\label{eq:ctcoperator}\mathfrak{ct}^p\mathfrak{c}_Q^{(ij)}T&\equiv\mathfrak{c}_{Q_{(i)}}^{(j)}(\text{tr}_{Q_{(i)}})^p\mathfrak{c}_Q^{(i)}T+\mathfrak{c}_{Q_{(j)}}^{(i)}(\text{tr}_{Q_{(j)}})^p\mathfrak{c}_Q^{(j)}T,
\end{align}
and we abbreviate, e.g., $\mathfrak{ctc}_Q^{(ij)}\equiv \mathfrak{ct}^1\mathfrak{c}_Q^{(ij)}$, and of course $\mathfrak{cc}_Q^{(ij)}\equiv \mathfrak{ct}^0\mathfrak{c}_Q^{(ij)}$. Both operations are symmetric wrst $i$ and $j$, and both restricts the remaining indices to a smaller range excluding $\{i,j\}$ (hence they induce tensors on $\Delta_{(ij)}$). While $\mathfrak{cc}_Q^{(ij)}$ brings $T$ down to a rank-$(0,k-2)$ tensor, $\mathfrak{ct}^p\mathfrak{c}_Q^{(ij)}$ brings it down to rank $(0,k-2p-2)$.

With the above notations the localization can be conveniently written as
\begin{equation}
\begin{split}
\int_\Delta T[X^k]
&=\sum_i\int_{\Delta_{(i)}}\frac{(\mathfrak{c}_Q^{(i)}T)[X_{(i)}^{k-1}]}{n+k-2}+\frac{k-1}{n+k-2}\int_\Delta(\text{tr}_{Q}T)[X^{k-2}]\\
&=\sum_{i<j}\int_{\Delta_{(ij)}}\frac{(\mathfrak{cc}_Q^{(ij)}T)[X_{(ij)}^{k-2}]}{(n+k-2)(n+k-4)}+\sum_i\int_{\Delta_{(i)}}\frac{(k-2)\,(\text{tr}_{Q_{(i)}}\mathfrak{c}_Q^{(i)}T)[X_{(i)}^{k-3}]}{(n+k-2)(n+k-4)}\\
&\quad+\sum_i\int_{\Delta_{(i)}}\frac{(k-1)\,(\mathfrak{c}_Q^{(i)}\text{tr}_QT)[X_{(i)}^{k-3}]}{(n+k-2)(n+k-4)}+\int_\Delta\frac{(k-1)(k-3)\,((\text{tr}_Q)^2T)[X^{k-4}]}{(n+k-2)(n+k-4)}\\
&=\cdots.
\end{split}
\end{equation}
In the above we avoided explicitly writing out the factor $\frac{\measure{X}{n-1}}{(XQX)^{\#}}$ or $\frac{\measure{X_{(ij)}}{n-3}}{(X_{(ij)}Q_{(ij)}X_{(ij)})^{\#}}$ in the integrand, which can be easily recovered from the context. Repeating the decomposition we can fully localize the original integral onto the codim-2 boundaries plus some remaining $e$ integral in higher dimensions. Specifically, when $k$ is even we have
\begin{equation}
\begin{split}
\int_\Delta T[X^k]=&\sum_{0\leq p_1\leq p_2\leq\frac{k-2}{2}}\frac{(k-1)!!(n+k-2p_2-6)!!\sum_{i<j}\int_{\Delta_{(ij)}}(\mathfrak{ct}^{p_2-p_1}\mathfrak{c}_Q^{(ij)}(\text{tr}_Q)^{p_1}T)[X_{(ij)}^{k-2p_2-2}]}{(k-2p_2-3)!!(n+k-2)!!(k-2p_1-1)}\\
&+\frac{(k-1)!!\,(n-2)!!}{(n+k-2)!!}\frac{(\text{tr}_Q)^{\frac{k}{2}}T}{\sqrt{q}}\,e_n,\qquad\text{even }k.
\end{split}
\end{equation}
The case with odd $k$ is a bit more involved, as the highest weight terms starts with $e_{n-1}$, which at least requires one localization. We have
\begin{equation}
\begin{split}
\int_\Delta T[X^k]=&\sum_{0\leq p_1\leq p_2\leq\frac{k-3}{2}}\frac{(k-1)!!(n+k-2p_2-6)!!\sum_{i<j}\int_{\Delta_{(ij)}}(\mathfrak{ct}^{p_2-p_1}\mathfrak{c}_Q^{(ij)}(\text{tr}_Q)^{p_1}T)[X_{(ij)}^{k-2p_2-2}]}{(k-2p_2-3)!!(n+k-2)!!(k-2p_1-1)}\\
&+\sum_{0\leq p\leq\frac{k-1}{2}}\frac{(k-1)!!\,(n-3)!!}{(n+k-2)!!(k-2p-1)}\sum_i\frac{((\text{tr}_{Q_{(i)}})^{\frac{k-1}{2}-p}\mathfrak{c}_Q^{(i)}(\text{tr}_Q)^{p}T)}{\sqrt{q_{(i)}}}e_{(i)},\qquad\text{odd }k.
\end{split}
\end{equation}

To proceed further we then leave the $e$ integrals aside and perform the same computation on the $E_{n-2,k-2p_2-2}$ integrals, landing on integrals schematically of the form
\begin{equation}
\int_{\Delta_{iji'j'}}\left(\mathfrak{ct}^{p'_2-p'_1}\mathfrak{c}_{Q_{(ij)}}^{(i'j')}(\text{tr}_{Q_{(ij)}})^{p'_1}\mathfrak{ct}^{p_2-p_1}\mathfrak{c}_Q^{(ij)}(\text{tr}_Q)^{p_1}T\right)[X_{(iji'j')}^{k-2p_2-2p'_2-4}]
\end{equation}
and so on. In the case when $n>k$ we in the end have both
\begin{equation}
\fixed
E_{n,k<n}\sim \begin{cases}e_{n}+e_{n-2}+\cdots +e_{n-k},&\text{even }k,\\ e_{n-1}+e_{n-3}+\cdots+e_{n-k},&\text{odd }k,\end{cases}
\end{equation}
where the explicit coefficients in the linear combination can be worked out using the above identities together with the discussion in Section \ref{sec:dif1}. When $k\geq n$, in the case of even $nk$ the recursion will in general ultimately land on some $E_{2,p}$ with $p\leq k-m+1$, whose localization generates a rational function (apart from possibly some log terms) and terminates; and in the case of odd $nk$ it terminate at some $E_{1,p}$, which is already a rational function. And so
\begin{equation}
\fixed
E_{n,k\geq n}\sim \begin{cases}e_{n}+e_{n-2}+\cdots +e_{2}+(\text{rational}),&\text{even }n,\text{ even }k,\\
e_{n-1}+e_{n-3}+\cdots +e_{3}+(\text{rational}),&\text{even }n,\text{ odd }k,\\
e_{n}+e_{n-2}+\cdots +e_{3}+(\text{rational}),&\text{odd }n,\text{ even }k,\\
e_{n-1}+e_{n-3}+\cdots +e_{2}+(\text{rational}),&\text{odd }n,\text{ odd }k.\end{cases}
\end{equation}

As a consequence, we see that the integrals with higher rank tensor numerators in general lead to a mixture of functions with different transcendental weights, with the highest weight related to the dimension of integrals we start with (in the same way as the $E_n$ and $E_{n,1}$ integrals).

In the situation when a rational piece is present its determination is a little bit different in the last step. The rational contributions can always be treated as arising from some one dimensional integral
\begin{equation}
\int_{\Delta}\frac{\langle X\mathrm{d}X\rangle\,T[X^k]}{(XQX)^{\frac{k+2}{2}}}=\frac{1}{k}\int_\Delta\mathrm{d}_X\!\!\left[\frac{\langle (Q^{-1}T)[X^{k-1}]\,X\rangle}{(XQX)^{\frac{k}{2}}}\right]+\frac{k-1}{k}\int_\Delta\frac{\langle X\mathrm{d}X\rangle\,(\text{tr}_QT)[X^{k-2}]}{(XQX)^{\frac{k}{2}}}.
\end{equation}
It suffices to focus just on the first term on RHS. Localizing this terms yields two contributions from the boundary points $V_1$ and $V_2$, which are
\begin{equation}\label{eq:rationalfrom1d}
\frac{1}{k}\left(\sum_{I=1,2}\frac{(Q^{-1})^{1I}T_{I2\ldots2}}{(Q_{22})^{\frac{k}{2}}}+(1\leftrightarrow2)\right).
\end{equation}

This seems to cause some confusion, because a generic scalar $n$-gon in $d$ dimensions has the form
\begin{equation}
\int_0^\infty\frac{\measure{X}{n-1}\,(LX)^{n-d}}{(XQX)^{\frac{n-d}{2}}},
\end{equation}
but we expect it to always have uniform transcendental weight $\frac{d}{2}$ for $n\geq d$ and even $d$, which is only the lowest weight we can get from the above analysis. Of course, the data $Q$ and $L$ we obtain from actual Feynman integrals cannot be random ones, and this indicates that they have to always kill the coefficients of the parts with higher transcendental weights. We will explain why this is true in detail later on. Before that we need to also understand what happens when the qaudric is indeed degenerate.

\subsection{An alternative analysis}

Instead of doing explicit decomposition of the tensor numerator, we can alternatively choose to relate a generic tensor integral to the elementary integrals $E_{n}$ and $E_{n,1}$ we studied in Section \ref{sec:dif1}. As usual this is achieved by observing
\begin{equation}
\verified{}
E_{n,k}=\begin{cases}{\displaystyle \frac{(-1)^{\frac{k}{2}}\Gamma(\frac{n+k}{2})}{\Gamma(\frac{n}{2})}\int_\Delta T[\partial_Q^{\frac{k}{2}}]\frac{\measure{X}{n-1}}{(XQX)^{\frac{n}{2}}}},&\text{even }k,\\
{\displaystyle \frac{(-1)^{\frac{k-1}{2}}\Gamma(\frac{n+k}{2})}{\Gamma(\frac{n+1}{2})}\int_\Delta T[\partial_Q^{\frac{k-1}{2}},X]\frac{\measure{X}{n-1}}{(XQX)^{\frac{n+1}{2}}}},&\text{odd }k.\end{cases}
\end{equation}
Here in actual computation we only require $T$ to be expressed in the form that is symmetric in each pair of indices that are contracted to the same $\partial_Q$. We can then pull the derivatives outside of the integral. Note that the differentiation directly applies to symbols as well; this allows us to first determine the symbol for the remaining $E_n$ or $E_{n,1}$ integral and they apply the derivative on it to further obtain the result for $E_{n,k}$.

When the quadric is non-degenerate, specifically the case with odd $k$ further decompose into
\begin{equation}
E_{n,k}=\frac{(-1)^{\frac{k-1}{2}}\Gamma(\frac{n+k}{2})}{\Gamma(\frac{n+1}{2})}\sum_{i=1}^n\frac{T[\partial_Q^{\frac{k-1}{2}},Q^{-1}H_i]}{n-1}\int_{\Delta_{(i)}}\frac{\measure{X_{(i)}}{n-2}}{(X_{(i)}Q_{(i)}X_{(i)})^{\frac{n-1}{2}}}.
\end{equation}
In the above particular attention should be paid that $\partial_Q$'s also acts on the vector $Q^{-1}H_i$ contracted with $T$ in each term.

\subsection{Integrals with a degenerate quadric}

In the discussions so far we have always assumed that the the quadric inside the integrals is non-degenerate, $q\equiv\det Q\neq0$. For the application of Feynman integrals, the quadric actually becomes degenerate every time when the particle number exceeds the spacetime dimensions by more than 2, $n>d+2$, as will be discussed in more detail in Section \ref{sec:generalscalar}. So we also need to understand how to handle this case.

In our discussion of the case $\det Q=0$ we always assume that $\mathrm{corank}Q=r>0$.

\subsubsection{General expectations}

When $\det Q=0$, $Q$ admits of a non-trivial null space $\mathfrak{N}_Q\cong\mathbb{CP}^{r-1}$. Clearly there is a risk for the integrals to run into divergence. Here we avoid this issue by requiring that the integration contour has empty overlap with this null space.

We can already make an estimation on some quantitative aspects of the resulting function. Since $\text{corank}(Q)=r$, we can always $\mathrm{PGL}(n)$ transform into a frame such that $XQX$ is manifestly free of the variables $\{x_{n-r+1},x_{n-r+2},\ldots,x_n\}$. So the integrals associated to these variables are just integration of a polynomial over certain bounded region, which can be straightforwardly completed,  resulting in another polynomial of the remaining variables with $r$ higher degrees. So an $E_{n,k}$ integral turns into
\begin{equation}
E_{n,k}=\sum_{\Delta'}\int_{\Delta'}\frac{\measure{X'}{n-r-1}\,T'[X^{k+r}]}{(X'Q'X')^{\frac{n+k}{2}}},
\end{equation}
where $Q'$ is now non-generic, so that the techniques we discussed before can be further applied. Note that even though we start from a single simplex, the integration region along the null directions can change depending on the values of the remaining variables, and so the result above is generically a summation over different simplices $\Delta'$ and the numerator $T'$ also varies with $\Delta'$. Despite of this, we immediately learn that any $E_{n,k}$ with $\mathrm{corank}Q=r$ cannot have transcendental weight higher than $\lfloor\frac{n-r}{2}\rfloor$.

\begin{figure}
\begin{center}
\begin{tikzpicture}
\coordinate (p3) at (0,0);
\coordinate (p1) at (5,0);
\coordinate (p2) at (0,5);
\draw [Orange,thick,dashed] (p3) -- (p1) arc [start angle=0, end angle=90,radius=5] -- cycle;
\draw [OrangeRed,thick,dotted] (p2) arc [start angle=90,end angle=0,radius=1] -- (1,0);
\draw [OrangeRed,thick,dotted] (p2) arc [start angle=90,end angle=0,radius=1.5] -- (1.5,0);
\draw [OrangeRed,thick,dotted] (p2) arc [start angle=90,end angle=0,radius=3] -- (3,0);
\draw [ProcessBlue,ultra thick] (p2) arc [start angle=90,end angle=0,radius=.4] -- (.4,0);
\draw [ProcessBlue,ultra thick] (p2) arc [start angle=90,end angle=0,radius=4] -- (4,0);
\coordinate [label=180:{\small $V_1$}] (v1) at (1,1.5);
\coordinate [label=180:{\small $V_2$}] (v2) at (1.5,.5);
\coordinate [label=0:{\small $V_3$}] (v3) at (3,2);
\fill [OrangeRed,opacity=.3] (v1) -- (v2) -- (v3) -- cycle;
\draw [OrangeRed,ultra thick] (v1) -- (v2) -- (v3) -- cycle;
\draw [OrangeRed,ultra thick] (v2) -- ($(v3)!.75!(v1)$);
\filldraw [OrangeRed] (v1) circle [radius=2pt];
\filldraw [OrangeRed] (v2) circle [radius=2pt];
\filldraw [OrangeRed] (v3) circle [radius=2pt];
\coordinate [label=-90:{\small $u_1$}] (u1) at (1,0);
\coordinate [label=-90:{\small $u_2$}] (u2) at (1.5,0);
\coordinate [label=-90:{\small $u_3$}] (u3) at (3,0);
\draw [OrangeRed,ultra thick] (u1) -- (u3);
\filldraw [OrangeRed] (u1) circle [radius=2pt];
\filldraw [OrangeRed] (u2) circle [radius=2pt];
\filldraw [OrangeRed] (u3) circle [radius=2pt];
\coordinate [label=-90:{\small $a$}] (a) at (.4,0);
\coordinate [label=-90:{\small $b$}] (b) at (4,0);
\coordinate [label=90:{\small $\mathfrak{N}_Q=\{[0:1:0]\}$}] (null) at (0,5);
\filldraw [black] (null) circle [radius=2pt];
\end{tikzpicture}
\end{center}
\caption{Projection through the null space $\mathfrak{N}_Q$ of $Q$.}
\label{fig:nullspaceprojection}
\end{figure}
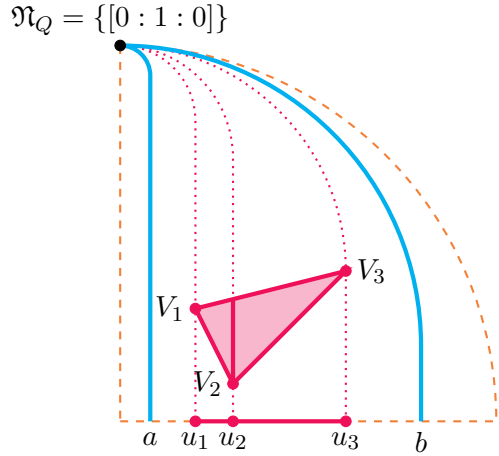

As a simple illustration, consider the following $E_{3,1}$ toy integral
\begin{equation}
\int_\Delta\frac{\measure{X}{2}\,(x_3)}{(x_1-a x_3)^2(x_1-b x_3)^2},
\end{equation}
where the simplex contour is defined by the three vertices $V_i=[u_i:v_i:1]$ ($i=1,2,3$), assuming $u_1<u_2<u_3$ and $v_2<v_1<v_3$. The quadric has corank-1 and its null space is a unique point $[0:1:0]$, which is obvious since the denominator is independent of $x_2$. These are depicted in Figure \ref{fig:nullspaceprojection}. Very conveniently we can fix $x_3=1$ and first perform the $x_2$ integral to obtain
\begin{equation}
\frac{\det(V_1V_2V_3)}{u_{12}u_{13}}\int_{u_1}^{u_2}\mathrm{d}x_1\frac{(x-u_1)}{(x_1-a)^2(x_1-b)^2}-\frac{\det(V_1V_2V_3)}{u_{31}u_{32}}\int_{u_2}^{u_3}\mathrm{d}x_1\frac{(x-u_3)}{(x_1-a)^2(x_1-b)^2},
\end{equation}
where $u_{ij}=u_i-u_j$. Hence the original integral is brought into two $E_{2,2}$ integrals (the fact we only see degree one in the numerator is merely a consequence of gauge-fixing), which can be easily performed.

More generally, the procedure of first integrating away the null directions is in correspondence to projecting through $\mathfrak{N}_Q$. In other words, we slice the space by $r$-planes passing through $\mathfrak{N}_Q$, and integrate over the intersection of the $r$-plane with the original simplex contour. In the above example the triangle projects into two simplices in $\mathbb{CP}^1$, and the resulting integrand differs between the two image simplices.

While in principle this procedure applies to higher dimensions and more complicated integrals, it can become tedious in practice, mainly due to the need of figuring out the image of the simplex under the projection (which is in general a particular trianglation of some complex). Moreover, the result it directly produces does not admits a nice geometric organization and can even hinder certain properties. For example, it is not at all obvious that the above toy integral is only a combination of log functions. Each of the $E_{2,2}$ integrals contains a rational part, and they turn out to cancel away between the two integrals. In fact, the entire result is as simple as
\begin{equation}
\frac{\det(V_1V_2V_3)}{(a-b)^3}\left(\frac{a+b-2 u_1}{u_{12}u_{13}}\log\left(\frac{a-u_1}{b-u_1}\right)+\text{permutations}\right).
\end{equation}
Altogether it decomposes into three terms, each of which seems to naturally locate on one of the codim-1 boundary of the triangle $(V_1V_2V_3)$.

Consequently, it is better to find out a more constructive procedure. 

\subsubsection{Summary of the analysis}

We consider an $E_{n,k}$ with $\mathrm{corank}Q=r$. In order to avoid too many technical details for a complete generality and to quickly move on to physical applications in the next section, here we only provide a summary for a slightly special situation that is sufficient for the examples we discuss later on. A complete discussion on the degenerate quadrics is contained in Appendix \ref{app:degenerateQ}.

So for the present discussion we assume that the numerator in $E_{n,k}$ satisfy a ``tangency'' condition (treating $T$ as a symmetric tensor)
\begin{equation}\label{eq:nullcondition}
\quad (TN)_{I_1I_2\ldots I_{k-1}}=0,\quad\forall N\in\mathfrak{N}_Q,
\end{equation}
which is a natural generalization of the condition \eqref{LNcondition} discussed before. In practice, to check this condition it suffices to pick up a basis for $\mathfrak{N}_Q$.  A convenient choice is to choose a arbitrary set of reference covectors  $\{R_1,\ldots,R_r\}$, and then
\begin{equation}\label{eq:nullspacebasis}
N_a:=
\parbox{5.5cm}{
\tikz{
\node [anchor=center] at (0,.2) {$R_a$};
\node [anchor=center] at (.6,0) {$Q$};
\node [anchor=center] at (2*.6,0) {$\ldots$};
\node [anchor=center] at (3*.6,0) {$Q$};
\node [anchor=center] at (4*.6,.2) {$R_1$};
\node [anchor=center] at (5*.6,.2) {$\ldots$};
\node [anchor=center] at (6*.6,.2) {$\widehat{a}$};
\node [anchor=center] at (7*.6,.2) {$\ldots$};
\node [anchor=center] at (8*.6,.2) {$R_{r}$};
\node [anchor=center] at (4*.6,-.2) {$R_1$};
\node [anchor=center] at (5*.6,-.2) {$\ldots$};
\node [anchor=center] at (6*.6,-.2) {$\widehat{a}$};
\node [anchor=center] at (7*.6,-.2) {$\ldots$};
\node [anchor=center] at (8*.6,-.2) {$R_{r}$};
\draw [black,thick] (0,.6) -- (8*.6,.6);
\draw [black,thick] (0,.6) -- (0,.45);
\draw [black,thick] (.6,.6) -- (.6,.25);
\draw [black,thick] (3*.6,.6) -- (3*.6,.25);
\draw [black,thick] (4*.6,.6) -- (4*.6,.45);
\draw [black,thick] (8*.6,.6) -- (8*.6,.45);
\draw [black,thick] (0,-.6) -- (8*.6,-.6);
\draw [black,thick] (0,-.6) -- (0,-.45);
\draw [black,thick] (.6,-.6) -- (.6,-.25);
\draw [black,thick] (3*.6,-.6) -- (3*.6,-.25);
\draw [black,thick] (4*.6,-.6) -- (4*.6,-.45);
\draw [black,thick] (8*.6,-.6) -- (8*.6,-.45);
}},\quad a=1,\ldots,r
\end{equation}
provide such a basis, where $\widehat{a}$ indicates that we omit $R_a$ in the corresponding contraction. The ladders denote contraction using Levi-Civita symbols. More details of the null space is discussed in Appendix \ref{app:Qnullspace}.

The same references allow us to define a modified determinant and inverse of $Q$
\begin{equation}
\mathring{q}=\parbox{4.2cm}{
\tikz{
\node [anchor=center] at (0,0) {$Q$};
\node [anchor=center] at (.6,0) {$Q$};
\node [anchor=center] at (1.2,0) {$\cdots$};
\node [anchor=center] at (1.8,0) {$Q$};
\node [anchor=center] at (2.4,.2) {$R_1$};
\node [anchor=center] at (3,.2) {$\cdots$};
\node [anchor=center] at (3.6,.2) {$R_r$};
\node [anchor=center] at (2.4,-.2) {$R_1$};
\node [anchor=center] at (3,-.2) {$\cdots$};
\node [anchor=center] at (3.6,-.2) {$R_r$};
\draw [black,thick] (0,.6) -- (3.6,.6);
\draw [black,thick] (0,.6) -- (0,.25);
\draw [black,thick] (.6,.6) -- (.6,.25);
\draw [black,thick] (1.8,.6) -- (1.8,.25);
\draw [black,thick] (2.4,.6) -- (2.4,.45);
\draw [black,thick] (3.6,.6) -- (3.6,.45);
\draw [black,thick] (0,-.6) -- (3.6,-.6);
\draw [black,thick] (0,-.6) -- (0,-.25);
\draw [black,thick] (.6,-.6) -- (.6,-.25);
\draw [black,thick] (1.8,-.6) -- (1.8,-.25);
\draw [black,thick] (2.4,-.6) -- (2.4,-.45);
\draw [black,thick] (3.6,-.6) -- (3.6,-.45);
}},\qquad\mathring{Q}^{-1}=\parbox{4cm}{
\tikz{
\node [anchor=center] at (.6,0) {$Q$};
\node [anchor=center] at (1.2,0) {$\cdots$};
\node [anchor=center] at (1.8,0) {$Q$};
\node [anchor=center] at (2.4,.2) {$R_1$};
\node [anchor=center] at (3,.2) {$\cdots$};
\node [anchor=center] at (3.6,.2) {$R_r$};
\node [anchor=center] at (2.4,-.2) {$R_1$};
\node [anchor=center] at (3,-.2) {$\cdots$};
\node [anchor=center] at (3.6,-.2) {$R_r$};
\draw [black,thick] (0,.6) -- (3.6,.6);
\draw [black,thick] (0,.6) -- (0,.25);
\draw [black,thick] (.6,.6) -- (.6,.25);
\draw [black,thick] (1.8,.6) -- (1.8,.25);
\draw [black,thick] (2.4,.6) -- (2.4,.45);
\draw [black,thick] (3.6,.6) -- (3.6,.45);
\draw [black,thick] (0,-.6) -- (3.6,-.6);
\draw [black,thick] (0,-.6) -- (0,-.25);
\draw [black,thick] (.6,-.6) -- (.6,-.25);
\draw [black,thick] (1.8,-.6) -- (1.8,-.25);
\draw [black,thick] (2.4,-.6) -- (2.4,-.45);
\draw [black,thick] (3.6,-.6) -- (3.6,-.45);
}}\Big/\mathring{q}.
\end{equation}
Under the condition \eqref{eq:nullcondition} these objects make it manifest that the integrand of $E_{n,k}$ can be decomposed into a part with a lower degree numerator plus a part that is exact. This leads to the following decomposition
\begin{equation}\label{eq:degenerateQtensorderivative}
\begin{split}
E_{n,k}=&\frac{(n-r)}{n+k-2}\sum_{i=1}^n\int_{\Delta_{(i)}}\frac{\measure{X_{(i)}}{n-2}\,(H_i\mathring{Q}^{-1}T)[(P(i)X_{(i)})^{k-1}]}{(X_{(i)}Q_{(i)}X_{(i)})^{\frac{n+k-2}{2}}}\\
&+\frac{(k-1)(n-r)}{n+k-2}\int_\Delta\frac{\measure{X}{n-1}\,(\text{tr}_{\mathring{Q}^{-1}}T)[X^{k-2}]}{(XQX)^{\frac{n+k-2}{2}}},
\end{split}
\end{equation}
which is structurally the same as the decomposition \eqref{eq:nondegeneratetensordecomposition} of $T_{n,k}$ in the case of non-degenerate quadric, except for the difference in coefficients. So the discussions there can equally apply in this case. In particular, we should expect to repeat this decomposition for the resulting integrals with lower-degree numerators. The only subtlety here is that it is not obvious that the numerators in the induced integrals on the boundaries still satisfy the condition \eqref{eq:nullcondition}. As we show in Appendix \ref{app:degenerateQintegralspecial}, this actually holds after any sequence of repeated decomposition using \eqref{eq:degenerateQtensorderivative}, as long as the quadric in the induced integral is still degenerate.

\subsection{Integrals with additional linear factors in the denominator}\label{sec:linearfactorsindenominator}

We can also study integrals which involves linear factors in the denominator, in addition to the quadric. These are essentially the same type of integrals as we have studied in previous subsections. To illustrate it, let us consider the situation with a single linear factor
\begin{equation}
\int_0^\infty\frac{\measure{X}{n}\,T[X^k]}{(LX)(XQX)^{\frac{n+k}{2}}},
\end{equation}
where we assume an arbitrary tensor in the numerator, and to simplify the discussion we already pick up a frame where the contour becomes the standard simplex. Here we can work in the affine patch, say $x_1=1$, and lift the integral to one higher dimension
\begin{equation}
\int_0^\infty\frac{\prod_{a=2}^{n+1}\mathrm{d}x_a\,T[X^k]}{(LX)(XQX)^{\frac{n+k}{2}}}=-\frac{2}{n+k}\int_0^\infty\frac{\prod_{a=2}^{n+2}\mathrm{d}x_a\,T[X^k]}{(XQX+(LX)x_{n+2})^{\frac{n+k}{2}+1}}.
\end{equation}
The new integral above is just a projective integral in $\bar{X}=[x_1:x_2:\ldots:x_{n+1}:x_{n+2}]\in\mathbb{CP}^{n+1}$ as expressed in the affine patch $x_1=1$
\begin{equation}
\int_0^\infty\frac{\langle \bar{X}\mathrm{d}^{n+1}\bar{X}\rangle\,\bar{T}[\bar{X}^{k}]}{(\bar{X}\bar{Q}\bar{X})^{\frac{n+k}{2}+1}},
\end{equation}
where $\bar{T}_{a_1\ldots a_k}=T_{a_1\ldots a_k}$ if $a_i<n+2$ ($\forall i$) and zero otherwise, and
\begin{equation}
\bar{Q}=\left(\begin{matrix}Q&L^{\mathrm{T}}\\ L&0\end{matrix}\right).
\end{equation}
The procedure is directly generalized to more general situations when we have several linear factors $(XQX)\prod_{i=1}^{l}(L_iX)$, where we lift the integral in $\mathbb{CP}^n$ to an integral in $\mathbb{CP}^{n+l}$, with the new quadric
\begin{equation}
\bar{Q}=\left(\begin{matrix}Q&L_1^{\mathrm{T}}&\cdots&L_l^{\mathrm{T}}\\ L_1&0&\cdots&0\\\vdots&\vdots&&\vdots\\L_l&0&\cdots&0\end{matrix}\right).
\end{equation}

\subsection{Some general aspects of the symbols}

Let us summarize some general properties of the structure of symbols for the quadric integrals $E_{n,k}$, where the quadric may or may not be degenerate, $\mathrm{corank}Q\geq0$. We have seen from previous analysis that in general the integral is mixture of parts with different transcendental weights, each of which can associates to a linear summation of $e$ integrals. 

By extracting exact rational forms from the integrand and Stokes' theorem, any $E_{n,k}$ can always be fully decomposed into $e_m$ integrals. Hence in a sense the $e$ are minimal. Especially, if $Q$ is non-degenerate, $r=0$, then there exists a canonical decomposition using the operations defined in \eqref{eq:troperator}, \eqref{eq:ccoperator} and \eqref{eq:ctcoperator}. There are some quantitative properties that holds for any decomposition
\begin{itemize}
\item For a given triple $(n,k,r)$ the highest transcendental weight of $E_{n,k}$ cannot be greater than $\lfloor\frac{n-r}{2}\rfloor$. While this bound is generically saturated, special configuration of the integrand numerator can actually turn it smaller.
\item If $k$ is even, the decomposition of $E_{n,k}$ into $e$ starts with $e_{n-r}$, plus $e$'s of lower degrees; while if $k$ is odd, it begins with $e_{n-r-1}$.
\item Regardless of the value of $r$, when $n+k$ is even, the decomposition only involves $e_m$'s with even $m$ (apart from possibly a rational piece); while when $n+k$ is odd, it only involves $e_m$'s with odd $m$. 
\end{itemize}

From the specific algorithm for the determination of symbols, it is also obvious that the $e$ integrals are in a sense maximal, when $r=0$. First of all, every $Q$ uniquely defines an $e$ integral. Then in the case $k>0$, all the $e$ integrals in the decomposition with lower degrees are uniquely associated to the induced quadric from $Q$ on some boundaries. Consequently their symbols are embedded as part of the symbol of the top $e$ associated to $Q$. When viewing $E_{n,k}$ as a single object, we see that every term in $\mathcal{S}E_{n,k}$ comes from the $e_n$ integral defined by its quadric, while the only difference from $\mathcal{S}e_n$ lies in the specific coefficients in front of the symbol terms.

This is a manifestation of the general expectation that all the singularities of an integral are tied to the singularities of its integrand. The advantage here is that the structure of the symbols terms are always completely fixed by the quadric itself, and for a specific integral $E_{n,k}$ the only remaining task is to work out the coefficients from the integrand numerator. 

There exists a linear map from the integrand numerator to the coefficients of the term, which can be worked out systematically following the analysis prescribed before, but in general this map can be very non-trivial. Despite of this, there always exist specific numerators such that the entire integral is restricted to some uniform transcendental weight, or even to a specific symbol term (as we will also see in some examples). Such configurations can be highly non-unique. Anyway it could potentially be interesting to work out the explicit constraints on the numerator for such special situations. We leave this for the future.

Regarding the elementary quadric integrals $e_n$ themselves, we observed a essential structural difference between the $e_n$ with even $n$ and the ones with odd $n$. In the case of odd $n$ there is always a specific label which does not enter the partition labeling a specific symbol term, although any symbol term itself appears to be a term belonging to some $\mathcal{S}e_{n-1}$. While not obvious from the differentiation analysis, it is in fact true that
\begin{equation}
e_n=\sum_{i=1}^ne_{n}^{(i)},\qquad\text{odd }n,
\end{equation}
where $e_{n}^{(i)}$ is some $e$ integral defined on $\Delta_{(i)}$. This relation is going to be discussed in more detail in Section \ref{sec:oddspacetime}.

As a consequence, any $\mathcal{S}E_{n,k}$ is a linear summation of $\mathcal{S}e_m$'s with $m$ valued in even integers, and all the $e_m$'s appearing are induced from the unique $e_n$ defined by the quadric in $E_{n,k}$ on the boundaries of the integration contour. Hence any quadric integral can be equally represented by the unique $e_n$ together with a set of rational functions specified for each boundary of $\Delta$, which indicates the coefficients in the expansion. This imposes non-trivial constraints on the structure of quadric integrals.

\section{More Feynman Integrals}\label{sec:feynman2}

Now let us apply the discussions in the previous section to general one-loop Feynman integrals.

We almost do not put any assumptions on the Feynman integrals we look at. This can be in any integral spacetime dimensions, involve any number of external line which are either (massless or massive) on-shell or off-shell, include loop propagators with no assumptions on their mass. For the time being, in order to make the discussion clean we only assume that we are studying integrals in \emph{even dimensions} and are \emph{finite}. We also only focus on planar objects; since at one loop everything can directly decompose into planar integrals, this does not put any extra restriction.

\subsection{Scalar integrals}\label{sec:generalscalar}

\subsubsection{Case of $n=d+2$}

One-loop scalar diagrams $I^\phi_{n,d}$ with $n=d+2$ are the first non-trivial case where we have to take care, as they are $E_{d+2,2}$ integrals, and
\begin{equation}
E_{d+2,2}\sim e_d+e_{d+2}.
\end{equation}
At first sight this object has mixed weights. However, the $e_{d+2}$ part is dressed with a vanishing coefficient $(LQ^{-1}L)$, and so that they again uniform weight $\frac{d}{2}$.

To confirm that indeed $LQ^{-1}L=0$, note that
\begin{equation}\label{detrelation}
LQ^{-1}L=\det\mathcal{Q}\equiv\det\left(\begin{matrix}0&L\\ L^{\mathrm{T}}&Q\end{matrix}\right),
\end{equation}
where $\mathcal{Q}$ is now a $(d+3)\times(d+3)$ matrix. Let us label the extra row and column by $0$, and define $\mathcal{Y}_0=Y_\infty$, then we have
\begin{equation}
\mathcal{Q}_{IJ}=\mathcal{Y}_I\cdot\mathcal{Y}_J,\quad I,J=0,1,\ldots,n.
\end{equation}
By the same argument as the one for rank$Q$ in Section \ref{sec:feynman1}, we immediately see that $\det\mathcal{Q}=0$ when $n>d+1$.

As a side comment, if we view $L_IL_J$ as a symmetric rank-$(0,2)$ tensor, then the above identity is nothing but the statement that the numerator tensor is traceless wrst $Q^{-1}$, i.e., $\mathrm{tr}_{Q^{-1}}(LL)=0$.

\subsubsection{Case of $n>d+2$}

When the multiplicity $n$ is even bigger, $I^\phi_{n,d}$ is of the type $E_{n,n-d}$, and by recursively applying \eqref{eq:degenerateQtensorderivative} it should again decompose into
\begin{equation}\label{degenerateQtensordecomposition}
e_d+e_{d+2}+\cdots+e_n.
\end{equation}
Pay attention that rank$Q=d+2<n$, so the subscripts above do not necessarily indicates the actual transcendental weights (here the weight has to be $\lfloor\frac{d+2}{2}\rfloor$ at most). Recall from \eqref{eq:nullcondition} that in order the decomposition \eqref{eq:degenerateQtensorderivative} is applicable we need an extra tangency condition for the tensor in the numerator. For the integrals $I^\phi_{n>d+2,d}$ this condition explicitly reduces to
\begin{equation}
(QN)_I=0\quad\Longrightarrow\quad LN=0\quad\forall N.
\end{equation}
Let us first show this is true. Again it is sufficient to introduce a basis for the null space of $Q$ (now it has dimension $r=n-d-2$) as what we did before, and the condition is equivalent to
\begin{equation}
\parbox{5.5cm}{
\tikz{
\node [anchor=center] at (0,.2) {$R_a$};
\node [anchor=center] at (0,-.2) {$L$};
\node [anchor=center] at (.6,0) {$Q$};
\node [anchor=center] at (2*.6,0) {$\cdots$};
\node [anchor=center] at (3*.6,0) {$Q$};
\node [anchor=center] at (4*.6,.2) {$R_1$};
\node [anchor=center] at (4*.6,-.2) {$R_1$};
\node [anchor=center] at (5*.6,.2) {$\cdots$};
\node [anchor=center] at (5*.6,-.2) {$\cdots$};
\node [anchor=center] at (4*.6,.2) {$R_1$};
\node [anchor=center] at (4*.6,-.2) {$R_1$};
\node [anchor=center] at (5*.6,.2) {$\cdots$};
\node [anchor=center] at (5*.6,-.2) {$\cdots$};
\node [anchor=center] at (6*.6,.2) {$\widehat{a}$};
\node [anchor=center] at (6*.6,-.2) {$\widehat{a}$};
\node [anchor=center] at (7*.6,.2) {$\cdots$};
\node [anchor=center] at (7*.6,-.2) {$\cdots$};
\node [anchor=center] at (8*.6,.2) {$R_r$};
\node [anchor=center] at (8*.6,-.2) {$R_r$};
\draw [black,thick] (0,.6) -- (8*.6,.6);
\draw [black,thick] (0,.6) -- (0,.45);
\draw [black,thick] (.6,.6) -- (.6,.25);
\draw [black,thick] (3*.6,.6) -- (3*.6,.25);
\draw [black,thick] (4*.6,.6) -- (4*.6,.45);
\draw [black,thick] (8*.6,.6) -- (8*.6,.45);
\draw [black,thick] (0,-.6) -- (8*.6,-.6);
\draw [black,thick] (0,-.6) -- (0,-.45);
\draw [black,thick] (.6,-.6) -- (.6,-.25);
\draw [black,thick] (3*.6,-.6) -- (3*.6,-.25);
\draw [black,thick] (4*.6,-.6) -- (4*.6,-.45);
\draw [black,thick] (8*.6,-.6) -- (8*.6,-.45);
}}=0,\quad\forall a=1,2,\ldots,r,
\end{equation}
where $R$'s are some independent random covectors. 

To see why this is true, we square the LHS, and after applying a Schouten identity the expression becomes
\begin{equation}\label{expr01}
\begin{split}
&\parbox{3.6cm}{
\tikz{
\node [anchor=center] at (.6,0) {$Q$};
\node [anchor=center] at (2*.6,0) {$\ldots$};
\node [anchor=center] at (3*.6,0) {$Q$};
\node [anchor=center] at (4*.6,.2) {$N_1$};
\node [anchor=center] at (5*.6,.2) {$\ldots$};
\node [anchor=center] at (6*.6,.2) {$N_r$};
\node [anchor=center] at (4*.6,-.2) {$N_1$};
\node [anchor=center] at (5*.6,-.2) {$\ldots$};
\node [anchor=center] at (6*.6,-.2) {$N_r$};
\draw [black,thick] (.6,.6) -- (6*.6,.6);
\draw [black,thick] (.6,.6) -- (.6,.25);
\draw [black,thick] (3*.6,.6) -- (3*.6,.25);
\draw [black,thick] (4*.6,.6) -- (4*.6,.45);
\draw [black,thick] (6*.6,.6) -- (6*.6,.45);
\draw [black,thick] (.6,-.6) -- (6*.6,-.6);
\draw [black,thick] (.6,-.6) -- (.6,-.25);
\draw [black,thick] (3*.6,-.6) -- (3*.6,-.25);
\draw [black,thick] (4*.6,-.6) -- (4*.6,-.45);
\draw [black,thick] (6*.6,-.6) -- (6*.6,-.45);
}}\times
\parbox{5.5cm}{
\tikz{
\node [anchor=center] at (0,.2) {$L$};
\node [anchor=center] at (0,-.2) {$L$};
\node [anchor=center] at (.6,0) {$Q$};
\node [anchor=center] at (2*.6,0) {$\ldots$};
\node [anchor=center] at (3*.6,0) {$Q$};
\node [anchor=center] at (4*.6,.2) {$N_1$};
\node [anchor=center] at (5*.6,.2) {$\ldots$};
\node [anchor=center] at (6*.6,.2) {$\widehat{a}$};
\node [anchor=center] at (7*.6,.2) {$\ldots$};
\node [anchor=center] at (8*.6,.2) {$N_r$};
\node [anchor=center] at (4*.6,-.2) {$N_1$};
\node [anchor=center] at (5*.6,-.2) {$\ldots$};
\node [anchor=center] at (6*.6,-.2) {$\widehat{a}$};
\node [anchor=center] at (7*.6,-.2) {$\ldots$};
\node [anchor=center] at (8*.6,-.2) {$N_r$};
\draw [black,thick] (0,.6) -- (8*.6,.6);
\draw [black,thick] (0,.6) -- (0,.45);
\draw [black,thick] (.6,.6) -- (.6,.25);
\draw [black,thick] (3*.6,.6) -- (3*.6,.25);
\draw [black,thick] (4*.6,.6) -- (4*.6,.45);
\draw [black,thick] (8*.6,.6) -- (8*.6,.45);
\draw [black,thick] (0,-.6) -- (8*.6,-.6);
\draw [black,thick] (0,-.6) -- (0,-.45);
\draw [black,thick] (.6,-.6) -- (.6,-.25);
\draw [black,thick] (3*.6,-.6) -- (3*.6,-.25);
\draw [black,thick] (4*.6,-.6) -- (4*.6,-.45);
\draw [black,thick] (8*.6,-.6) -- (8*.6,-.45);
}}+\\&+
\parbox{10.3cm}{
\tikz{
\node [anchor=center] at (.6,0) {$Q$};
\node [anchor=center] at (2*.6,0) {$\ldots$};
\node [anchor=center] at (3*.6,0) {$Q$};
\node [anchor=center] at (4*.6,.2) {$N_1$};
\node [anchor=center] at (5*.6,.2) {$\ldots$};
\node [anchor=center] at (6*.6,.2) {$\widehat{a}$};
\node [anchor=center] at (7*.6,.2) {$\ldots$};
\node [anchor=center] at (8*.6,.2) {$N_r$};
\node [anchor=center] at (4*.6,-.2) {$N_1$};
\node [anchor=center] at (5*.6,-.2) {$\ldots$};
\node [anchor=center] at (6*.6,-.2) {$\widehat{a}$};
\node [anchor=center] at (7*.6,-.2) {$\ldots$};
\node [anchor=center] at (8*.6,-.2) {$N_r$};
\node [anchor=center] at (9*.6,.2) {$L$};
\node [anchor=center] at (10*.6,0) {$Q$};
\node [anchor=center] at (11*.6,-.2) {$L$};
\node [anchor=center] at (12*.6,0) {$Q$};
\node [anchor=center] at (13*.6,0) {$\ldots$};
\node [anchor=center] at (14*.6,0) {$Q$};
\node [anchor=center] at (15*.6,.2) {$N_1$};
\node [anchor=center] at (16*.6,.2) {$\ldots$};
\node [anchor=center] at (17*.6,.2) {$N_r$};
\node [anchor=center] at (15*.6,-.2) {$N_1$};
\node [anchor=center] at (16*.6,-.2) {$\ldots$};
\node [anchor=center] at (17*.6,-.2) {$N_r$};
\draw [black,thick] (.6,.6) -- (9*.6,.6);
\draw [red,thick] (.6,.6) -- (.6,.25);
\draw [red,thick] (3*.6,.6) -- (3*.6,.25);
\draw [red,thick] (4*.6,.6) -- (4*.6,.45);
\draw [red,thick] (8*.6,.6) -- (8*.6,.45);
\draw [red,thick] (9*.6,.6) -- (9*.6,.45);
\draw [black,thick] (.6,-.6) -- (10*.6,-.6);
\draw [black,thick] (.6,-.6) -- (.6,-.25);
\draw [black,thick] (3*.6,-.6) -- (3*.6,-.25);
\draw [black,thick] (4*.6,-.6) -- (4*.6,-.45);
\draw [black,thick] (8*.6,-.6) -- (8*.6,-.45);
\draw [black,thick] (10*.6,-.6) -- (10*.6,-.25);
\draw [black,thick] (10*.6,.6) -- (17*.6,.6);
\draw [red,thick] (10*.6,.6) -- (10*.6,.25);
\draw [black,thick] (12*.6,.6) -- (12*.6,.25);
\draw [black,thick] (14*.6,.6) -- (14*.6,.25);
\draw [black,thick] (15*.6,.6) -- (15*.6,.45);
\draw [black,thick] (17*.6,.6) -- (17*.6,.45);
\draw [black,thick] (11*.6,-.6) -- (17*.6,-.6);
\draw [black,thick] (11*.6,-.6) -- (11*.6,-.45);
\draw [black,thick] (12*.6,-.6) -- (12*.6,-.25);
\draw [black,thick] (14*.6,-.6) -- (14*.6,-.25);
\draw [black,thick] (15*.6,-.6) -- (15*.6,-.45);
\draw [black,thick] (17*.6,-.6) -- (17*.6,-.45);
}}.
\end{split}
\end{equation}
Let us first look at the second term. Note that there are in total $d+2$ $Q$'s in the front. By applying a Schouten identity on the legs colored red, each of the resulting terms factorize into two parts, of which the first is always a contraction involving $d+3$ $Q$'s, thus vanishing due to the rank of $Q$. 

In the first term above, its second factor involves the substructure
\begin{equation}
\parbox{4cm}{
\tikz{
\node [anchor=center] at (0,.2) {$L$};
\node [anchor=center] at (0,-.2) {$L$};
\node [anchor=center] at (.6,0) {$Q$};
\node [anchor=center] at (2*.6,0) {$\ldots$};
\node [anchor=center] at (3*.6,0) {$Q$};
\node [anchor=center] at (5*.6,0) {$\ldots$};
\draw [black,thick] (0,.6) -- (6*.6,.6);
\draw [black,thick] (0,.6) -- (0,.45);
\draw [black,thick] (.6,.6) -- (.6,.25);
\draw [black,thick] (3*.6,.6) -- (3*.6,.25);
\draw [black,thick] (4*.6,.6) -- (4*.6,.45);
\draw [black,thick] (6*.6,.6) -- (6*.6,.45);
\draw [black,thick] (0,-.6) -- (6*.6,-.6);
\draw [black,thick] (0,-.6) -- (0,-.45);
\draw [black,thick] (.6,-.6) -- (.6,-.25);
\draw [black,thick] (3*.6,-.6) -- (3*.6,-.25);
\draw [black,thick] (4*.6,-.6) -- (4*.6,-.45);
\draw [black,thick] (6*.6,-.6) -- (6*.6,-.45);
}},
\end{equation}
which has $2(r-1)$ upper indices and contains $d+2$ $Q$'s. This tensor is zero in every component. We can actually prove a stronger statement: a rank-$2r$ tensor constructed in the same way as above but with $d+1$ $Q$ is zero. The reason is that
\begin{equation}
L_{I_1}L_{J_1}Q_{I_2J_2}\cdots Q_{I_{d+2}J_{d+2}}\epsilon^{I_1\ldots I_n}\epsilon^{J_1\ldots J_n}\propto \det\left(\mathcal{Q}_{\widehat{I_{d+3}}\widehat{I_{d+4}}\ldots\widehat{I_n}}^{\widehat{I_{d+3}}\widehat{I_{d+4}}\ldots\widehat{I_n}}\right),\quad\forall I_{d+3},\ldots,I_n,J_{d+3},\ldots,J_n,
\end{equation}
where the factor of proportionality is just some none-zero constant. In the above the hats indicate the rows and columns deleted from $\mathcal{Q}$. Recall that by our construction \eqref{detrelation} the indices of $\mathcal{Q}$ runs from $0$ to $n$, and so the submatrix $\mathcal{Q}_{\widehat{I_{d+3}}\widehat{I_{d+4}}\ldots\widehat{I_n}}^{\widehat{I_{d+3}}\widehat{I_{d+4}}\ldots\widehat{I_n}}$ always keeps the row and column with entries $L$. Since this submatrix is of size $(d+3)\times(d+3)$ but at most rank $d+2$, our tensor vanishes.

As a result \eqref{expr01} vanishes, and hence the entire null space of $Q$ is contained in the hyperplane defined by $L$, verifying the condition under which we perform the decomposition \eqref{eq:degenerateQtensorderivative}.

In fact, the proof above has a further remarkable consequence. First note it directly indicates that 
\begin{equation}
L\mathring{Q}^{-1}L=0.
\end{equation}
From the explicit decomposition procedure, it is not hard to observe that apart from the first term $e_d$ each of the rest terms in \eqref{degenerateQtensordecomposition} is dressed with a coefficient that is proportional to the numerator tensor $T_{I_1\ldots I_{n-d}}\equiv L_{I_1}L_{I_2}\cdots L_{I_{n-d}}$ contracted with one or more $\mathring{Q}^{-1}$ (with both indices of $\mathring{Q}^{-1}$ contracted). Since this is always proportional to $L\mathring{Q}^{-1}L$, we conclude that the scalar integrals in the case $n>d+2$
\begin{equation}
I^\phi_{n>d+2,d}\sim e_d,
\end{equation}
and so have uniform transcendental weight $\lfloor\frac{d}{2}\rfloor$.

In actual computations, what the above conclusion specifically mean is that in every step of the decomposition procedure \eqref{eq:degenerateQtensorderivative} we directly through away the trace term, and so, e.g., in the first step
\begin{equation}
I^\phi_{n>d+2,d}=\frac{(n-1)(d+2)}{2n-d-2}\sum_{H_i\in\partial\Delta}(H_i\mathring{Q}^{-1}L)\int_{\Delta_{(i)}}\frac{\langle X_{(i)}\mathrm{d}X_{(i)}^{n-2}\rangle\,(L_{(i)}X_{(i)})^{n-d-1}}{(X_{(i)}Q_{(i)}X_{(i)})^{n-\frac{d}{2}-1}}.
\end{equation}

\subsection{Integrals with tensor numerator}

When the Feynman diagrams themselves involves tensor numerators depending on the loop momentum, as usual they can be traded off by replacing each loop momentum with a derivative. Without loss of generality let us assume a numerator of degree $k$ in the loop momentum; specifically in our setup for the Feynman parametrization it has the form
\begin{equation}
\int_0^\infty\langle X\mathrm{d}^{n-1}X\rangle\int\mathrm{d}^dY\frac{N_k[Y]\,(Y\cdot Y_\infty)^{n-k-d}}{(Y\cdot W)^n}=
\int_0^\infty N_k\left[\partial_W\right](Y_\infty^I\partial_{W^I})^{n-k-d}\frac{\langle X\mathrm{d}^{n-1}X\rangle}{(W\cdot W)^{\frac{d}{2}}}.
\end{equation}
We do not consider the case with $n-k-d<0$ as it has a bad power counting of loop momentum at infinity. It is easy to see that the result is always a linear combination of the following integrals
\begin{equation}
I_{n,g}^{(m,t)}[U]=\int_0^\infty\langle X\mathrm{d}^{n-1}X\rangle\frac{U[W^{d-n+2m-t}]\,(Y_\infty\cdot W)^t}{(W\cdot W)^{\frac{d}{2}+m}},\quad 0\leq t\leq d-n+2m.
\end{equation}
This defines a quadric by $W\cdot W\equiv XQX$, i.e., $Q_{IJ}=\mathcal{Y}_I\cdot\mathcal{Y}_J$ as before. We can treat the entire numerator as a symmetric rank-$(d-n+2m)$ tensor, but here it is more convenient to study it directly. In particular we would like to check whether the numerator as a symmetric tensor is again traceless wrst $Q^{-1}$ (if it is non-degenerate) or $\mathring{Q}^{-1}$ (if it is degenerate). The explicit expression above suggests that this condition is equivalent to the following conditions
\begin{equation}
Y^MQ^{-1}Y^N=Y^MQ^{-1}L=LQ^{-1}L=0,\quad\forall M,N.
\end{equation}
Note that in the above again $L=[1:1:\ldots:1]$, and $Y^M$ is the covector made from the $M^{\rm th}$ Lorentz component of each $\mathcal{Y}_I$ ($I=1,\ldots,n$). The last condition $LQ^{-1}L=0$ is exactly the same as the case of scalar integrals, which we have verified before. Following the same logic there using the rank of an extended matrix $\mathcal{Q}$, we have
\begin{equation}
Y^MQ^{-1}Y^N\propto\det\left(\begin{matrix}0&Y^N\\(Y^M)^{\rm T}&Q\end{matrix}\right),\qquad
Y^MQ^{-1}L\propto\det\left(\begin{matrix}0&L\\(Y^M)^{\rm T}&Q\end{matrix}\right).
\end{equation}
Hence we only need to check whether each $Y^M$ can be written as
\begin{equation}
(Y^M)_I=Y_{(M)}\cdot \mathcal{Y}_I,\quad\forall I=1,2,\ldots,n
\end{equation}
for some embedding space vector $Y_{(M)}$. This is obviously true since (with the light cone metric) we can assign
\begin{equation}
Y_{(1)}^N=-2\delta_{N,2},\qquad Y_{(2)}^N=-2\delta_{N,1},\qquad Y_{(M)}^N=\delta_{M,N}\quad N>2.
\end{equation}
In particular $Y_{(1)}\cdot \mathcal{Y}_I=L$. Similar arguments also hold for degenerate $Q$'s. Hence we see the numerator tensor is again always traceless wrst the inversed quadric.

\subsection{More examples}\label{sec:moreintegralsdif}

\flag{No further revisions.}

Now let us study several explicit examples of diagrams with non-trivial numerators, to further illustrate the discussion so far. The first three examples, together with the notations used, are from \cite{ArkaniHamed:2010gh}.

\subsubsection{E.g.~4: a finite hexagon in 4d}

In the fourth example we study the finite hexagon in 4d (expressed in terms of momentum twistors)
\begin{equation}
I_{\text{e.g.}4}=\underset{AB}{\int}\frac{\langle AB13\rangle\langle AB46\rangle\langle 5612\rangle\langle 2345\rangle}{\langle AB12\rangle\langle AB23\rangle\langle AB34\rangle\langle AB45\rangle\langle AB56\rangle\langle AB61\rangle}.
\end{equation}
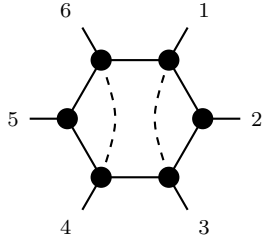
\begin{center}
\begin{tikzpicture}
\coordinate [label=60:{\scriptsize $1$}] (e1) at (60:1.4cm);
\coordinate [label=0:{\scriptsize $2$}] (e2) at (0:1.4cm);
\coordinate [label=-60:{\scriptsize $3$}] (e3) at (-60:1.4cm);
\coordinate [label=-120:{\scriptsize $4$}] (e4) at (-120:1.4cm);
\coordinate [label=180:{\scriptsize $5$}] (e5) at (180:1.4cm);
\coordinate [label=120:{\scriptsize $6$}] (e6) at (120:1.4cm);
\coordinate (v1) at (60:.9cm);
\coordinate (v2) at (0:.9cm);
\coordinate (v3) at (-60:.9cm);
\coordinate (v4) at (-120:.9cm);
\coordinate (v5) at (180:.9cm);
\coordinate (v6) at (120:.9cm);
\draw [black,thick] (v1) -- (v2) -- (v3) -- (v4) -- (v5) -- (v6) -- cycle;
\draw [black,thick] (v1) -- (e1);
\draw [black,thick] (v2) -- (e2);
\draw [black,thick] (v3) -- (e3);
\draw [black,thick] (v4) -- (e4);
\draw [black,thick] (v5) -- (e5);
\draw [black,thick] (v6) -- (e6);
\draw [black,thick,dashed] (v1) .. controls (.2cm,0) .. (v3);
\draw [black,thick,dashed] (v4) .. controls (-.2cm,0) .. (v6);
\fill [black] (v1) circle(4pt);
\fill [black] (v2) circle(4pt);
\fill [black] (v3) circle(4pt);
\fill [black] (v4) circle(4pt);
\fill [black] (v5) circle(4pt);
\fill [black] (v6) circle(4pt);
\end{tikzpicture}
\end{center}
In the embedding formalism and turn into Feynman parametrization we have
\begin{equation}
\begin{split}
\frac{I_{\text{e.g.}4}}{\langle 5612\rangle\langle 2345\rangle}&=\int_0^\infty\measure{X}{5}(Y_{[13]}\cdot\partial_W)(Y_{[46]}\cdot\partial_W)\frac{1}{(W\cdot W)^2}\\
&=4(Y_{[13]}\cdot Y_{[46]})\underbrace{\int_0^\infty\frac{\measure{X}{5}}{(W\cdot W)^3}}_{E_6}-12\underbrace{\int_0^\infty\measure{X}{5}\frac{2(Y_{[13]}\cdot W)(Y_{[46]}\cdot W)}{(W\cdot W)^4}}_{E_{6,2}}.
\end{split}
\end{equation}
This is a summation of an $E_6$ and an $E_{6,2}$. In both integrals we have the same quadric which in terms of components is $Q_{IJ}=Y_I\cdot Y_J=\frac{\langle I-1\,I\,J-1\,J\rangle}{\langle I-1\,I\rangle\langle J-1\,J\rangle}$, and the tensor in $E_{6,2}$ is $T_{IJ}=(Y_{[13]}\cdot Y_I)(Y_{[46]}\cdot Y_J)+(Y_{[13]}\cdot Y_J)(Y_{[46]}\cdot Y_I)$.

We first check the $Q^{-1}$ trace of $T$. In the case of a hexagon this is particularly simple. regarding $Y_I^{\;M}$ as a non-degenerate matrix $\mathbf{Y}$ and the flat metric $\mathbf{\eta}$, we have $\mathbf{Q}=\mathbf{Y}\mathbf{\eta}\mathbf{Y}^{\rm T}$, and so
\begin{equation}
\text{tr}_{Q^{-1}}T=2\,\vec{Y}_{[13]}\mathbf{\eta}\mathbf{Y}^{\rm T}\mathbf{Q}^{-1}\mathbf{Y}\mathbf{\eta}\vec{Y}_{[46]}=2\,Y_{[13]}\cdot Y_{[46]}.
\end{equation}
Recall that in the decomposition of an $E_{6,2}$ integral there is an extra constant factor $\frac{1}{6}$ for the trace part (which reduces to an $E_6$), we see that the weight-3 part of the above $E_{6,2}$ exactly cancels the $E_6$ in the first term, hence $I_{\text{e.g.}4}$ has a uniform weight 2.

To work out the explicitly result, let us write $E_{6,2}$ back into the momentum twistor space
\begin{equation}
-\frac{1}{12}I_{\text{e.g.}4}\Big|_{E_{6,2}}=\langle5612\rangle\langle2345\rangle\int_0^\infty\measure{X}{5}\frac{2(\langle1345\rangle x_5+\langle1356\rangle x_6)(\langle1246\rangle x_2+\langle 2346\rangle x_3)}{(XQX)^4},
\end{equation}
with $Q_{IJ}=\langle I-1\,I\,J-1\,J\rangle$. The integral $I_{\text{e.g.}4}$ is just the weight-2 part of this integral, which can be decomposed into localized contribution from each of the codim-2 boundaries, i.e., a linear combination of $E_4$ integrals
\begin{equation}
I_{\text{e.g.}4}=\sum_{i<j}C_{(ij)}\int_0^\infty\frac{\measure{X_{(ij)}}{3}\,\sqrt{\det Q_{(ij)}}}{(X_{(ij)}Q_{(ij)}X_{(ij)})^2}.
\end{equation}

We can simplify the problem by rescaling the variables to bring the integral into the form
\begin{equation}
-\frac{1}{12}I_{\text{e.g.}4}\Big|_{E_{6,2}}=\frac{1}{16}\int_0^\infty\frac{\measure{X}{5}\left(u_2(u_1-1)x_2x_5+u_2^2(u_4-1)x_3x_5+\frac{u_1u_3-u_2u_4}{u_4}x_2x_6+u_2(u_3-1)x_3x_6\right)}{(x_1x_3+x_1x_4+u_1x_2x_4+x_1x_5+x_2x_5+u_2x_3x_5+x_2x_6+x_3x_6+u_3x_4x_6)^4}.
\end{equation}
with the crossratios
\begin{equation}
u_1=\frac{\langle1234\rangle\langle4561\rangle}{\langle1245\rangle\langle3461\rangle},\quad
u_2=\frac{\langle2345\rangle\langle5612\rangle}{\langle2356\rangle\langle4512\rangle},\quad
u_3=\frac{\langle3456\rangle\langle6123\rangle}{\langle3461\rangle\langle5623\rangle},\quad
u_4=\frac{\langle1234\rangle\langle3456\rangle}{\langle3461\rangle\langle2345\rangle}.
\end{equation}
Depending on specific data $u_4=u_1u_3x_\pm$. Here the crossratios $\{u_1,u_2,u_3\}$ and $x_\pm$ are the same as those defined in the hexagon in 6d. We again assume to work in the region where $0<u_i<1$ ($i=1,2,3$). 

To work out the coefficients, following the standard procedure we described before, they are given by the following matrix operations
\begin{equation}
C_{(ij)}=(\det \mathbf{Q}_{\widehat{\{i,j\}},\widehat{\{i,j\}}})^{-\frac{1}{2}}\left[((\mathbf{Q}^{-1}\mathbf{T})_{i,\hat{i}}(\mathbf{Q}_{\widehat{\{i\}},\widehat{\{i\}}})^{-1})_j+(i\leftrightarrow j)\right].
\end{equation}
In this example it turns out that
\begin{equation}
\begin{split}
&C_{(12)}=C_{(16)}=C_{(26)}=C_{(34)}=C_{(35)}=C_{(45)}=0,\\
&C_{(13)}=-C_{(14)}=C_{(15)}=-C_{(23)}=C_{(24)}=-C_{(25)}=-C_{(36)}=C_{(46)}=-C_{(56)}=8,\\
\end{split}
\end{equation}
and so $I_{\text{e.g.}4}$ is also a pure function.

Since we are starting from an integral in $\mathbb{CP}^5$, this means that in studying the above induced $E_{4}$ integrals on the codim-2 boundaries we can directly borrow the symbol for the unique $E_6$ integral defined by the same quadric and truncate it. In the present example this $E_6$ integral is exactly identical to the case of scalar hexagon in 6d that we investigated before, the detailed result of whose symbol is collected in Appendix \ref{app:hexagon}.

As an example, consider the integral $I^{\text{e.g.}4}_{(12)}$ induced on the boundary $\Delta_{(12)}$. Its symbol is obtained by collecting terms in $\mathcal{S}I_{\text{e.g.}2}$ ended with $\underline{12}$ and chop off the last entry, i.e.,
\begin{equation}
\begin{split}
\mathcal{S}I^{\text{e.g.}4}_{(12)}&=\underline{35}\otimes\underline{46}\otimes\widehat{\underline{12}}+\underline{46}\otimes\underline{35}\otimes\widehat{\underline{12}}\\
&=8\,u_2\otimes u_2+8\,u_3\otimes u_3+4\,u_2\otimes u_3+4\,u_3\otimes u_2.
\end{split}
\end{equation}
In fact, since the symbol terms associated with induced integrals on different codim-2 boundaries have no overlaps at all, all what we need is to sum up all the symbol terms of $I_{\text{e.g.2}}$, dropping the last entry in each term and dressing it with the coefficient indicated by the two labels therein
\begin{equation}\label{eg4decomposition}
\mathcal{S}I_{\text{e.g.}4}=\sum_\rho C_{(\rho_5\rho_6)}\,\underline{\rho_1\rho_2}\otimes\underline{\rho_3\rho_4}\otimes\widehat{\underline{\rho_5\rho_6}}.
\end{equation}
This reduces the expression to 11 symbol terms
\begin{equation}
\begin{split}
&\underline{24}\otimes(-\underline{13}\otimes\widehat{\underline{56}}-
\underline{15}\otimes\widehat{\underline{36}}+
\underline{36}\otimes\widehat{\underline{15}}+
\underline{56}\otimes\widehat{\underline{13}})\\
+&\underline{35}\otimes(\underline{12}\otimes\widehat{\underline{46}}+
\underline{16}\otimes\widehat{\underline{24}}-
\underline{26}\otimes\widehat{\underline{14}})\\
+&\underline{46}\otimes(-\underline{13}\otimes\widehat{\underline{25}}-
\underline{15}\otimes\widehat{\underline{23}}+
\underline{23}\otimes\widehat{\underline{15}}+
\underline{25}\otimes\widehat{\underline{13}}).
\end{split}
\end{equation}
Substituting the result for each symbol term collected in Appendix \ref{app:hexagon}, the expression reduces to
\begin{equation}
-4\left(u_1\otimes u_3+u_3\otimes u_1-\sum_{i=1}^3u_i\otimes(1-u_i)\right),
\end{equation}
which matches the previous result in \cite{ArkaniHamed:2010gh}.

\subsubsection{E.g.~5: another finite hexagon in 4d}

The fifth example is another finite hexagon in 4d, defined as
\begin{equation}
I_{\text{e.g.}5}=\underset{AB}{\int}\frac{\langle AB(612)\cap(234)\rangle\langle AB46\rangle}{\langle AB12\rangle\langle AB23\rangle\langle AB34\rangle\langle AB45\rangle\langle AB56\rangle\langle AB61\rangle}.
\end{equation}
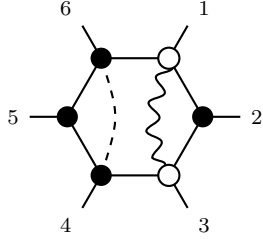
\begin{center}
\begin{tikzpicture}[decoration=snake]
\coordinate [label=60:{\scriptsize $1$}] (e1) at (60:1.4cm);
\coordinate [label=0:{\scriptsize $2$}] (e2) at (0:1.4cm);
\coordinate [label=-60:{\scriptsize $3$}] (e3) at (-60:1.4cm);
\coordinate [label=-120:{\scriptsize $4$}] (e4) at (-120:1.4cm);
\coordinate [label=180:{\scriptsize $5$}] (e5) at (180:1.4cm);
\coordinate [label=120:{\scriptsize $6$}] (e6) at (120:1.4cm);
\coordinate (v1) at (60:.9cm);
\coordinate (v2) at (0:.9cm);
\coordinate (v3) at (-60:.9cm);
\coordinate (v4) at (-120:.9cm);
\coordinate (v5) at (180:.9cm);
\coordinate (v6) at (120:.9cm);
\draw [black,thick] (v1) -- (v2) -- (v3) -- (v4) -- (v5) -- (v6) -- cycle;
\draw [black,thick] (v1) -- (e1);
\draw [black,thick] (v2) -- (e2);
\draw [black,thick] (v3) -- (e3);
\draw [black,thick] (v4) -- (e4);
\draw [black,thick] (v5) -- (e5);
\draw [black,thick] (v6) -- (e6);
\draw [black,thick,decorate] (v1) .. controls (.2cm,0) .. (v3);
\draw [black,thick,dashed] (v4) .. controls (-.2cm,0) .. (v6);
\path [draw=black,fill=white,thick] (v1) circle(4pt);
\fill [black] (v2) circle(4pt);
\path [draw=black,fill=white,thick] (v3) circle(4pt);
\fill [black] (v4) circle(4pt);
\fill [black] (v5) circle(4pt);
\fill [black] (v6) circle(4pt);
\end{tikzpicture}
\end{center}
Its Feynman parametrization is the same as the one for the previous example, but with out the prefactor $\langle5612\rangle\langle2345\rangle$ and with the vector $Y_{[13]}$ substituted by $Y_{(612)\cap(234)}$. Following the same logic this immediately indicates that the weight-3 parts cancel away, and we conclude again that the function has uniform weight 2.

The numerator in the $E_{6,2}$ part of this integral is proportional to
\begin{equation}
(\langle6124\rangle\langle2345\rangle x_5+\langle6125\rangle\langle2346\rangle x_6)(\langle1246\rangle x_2+\langle2346\rangle x_3)
\end{equation}
In analogy to the previous case, we can first rescale the Feynman parameters to bring the quadric into the same form, and then compute the coefficients $C_{(ij)}$'s in the decomposition \eqref{eg4decomposition}, where it turns out that $C_{(14)}=C_{(25)}=C_{(36)}=0$. We can re-organize the remaining terms into
\begin{equation}
\begin{split}
&2(4C_{(56)}-C_{(35)}+4C_{(16)}+3C_{15}-C_{(13)})u_1\otimes u_1\\
+&2(-C_{(46)}+3C_{(26)}-C_{(24)}+4C_{(16)}+4C_{(12)})u_2\otimes u_2\\
+&2(-C_{(35)}+4C_{(23)}-C_{(15)}+3C_{(13)}+4C_{(12)})u_3\otimes u_3\\
+&4C_{(16)}(u_1\otimes u_2+u_2\otimes u_1)\\
+&4C_{(12)}(u_2\otimes u_3+u_3\otimes u_2)\\
+&2(C_{(35)}-C_{(15)}-C_{(13)})(u_3\otimes u_1+u_1\otimes u_3).
\end{split}
\end{equation}
Explicit computation shows that the coefficients in front of the $u_i\otimes u_i$ terms all vanish. Furthermore we have
\begin{align}
4C_{(16)}&=16\frac{\langle1234\rangle}{\langle1345\rangle\langle1235\rangle},\\
4C_{(12)}&=16\frac{\langle1236\rangle}{\langle1235\rangle\langle1356\rangle},\\
2(C_{(35)}-C_{(15)}-C_{(13)})&=16\frac{\langle3461\rangle}{\langle1345\rangle\langle1356\rangle}.
\end{align}
This also perfectly matches the existing result in \cite{ArkaniHamed:2010gh}.

\subsubsection{E.g.~6: parity-odd diagrams}

Parity-odd diagrams has to integrate to zero. It is interesting to see how this occur explicitly in Feynman parametrization.

The simplest case is parity-odd pentagon in 4d, which is
\begin{equation}
I_{\text{e.g.}6a}=\underset{AB}{\int}\frac{\langle AB13\rangle\langle2345\rangle\langle4512\rangle-\langle AB(512)\cap(234)\rangle\langle 3451\rangle}{\langle AB12\rangle\langle AB23\rangle\langle AB34\rangle\langle AB45\rangle\langle AB51\rangle}.
\end{equation}
Its Feynman parametrization, working again in the ambient space, is
\begin{equation}
\begin{split}
&\int_0^\infty\measure{X}{4}\frac{(Y_3\cdot Y_5)(Y_2\cdot Y_5)(Y_{[13]}\cdot W)-(Y_1\cdot Y_4)(Y_{(512)\cap(234)}\cdot W)}{(W\cdot W)^3}.
\end{split}
\end{equation}
Note that the numerator is proportional to
\begin{equation}
\langle 2345\rangle\langle4512\rangle(\langle 1345\rangle x_4)-\langle 3451\rangle (\langle4512\rangle\langle5234\rangle x_4)=0,
\end{equation}
hence the parity-odd pentagon has a trivial Feynman parametrization.

We can also check the parity-odd combination of hexagons in 4d. For example
\begin{equation}
\begin{split}
I_{\text{e.g.}6b}&=\underset{AB}{\int}\frac{\langle AB(612)\cap(234)\rangle\langle AB(345)\cap(561)\rangle-\langle AB13\rangle\langle AB46\rangle\langle5612\rangle\langle2345\rangle}{\langle AB12\rangle\langle AB23\rangle\langle AB34\rangle\langle AB45\rangle\langle AB56\rangle\langle AB61\rangle}\\
&=\left(\langle1234\rangle\langle3456\rangle\langle5612\rangle-\langle2345\rangle\langle4561\rangle\langle6123\rangle\right)\int_0^\infty\measure{X}{5}\frac{\langle1256\rangle x_2x_6-\langle 2345\rangle x_3x_5}{(\sum_{i<j}\langle i-1\,i\,j-1\,j\rangle x_ix_j)^4},
\end{split}
\end{equation}
It turns out that their Feynman parametrizations are non-trivial. What one can observe there is that the weight-3 part of the expression is again trivially zero due to the tracelessness of the numerator, so that the integrand is again a total derivative. When localized to the codim-1 boundaries the integrals still have a non-trivial integrand, but the further localized contributions on each codim-2 boundary cancel away (between each pair of codim-1 boundaries adjacent to it). Hence this is actually a non-trivial expression that evaluates to zero.

\subsubsection{E.g.~7: a box in 4d with mixed weights}

We use a seventh example to illustrate tensor integrals that give rise to functions with mixed weights. These could arise even from scalar Feynman diagram, as long as the degrees of some propagators are higher than one. For example, consider again the massive box diagram in 4d but we assign the integral
\begin{equation}
\begin{split}
I_{\text{e.g.}7}&=\int\mathrm{d}^4y\frac{m^8}{((y-y_1)^2+m^2)^2((y-y_2)^2+m^2)((y-y_3)^2+m^2)^2((y-y_4)^2+m^2)}\\
&=\int\frac{\mathrm{d}^6Y\,\delta(Y\cdot Y)}{\text{vol.}\mathrm{GL}(1)}\frac{m^8\,(Y\cdot Y_\infty)^2}{(Y\cdot Y_1)^2(Y\cdot Y_2)(Y\cdot Y_3)^2(Y\cdot Y_4)}.
\end{split}
\end{equation}
Since the first propagator has degree 2, its Feynman parametrization acquires an extra factor $x_1$. As a result
\begin{equation}
\verified{}
I_{\text{e.g.}7}=\int_0^\infty\frac{\measure{X}{3}\,x_1x_3\,(x_1+x_2+x_3+x_4)^2}{(\frac{4}{u}x_1x_3+\frac{4}{v}x_2x_4+(x_1+x_2+x_3+x_4)^2)^4},
\end{equation}
where we have performed the same rescaling $x_i\mapsto x_i/m$ so as to bring the quadric $Q$ into the same form as in our first example.

Let us denote the rank-4 tensor in the numerator as $T[X^4]$. It can be easily verified that this integral contains a weight-2 piece whose symbol is exactly the same as $\mathcal{S}I_{\text{e.g.}1}$, dressed with a coefficient
\begin{equation}
\verified{}
\frac{4-1}{4+4-2}\frac{2-1}{4+2-2}\frac{(\text{tr}_Q)^2T}{\sqrt{\det Q}}=\frac{u^2v(4u+u^2+2v+3uv+2v^2)}{384(1+u+v)^{\frac{5}{2}}}.
\end{equation}

Apart from the above highest-weight piece, the fact that we starts with a rank-4 tensor indicates there is in general also a weight-1 piece and a rational piece. The weight-1 piece comes from three contributions, as follows
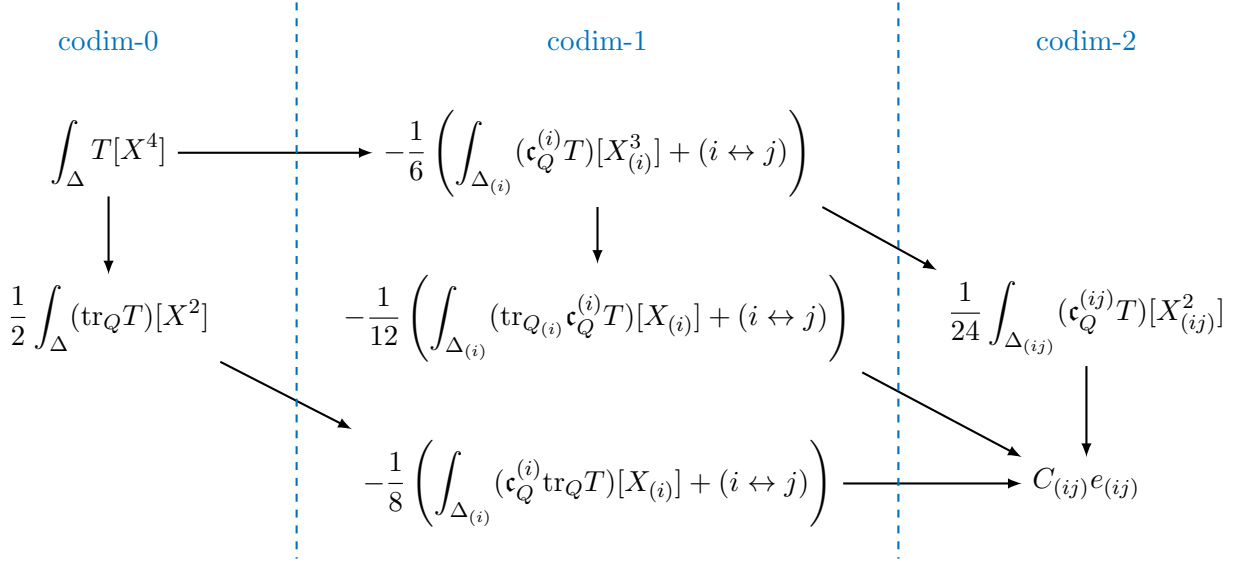
\begin{center}
\begin{tikzpicture}
\node [anchor=center] (n1) at (0,0) {\small $\displaystyle \int_\Delta T[X^4]$};
\node [anchor=center] (n2) at (0,-2.2) {\small $\displaystyle \frac{1}{2}\int_{\Delta}(\text{tr}_{Q}T)[X^2]$};
\node [anchor=center] (n3) at (6.5,0) {\small $\displaystyle -\frac{1}{6}\left(\int_{\Delta_{(i)}}(\mathfrak{c}_Q^{(i)}T)[X_{(i)}^3]+(i\leftrightarrow j)\right)$};
\node [anchor=center] (n4) at (6.5,-2.2) {\small $\displaystyle -\frac{1}{12}\left(\int_{\Delta_{(i)}}(\text{tr}_{Q_{(i)}}\mathfrak{c}_Q^{(i)}T)[X_{(i)}]+(i\leftrightarrow j)\right)$};
\node [anchor=center] (n5) at (6.5,-4.4) {\small $\displaystyle -\frac{1}{8}\left(\int_{\Delta_{(i)}}(\mathfrak{c}_Q^{(i)}\text{tr}_QT)[X_{(i)}]+(i\leftrightarrow j)\right)$};
\node [anchor=center] (n6) at (13,-2.2) {\small $\displaystyle \frac{1}{24}\int_{\Delta_{(ij)}}(\mathfrak{c}_Q^{(ij)}T)[X_{(ij)}^2]$};
\node [anchor=center] (n7) at (13,-4.4) {\small $\displaystyle C_{(ij)}e_{(ij)}$};
\draw [black,thick,-latex] (n1.south) to (n2.north);
\draw [black,thick,-latex] (n1.east) to (n3.west);
\draw [black,thick,-latex] (n2.south east) to (n5.north west);
\draw [black,thick,-latex] (n3.south) to (n4.north);
\draw [black,thick,-latex] (n3.south east) to (n6.north west);
\draw [black,thick,-latex] (n4.south east) to (n7.north west);
\draw [black,thick,-latex] (n5.east) to (n7.west);
\draw [black,thick,-latex] (n6.south) to (n7.north);
\node [RoyalBlue,anchor=center] at (0,1.5) {\small codim-0};
\node [RoyalBlue,anchor=center] at (6.5,1.5) {\small codim-1};
\node [RoyalBlue,anchor=center] at (13,1.5) {\small codim-2};
\draw [RoyalBlue,thick,dashed] (2.5,2) -- (2.5,-5.4);
\draw [RoyalBlue,thick,dashed] (10.5,2) -- (10.5,-5.4);
\end{tikzpicture}
\end{center}
Altogether the coefficient $C_{(ij)}$ above reads
\begin{equation}
\verified{}
C_{(ij)}=\frac{1}{\sqrt{q_{(ij)}}}\left(\frac{\mathfrak{c}_Q^{(ij)}\text{tr}_QT}{16}+\frac{\mathfrak{ctc}_Q^{(ij)}T}{24}+\frac{\text{tr}_{Q_{(ij)}}\mathfrak{c}_Q^{(ij)}T}{48}\right).
\end{equation}
Since $\mathcal{S}I_{\text{e.g.}1}$ receives non-trivial contributions from only $\Delta_{(13)}$ and $\Delta_{(24)}$, again we only need to consider these two boundaries. Hence the weight-1 piece is
\begin{equation}
\verified{}
\mathcal{S}I_{\text{e.g.}7}\big|_{\text{weight }1}=C_{(13)}\,\underline{24}\otimes\widehat{\underline{13}}+C_{(24)}\,\underline{13}\otimes\widehat{\underline{24}},
\end{equation}
where
\begin{align}
\label{eq:mixedweightC13}\verified{}C_{(13)}&=-\frac{i\,u^2v\sqrt{1+v}(2-u-v)}{384(1+u+v)^2},\\
\label{eq:mixedweightC24}\verified{}C_{(24)}&=-\frac{i\,u^2v(2+u+u^3+2v+2uv+u^2v)}{384(1+u)^{\frac{3}{2}}(1+u+v)^2}.
\end{align}

To work out the rational piece, we just need to further localize each of the integrals $\int_{\Delta_{(ij)}}(\mathfrak{c}_Q^{(ij)}T)[X_{(ij)}^2]$ and further localize according to \eqref{eq:rationalfrom1d}, i.e.,
\begin{equation}
-\frac{1}{48}\left(\sum_{I=2,4}\frac{(Q_{(13)}^{-1})^{2I}(\mathfrak{c}_Q^{(13)}T)_{I4}}{((Q_{(13)})_{44})^2}-(2\leftrightarrow4)\right)-\frac{1}{48}\left(\sum_{I=1,3}\frac{(Q_{(24)}^{-1})^{1I}(\mathfrak{c}_Q^{(24)}T)_{I3}}{((Q_{(24)})_{33})^2}-(1\leftrightarrow3)\right).
\end{equation}
This yields
\begin{equation}
\mathcal{S}I_{\text{e.g.}7}\big|_{\text{weight }0}=\frac{u^2v}{192(1+u)(1+u+v)}.
\end{equation}

\section{Discontinuity I: Spherical Contours and $e_n$}\label{sec:dis1}

\flag{No further revisions.}

We can study the functions produced by Feynman integrals by analyzing their branching points and the corresponding discontinuities. When represented as contour integrals in the complexified space (no matter whether its is the original loop momentum space or the Feynman parameter space or some other space) these properties are always encoded by the relations between the contour and the singularity points of the integrand under the deformation of data defining the integral. An actual singular point (in the data space) of the function is produced whenever it forces the contour to be pinched by some singular points of the integrand.

\subsection{A toy example}

This relation becomes very explicit for iterated integrals. As the simplest example, let us consider the log function
\begin{equation}
\fixed
\log(z)=\int_0^\infty\frac{(z-1)\,\mathrm{d}x_1}{(x_1+z)(x_1+1)}.
\end{equation}
The contour is the positive real axis in the $x_1$ plane, and the integrand has a pole at $x_1=-z$. We can freely deform $z$ so as to move this pole, and let it travel along a $S^1$ contour on the complex plane. At long as its path has a trivial winding number around $z=0$ or $z=\infty$, we can always deform the integration contour to avoid hitting the pole. If the path winds around $z=0$ once, however, the final integral we obtain can only be identical to the original one by a residue at $x_1=z$. This tells us that the function has branch points at $z=0$ and at $z=\infty$, and that the discontinuity across the cut stretching between these two points is
\begin{equation}\label{eq:ordinarydiscontinuity}
\fixed
\text{Disc}\log(z)=-2\pi i\,\underset{x_1=-z}{\text{Res}}\frac{(z-1)\mathrm{d}x_1}{(x_1+z)(x_1+1)}=2\pi i.
\end{equation}
Note that the extra minus sign in front of the residue is because as we move $z$ counter-clockwise around 0 the extra contour for $x_1$ winds clockwise around $z$.

The situation with integrals involving a single quadric denominator is different. Already, we see that for iterated integrals every time we compute a discontinuity (thus reducing the transcendentality by one) we obtain an integration in one lower dimensions, and correspondingly the transcendental weight of the function is the same as the dimensions in which the integral is defined. But this is no longer true for the integrals with a quadric. To seek for the right prescription for the discontinuity contour, here let us use the same $\log(z)$ example but first lift it to one higher dimensions
\begin{equation}\label{eq:logasquadric}
\fixed
\log(z)=\int_0^\infty\mathrm{d}x_1\int_0^\infty\mathrm{d}x_2\frac{z-1}{(x_2(x_1+1)+(x_1+z))^2}=\int_0^\infty\frac{\langle X\mathrm{d}^2X\rangle\,(LX)}{(XQX)^2},
\end{equation}
which now is explicitly identified with an $E_1$ integral, with
\begin{equation}
\fixed
L=[0:0:z-1],\quad Q=\frac{1}{2}\left(\begin{matrix}0&1&1\\1&0&1\\1&1&2z\end{matrix}\right).
\end{equation}
Note that the denominator can be written into the form $(x_1+x_3)(x_2+x_3)+(z-1)x_3^2$, and so we first do a shift $x_1=y_+-x_3$, $x_2=y_--x_3$ to bring it into the form $(y_+y_-+(z-1)x_3^2)$ and then integrate over the two dimensional contour
\begin{equation}
\fixed
y_+=w,\quad y_-=\overline{w},\quad\text{the entire }w\text{ plane.}
\end{equation}
With a point compactification this is just an $S^2$ contour; explicitly we can also assign $y_\pm=r\, e^{\pm i\phi}$, $r\in[0,+\infty]$, $\phi\in[0,2\pi]$. Integration over this contour also reproduces the same discontinuity
\begin{equation}
\fixed
-\int_0^{+\infty}\mathrm{d}r\,(-2ir)\int_0^{2\pi}\mathrm{d}\phi\,\frac{(z-1)x_3^2}{(r^2+(z-1)x_3^2)^2}=2\pi i.
\end{equation}
Here apart from the above definition of the contour we have inserted an extra factor $(-1)$ by hand, in accord with the extra minus sign we put for the ordinary residue computation in \eqref{eq:ordinarydiscontinuity}, and the results match.

We call the two-dimensional contour used in the above computation a \textit{spherical contour}. Though different in natural for an ordinary residue contour (as we observed in computing discontinuities of a multiple polylogarithm as represented by iterated integrals, or Aomoto polylogarithms, see Appendix \ref{app:sec:aomotodisc}), it manages to extract the desired discontinuities of $\log(z)$ starting from a representation using quadric integrals \eqref{eq:logasquadric}.   While the way it works here may look a bit accidental, it turns out to work for discontinuities in generic quadric integrals, as will be presented in the next subsection.  The connection between this contour and the more familiar residue contours will be further explained in Section \ref{sec:S2fibration}.

\subsection{Discontinuities of $e$ integrals and spherical contours}

The computation for discontinuities presented above is in fact naturally expected from our understanding of the symbol structure as learned from the differentiation method. There the symbol of an $e$ integral is organized as a summation over ordered partition of the Feynman parameter label sets into symmetric pairs. In particular, each of the first symbol entry, which encodes the location of the branch points, associates to certain pair of labels. Hence each computation of the discontinuity across certain branch cut should be a two-fold contour integral.

This expectation leads to a prescription for extracting the symbols of the functions purely from the study of discontinuities. Of course this involves two tasks
\begin{enumerate}
\item Derive the explicit expression in each symbol entry;
\item Identify a new integral whose symbol is the same as that obtained by chopping off the first symbol entry.
\end{enumerate}

For computational convenience the procedure is best illustrated in the canonical frame. The general picture will be discussed in detail later.

In this subsection we focus on the $e_n$ integrals with even $n$. The case of odd $n$ has some qualitative difference, and we postpone their analysis to later discussions Section \ref{sec:oddspacetime}.

For these integrals the symbols are again organized into ordered partitions
\begin{equation}\label{eq:symbolfromdiscontinuity}
\fixed
\mathcal{S}e_n=C_n\sum_{\rho}\overline{\rho_1\rho_2}\otimes\overline{\rho_3\rho_4}\otimes\cdots\otimes\overline{\rho_{n-1}\rho_{n}},\qquad\text{even }n.
\end{equation}
However, the labels $\{\rho_i\}$ here are labels for vertices (instead of the codim-1 boundaries as in the differentiation method), and correspondingly we switch the notation into an overline. There is also an overall factor $C_n$ which is going to be determined. 

Given the intuition from the previous subsection, it is straightforward to first deal with the second target above. We begin with the integral $e_n$ to determine the first symbol entries. The objects playing the same role for the second symbol entries are the various discontinuities computed from $e_n$, and similarly for the third entries and so on. To compute the discontinuity of $e_n$ across a branch cut associated to a pair of labels, say $(ij)$, we introduce a corresponding spherical contour. This is done by first performing an affine transformation to the variables $\{x_i,x_j\}\mapsto\{w_i,w_j\}$
\begin{equation}\label{affinetransformation}
\fixed
\left(\begin{matrix}x_i\\x_j\end{matrix}\right)=R\left(\begin{matrix}w_i\\w_j\end{matrix}\right)-Q_{\{i,j\},\{i,j\}}^{-1}Q_{\{i,j\},\widehat{\{i,j\}}}X_{(ij)},
\quad\text{with }R^{\rm T}Q_{\{i,j\},\{i,j\}}R=\left(\begin{matrix}
0&\frac{1}{2}\\\frac{1}{2}&0
\end{matrix}\right).
\end{equation}
The specific form of the transformation matrix $R$ depends on the configuration of $Q$ and there is not a unique choice, but this does not affect the result yielded by the $S^2$ contour integration described below.

Note that the original matrix $Q$ naturally decomposes into four blocks: $Q_{\{i,j\},\{i,j\}}$, $Q_{\{i,j\},\widehat{\{i,j\}}}$, $Q_{\widehat{\{i,j\}},\{i,j\}}\equiv(Q_{\{i,j\},\widehat{\{i,j\}}})^{\mathrm{T}}$, and $Q_{\widehat{\{i,j\}},\widehat{\{i,j\}}}$. Here the subscripts indicate the labels to be included or excluded (when with a hat) in the range of the indices. The above brings the quadric to the form
\begin{equation}\label{spherical}
\fixed
w_iw_j+X_{(ij)}Q^{(ij)}X_{(ij)},
\end{equation}
with
\begin{equation}\label{Qijprojection}
\fixed
Q^{(ij)}=Q_{\widehat{\{i,j\}},\widehat{\{i,j\}}}-Q_{\widehat{\{i,j\}},\{i,j\}}(Q_{\{i,j\},\{i,j\}})^{-1}Q_{\{i,j\},\widehat{\{i,j\}}}.
\end{equation}

In the integration, we preserve the contour for the remaining $x$ variables (i.e., over $[0,+\infty)$ in the canonical frame), but switch the $\{w_i,w_j\}$ integration to the same $S^2$ contour: $w_i=w$, $w_j=\bar{w}$, and integrate over the entire $w$ plane (in correspondence to the usual definition of residues, we further divide by a factor $\frac{1}{2\pi i}$). Completing this $S^2$ integration yields a new integral in two lower dimensions
\begin{equation}
e_n^{(ij)}=\frac{\sqrt{-1}}{2(n-2)}\int_{\Delta^{(ij)}}\frac{\sqrt{-\det Q^{(ij)}}\,\measure{X_{(ij)}}{n-3}}{(X_{(ij)}Q^{(ij)}X_{(ij)})^{\frac{n-2}{2}}},
\end{equation}
with the $Q^{(ij)}$ given in \eqref{Qijprojection}, and $\Delta^{(ij)}$ the codim-2 boundray excluding vertices $V_i$ and $V_j$ (note the difference from $\Delta_{(ij)}$). This is exactly an $e_{n-2}$ integral. 
We then have
\begin{equation}
\fixed
\mathcal{S}e_n^{(ij)}=C_n\sum_{\rho:\rho_1=i,\rho_2=j}\widehat{\overline{ij}}\otimes\overline{\rho_3\rho_4}\otimes\cdots\otimes\overline{\rho_{n-1}\rho_n}.
\end{equation}

Given the $e_n^{(ij)}$ integral for each choice of $(ij)$, we can then go to compute its own discontinuities $e_n^{(ij;kl)}$ in the same way, with $(kl)$ a pair chosen from the remaining labels. This is iterated $\frac{n-2}{2}$ times, until we land on a one-fold integral that produces a log corresponding to the last symbol entry. There is no more spherical contour applicable to this remaining integral, but it can be straightforwardly computed anyway.

At this stage, its is obvious what the overall factor is. It equals the product of the factors from each previous $S^2$ integral, times an additional factor from the last integral (which is the same as for a generic $e_2$ integral shown in \eqref{eq:e2result}), and so
\begin{equation}
C_n=\frac{-(\sqrt{-1})^{\frac{n-2}{2}}}{2^{\frac{n}{2}}(n-2)!!}.
\end{equation}
Along the way we also have acquired a list of objects for the determination of each symbol entry
\begin{equation}
\fixed
\mathcal{S}e_n=C_n\sum_{\rho}\underset{\substack{\uparrow\\ e_n}}{\overline{\rho_1\rho_2}}\otimes\underset{\substack{\uparrow\\ e_n^{(\rho_1\rho_2)}}}{\overline{\rho_3\rho_4}}\otimes\cdots\otimes\underset{\substack{\uparrow\\ e_n^{(\rho_1\rho_2;\ldots;\rho_{n-5}\rho_{n-4})}}}{\overline{\rho_{n-3}\rho_{n-2}}}\otimes\overline{\rho_{n-1}\rho_{n}},\qquad\text{even }n.
\end{equation}

\subsection{Determinantion of symbol entries}\label{sec:symboldetermination}

The remaining question is how to actually determine each symbol entry. Already from the last symbol entries we see that they are naturally associated with an $e_2$ integral. How can we identify such $e_2$ integral for every of the previous entries?

Take the first entry $\overline{ij}_{\text{(first entry)}}=0$ as an example. Observe that when we compute the discontinuity, the structure of the spherical contour that we do in the $(ij)$ direction does not rely at all on what other directions we have, or even the total dimensions of the space. This indicates that the existence of this particular branch cut and its local behavior is a property that solely arises from the the system as restricted to the (complex) line $\overline{V_iV_j}$. This suggests that we focus on the $e_2$ integral obtained by setting $x_k=0$ ($\forall k\neq i,j$)
\begin{equation}
\fixed
\int_{\overline{V_iV_j}}\frac{\sqrt{-\det Q_{\{i,j\},\{i,j\}}}\,\langle X_{\{i,j\}}\mathrm{d}X_{\{i,j\}}\rangle}{X_{\{i,j\}}Q_{\{i,j\},\{i,j\}}X_{\{i,j\}}}.
\end{equation}
Here $Q_{\{i,j\},\{i,j\}}$ is the submatrix of $Q$ labeled by $\{i,j\}$ for both columns and rows. The quadric in this 1d system factorizes into two linear factors as usual, and the effect of the spherical contour integration we performed above can be understood exactly in the same way as we did for the toy integral in the previous section. This analogy suggests the first symbol entry is again computed by
\begin{equation}\label{eq:symbolentryfromdiscontinuity}
\fixed
\overline{ij}_{\text{(first entry)}}=r(Q_{\{i,j\},\{i,j\}}^{-1}),
\end{equation}
where the ratio of roots $r(\mathbf{M})$ follows the same definition as in \eqref{ratioofroots}. This expression indicates a branch point at $\overline{ij}_{\text{(first entry)}}=0$ and at $1/\overline{ij}_{\text{(first entry)}}=0$. 

The other intermediate symbol entries follows exactly the same prescription \eqref{eq:symbolentryfromdiscontinuity}, since they all associate to some $e$ integral. The only difference is that we need to submit the quadric $Q$ by the corresponding projected quadric.

Here let us point out one simplification. Note that any $e$ is uniquely specified by the quadric in it. This means that for the $e$ integrals the $S^2$ contour integration is equivalent to a purely algebraic operation, e.g., in the first step, 
\begin{equation}
\fixed
\mathfrak{p}_{(ij)}:Q\mapsto Q^{(ij)},
\end{equation}
which is defined by \eqref{Qijprojection}. The next spherical contour is equivalent to another operation, say $Q^{(ij)}\mapsto\mathfrak{p}_{(kl)}Q^{(ij)}=Q^{(ij;kl)}$. The operator $\mathfrak{p}$ is actually commutative in the sense that $Q^{(ij;kl)}=Q^{(kl;ij)}$, or in other words, $\mathfrak{p}_{(kl)}\mathfrak{p}_{(ij)}=\mathfrak{p}_{(ij)}\mathfrak{p}_{(kl)}$ for any $\{ij,kl\}$. In fact we have a larger identity
\begin{equation}\label{eq:pijcommutativity}
\fixed
\mathfrak{p}_{(kl)}\mathfrak{p}_{(ij)}=\mathfrak{p}_{(jl)}\mathfrak{p}_{(ik)}=\mathfrak{p}_{(jk)}\mathfrak{p}_{(il)}=\mathfrak{p}_{(ij)}\mathfrak{p}_{(kl)}=\mathfrak{p}_{(ik)}\mathfrak{p}_{(jl)}=\mathfrak{p}_{(il)}\mathfrak{p}_{(jk)},\quad\forall i,j,k,l.
\end{equation}
Hence after completing any sequence of two $S^2$ contour integrals, as long as they together involve the same set of four variables $\{x_i,x_j,x_k,x_l\}$, we land on a unique quadric depending on the remaining $(n-4)$ variables, which we can denote as $Q^{(ijkl)}$ (where the parentheses indicate that the ordering is irrelevant).

This holds more generally for any sequence of $S^2$ contours (of any size). Starting with the original quadric $Q$, suppose we have completed a specific sequence of such integrals where the set of variables we get rid of is $S$, then we have
\begin{equation}\label{operationp}
\fixed
Q^{(S)}\equiv \mathfrak{p}_{(S)}Q=Q_{\widehat{S},\widehat{S}}-Q_{\widehat{S},S}(Q_S)^{-1}Q_{S,\widehat{S}},\qquad\text{even }|S|.
\end{equation}
This $Q^{(S)}$ can then be directly used to determine the next symbol entry. Explicitly, in \eqref{eq:symbolfromdiscontinuity} each entry is computed by
\begin{equation}
\fixed
\overline{\rho_{2k-1}\rho_{2k}}=r\left((Q^{(\rho_1\ldots\rho_{2k-2})})_{\{\rho_{2k-1},\rho_{2k}\},\{\rho_{2k-1},\rho_{2k}\}}^{-1}\right).
\end{equation}
In particular, this applies to the last symbol entries as well, thus putting all entries on an equal footing.

The above prescription is for a generic $Q$ whose diagonals are all none-zero. When some of the diagonal elements are zero, again we face the issue that the ratio of roots is not well-defined. Geometrically, $Q_{ii}=0$ indicates that the vertex $V_{i}$ resides on the quadric, and so from our experience with the differentiation method we know there are naive divergences in the contributions from the boundaries that have to cancel out after summation.

Here we directly work on this ``singular'' case by a limiting procedure. Consider, say $Q_{ii}=0$. We approach this by taking $Q_{ii}\to0$ so that the first entry $\overline{ij}$ approximates to
\begin{equation}
\fixed
\overline{ij}_{\text{first entry}}\longrightarrow\left(\frac{Q_{ii}Q_{jj}}{4Q_{ij}^2}\right)^{\text{sign}(Q_{ij})},
\end{equation}
and similarly if $Q_{jj}\to0$ or both of them vanish. Due to the algebraic properties of symbols we can now unambiguously separate the divergence piece
\begin{equation}
\fixed
Q_{ii}^{\text{sign}(Q_{ij})}\otimes\cdots
\end{equation}
(and similarly if $Q_{jj}\to0$) from the rest. The remaining finite piece is the actual contribution to the symbol of the integral. In consequence, the general prescription for the first symbol entries of a non-degenerate quadric integral is
\begin{equation}\label{eq:discentryformula}
\fixed
\overline{ij}_{\text{first entry}}=
\begin{cases}
r(Q_{\{i,j\},\{i,j\}}^{-1}),&Q_{ii}\neq 0,Q_{jj}\neq0,\\
\left(\frac{Q_{ij}^2}{Q_{jj}}\right)^{-\text{sign}(Q_{ij})},&Q_{ii}=0,Q_{jj}\neq0,\\
\left(\frac{Q_{ij}^2}{Q_{ii}}\right)^{-\text{sign}(Q_{ij})},&Q_{ii}\neq0,Q_{jj}=0,\\
Q_{ij}^{-2\,\text{sign}(Q_{ij})},&Q_{ii}=Q_{jj}=0.
\end{cases}
\end{equation}
Its generalization to other symbol entries is obvious.

Of course we need to confirm that the divergent parts indeed cancel away. From the above discussion we see that the cancellation explicitly require the following identity to hold, e.g., associating to the first symbol entry
\begin{equation}
\sum_{j\neq i}\text{sign}(Q_{ij})e_n^{(ij)}=0,\qquad \text{iff }Q_{ii}=0.
\end{equation}
Structurally this resembles the cancellation identity \eqref{spuriousterms} we encountered in the differentiation method. We do not provide a complete proof for this cancellation in this paper, but just point out that this has to come as a consequence of the equivalence between the discontinuity and differentiation methods that we are going to discuss in the next subsection.

\subsection{Equivalence of the discontinuity and the differentiation methods}\label{subsec:eqdis2dif}

So far we have introduced two different methods to determine the symbol of any $e_n$ integral with even $n$, which is a pure function of transcendental weight $\frac{n}{2}$. The result from both methods are organized into ordered partitions of the label sets into symmetric pairs
\begin{equation}
\fixed
\mathcal{S}e_n=\sum_{\rho}(\rho_1\rho_2)\otimes(\rho_3\rho_4)\otimes\cdots\otimes(\rho_{n-1}\rho_{n}).
\end{equation}
In the differentiation method we have $(\rho_{2k-1}\rho_{2k})=\underline{\rho_{2k-1}\rho_{2k}}$ and in the discontinuity method $(\rho_{2k-1}\rho_{2k})=\overline{\rho_{2k-1}\rho_{2k}}$. We now show that
\begin{equation}\label{overunderid}
\fixed
\underline{\rho_{2k-1}\rho_{2k}}=\overline{\rho_{2k-1}\rho_{2k}},\qquad\forall k.
\end{equation}

In the differentiation method $\underline{\rho_{2k-1}\rho_{2k}}$ is obtained by computing the inverse of the submatrix $Q_{\{1,\ldots,2k\},\{1,\ldots,2k\}}$ and extracting its $2\times2$ submatrix $\underline{M}$ labeled by $\{2k-1,2k\}$, and then computing the ratio of its two associated roots. Note that
\begin{equation}
\begin{split}\fixed
&\underline{M}\,\det Q_{\{1,\ldots,2k\},\{1,\ldots,2k\}}\equiv \underline{M}'=\\
&\quad\left(\begin{matrix}\det Q_{\{1,\ldots,2k-2,2k\},\{1,\ldots,2k-2,2k\}}&-\det Q_{\{1,\ldots,2k-2,2k\},\{1,\ldots,2k-2,2k-2\}}\\-\det Q_{\{1,\ldots,2k-2,2k-1\},\{1,\ldots,2k-2,2k\}}&\det Q_{\{1,\ldots,2k-2,2k-1\},\{1,\ldots,2k-2,2k-1\}}\end{matrix}\right),
\end{split}
\end{equation}
and so
\begin{equation}
\fixed
\underline{\rho_{2k-1}\rho_{2k}}=r(\underline{M})=r(\underline{M}')^{\text{sign}(\det Q_{\{1,\ldots,2k\},\{1,\ldots,2k\}})}.
\end{equation}

In the discontinuity method we first compute the projected matrix $Q^{(1,2,\ldots,2k-2)}$ and then extract its submatrix $\overline{M}$ labeled by $\{2k-1,2k\}$
\begin{equation}
\fixed
\overline{M}=\left(\begin{matrix}Q^{(1,2,\ldots,2k-2)}_{2k-1,2k-1}&Q^{(1,2,\ldots,2k-2)}_{2k-1,2k}\\Q^{(1,2,\ldots,2k-2)}_{2k,2k-1}&Q^{(1,2,\ldots,2k-2)}_{2k,2k}\end{matrix}\right),
\end{equation}
and $\overline{\rho_{2k-1}\rho_{2k}}=r(\overline{M}^{-1})$. 

Note that, for example
\begin{equation}
\begin{split}\verified
&Q^{(1,2,\ldots,2k-2)}_{2k-1,2k-1}\det Q_{\{1,\ldots,2k-2\},\{1,\ldots,2k-2\}}\\
&\qquad=Q_{2k-1,2k-1}\det Q_{\{1,\ldots,2k-2\},\{1,\ldots,2k-2\}}\\
&\qquad\quad-\sum_{I,J=1}^{2k-2}(-1)^{I+J}Q_{2k-1,I}(\det Q_{\{1,\ldots,\hat{I},\dots,2k-2\},\{1,\ldots,\hat{J},\ldots,2k-2\}})Q_{J,2k-1}\\
&\qquad=\det Q_{\{1,\ldots,2k-2,2k-1\},\{1,\ldots,2k-2,2k-1\}}.
\end{split}
\end{equation}
Similar identities hold also for the other three entries. Altogether we obtain
\begin{equation}
\fixed
\overline{M}^{-1}\,\det\overline{M}\equiv\left(\begin{matrix}0&-1\\1&0\end{matrix}\right)\times\overline{M}\times\left(\begin{matrix}0&1\\-1&0\end{matrix}\right)=\frac{\underline{M}'}{\det Q_{\{1,\ldots,2k-2\},\{1,\ldots,2k-2\}}}.
\end{equation}
Recall the basic property of determinant
\begin{equation}
\fixed
\det Q_{\{1,\ldots,2k\},\{1,\ldots,2k\}}=\det Q_{\{1,\ldots,2k-2\},\{1,\ldots,2k-2\}}\times\det\overline{M},
\end{equation}
we thus have
\begin{equation}\label{eq:MMequality}
\fixed
\overline{M}^{-1}=\underline{M}.
\end{equation}
This indicates the validity of \eqref{overunderid} for every symbol entry, and so the two methods are equivalent for $e_n$ integrals with even $n$.

Before we end this subsection, let us quickly point out that obviously the proof of the matrix identity \eqref{eq:MMequality} also holds for half-integral $k$ and $n$ being any integer. This in particular indicates an algebraic way to determine the symbol $\mathcal{S}e_n$ with odd $n$, much like the discontinuity method prescribed before. This is going to be discussed in more detail in Section \ref{sec:oddspacetime}.

\subsection{A duality among $e_n$ integrals with even $n$}

To better understand the relation between the differentiation and the discontinuity operations on the $e_n$ integrals with even $n$, let us present a duality among these objects, as revealed by these methods. This ``duality'' refers to the correspondence between an object in the original space and its counterpart in the dual space. The precise statement is the following.

Consider an $e_n$ integral with even $n$, specified by an arbitrary non-degenerate quadric $Q$ and an arbitrary simplex contour $\Delta$ (defined by its vertices $\{V_i\}$ or boundaries $\{H_i\}$). In the dual space we construct another integral
\begin{equation}
\widetilde{e}_n=\int_{\widetilde{\Delta}}\frac{\sqrt{\det\widetilde{Q}}\measure{\widetilde{X}}{n-1}}{(\widetilde{X}\widetilde{Q}\widetilde{X})^{\frac{n}{2}}}.
\end{equation}
where $\widetilde{Q}=Q^{-1}$, and we define the contour $\widetilde{\Delta}$ by specifying its vertices to be $\{H_i\}$ or equivalently its boundaries to be $\{V_i\}$. Then we have the following identity between the symbols
\begin{equation}\label{reversionid}
\mathcal{S}e_n=(\mathcal{S}\widetilde{e}_n)^{\mathrm{R}},
\end{equation}
where $\mathrm{R}$ is the operation that reverses the entries in each symbol term
\begin{equation}
\mathrm{R}:w_1\otimes w_2\otimes\cdots\otimes w_{\frac{n}{2}}\longmapsto w_{\frac{n}{2}}\otimes w_{\frac{n}{2}-1}\otimes\cdots\otimes w_1.
\end{equation}

Since $Q$ is non-degenerate, we can regard it as a $\mathrm{PGL}(n)$ transformation, and so the $e_n$ integral can as well be defined by setting $\widetilde{Q}=Q$, and setting the contour's vertices to be $\{Q^{-1}H_i\}$ or its boundaries to be $\{QV_i\}$.

We provide a proof in Appendix \ref{app:duality}.

\section{Discontinuity II: Tensor Integrals}\label{sec:dis2}

\flag{No further revisions.}

The spherical contour that we introduced in the previous section has the advantage that it directly applies to integrals with arbitrary tensor numerators as well.

\subsection{Spherical contour on tensor integrals}

Let us now work out the explicit result for the action of the spherical contour on a generic tensor integral
\begin{equation}
E_{n,k}=\int_\Delta\frac{\measure{X}{n-1}\,T[X^k]}{(XQX)^{\frac{n+k}{2}}}.
\end{equation}
In this whole section we assume $n+k$ is even.

It is sufficient to look at a specific component ($R$ being the transformation in \eqref{affinetransformation})
\begin{equation}
\begin{split}
E_{n,\{i,j\}}&=\int_{\Delta^{(ij)}}\measure{X_{(ij)}}{n-3}\underset{S^2}{\int}\mathrm{d}x_i\mathrm{d}x_j\frac{x_i^{p_i}x_j^{p_j}}{(XQX)^{\frac{n+k}{2}}}\\
&=\sum_{m=0}^{\frac{p_i+p_j}{2}}\left(\begin{matrix}p_i\\m\end{matrix}\right)\left(\begin{matrix}p_j\\\frac{p_i+p_j}{2}-m\end{matrix}\right)\frac{\pi\,R_{ii}^mR_{ij}^{p_i-m}R_{ji}^{\frac{p_i+p_j}{2}-m}R_{jj}^{\frac{p_j-p_i}{2}+m}}{\sqrt{Q_{ij}^2-Q_{ii}Q_{jj}}}\times\\
&\qquad\qquad\times\int_{\Delta^{(ij)}}\measure{X_{(ij)}}{n-3}\int_0^\infty\frac{\mathrm{d}r\,r^{p_i+p_j+1}}{(r^2+X_{(ij)}Q^{(ij)}X_{(ij)})^{\frac{n+k}{2}}}
\end{split}
\end{equation}
with $p_i+p_j\leq k$. Completing the $r$ integral yields
\begin{equation}
\begin{split}
E_{n,\{i,j\}}&=\pi\sum_{m=0}^{\frac{p_i+p_j}{2}}\left(\begin{matrix}p_i\\m\end{matrix}\right)\left(\begin{matrix}p_j\\\frac{p_i+p_j}{2}-m\end{matrix}\right)\frac{\Gamma(\frac{p_i+p_j+2}{2})\Gamma(\frac{n+k-p_i-p_j+2}{2})}{\Gamma(\frac{n}{2})\,\sqrt{Q_{ij}^2-Q_{ii}Q_{jj}}}R_{ii}^mR_{ij}^{p_i-m}R_{ji}^{\frac{p_i+p_j}{2}-m}R_{jj}^{\frac{p_j-p_i}{2}+m}\times\\
&\qquad\qquad\times\int_{\Delta^{(ij)}}\frac{\measure{X_{(ij)}}{n-3}}{(X_{(ij)}Q^{(ij)}X_{(ij)})^{\frac{n+k-p_i-p_j}{2}-1}}.
\end{split}
\end{equation}

\subsection{General integrals}

The discontinuity analysis directly generalizes to integrals with arbitrary tensor numerators, and also to the case when $Q$ itself is degenerate. In general, as concluded in the differentiation discussion, the functions produced by these integrals have mixed transcendental weights, and even when they turn out to have a uniform weight they can be a linear combination of pure functions with non-trivial coefficients.

In such situations the discontinuity analysis does not directly obtain the information about the entire singularity structure of the integral, mainly due to the fact that the computation of a discontinuity vanishes if a potential branch point turns out not to be an actual one. Here what one can expect to directly extract from the integral representation using by the method are
\begin{enumerate}
\item Highest transcendental weight of the function.
\item Symbols of the highest-weight part, including the coefficients in front.
\end{enumerate}
More precisely, starting from a generic $E_{n,k}$ integral, one applies the following algorithm:
\begin{enumerate}
\item Bring the integral to the frame where the contour is canonical.
\item List out all possible pairs of labels in the set $\{1,2,\ldots,n\}$.
\item For each pair $(i_1i_2)$, do an affine transformation to the variables $\{x_{i_1},x_{i_2}\}$ according to \eqref{affinetransformation} to bring the quadric to the standard form \eqref{spherical}, and perform the corresponding $S^2$ contour integral, which in general yields another integral $E_{n,k}^{(i_1i_2)}$ in two lower dimensions and depends on the remaining $x$ variables.
\item It can happen that $E_{n,k}^{(i_1i_2)}$ already vanishes, or that the affine transformation to the standard form does not exist. In this case we do nothing, as it just means that a branch point associated to $\overline{i_1i_2}$ does not exist. However, in the case when it is non-zero, it indicates that we should include symbol terms that starts with $\overline{i_1i_2}$ in the first entry
\begin{equation}
\mathcal{S}E_{n,k}\supset \overline{i_1i_2}\otimes(\mathcal{S}E_{n,k}^{(i_1i_2)}).
\end{equation}
At this stage the coefficients in front are not yet determined.
\item For nonvanishing $E_{n,k}^{(i_1i_2)}$, we repeat the analysis 1 to 4 on it, extracting the expressions in latter symbol entries.
\item For every sequence of $S^2$ contour integral computations, it has to terminate after a finite number of steps, in the sense that in the last step for whatever choice of pairs from the remaining label set the $S^2$ integral always generates zero or does not exist. Suppose this occurs at the $s^{\rm th}$ step, then this indicates that there exist a specific symbol term
\begin{equation}
\mathcal{S}E_{n,k}\ni E_{n,k}^{(\text{last})}\left(\overline{i_1i_2}\otimes\overline{i_3i_4}\otimes\cdots\otimes\overline{i_{2s-3}i_{2s-2}}\right).
\end{equation}
Here the coefficient $E_{n,k}^{(\text{last})}$ is exactly the integral expression we obtain before the last $S^2$ integral, which is guaranteed to be rational.
\item In the end, we collect all these terms and simply sum them up.
\end{enumerate}

The result we obtain from this algorithm might possibly contain symbol terms of various lengths. It is obvious from the above discussion that, suppose there is a term of length $s_1$ that is not the maximal length in the entire result, it can never be embedded in any other symbol term of greater length as the first $s_1$ entries. Here the phrase ``embedded'' has to be understood in a generalized sense. For example, from some integral in $\mathbb{CP}^7$ we might extract some terms with the structure
\begin{equation}
c_1\,\overline{12}\otimes\overline{34}\otimes\overline{56}+c_2\,\overline{13}\otimes\overline{57}+\cdots.
\end{equation}
In this case, of course we should not expect to see other terms with the structure, e.g., $\overline{13}\otimes\overline{57}\otimes\overline{48}$, in which the second term above can be directly embedded, since this obviously violates the above algorithm. In fact, we cannot expect any other terms with length $k>2$ such that the $2k$ labels in it contains $\{1,3,5,7\}$ as a subset, e.g., $\overline{14}\otimes\overline{35}\otimes\overline{78}$. This is because if there is such a term, then the fact that the symbol of an integral always decomposes into those of $e$ integrals indicates there necessarily exist other terms related to this by permutation of labels (unless those terms trivially vanishes due to trivial entries), which in particular means that the term $\overline{13}\otimes\overline{57}\otimes\cdots$ cannot terminate at length 2.

In consequence, the result from this algorithm is organized in terms of a linear combination of $\mathcal{S}e_m$'s with possibly various $m$ values, each of which corresponds to an induced quadric integral on some codim-$(n-m)$ boundaries, and furthermore, each lower-dim boundary is not a boundary of any other higher-dim boundaries that appear. This result as a whole is contained in the symbol of the $E_{n,k}$ integral under study.

However, it is not necessary that this produces the entire $\mathcal{S}E_{n,k}$. For example, consider again the above integral in $\mathbb{CP}^7$. In principle there could possibly exist a term $\overline{12}\otimes\overline{34}$, but such term cannot be detected as a further spherical integration in the $(56)$ direction will ignore it.

In order to analyze the terms ignored by the above algorithm, we need to first reconstruct a function corresponding to the symbol terms (as well as their coefficients) that has been worked out and subtract it from $E_{n,k}$ and repeat the same analysis again. In principle the construction of this function can be directly read off from the symbol terms already worked out, since their corresponding $e$ integrals always descend from the same quadric in the original integral.

Even though the discontinuity analysis is not very straightforward in fully extracting the singularity structure in the case of mixed transcendental weights, it can become very convenient in special situations. It is in particular useful when the quadric is highly degenerate, where the differentiation analysis could potentially becomes complicated.  Also, if the integral vanishes on every choice of $S^2$ contour, one immediately concludes it has to be a rational function. Furthermore, if a function has a uniform transcendental weight but is not pure, the non-trivial coefficients can be easily obtained by the iterated $S^2$ contour integrals. These will be illustrated in the Feynman integral examples in the following.

\subsection{Application to Feynman integrals}

Since we have shown that for $e_n$ integrals with even $n$ the discontinuity analysis is equivalent to the differentiation method, we do not provide any further example of this type. Here we illustrate how the discontinuity method works in the last three examples of tensor integrals discussed in Section \ref{sec:moreintegralsdif}.

\subsubsection{Finite hexagon in 4d}

First let us return to our previous example 5
\begin{equation}
I_{\text{e.g.}5}=\underset{AB}{\int}\frac{\langle AB(612)\cap(234)\rangle\langle AB46\rangle}{\langle AB12\rangle\langle AB23\rangle\langle AB34\rangle\langle AB45\rangle\langle AB56\rangle\langle AB61\rangle}.
\end{equation}
There using the differentiation analysis we observe that it has uniform weight 2 but is not pure. We thus should expect to extract the non-trivial coefficients after computing two spherical integrals.

The total numerator in the Feynman parametrization of this integral is
\begin{equation}
\langle6124\rangle\langle2346\rangle\left(\sum_{i<j}\langle i-1\,i\,j-1\,j\rangle x_ix_j\right)-6(\langle6124\rangle\langle2345\rangle x_5+\langle6125\rangle\langle2346\rangle x_6)(\langle1246\rangle x_2+\langle2346\rangle x_3).
\end{equation}
The determination of each specific symbol entry can be worked out purely algebraically and was previously shown to yield the same result as is obtained from differentiation, and so we skip this step and merely refer to the result listed in Appendix \ref{app:hexagon}. In particular, this indicates that in the first step we only need to consider $S^2$ contour in the direction $(24)$, $(35)$ and $(46)$. Take the direction $(46)$ as an example; the spherical contour yields an $E_{4,2}$ on $\Delta^{(46)}$. Upon that we can further choose an $S^2$ contour. It turns out that the contour along $(13)$ yields
\begin{equation}
I_{\text{e.g.}5}^{(4613)}=0,
\end{equation}
indicating there is not contribution from $\overline{46}\otimes\overline{13}\otimes\cdots$. This is desired since this symbol term generates $u_3\otimes(1-u_3)$, which only resides in this term but is absent in $\mathcal{S}I_{\text{e.g.}4}$. On the other hand, e.g., the contour along $(12)$ leads to
\begin{equation}
\begin{split}
I_{\text{e.g.}5}^{(4612)}=&\frac{-\langle1246\rangle\langle2346\rangle}{8\langle6134\rangle\langle1245\rangle\langle2356\rangle}\times\\&\int_{\Delta^{(4612)}}\frac{(u_1x_3-x_5+u_1x_5+u_5x_5)(u_5^{-1}u_2x_3-x_3+u_3x_3+u_3x_5)}{(u_1(u_3-u_1)x_3^2+(1-u_1-u_2-u_3+2u_1u_3)x_3x_5+u_3(u_1-1)x_5^2)^2},
\end{split}
\end{equation}
with $u_5=\frac{\langle1246\rangle\langle2345\rangle}{\langle2346\rangle\langle1245\rangle}$. This remaining one-dimensional integral turns out to be a rational function
\begin{equation}
I_{\text{e.g.}4}^{(4612)}=\frac{1}{4}\frac{\langle1346\rangle}{\langle1345\rangle\langle1356\rangle}.
\end{equation}
Similarly, we also have
\begin{equation}
I_{\text{e.g.}4}^{(4615)}=-I_{\text{e.g.}4}^{(4612)},\quad
I_{\text{e.g.}4}^{(4623)}=I_{\text{e.g.}4}^{(4635)}=\frac{1}{4}\frac{\langle1236\rangle}{\langle1235\rangle\langle1356\rangle},\quad
I_{\text{e.g.}4}^{(4625)}=-\frac{1}{4}\frac{\langle1234\rangle}{\langle1345\rangle\langle1235\rangle}.
\end{equation}
All these integrals account for symbol terms starting with $u_3\otimes\cdots$ and they confirm this part uniformly has length 2. Note that the rational functions generated in the end already match those worked out in the differentiation analysis. Here again we need to observe the cancellation of $u_3\otimes u_3$, which is guaranteed by the identity
\begin{equation}
\frac{\langle1346\rangle}{\langle1345\rangle\langle1356\rangle}+\frac{\langle1236\rangle}{\langle1235\rangle\langle1356\rangle}+\frac{\langle1234\rangle}{\langle1345\rangle\langle1235\rangle}=0.
\end{equation}
The other sequences of $S^2$ can be analyzed similarly and we do not go into further details.

\subsubsection{The parity-odd diagrams}

The spherical contours can usually help quickly detect the highest weight of the integral. An extreme case are the parity-odd diagram in our example 6, which are simply zero. Here the spherical contours cannot tell that the result is exactly zero. However, it has a sharp characterization of these integrals being rational: in any choice of directions $(ij)$ the spherical contour either is not well-defined, or leads to zero.

\subsubsection{The massive box with mixed weights}

We now illustrate how the discontinuity method works in the case when the integral involves mixed weights. We again take the previous example
\begin{equation}
\verified{}
I_{\text{e.g.}7}=\int_0^\infty\frac{\measure{X}{3}\,x_1x_3\,(x_1+x_2+x_3+x_4)^2}{(\frac{4}{u}x_1x_3+\frac{4}{v}x_2x_4+(x_1+x_2+x_3+x_4)^2)^4},
\end{equation}
so as to compare with the differentiation method.

First of all the spherical contours are well-defined only in the directions $(13)$ and $(24)$, as the $2\times2$ submatrix of $Q$ in all the other choices are singular. Correspondingly the first symbol entries only choose between
\begin{equation}
\overline{13}=r(Q_{\{1,3\}})=\left(\frac{\sqrt{1+u}-1}{\sqrt{1+u}+1}\right)^{2},\qquad
\overline{24}=r(Q_{\{2,4\}})=\left(\frac{\sqrt{1+v}-1}{\sqrt{1+v}+1}\right)^{2}.
\end{equation}
Now we compute the $S^2$ contour, say in the $(13)$ direction. This yields an $E_{2,4}$ integral on $\Delta^{(13)}$, whose quadric reads
\begin{equation}
vx_2^2+(4+4u+2v)x_2x_4+vx_4^2.
\end{equation}
This is a one-dimensional integral and the spherical contours no longer applies. But here the discontinuity problem reduces to the usual one as that for a logarithm, so all what we need is a further $S^1$ wrapping one of the two roots of the quadric. Thus the corresponding second symbol entry is again just the ratio of the two roots (as is for an ordinary log) 
\begin{equation}
\overline{24}=\left(\frac{\sqrt{1+u+v}-\sqrt{1+u}}{\sqrt{1+u+v}+\sqrt{1+u}}\right)^2.
\end{equation}
The result of the two contour integrals yields
\begin{equation}
-\frac{\pi^2}{2}\frac{u^2v(4u+u^2+2v+3uv+2v^2)}{384(1+u+v)^{\frac{5}{2}}}.
\end{equation}
Taking into consideration the extra factors coming from the integrations, this produces the correct coefficient of the weight-2 part of the integral. The same result can be obtained by starting with the $(24)$ direction as well. With these two computations we also manage to recover the correct symbols for this part.

From the above, we explicitly see that in the case of mixed transcendental weights the direct application of the discontinuity method can only reveal information of the highest-weight part of the integral, while all the lover-weight pieces are discarded by the intermediate computations.

In order to recover the lower-weight pieces, a simple (though indirect) practice to follow is to subtract from the original integral the $e$ integrals for the highest-weight piece dressed by its corresponding coefficient. In this particular example we thus go to
\begin{equation}
\verified{}
I'_{\text{e.g.}7}=I_{\text{e.g.}7}-\frac{u^2v(4u+u^2+2v+3uv+2v^2)}{384(1+u+v)^{\frac{5}{2}}}\int_\Delta\frac{\sqrt{q}\,\langle X\mathrm{d}X^3\rangle}{(XQX)^2}.
\end{equation}
Combine the two terms yields a new tensor numerator $T'[X^4]$. Now we can repeat the same discontinuity analysis in exactly the same way as before: compute the spherical contour integrals in the $(13)$ and $(24)$ and detect the first symbol entries. However, as we try to further apply an $S^1$ contour integral on the induced integrals, say $I'^{(13)}_{\text{e.g.}7}$, the result is simply zero. This indicates that $I'^{(13)}_{\text{e.g.}7}$ is purely rational. Moreover, this rational function (after chopping of the extra overall factor from the $S^2$ contour) is exactly the coefficient in front of the corresponding weight-1 piece. It can be easily worked out that
\begin{align}
I'^{(13)}_{\text{e.g.}7}&=\frac{\pi\,u^2v(2+u+u^3+2v+2uv+u^2v)}{384(1+u)^{\frac{3}{2}}(1+u+v)^2},\\
I'^{(24)}_{\text{e.g.}7}&=\frac{\pi\,u^2v\sqrt{1+v}(2-u-v)}{384(1+u+v)^2}.
\end{align}
We see these match \eqref{eq:mixedweightC13} and \eqref{eq:mixedweightC24} perfectly.

In order to work out the rational piece, we need to find out integrals $E_{4,2}$ that purely yields a logarithm matching $\overline{13}$ and $\overline{24}$, respectively, and then dress them with $I'^{(13)}_{\text{e.g.}7}$ and $I'^{(24)}_{\text{e.g.}7}$ and subtract from $I'^{(13)}$ to obtain yet another new integral $I''^{(13)}$. The choice is not unique, but for any such choice $I''^{(13)}$ is guaranteed to be rational (and the way to observe it is to see that the $S^2$ contours in $(13)$ and $(24)$ both leads to zero). This rational function is exactly the weight-0 piece of $I_{\text{e.g.}7}$.

To make this explicit, for example, we can choose $E_{4,2}$ integrals
\begin{align}
U_{24}&=\int_{\Delta}\frac{8\sqrt{q_{(24)}}}{(XQX)^3}\left(-\frac{4}{u}x_1x_3+\frac{2+3v}{v}(x_1+x_3+x_4)x_4+\frac{4+3v}{v}x_2x_4\right),\\
U_{13}&=\int_{\Delta}\frac{8\sqrt{q_{(13)}}}{(XQX)^3}\left(-\frac{4(1+u)}{u^2}x_1x_3+\frac{2+3u}{u}(x_1+x_3+x_4)x_4+\frac{4+2u+2v+3uv}{uv}x_2x_4\right).
\end{align}
Though not very obvious, it can be verified that
\begin{equation}
\mathcal{S}U_{24}=\overline{13},\qquad\mathcal{S}U_{13}=\overline{24}.
\end{equation}
So we can construct a new integral bu subtraction
\begin{equation}
I''_{\text{e.g.}7}=I'_{\text{e.g.}7}-\frac{1}{\pi\,i}I'^{(13)}_{\text{e.g.}7}U_{24}-\frac{1}{\pi\,i}I'^{(24)}_{\text{e.g.}7}U_{13}.
\end{equation}
As we perform the spherical contour integral in the $(13)$ and $(24)$ directions, although the resulting integral may appear to be non-trivial and dependent on the choice of the $U$ integrals, it actually vanishes (as can be seen by explicitly doing the remaining one-fold integration). Hence $I''_{\text{e.g.}7}$ is identical to the rational part of the original integral
\begin{equation}
I''_{\text{e.g.}7}=I_{\text{e.g.}7}\big|_{\text{rational}}.
\end{equation}

\section{Discontinuity III: Hypercontours}\label{sec:contour}

\flag{No further revisions.}

In Section \ref{sec:dis1} we introduced a method based on discontinuities to determine the singularity structure of non-degenerate $e_n$ integrals with even $n$, and in Section \ref{sec:dis2} showed its applicability to a generic $E_{n,k}$ integral such that $n+k$ is even. As we are going to explain in this section, this prescription has an intrinsic geometric interpretation as a projection, which is manifestly frame-independent. We also generalize this prescription to arbitrary projections, so as to account for arbitrary integrals.


\subsection{Geometries of the spherical contour}

Previously we showed that for an $e_n$ integral, the $S^2$ integral associated to labels, say $(ij)$, generates $e_n^{(ij)}$, which is an $e_{n-2}$ integral associated to a quadric $Q^{(ij)}$. This was derived in the canonical frame. 

In fact the new quadric $Q^{(ij)}$ admits a frame-independent definition, geometrically, as a projection of the original quadric $Q$ through the line that passes the two vertices $V_i$ and $V_j$ of the simplex contour in $e_n$.

\begin{figure}[ht]
\begin{center}
\begin{tikzpicture}
\coordinate [label=-135:$V_i(V_j)$] (vij) at (0,0);
\coordinate (v1) at (5,0);
\coordinate (v2) at (0,5);
\fill [OrangeRed,opacity=.1] (vij) -- (v1) arc [start angle=0, end angle=90,radius=5] -- cycle;
\draw [OrangeRed,very thick] (v1) -- (vij) -- (v2);
\draw [OrangeRed,very thick] (v1) arc [start angle=0, end angle=90,radius=5];
\draw [ProcessBlue,thick] (45:3) circle [radius=1.5];
\shade[ball color = ProcessBlue!40, opacity = 0.4] (45:3) circle [radius=1.5];
\draw [Violet,thick,dotted] (vij) -- +(15:5.5);
\draw [Violet,thick,dotted] (vij) -- +(75:5.5);
\draw [ProcessBlue,thick] (15:5.04) arc [start angle=15, end angle=75,radius=5.04];
\fill [ProcessBlue] (15:2.6) circle(2pt);
\fill [ProcessBlue] (75:2.6) circle(2pt);
\fill [ProcessBlue] (15:5) circle(2pt);
\fill [ProcessBlue] (75:5) circle(2pt);
\end{tikzpicture}
\caption{Spherical contour as a projection, viewed through the line $H_{(ij)}$.}
\label{fig:S2projection}
\end{center}
\end{figure}
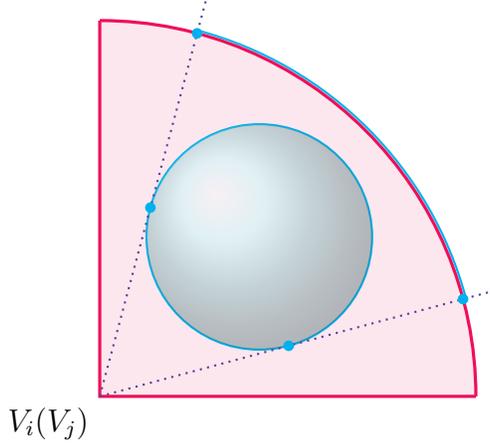

To be more explicit, let us denote $H_{(P_1\ldots P_l)}$ has an $(l-1)$-plane $\mathbb{CP}^{l-1}\in\mathbb{CP}^{n-1}$ which is specified by $l$ non-degenerate points $\{P_1,\ldots,P_l\}$ contained in it. In this notation, for example the codim-1 boundaries of the contour $H_i\equiv H_{(12\ldots\hat{i}\ldots n)}$. $H_{(ij)}$ is the line passing through $V_i$ and $V_j$. To do the projection through $H_{(ij)}$, we consider the collection of all planes (dim-2) $H_{(ijP)}$, which are indexed by certain point $P$ up to equivalence. In particular we can consider $P$ to be inside some codim-2 hyperplane $H'\simeq\mathbb{CP}^{n-3}$ as long as $H'$ does not contain $H_{(ij)}$. Also note that $H'\cap H_{(ijP)}=\{P\}$ ($\forall P\in H'$) (just by counting of equations), hence there is a one-to-one correspondence between $P\in H'$ and $H_{(ijP)}$. Now the new quadric $Q^{(ij)}$ is
\begin{equation}
\fixed
Q^{(ij)}:\quad\{P\in H'|H_{(ijP)}\text{ is tangent to }Q\}.
\end{equation}
This is visualized in Figure \ref{fig:S2projection}.

At first sight it may not even be obvious why the set of $P$ in $H'$ satisfy the tangent condition has to form a quadric. To confirm this, note that we can parametrize points on $H_{(ijP)}$ by
\begin{equation}
\fixed
X=P+\alpha_iV_i+\alpha_jV_j.
\end{equation}
Then the tangent condition is equivalent to
\begin{equation}
\fixed
XQX=\partial_{\alpha_i}(XQX)=\partial_{\alpha_j}(XQX)=0,
\end{equation}
for some choice of $\{\alpha_i,\alpha_j\}$. Eliminating $\{\alpha_i,\alpha_j\}$ results in a unique quadratic equation of $P$, which is
\begin{equation}\label{eq:Qinprojectioninvariant}
\fixed
PQ'P\equiv P^I\left(Q_{IJ}-\left(\begin{matrix}(V_iQ)_I\\(V_jQ)_I\end{matrix}\right)^{\rm T}\left(\begin{matrix}V_iQV_i&V_iQV_j\\V_jQV_i&V_jQV_j\end{matrix}\right)^{-1}\left(\begin{matrix}(V_iQ)_J\\(V_jQ)_J\end{matrix}\right)\right)P_J=0.
\end{equation}
Hence the projection leads to a quadric in $H'$. 

Correspondingly, we define a new simplex integration contour $\Delta'$ in $H'$, by the set of points $P$ such that $H_{(ijP)}$ intersects $\Delta$ at points other than $H_{(ij)}$. This together with $Q'$ uniquely defines an $e_{n-2}$ integral in $H'$, which is related to $e_n^{(ij)}$ by some $\mathrm{PGL}(n-2)$ transformation, thus identical.

The above $\mathrm{PGL}(n-2)$ transformation is tied to different choices of $H'$. In particular, the choice $H'=H_{(1\ldots\hat{i}\ldots\hat{j}\ldots n)}$ sets the system in the canonical frame, where the new quadric exactly becomes the expression $Q^{(ij)}$ we obtained in \eqref{Qijprojection}, and the new contour is the facet $\Delta_{(ij)}$.

As a result, we see that for the $e_n$ integrals with even $n$ the discontinuity is literally a projection through a line in the Feynman parameter space.

\subsection{An alternative $S^1$ contour}

In Section \ref{sec:symboldetermination} we showed that the spherical contours are commutative to each other. There are two aspects of this property:
\begin{enumerate}
\item At the level of the quadrics, a spherical contour in $(ij)$ direction corresponds to an algebraic operation $\mathfrak{p}_{(ij)}$ on the quadric, as defined in \eqref{Qijprojection} and more invariantly in \eqref{eq:Qinprojectioninvariant}. These algebraic operations commute to each other.
\item At the level of the integrals, for any generic $E_{n,k}$ integral, if we apply spherical contour integrations twice, say in the $(i_1j_1)$ and $(i_2j_2)$ directions respectively, the result does not rely on in which direction we perform the integration first. Hence we are allowed to denote these integral-level operations abstractly using the same notation $\mathfrak{p}$ as above.
\end{enumerate}

While as originally defined the projection associated to the algebraic operation $\mathfrak{p}$ is always through some $m$-plane with $m$ odd, projections through even planes are equally natural, hence in general
\begin{equation}\label{eq:QSprojection}
\fixed
Q^{(S)}\equiv \mathfrak{p}_{(S)}Q=Q_{\widehat{S},\widehat{S}}-Q_{\widehat{S},S}(Q_S)^{-1}Q_{S,\widehat{S}}
\end{equation}
is well-defined for any cardinality $|S|$. In particular $S$ can be a single point. Meanwhile, as mentioned in \eqref{eq:pijcommutativity} the commutavitity actually holds for arbitrary permutation of individual labels but not just permutation of the label pairs. These suggests there should exist some corresponding notion of $S^1$ contours respecting the above mentioned properties, and it is interesting and important to understand their effects.

In fact such $S^1$ contour can be defined in natural analogy to the $S^2$ contour we did in Section \ref{sec:dis1}: starting from an integral as expressed in the canonical frame, for a given label $i$ we affine transform the variable $x_i$ to $\chi_i$ such that
\begin{equation}
XQX\longmapsto \chi_i^2+X_{(i)}Q^{(i)}X_{(i)},
\end{equation}
with $Q^{(i)}$ given in \eqref{eq:QSprojection}, and then we integrate $w_i$ over $(-\infty,+\infty)$. The ``$S^1$'' is, as was the $S^2$ contour, the point compactification of the real axis, or in other words, the real slice of $\mathbb{CP}^1$ for $w_i$. Note that the second term above is also identical to the discriminant of the original quadric wrst $x_i$
\begin{equation}
X_{(i)}Q^{(i)}X_{(i)}=\text{Dis}(XQX,x_i).
\end{equation}

This $S^1$ integration can be generically written into a linear summation of the form
\begin{equation}
\int_{-\infty}^{+\infty}\frac{\mathrm{d}\chi_i\,\chi_i^m}{(\chi_i^2+X_{(i)}Q^{(i)}X_{(i)})^{\frac{n+k}{2}}}\equiv(X_{(i)}Q^{(i)}X_{(i)})^{\frac{n+k-m-1}{2}}\int_{-\infty}^{+\infty}\frac{\mathrm{d}\chi_i\,\chi_i^m}{(\chi_i^2+1)^{\frac{n+k}{2}}}.
\end{equation}
(Here we assume $\sqrt{X_{(i)}Q^{(i)}X_{(i)}}$ is real.) The remaining integral on RHS is merely some constant, and vanishes for odd $m$. The case of even $m$ divides into two situations. When $n+k$ is even it is a rational number times $\pi$. When $n+k$ is odd it is merely a rational constant. These seem to indicate that the $S^1$ contour integration is computing certain discontinuity in the case of even $n+k$, but some part of the original integral in the case of odd $n+k$.

By definition the action of the $S^1$ contour on the quadric is guaranteed to be the projection of $Q$ through the vertex $V_i$, following similar argument as in the previous subsection. It is also true in general that
\begin{equation}\label{eq:pipjcomposition}
\mathfrak{p}_{(i)}\mathfrak{p}_{(j)}=\mathfrak{p}_{(j)}\mathfrak{p}_{(i)}=\mathfrak{p}_{(ij)}=\mathfrak{p}_{(ji)},\qquad\forall i,j,
\end{equation}
either as algebraic operations on the matrix or as geometric projections of the curve. It turns out that this composition relation also holds as operations on the integrals. In other words, for a generic $E_{n,k}$ integral the result from first applying a $S^1$ contour integration in the $(i)$ direction and then another in the $(j)$ direction (or alternatively first $(j)$ and then $(i)$) is the same as that from a single $S^2$ contour in the $(ij)$ direction. In particular, this implies the following relations for the symbols of the elementary integrals
\begin{equation}
\begin{cases}
\mathcal{S}e_n\supset\underbrace{\mathcal{S}e_{n}^{(i)}\supset\mathcal{S}e_{n}^{(ij)}}_{\text{same length}},&\text{even }n,\\
\underbrace{\mathcal{S}e_n\supset\mathcal{S}e_{n}^{(i)}}_{\text{same length}}\supset\mathcal{S}e_{n}^{(ij)},&\text{odd }n,
\end{cases}
\end{equation}
where $e_{n}^{(i)}\equiv\mathfrak{p}_{(i)}e_n$. 

\begin{figure}[h]
\begin{center}
\begin{tikzpicture}
\begin{scope}[xshift=-4cm]
\draw [black,thick,dotted] (0,0) -- (5,0);
\coordinate [label=-90:$0$] (u1) at (3,0);
\coordinate [label=-90:$-z$] (u2) at (4,0);
\coordinate [label=-90:$-1$] (u3) at (1,0);
\filldraw [black] (u1) circle [radius=1pt];
\filldraw [ProcessBlue] (u2) circle [radius=1.5pt];
\filldraw [ProcessBlue] (u3) circle [radius=1.5pt];
\draw [OrangeRed,very thick] (u2) circle [radius=.2];
\draw [OrangeRed,very thick,->] ($(u2)-(.2,0)$) arc [start angle=180, end angle=80,radius=.2];
\end{scope}
\begin{scope}[xshift=4cm]
\draw [black,thick,dotted] (0,0) -- (5,0);
\coordinate [label=-90:$0$] (v1) at (3,0);
\coordinate [label=-90:$-z$] (v2) at (4,0);
\coordinate [label=-90:$-1$] (v3) at (1,0);
\filldraw [black] (v1) circle [radius=1pt];
\filldraw [ProcessBlue] (v2) circle [radius=1.5pt];
\filldraw [ProcessBlue] (v3) circle [radius=1.5pt];
\draw [OrangeRed,very thick,dashed] (5,0) arc [start angle=0,end angle=-180, radius=2.5];
\draw [OrangeRed,very thick] (0,0) -- ($(v3)-(.2,0)$) arc [start angle=180, end angle=360, radius=.2] -- ($(v2)-(.2,0)$) arc [start angle=180, end angle=0, radius=.2] -- (5,0);
\draw [OrangeRed,very thick,->] ($(v3)+(.2,0)$) -- +(1,0);
\end{scope}
\node [anchor=center] at (2.5,0) {\huge $\Rightarrow$};
\end{tikzpicture}
\caption{Deformation of the ordinary residue contour into the new $S^1$ contour.}
\label{fig:S1deformation}
\end{center}
\end{figure}
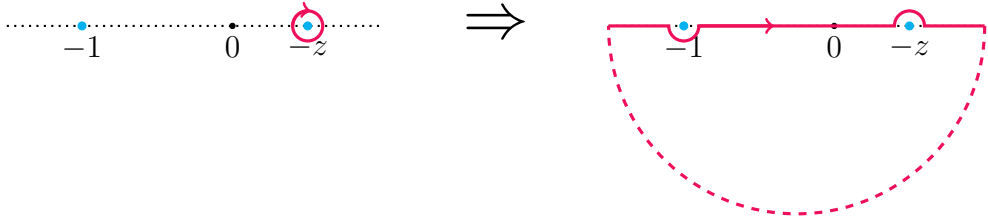

Let us now have a more careful look at what this $S^1$ really does. This can be very explicitly illustrated by the toy example of $\log(z)$ we started with in Section \ref{sec:dis1}. Recall that there the integrand has two poles, located at $x_1=-1$ and $x_1=-z$ respectively, and the discontinuity $2\pi i$ can be computed, e.g., by an ordinary $S^1$ contour wrapping around $x_1=-z$ clockwise. As illustrated in figure \ref{fig:S1deformation}, in this simple example we can in fact continuously deform the contour into a semi-circle in the lower half of the $x_1$ complex plane. Since the integrand falls off as $|x_1|^{-2}$ as $|x_1|\to\infty$, the contribution from the infinity vanishes, and so the original integration is identical to a contour along the real axis. Of course since the poles of the integration locates on the real axis as well, we need to take care of what side of the poles the contour passes by, which is analogous to the role of the $i\epsilon$ prescription in the usual Feynman integrals in momentum space.

In actual computation this contour is very similar to that of the original simplex contour, since we are just evaluating
\begin{equation}
\log\left(\frac{x_1+1}{x_1+z}\right)\bigg|_{-\infty}^{+\infty},
\end{equation}
but the non-trivial result arises from the monodromy that the deformed contour picks up. Now it is obvious that this contour is equivalent to the $S^1$ contour we defined previously, in this particular case of $e_2$ integral, since they are just related by a further shift of the variable $x_1\mapsto x_1-\frac{z+1}{2}$.

The virtue of our prescribed $S^1$ contour is that in practice it allows us to ignore the detailed profile of the poles of the integrand as well as the slight deformation of the contour around them, as these are automatically taken careful of by the property of a resulting $\arctan$ function, so that the computation can be performed very straightforwardly (assuming that the parameters are in a proper range). This also make the contour specification less confusing as we deals with integrals in higher dimensions. Furthermore, as we already observed in the explicit definition of the $S^2$ contours, the resulting prescription directly generalizes.

\begin{figure}[h]
\begin{center}
\begin{tikzpicture}
\draw [black,thick,dotted] (0,0) -- (5,0);
\coordinate [label=-90:$0$] (u1) at (3,0);
\coordinate [label=-90:$-z$] (u2) at (4,0);
\coordinate [label=-90:$-1$] (u3) at (1,0);
\coordinate [label=180:$\infty$] (inf1) at (-.5,-.1);
\coordinate [label=0:$\infty'$] (inf2) at (5.5,.1);
\filldraw [black] (u1) circle [radius=1pt];
\filldraw [ProcessBlue] (u2) circle [radius=1.5pt];
\filldraw [ProcessBlue] (u3) circle [radius=1.5pt];
\draw [OrangeRed,very thick] (inf1) -- ($(u3)+(0,-.1)$) .. controls ($(u3)+(.5,-.1)$) and (2.5,0) .. (2.5,0);
\draw [OrangeRed,very thick,dashed] (2.5,0) .. controls (2.5,0) and ($(u2)+(-.5,.1)$) .. ($(u2)+(0,.1)$) -- (inf2);
\filldraw [OrangeRed] (inf1) circle [radius=1.5pt];
\filldraw [OrangeRed] (inf2) circle [radius=1.5pt];
\draw [ProcessBlue,thick,decoration={zigzag,segment length=4pt, amplitude=1.5pt},decorate] (u3) -- (u2);
\end{tikzpicture}
\caption{$S^1$ contour in the case of odd $n+k$.}
\label{fig:S1odd}
\end{center}
\end{figure}
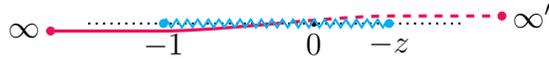

For integrals where $n+k$ is odd, in analogy with the above example let us assume the integrand again contains singular points at $x_1=-1$ and at $x_1=-z$. But these are not branch points instead of the poles. The resulting analogue of the $S^1$ contour is not an actual $S^1$ but a curve whose ends are anchored at the infinity located at different Riemann sheets, as shown in Figure \ref{fig:S1odd}. Correspondingly the integration yields another algebraic function.

\subsection{$S^2$ fibrated over $S^1$}\label{sec:S2fibration}

As mentioned in \eqref{eq:pipjcomposition} a sequence of two $S^1$ contours in general leads to the same result as produced by a single $S^2$ contour. This is simple to understand by showing how the contours are connected.

Consider an $S^1$ in direction $(i)$ followed by an $S^1$ in direction $(j)$. This is equivalent to first affine transform $x_i,x_j$ to bring the integral into
\begin{equation}\label{eq:twochi}
\int_{\Delta_{(ij)}}\measure{X_{(ij)}}{n-3}\int\mathrm{d}w_i\mathrm{d}w_j\frac{T[\chi_i,\chi_j,X_{(ij)}]}{(\chi_i^2+\chi_j^2+X_{(ij)}Q^{(ij)}X_{(ij)})^{\frac{n+k}{2}}}.
\end{equation}
And then the total integration contour for $\chi_i,\chi_j$, according to the definition for the $S^1$ contours, is just the entire real slice associated to these two variables. Of course in the region where $X_{(ij)}Q^{(ij)}X_{(ij)}<0$ one has to avoid singularities along the contour using the prescription provided in the previous subsection; this is also equivalent to introducing an $i\epsilon$ deformation to the integration denominator
\begin{equation}
\chi_i^2+\chi_j^2+X_{(ij)}Q^{(ij)}X_{(ij)}-i\epsilon.
\end{equation}
As directly defined, the topology of this contour is $S^1\times I$ (since the second ``$S^1$'' does not really close into a circle as pointed out previously). However, it is obvious that the contribution from the infinity is trivial, and so we can identify the entire region where $\chi_i,\chi_j\to\infty$, resulting in an $S^2$. From this point of view, the two $S^1$ contours exactly provides a fibration of the $S^2$, with a fixed point at infinity.

By a further coordinate transformation $\chi_i=\frac{w_i+w_j}{2}$, $\chi_j=\frac{w_i-w_j}{2i}$, the above expression obvious equals that for our original definition of the $S^2$ contour.

\subsection{Feynman integrals in odd dimensional spacetime}\label{sec:oddspacetime}

The discussion in the previous subsection immediately allows us to generalize the discontinuity analysis to Feynman integrals in odd spacetime dimensions, or more generally, any $E_{n,k}$ integral with odd $n+k$.

Here let us illustrate this for $e_n$ integrals with odd $n$. Apart from the scalar $d$-gons in $d$ dimensions, all the other cases necessarily involve a non-trivial tensor numerator in the Feynman parametrization. But this adds no more difficulty, because the decomposition procedure we discussed before applies regardless of the specific exponent in the denominator, and hence the decomposition on to $o$ integrals is always valid.

The strategy is to start by applying a single $S^1$ contour in some direction $(i)$. We have confirmed previously that this leads to a new integral $e_{n}^{(i)}$ whose symbol $\mathcal{S}e_{n-1}^{(i)}$ can be determined using the method prescribed in Section \ref{sec:dis1}. The quadric $Q^{(i)}=\mathfrak{p}_{(i)}Q$ comes as the projection of $Q$ through $V_i$. Recall that the equivalence relation $\overline{\rho_{2k-1}\rho_{2k}}=\underline{\rho_{2k-1}\rho_{2k}}$ proved in Section \ref{subsec:eqdis2dif} obviously holds for half-integral $k$ as well. This indicates that $\mathcal{S}e_{n}^{(i)}$ has to be contained inside $\mathcal{S}e_n$. Hence we conclude that
\begin{equation}\label{eq:enodddecomposition}
\mathcal{S}e_n=\frac{1}{n-1}\sum_{i=1}^n\mathcal{S}e_{n}^{(i)},\qquad\text{odd }n.
\end{equation}

Very explicitly, we observe that the $S^1$ contour in this case not only computes no discontinuities but also performs the role as extracting part of the original integral. This relation strongly suggests that there should be a corresponding geometric identity between the original simplex contour and the summation of the $n$ $S^1$ contours. 

\begin{figure}[h]
\begin{center}
\begin{tikzpicture}
\begin{scope}[xshift=3cm,yshift=2.5cm]
\fill [Orange,opacity=.2] (0,0) -- (2,0) arc [start angle=0,end angle=90,radius=2];
\draw [black,very thick,dashed] (110:2) .. controls (110:1.5) and ($(200:.5)+(110:.5)$) .. (200:.5) .. controls ($(200:.5)+(290:.5)$) and (-40:1.5) .. (-40:2);
\draw [black,very thick,dashed] (140:2) .. controls (140:1.5) and ($(200:1)+(110:.5)$) .. (200:1) .. controls ($(200:1)+(290:.5)$) and (-70:1.5) .. (-70:2);
\draw [Green] (0,1.8) -- +(2,0);
\draw [Green] (0,1.4) -- +(2,0);
\draw [Green] (0,1) -- +(2,0);
\draw [Green] (0,.6) -- +(2,0);
\draw [Green] (0,.2) -- +(2,0);
\draw [Green] (0,-.2) -- +(-2,0);
\draw [Green] (0,-.6) -- +(-2,0);
\draw [Green] (0,-1) -- +(-2,0);
\draw [Green] (0,-1.4) -- +(-2,0);
\draw [Green] (0,-1.8) -- +(-2,0);
\draw [ProcessBlue] (.8,2) -- (2,.8);
\draw [ProcessBlue] (.4,2) -- (2,.4);
\draw [ProcessBlue] (0,2) -- (2,0);
\draw [ProcessBlue] (0,1.6) -- (2,-.4);
\draw [ProcessBlue] (0,1.2) -- (2,-.8);
\draw [ProcessBlue] (0,.8) -- (2,-1.2);
\draw [ProcessBlue] (0,.4) -- (2,-1.6);
\draw [ProcessBlue] (0,0) -- (2,-2);
\draw [ProcessBlue] (0,-.4) -- (1.6,-2);
\draw [ProcessBlue] (0,-.8) -- (1.2,-2);
\draw [ProcessBlue] (0,-1.2) -- (.8,-2);
\draw [ProcessBlue] (0,-1.6) -- (.4,-2);
\draw [OrangeRed] (-2,.8) -- (-.8,2);
\draw [OrangeRed] (-2,.4) -- (-.4,2);
\draw [OrangeRed] (-2,0) -- (0,2);
\draw [OrangeRed] (-1.6,0) -- (.4,2);
\draw [OrangeRed] (-1.2,0) -- (.8,2);
\draw [OrangeRed] (-.8,0) -- (1.2,2);
\draw [OrangeRed] (-.4,0) -- (1.6,2);
\draw [OrangeRed] (0,0) -- (2,2);
\draw [OrangeRed] (.4,0) -- (2,1.6);
\draw [OrangeRed] (.8,0) -- (2,1.2);
\draw [OrangeRed] (1.2,0) -- (2,.8);
\draw [OrangeRed] (1.6,0) -- (2,.4);
\fill [white] (2,2) -- (0,2) arc [start angle=90,end angle=0,radius=2] -- cycle; 
\fill [white] (-2,2) -- (-2,0) arc [start angle=180,end angle=90,radius=2] -- cycle; 
\fill [white] (-2,-2) -- (-2,0) arc [start angle=180,end angle=270,radius=2] -- cycle; 
\fill [white] (2,-2) -- (2,0) arc [start angle=0,end angle=-90,radius=2] -- cycle;
\draw [black,thick] (0,0) circle [radius=2];
\draw [black,thick] (-2,0) -- (2,0);
\draw [black,thick] (0,-2) -- (0,2);
\draw [black,thick,->>] (2,0) arc [start angle=0,end angle=45,radius=2]; 
\draw [black,thick,->>>] (0,2) arc [start angle=90,end angle=135,radius=2]; 
\draw [black,thick,->>] (-2,0) arc [start angle=180,end angle=225,radius=2]; 
\draw [black,thick,->>>] (0,-2) arc [start angle=270,end angle=315,radius=2]; 
\end{scope}
\begin{scope}[xshift=-3cm,yshift=2.5cm]
\fill [Orange,opacity=.2] (0,0) -- (2,0) arc [start angle=0,end angle=90,radius=2];
\draw [black,very thick,dashed] (110:2) .. controls (110:1.5) and ($(200:.5)+(110:.5)$) .. (200:.5) .. controls ($(200:.5)+(290:.5)$) and (-40:1.5) .. (-40:2);
\draw [black,very thick,dashed] (140:2) .. controls (140:1.5) and ($(200:1)+(110:.5)$) .. (200:1) .. controls ($(200:1)+(290:.5)$) and (-70:1.5) .. (-70:2);
\draw [OrangeRed] (-2,.8) -- (-.8,2);
\draw [OrangeRed] (-2,.4) -- (-.4,2);
\draw [OrangeRed] (-2,0) -- (0,2);
\draw [OrangeRed] (-1.6,0) -- (.4,2);
\draw [OrangeRed] (-1.2,0) -- (.8,2);
\draw [OrangeRed] (-.8,0) -- (1.2,2);
\draw [OrangeRed] (-.4,0) -- (1.6,2);
\draw [OrangeRed] (0,0) -- (2,2);
\draw [OrangeRed] (.4,0) -- (2,1.6);
\draw [OrangeRed] (.8,0) -- (2,1.2);
\draw [OrangeRed] (1.2,0) -- (2,.8);
\draw [OrangeRed] (1.6,0) -- (2,.4);
\fill [white] (2,2) -- (0,2) arc [start angle=90,end angle=0,radius=2] -- cycle; 
\fill [white] (-2,2) -- (-2,0) arc [start angle=180,end angle=90,radius=2] -- cycle; 
\fill [white] (-2,-2) -- (-2,0) arc [start angle=180,end angle=270,radius=2] -- cycle; 
\fill [white] (2,-2) -- (2,0) arc [start angle=0,end angle=-90,radius=2] -- cycle;
\draw [black,thick] (0,0) circle [radius=2];
\draw [black,thick] (-2,0) -- (2,0);
\draw [black,thick] (0,-2) -- (0,2);
\draw [black,thick,->>] (2,0) arc [start angle=0,end angle=45,radius=2]; 
\draw [black,thick,->>>] (0,2) arc [start angle=90,end angle=135,radius=2]; 
\draw [black,thick,->>] (-2,0) arc [start angle=180,end angle=225,radius=2]; 
\draw [black,thick,->>>] (0,-2) arc [start angle=270,end angle=315,radius=2]; 
\draw [OrangeRed,ultra thick] (-2,0) -- (2,0) arc [start angle=0,end angle=180,radius=2];
\draw [OrangeRed,ultra thick,-latex] (1.5,.2) -- (.5,.2);
\draw [OrangeRed,ultra thick,-latex] (-.5,.2) -- (-1.5,.2);
\end{scope}
\begin{scope}[xshift=-3cm,yshift=-2.5cm]
\fill [Orange,opacity=.2] (0,0) -- (2,0) arc [start angle=0,end angle=90,radius=2];
\draw [black,very thick,dashed] (110:2) .. controls (110:1.5) and ($(200:.5)+(110:.5)$) .. (200:.5) .. controls ($(200:.5)+(290:.5)$) and (-40:1.5) .. (-40:2);
\draw [black,very thick,dashed] (140:2) .. controls (140:1.5) and ($(200:1)+(110:.5)$) .. (200:1) .. controls ($(200:1)+(290:.5)$) and (-70:1.5) .. (-70:2);
\draw [Green] (0,1.8) -- +(2,0);
\draw [Green] (0,1.4) -- +(2,0);
\draw [Green] (0,1) -- +(2,0);
\draw [Green] (0,.6) -- +(2,0);
\draw [Green] (0,.2) -- +(2,0);
\draw [Green] (0,-.2) -- +(-2,0);
\draw [Green] (0,-.6) -- +(-2,0);
\draw [Green] (0,-1) -- +(-2,0);
\draw [Green] (0,-1.4) -- +(-2,0);
\draw [Green] (0,-1.8) -- +(-2,0);
\fill [white] (2,2) -- (0,2) arc [start angle=90,end angle=0,radius=2] -- cycle; 
\fill [white] (-2,2) -- (-2,0) arc [start angle=180,end angle=90,radius=2] -- cycle; 
\fill [white] (-2,-2) -- (-2,0) arc [start angle=180,end angle=270,radius=2] -- cycle; 
\fill [white] (2,-2) -- (2,0) arc [start angle=0,end angle=-90,radius=2] -- cycle;
\draw [black,thick] (0,0) circle [radius=2];
\draw [black,thick] (-2,0) -- (2,0);
\draw [black,thick] (0,-2) -- (0,2);
\draw [black,thick,->>] (2,0) arc [start angle=0,end angle=45,radius=2]; 
\draw [black,thick,->>>] (0,2) arc [start angle=90,end angle=135,radius=2]; 
\draw [black,thick,->>] (-2,0) arc [start angle=180,end angle=225,radius=2]; 
\draw [black,thick,->>>] (0,-2) arc [start angle=270,end angle=315,radius=2]; 
\draw [Green,ultra thick] (0,2) -- (0,-2);
\draw [Green,ultra thick] (-2,0) -- (2,0);
\draw [Green,ultra thick,-latex] (.5,.2) -- (1.5,.2);
\draw [Green,ultra thick,-latex] (-1.5,-.2) -- (-.5,-.2);
\end{scope}
\begin{scope}[xshift=3cm,yshift=-2.5cm]
\fill [Orange,opacity=.2] (0,0) -- (2,0) arc [start angle=0,end angle=90,radius=2];
\draw [black,very thick,dashed] (110:2) .. controls (110:1.5) and ($(200:.5)+(110:.5)$) .. (200:.5) .. controls ($(200:.5)+(290:.5)$) and (-40:1.5) .. (-40:2);
\draw [black,very thick,dashed] (140:2) .. controls (140:1.5) and ($(200:1)+(110:.5)$) .. (200:1) .. controls ($(200:1)+(290:.5)$) and (-70:1.5) .. (-70:2);
\draw [ProcessBlue] (.8,2) -- (2,.8);
\draw [ProcessBlue] (.4,2) -- (2,.4);
\draw [ProcessBlue] (0,2) -- (2,0);
\draw [ProcessBlue] (0,1.6) -- (2,-.4);
\draw [ProcessBlue] (0,1.2) -- (2,-.8);
\draw [ProcessBlue] (0,.8) -- (2,-1.2);
\draw [ProcessBlue] (0,.4) -- (2,-1.6);
\draw [ProcessBlue] (0,0) -- (2,-2);
\draw [ProcessBlue] (0,-.4) -- (1.6,-2);
\draw [ProcessBlue] (0,-.8) -- (1.2,-2);
\draw [ProcessBlue] (0,-1.2) -- (.8,-2);
\draw [ProcessBlue] (0,-1.6) -- (.4,-2);
\fill [white] (2,2) -- (0,2) arc [start angle=90,end angle=0,radius=2] -- cycle; 
\fill [white] (-2,2) -- (-2,0) arc [start angle=180,end angle=90,radius=2] -- cycle; 
\fill [white] (-2,-2) -- (-2,0) arc [start angle=180,end angle=270,radius=2] -- cycle; 
\fill [white] (2,-2) -- (2,0) arc [start angle=0,end angle=-90,radius=2] -- cycle;
\draw [black,thick] (0,0) circle [radius=2];
\draw [black,thick] (-2,0) -- (2,0);
\draw [black,thick] (0,-2) -- (0,2);
\draw [black,thick,->>] (2,0) arc [start angle=0,end angle=45,radius=2]; 
\draw [black,thick,->>>] (0,2) arc [start angle=90,end angle=135,radius=2]; 
\draw [black,thick,->>] (-2,0) arc [start angle=180,end angle=225,radius=2]; 
\draw [black,thick,->>>] (0,-2) arc [start angle=270,end angle=315,radius=2]; 
\draw [ProcessBlue,ultra thick] (0,2) -- (0,-2) arc [start angle=-90,end angle=90,radius=2];
\end{scope}
\end{tikzpicture}
\caption{Relation between $\Delta$ and the $S^1$ contours}
\label{fig:contouridentity}
\end{center}
\end{figure}
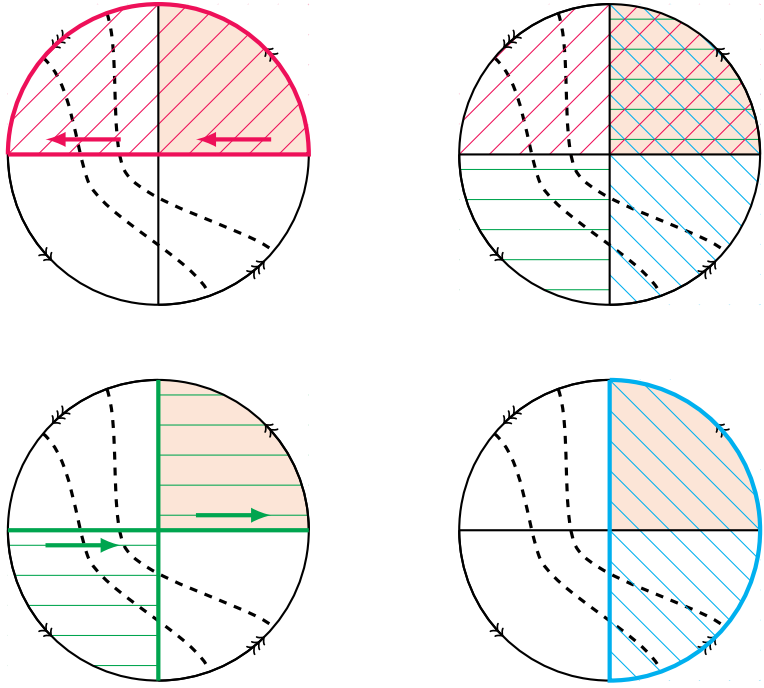

Here we provide an intuitive illustration for how this happens (without being completely rigorous), in the simple case of integrals in $\mathbb{CP}^2$. Figure \ref{fig:contouridentity} shows the entire real slice of $\mathbb{CP}^2$, which divides into four regions according to the relative signs of the three homogeneous variables $[x_1:x_2:x_3]$, and the original simplex contour covers the region with $[+:+:+]$.

Let us consider an $e_3$ integral. In order to make the integrals itself straightforwardly well-defined we assume that the quadric $Q$ imposes no singularities in the region $[+;+;+]$. For simplicity we temporarily focus on the case such that the quadric intersects every real axis at two generic positions. Without loss of generality this means that the real slice of the singularities should look like the dashed curves in Figure \ref{fig:contouridentity}.

Obviously the red region denotes the $S^1$ contour in $x_1$ (together with the original contour for the remaining variables), and the blue one for $x_2$ and the green one for $x_3$ respectively. The aim is to show that their summation is equivalent to the original simplex contour, which is exactly their overlap.

To verify this, let us first obverse that the integrand of $e_n$ with odd $n$ is actually the differentiation of a lower form that is algebraic. This is essentially because a quadric is always rational, which mean we can always find a reparametrization of the coordinates such that the integrand is rationalized. Since the denominator of this new integrand has an integral power, it necessarily has the form
\begin{equation}
\frac{\measure{X}{n-1}\,(L'X)}{(XQ'X)^{\frac{n+1}{2}}},
\end{equation}
and so it is explicitly exact. Back in the original coordinate the lower form contains square root again imposed by the same quadric $Q$, and so still has branch cuts. Nevertheless, by Stokes' theorem we can localized the integrals on to the closed boundary of each colored region.

By a careful inspection of the boundary contours, we can easily observe, e.g., the overlap of the red boundary contour and the green one in region $[-:+:+]$ (i.e., the negative $x_1$ real axis) share the same orientation. Naively this two integrals does not cancel each other. However, note that the they are continued from the region $[+:+:+]$ from different directions: the red one is continued from $x_1>0$ to $x_1<0$, while the green one from $x_3>0$ to $x_3<0$, as indicated by the red and green arrows respectively. In the region $[-:+:+]$ as we pass through the singularities the integrand travels to a different Riemann sheet. Hence it is obvious that at every point of the overlap between the red and green contours in this region the integrand can be tuned to have exactly the opposite phase. As a result, although the two contours share the same orientation, the integrals turns out to cancel by phase.

The same phenomena holds similarly for all the other overlaps in $[-:+:+]$, $[-:-:+]$ and $[+:-:+]$. And so as we sum up the three terms the only contribution that remains are the boundary contours to the region $[+:+:+]$, which doubles the boundary for the original simplex contour. Thus we verify the relation \eqref{eq:enodddecomposition}.

The above arguments obviously generalizes to $e_n$ with higher $n$, where the prefactor $\frac{1}{n-1}$ can be easily understood. In the more general situation when the real slice of the quadric does not necessarily intersect with the real axis, the branch cuts still exist in the corresponding region, and so the above arguments remain valid.

\subsubsection{An example}

To illustrate the above observations let us study an explicit $e_3$ example
\begin{equation}
I=\int_\Delta\frac{\sqrt{-q}\,\measure{X}{2}}{(x_1^2+x_2^2+x_3^2+u_1x_2x_3+u_2x_3x_1+u_3x_1x_2)^{\frac{3}{2}}},
\end{equation}
where $q=\frac{1}{4}(4-u_1^2-u_2^2-u_3^2+u_1u_2u_3)$. Here we choose values $u_1,u_2,u_3>2$ such that $q<0$, e.g., $u_1=29,u_2=3,u_3=11$. The result is
\begin{equation}
I=\frac{1}{2}\left(\log\left(\frac{-2u_1+u_2u_3-4\sqrt{-q}}{-2u_1+u_2u_3+4\sqrt{-q}}\right)+\text{permutations}\right),
\end{equation}
as one can check numerically.

Let us try the $S^1$ contours. An $S^1$ contour in the $(1)$ direction leads to
\begin{equation}
\begin{split}
\int_{\Delta_{(1)}}\langle X_{(1)}\mathrm{d}X_{(1)}\rangle&\int_{-\infty}^{+\infty}\mathrm{d}x_1\frac{\sqrt{-q}}{(x_1^2+(1-\frac{u_3^2}{4})x_2^2+(1-\frac{u_2^2}{4})x_3^2+(u_1-\frac{u_2u_3}{2})x_2x_3)^{\frac{3}{2}}}\\
&=\log\left(\frac{-2u_1+u_2u_3-4\sqrt{-q}}{-2u_1+u_2u_3+4\sqrt{-q}}\right).
\end{split}
\end{equation}
We have similar results for the other two choices of $S^1$ contours. Explicitly, we find in this case that
\begin{equation}
e_3=\frac{1}{2}\sum_{i=1}^3e_{3}^{(i)}.
\end{equation}

\subsection{Generalized projection contours}

The previous discussions on the $S^2$ and $S^1$ contours immediately suggest an investigation of the generalization of projections $\mathfrak{p}$ to that through any boundaries of the simplex contour with codimension higher than 1. 

\subsubsection{As algebraic operations}

Focusing on the $e_n$ integrals, the generalized projection contours continue to play as algebraic operations on the quadric $Q$. In general we specify a subset of labels $S\subset\{1,2,\ldots,n\}$, of cardinality $|S|<n-1$, and consider a projection through the boundary $H_{(S)}$, which has dimension $|S|-1$. For convenience we can use the hyperplane defined by the remaining vertices, i.e., $H_{(S^{\rm c})}$, to parametrize the space after projection, which is of dimension $n-|S|-1$. Again the projection leads to a configuration where the new contour $\Delta^{(S)}$ is just the corresponding boundary of the original simplex in $H_{(S^{\rm c})}$, and the new quadric
\begin{equation}
\mathfrak{p}_{(S)}Q\equiv Q^{(S)}:\quad\{P\in H_{(S^{\rm c})}|H_{(S\cup\{P\})}\text{ is tangent to }Q\}.
\end{equation}
Here we use $\mathfrak{p}_{(S)}$ to refer to both the projection and the corresponding algebraic operation \eqref{operationp} on the matrix $Q$ (in the canonical frame). This leads to an integral
\begin{equation}
e_n^{(S)}\equiv \mathfrak{p}_{(S)}e_n=\int_{\Delta^{(S)}}\frac{\measure{X_{(S^{\rm c})}}{n-1-|S|}}{(X_{(S^{\rm c})}Q^{(S)}X_{(S^{\rm c})})^{\frac{n-|S|}{2}}},
\end{equation}
where we use $X_{(S^{\rm c})}$ to denote coordinates on $H_{(S^{\rm c})}$. Obviously, the projection $\mathfrak{p}$ enjoys the property
\begin{equation}\label{eq:generalizedprojectioncommutation}
\mathfrak{p}_{(S_1)}\mathfrak{p}_{(S_2)}=\mathfrak{p}_{(S_2)}\mathfrak{p}_{(S_1)}=\mathfrak{p}_{(S_1\cup S_2)},
\end{equation}
for arbitrary subsets $S_1$ and $S_2$ such that $S_1\cap S_2=\varnothing$.

At the level of explicit symbols the effects of these projections are all certain truncation of $\mathcal{S}e_n$. When $n$ is even
\begin{equation}
\mathcal{S}\mathfrak{p}_{(S)}e_n=\begin{cases}\underset{\rho:\{\rho_1,\ldots,\rho_{|S|}\}=S}{\sum}\underline{\rho_{|S|+1}\rho_{|S|+2}}\otimes\cdots\otimes\underline{\rho_{n-1}\rho_{n}}&|S|\text{ even},\\\underset{\rho:\{\rho_1,\ldots,\rho_{|S|+1}\}\supset S}{\sum}\underline{\rho_{|S|+2}\rho_{|S|+3}}\otimes\cdots\otimes\underline{\rho_{n-1}\rho_{n}}&|S|\text{ odd}.\end{cases}
\end{equation}
When $n$ is odd
\begin{equation}
\mathcal{S}\mathfrak{p}_{(S)}e_n=\begin{cases}\underset{\rho:\{\rho_1,\ldots\rho_{|S|}\}\supset S}{\sum}\underline{\rho_{|S|+2}\rho_{|S|+3}}\otimes\cdots\otimes\underline{\rho_{n-1}\rho_{n}}&|S|\text{ even},\\\underset{\rho:\{\rho_1,\ldots,\rho_{|S|}\}=S}{\sum}\underline{\rho_{|S|+1}\rho_{|S|+2}}\otimes\cdots\otimes\underline{\rho_{n-1}\rho_{n}}&|S|\text{ odd}.\end{cases}
\end{equation}

\subsubsection{As integral operations}

For a generic quadric integral also with a non-trivial numerator, generalized projection contours can be explicitly defined such that the commutation relation \eqref{eq:generalizedprojectioncommutation} is satisfied. The construction is straighforward given the definition for the $S^2$ contour prescribed before. Although in the case when we have $|S|>2$ variables there is no unique way to pick out complex conjugate pairs, we can always affine transform the variables $x_i$ ($i\in S$) such that
\begin{equation}
XQX\longrightarrow\sum_{i\in S}\chi_i^2+X_{(S)}Q^{(S)}X_{(S)}
\end{equation}
(as long as $Q$ is generic), and then the generalized projection contour is specified by integrating the $\chi$ variables over $\mathbb{R}^{|S|}$ while preserving the original contours for the remaining variables (i.e., $\Delta_{(S)}$). In the special case of $|S|=2$ this is identical to \eqref{eq:twochi} and thus explicitly reduces to the $S^2$ contours for the discontinuity computations.

This operation is particularly useful in extracting non-trivial coefficients in front of different symbol terms when the given integral has uniform transcendental weight but is not pure. For this purpose, in the case of generic $Q$, all what is needed is to study all possible sets of labels of maximal size such that the corresponding projection contours still produce a non-vanishing result. Note that although naively the maximal size is $n-1$, the integral can turns out to vanish, but it that case it just means the actual transcendental weight has to be lower, because as we saw in the differentiation analysis the naive highest weight part may vanish if the tensor numerator is traceless wrst $Q^{-1}$.

\subsubsection{Case of degenerate quadric}

When the quadric is degenerate, it is not possible to project it all the way through $n-1$ vertices, as the quadric is doomed to degenerate into lines at some intermediate steps, beyond which the projection contour is no longer well-defined. As a result there exists some non-trivial maximal set of labels for the generalized projections. The integrals arising from such maximal projections are guaranteed to give results that are rational, and similar to the case of generic $Q$ these are the coefficients appearing in the original integral. Readers interested in the detailed analysis can refer to Appendix \ref{app:projectdegenerateQ}.

\subsection{Geometry of individual symbol entries and integrability}\label{sec:symbolentrygeometry}

By definition each symbol entry reflects certain logarithmic singularity. This fact can manifestly be visualized as a single log in the geometry. Consider some $o_m$ integral, and assume that we just want to extract the expression at the $k^{\rm th}$ entry of some symbol term, labeled by $(i_1i_2)$, which is preceded by entries with labels in a subset $S$ ($|S|$ being even for integral $m$ and odd otherwise). For this purpose we do a $\mathfrak{p}_{(S)}$ projection as prescribed before, landing on a quadric $Q^{(S)}$ and simplex $\Delta^{(S)}$ which is the boundary of $\Delta$ in $H_{(S^{\rm c})}$. Upon this we further quotient the configuration to the line passing through the images $\mathfrak{p}_{(S)}V_{i_1}$ and $\mathfrak{p}_{(S)}V_{i_2}$ (which in the current setup are just $V_{i_1},V_{i_2}$ themselves when embedded in the original space), resulting in a configuration of four points $\{v_{i_1},v_{i_2},q_1,q_2\}$ in $\mathbb{CP}^1$. $v_{i_1},v_{i_2}$ are the positions of $\mathfrak{p}_{(S)}V_{i_1},\mathfrak{p}_{(S)}V_{i_2}$, which defines a simplex in $\mathbb{CP}^1$, and $q_1,q_2$ are the two intersection points of $Q^{(S)}$ with $H_{(\mathfrak{p}_{(S)}V_{i_1},\mathfrak{p}_{(S)}V_{i_2})}$, which defines a new quadric. We then have the identification
\begin{equation}
\log(i_1i_2)=\int_{v_{i_1}}^{v_{i_2}}\frac{\sqrt{\det(q_1q_2)}\,\langle X\mathrm{d}X\rangle}{(q_1X)(q_2X)}.
\end{equation}

Now we can provide a simple proof of the integrablity of the symbols as computed from either the differentiation or the discontinuity method, by associating it to quadric integrals in $\mathbb{CP}^3$. It suffices to just focus on the elementary integrals $e_n$. Recall that previously we pointed out in \eqref{seedintegrability} the integrability is actually seeded in a finer condition
\begin{equation}
\sum_{S_4}\mathrm{d}\log(\underline{\rho_{2k-1}\rho_{2k}})\wedge\mathrm{d}\log(\underline{\rho_{2k+2}\rho_{2k+2}})=0,
\end{equation}
for each choice of fixed set of labels $\{\rho_{2k-1},\rho_{2k},\rho_{2k+1},\rho_{2k+2}\}$. Note that for any such choice, the combination
\begin{equation}
\sum_{S_4}\mathrm{d}\underline{\rho_{2k-1}\rho_{2k}}\otimes\underline{\rho_{2k+2}\rho_{2k+2}}
\end{equation}
is exactly the symbol of an $e_4$ integral, induced by a projection $\mathfrak{p}{(S)}\equiv\mathfrak{p}_(\rho_1\ldots\rho_{2k-2})$ and then restricting to the subspace $H_{(\mathfrak{p}_{(S)}V_{\rho_{2k-1}},\mathfrak{p}_{(S)}V_{\rho_{2k}},\mathfrak{p}_{(S)}V_{\rho_{2k+1}},\mathfrak{p}_{(S)}V_{\rho_{2k+2}})}$. Hence the refined condition \eqref{seedintegrability} reduces to the integrability condition for a generic $e_4$ integral.

Work again in the canonical frame. Note that the symbols remain invariant under overall rescaling of any individual row and corresponding column (which is a simple consequence of the rescaling invariance of the ratio of roots), and so without loss of generality that the $Q_{ii}=1$ ($i=1,2,3,4$) for the quadric in a generic $e_4$. One can verify that
\begin{equation}
\begin{split}
\verified{}
\frac{1}{4}\mathrm{d}\log(\underline{12})\wedge\mathrm{d}\log(\underline{34})=&\mathrm{d}Q_{12}\wedge\Big(-\frac{\det Q_{\widehat{1},\widehat{4}}}{\det Q_{\widehat{4},\widehat{4}}}\mathrm{d}Q_{13}+\frac{\det Q_{\widehat{1},\widehat{3}}}{\det Q_{\widehat{3},\widehat{3}}}\mathrm{d}Q_{14}\\
&+\frac{\det Q_{\widehat{2},\widehat{4}}}{\det Q_{\widehat{2},\widehat{2}}}\mathrm{d}Q_{23}-\frac{\det Q_{\widehat{2},\widehat{3}}}{\det Q_{\widehat{3},\widehat{3}}}\mathrm{d}Q_{24}+\mathrm{d}Q_{34}\Big),
\end{split}
\end{equation}
while the other terms can be easily obtained by permuting the labels. It is then obvious that each $\mathrm{d}Q_{ij}\wedge\mathrm{d}Q_{kl}$ receives contribution from only two symbol terms, which cancel purely due to the antisymmetry of the wedge product. This confirms the integrability of the symbols for generic $e_4$ integrals, thus proving \eqref{seedintegrability}.

\section{Unitarity}\label{sec:unitarity}

In the last three sections we have been discussing the discontinuities of Feynman integrals as arising from spherical contours and their generalizations, as well as their corresponding geometric interpretation as projections in the Feynman parameter space. In this section we explore their connection to the (generalized) unitarity cut on Feyman diagrams. Here we focus on the scalar diagrams $I^\phi_{n,d}$.

\subsection{Feynman parametrization of the cut scalar $n$-gon in $d$ dimensions}\label{sec:FPofcuts}

We first derive the Feynman parametrization of an arbitrary scalar $n$-gon in $d$ dimensions, possibly with massive loop propagators. 

Working in the embedding space $\mathbb{R}^{1,d+1}$, we introduce one parameter for each un-cut propagator. Assume that we cut the propagators labeled by $i$ and $j$. Here we assume that the cut propagators are massless. The cut integral reads
\begin{equation}\label{eq:Indij}
\begin{split}
I_{n,d}^{(ij)}&=\int_0^\infty\langle X\mathrm{d}^{n-3}X\rangle\int\frac{\mathrm{d}^{d+2}Y}{\text{vol.}\mathrm{GL}(1)}\frac{(Y\cdot Y_\infty)^{n-d}\,\delta(Y\cdot Y)\delta(Y\cdot Y_i)\delta(Y\cdot Y_j)}{(Y\cdot \sum_{k\neq i,j}x_kY_k)^{n-2}}\\
&=\int_0^\infty\langle X\mathrm{d}^{n-3}X\rangle\left(Y_\infty^I\frac{\partial}{\partial W^I}\right)^{n-d}\dot{I}_{n,d}^{(ij)},
\end{split}
\end{equation}
where $W=\sum_{k\neq i,j}x_kY_k$ and
\begin{equation}
\dot{I}_{n,d}^{(ij)}=\int\frac{\mathrm{d}^{d+2}Y}{\text{vol.}\mathrm{GL}(1)}\frac{\delta(Y\cdot Y)\delta(Y\cdot Y_i)\delta(Y\cdot Y_j)}{(Y\cdot W)^{d-2}}.
\end{equation}
Note that the integrand in $\dot{I}_{n,d}^{(ij)}$ manifestly rescaling, and so it is expected that the result has to be a combination of $\{W\cdot W,W\cdot Y_1,W\cdot Y_2,Y_1^2,Y_1\cdot Y_2,Y_2^2\}$ with correct weights. 

To work out the detailed result, note that the operator
\begin{equation}
Y_i^IY_{iJ}+Y_j^IY_{iJ}-\delta^I_J\,Y_i\cdot Y_j
\end{equation}
projects any vector onto the subspace constrained by $\delta(Y\cdot Y_i)\delta(Y\cdot Y_j)$ (recall that $Y_i^2=Y_j^2=0$). So the above integral can be equally written as
\begin{equation}
\dot{I}_{n,d}^{(ij)}=\int\frac{\mathrm{d}^{d+2}Y}{\text{vol.}\mathrm{GL}(1)}\frac{\delta(Y\cdot Y)\,(Y_i\cdot Y_j)^{d-3}}{(Y\cdot Y_i\,Y_j\cdot W+Y\cdot Y_j\,Y_i\cdot W-Y\cdot W\,Y_i\cdot Y_j)^{d-2}},
\end{equation}
where the numerator as well as the exponent in the denominator is uniquely fixed by conformal weights.
Borrowing the result for a $d$-gon in $d$ dimensions we conclude
\begin{equation}
\dot{I}_{n,d}^{(ij)}=\frac{(Y_i\cdot Y_j)^{\frac{d-4}{2}}}{(Y_i\cdot Y_j\,W\cdot W-2W\cdot Y_i\,W\cdot Y_j)^{\frac{d-2}{2}}}.
\end{equation}
Then $I_{n,d}^{(ij)}$ is easily obtained by repeatedly applying differential operators on $\dot{I}_{n,d}^{(ij)}$.

More generally when the two cut propagators have arbitrary masses, correspondingly we use $\mathcal{Y}_i=(1,y_i^2+m_i^2,y_i^\mu)$ instead of $Y_i$ and similarly for the $j^{\rm th}$ propagator. In this case we $\mathcal{Y}_i^2$ and $\mathcal{Y}_j^2$ are non-zero. Correspondingly the projection operator has to be modified, and we have
\begin{equation}
\dot{I}_{n,d}^{(ij)}=\int\frac{\mathrm{d}^{d+2}Y}{\text{vol.}\mathrm{GL}(1)}\frac{\delta(Y\cdot Y)\,((\mathcal{Y}_i\cdot \mathcal{Y}_j)^2-\mathcal{Y}_i\cdot \mathcal{Y}_i\,\mathcal{Y}_j\cdot \mathcal{Y}_j)^{\frac{2d-5}{2}}}{(Y\cdot\widetilde{W})^{d-2}},
\end{equation}
where
\begin{equation}
\begin{split}
\widetilde{W}=&\mathcal{Y}_i\,\mathcal{Y}_i\cdot \mathcal{Y}_j\,\mathcal{Y}_j\cdot W-\mathcal{Y}_i\,\mathcal{Y}_j\cdot \mathcal{Y}_j\,\mathcal{Y}_i\cdot W+\mathcal{Y}_j\,\mathcal{Y}_i\cdot \mathcal{Y}_j\,\mathcal{Y}_i\cdot W\\
&-\mathcal{Y}_j\,\mathcal{Y}_i\cdot \mathcal{Y}_i\,\mathcal{Y}_j\cdot W+W\,\mathcal{Y}_i\cdot \mathcal{Y}_i\,\mathcal{Y}_j\cdot \mathcal{Y}_j-W\,(\mathcal{Y}_i\cdot \mathcal{Y}_j)^2.
\end{split}
\end{equation}
In the end we obtain
\begin{equation}
\dot{I}_{n,d}^{(ij)}=\frac{((\mathcal{Y}_i\cdot \mathcal{Y}_j)^2-\mathcal{Y}_i^2\,\mathcal{Y}_j^2)^{\frac{d-3}{2}}}{\left(W\cdot W\,((\mathcal{Y}_i\cdot \mathcal{Y}_j)^2-\mathcal{Y}_i^2\,\mathcal{Y}_j^2)+(W\cdot \mathcal{Y}_i)^2\mathcal{Y}_j^2+(W\cdot \mathcal{Y}_j)^2\mathcal{Y}_i^2-2\,W\cdot \mathcal{Y}_i\,W\cdot \mathcal{Y}_j\,\mathcal{Y}_i\cdot \mathcal{Y}_j\right)^{\frac{d-2}{2}}}.
\end{equation}

\subsection{Comparison with the projection of Feynman parametrization}

For simplicity let us return to the cases with massless cut propagators.

We can actually rewrite the above expression in a slightly better way, into
\begin{equation}
\dot{I}_{n,d}^{(ij)}=\frac{1}{(Y_i\cdot Y_j)(W^M\widetilde{\eta}_{MN}W^N)^{\frac{d-2}{2}}},
\end{equation}
with
\begin{equation}
\widetilde{\eta}_{MN}=\eta_{MN}-\frac{Y_{iM}Y_{jN}+Y_{iN}Y_{jM}}{Y_i\cdot Y_j},
\end{equation}
where $\eta_{MN}$ is the flat space metric, and $Y_{iM}=\eta_{MN}Y_i^N$.  Inserting this into \eqref{eq:Indij} we obtain the explicit Feynman parametrization of a generic scalar diagram
\begin{equation}
I_{n,d}^{(ij)}=\int_{\Delta_{(ij)}}\frac{\measure{X_{(ij)}}{n-3}}{Y_i\cdot Y_j}\sum_{k=0}^{\lfloor\frac{n-d}{2}\rfloor}\frac{\left(\substack{n-d\\k}\right)\left(\substack{n-d-k\\k}\right)}{2^k}\frac{\Gamma(n-\frac{d}{2}-k-1)}{\Gamma(\frac{d}{2}-1)}\frac{(\frac{2}{Y_i\cdot Y_j})^k(-Y_\infty\tilde\eta W)^{n-d-2k}}{(W\tilde\eta W)^{n-\frac{d}{2}-k-1}}.
\end{equation}

On the other hand, we start from the Feynman parametrization for the $n$-gon in $d$ dimensions, and perform an $S^2$ contour integral in the $(ij)$ directly. We first do the transformation
\begin{equation}
x_i\mapsto\frac{x_i}{\sqrt{Y_i\cdot Y_j}}-\sum_{k\neq i,j}\frac{Y_i\cdot Y_k}{Y_i\cdot Y_j}x_k,\qquad
x_j\mapsto\frac{x_j}{\sqrt{Y_i\cdot Y_j}}-\sum_{k\neq i,j}\frac{Y_j\cdot Y_k}{Y_i\cdot Y_j}x_k,
\end{equation}
which brings the Feynman parametrization into the form
\begin{equation}
\int_{\Delta_{(ij)}}\measure{X_{(ij)}}{n-3}\int\frac{\mathrm{d}x_i\mathrm{d}x_j}{(Y_i\cdot Y_j)}\frac{\left(\frac{x_i+x_j}{\sqrt{Y_i\cdot Y_j}}+Y_\infty\widetilde{\eta}W\right)^{n-d}}{(x_ix_j+W\widetilde{\eta}W)^{n-\frac{d}{2}}},
\end{equation}
where the vector $W$ is the same as before. For the spherical contour we do the parametrization $x_i=r e^{i\phi}$ and $x_j=r e^{-i\phi}$. Obviously the only dependence on $\phi$ in the integrand is in the term $\frac{x_i+x_j}{\sqrt{Y_i\cdot Y_j}}$. Integrating $\phi$ away we obtain
\begin{equation}
\begin{split}
&2\pi\int_{\Delta_{(ij)}}\frac{\measure{X_{(ij)}}{n-3}}{Y_i\cdot Y_j}\int_0^\infty\mathrm{d}(r^2)\frac{\sum_{k=0}^{\lfloor\frac{n-d}{2}\rfloor}\left(\substack{n-d\\2k}\right)\left(\substack{2k\\k}\right)\frac{(Y_\infty\widetilde{\eta}W)^{n-d-2k}}{(Y_i\cdot Y_j)^k}r^{2k}}{(r^2+W\widetilde{\eta}W)^{n-\frac{d}{2}}}\\
=&2\pi\int_{\Delta_{(ij)}}\frac{\measure{X_{(ij)}}{n-3}}{Y_i\cdot Y_j}                                                                                                                                                                                                                                                                                                                                                                                                                                                                                                                                                                                                                                                                                                                                                                                                                                                                                                                                                                                                                                                                                                                                                                                                                                                                                                                                                                                                                                                                                                                                                                                                                                                                                                                                                                                                                                                                                                                                                                                                                                                                                                                                                                                                                                                                                                                                                                                                                                                                                                                                                                                                                                                                                                                                                                                                                                                                                                                                                                                                                                                                                                                                                                                                                                                                                             \sum_{k=0}^{\lfloor\frac{n-d}{2}\rfloor}\left(\substack{n-d\\2k}\right)\left(\substack{2k\\k}\right)\frac{\Gamma(n-\frac{d}{2}-k-1)\Gamma(k+1)}{\Gamma(n-\frac{d}{2})}\frac{(Y_\infty\widetilde{\eta}W)^{n-d-2k}}{(Y_i\cdot Y_j)^k(W\tilde\eta W)^{n-\frac{d}{2}-k-1}}\\
=&\frac{(-1)^{n-d}2\pi\,\Gamma(\frac{d}{2}-1)}{\Gamma(n-\frac{d}{2})}I_{n,d}^{(ij)}.
\end{split}
\end{equation}
So up to an overall constant we explicitly verified that the discontinuity computed by the $S^2$ contour matches the Feynman parametrization of the cut diagram for any choice of unitarity cut.

\acknowledgments
We thank Tim Adamo, Jacob Bourjaily, Yvonne Geyer, Sasha Goncharov, Song He, Enrico Herrmann, Andrew Hodges, Yu-tin Huang, William Linch III, Julio Parra Martinez, Mauricio Romo, Jaroslav Trnka and Akshay Yelleshpur for useful discussions. NA-H and EYY are supported by the DOE under grant DE-SC0009988. EYY is in addition supported by a Carl P.~Feinberg Founders’ Circle Membership.

\appendix

\section{Notations and Conventions}\label{app:convention}

\flag{No further revisions.}

In this appendix we collect some notations and conventions that are used throughout this paper.

The integrals under study are in a complex projective space of arbitrary dimensions, $\mathbb{CP}^{n-1}$, parametrized by the homogeneous coordinates  $X\equiv[x_1:x_2:\ldots:x_n]$.  We also frequently consider subspace formed by intersections of hyperplanes $x_i=0$, $i\in S\subset\{1,2,\ldots,n\}$, with $0<|S|<n$; such subspace is a $\mathbb{CP}^{n-|S|-1}$, whose coordinates are naturally induced from the ambient space
\begin{equation}\label{app:subspaceS}
\fixed
X_{(S)}\equiv[x_1:\ldots:\widehat{x_{i_1}}:\ldots:\widehat{x_{i_{|S|}}}:\ldots:x_n],\quad S\equiv\{i_1,\ldots,i_{|S|}\},
\end{equation}
where the hats indicate the absence of the corresponding entries.

A top form in $\mathbb{CP}^{n-1}$ is always proportional to
\begin{equation}
\fixed
\measure{X}{n-1}\equiv\frac{\epsilon_{I_1,\ldots,I_n}}{(n-1)!}\,X^{I_1}\,\mathrm{d}X^{I_2}\wedge\cdots\wedge\mathrm{d}X^{I_n},\quad \epsilon_{12\ldots n}=1.
\end{equation}
More generally, to construct lower forms we also use a similar notation
\begin{equation}
\fixed
\form{W_1\ldots W_k}{X}{n-k}\equiv\frac{\epsilon_{I_1\ldots I_n}}{(n-k)!}\,W_1^{I_1}\cdots W_k^{I_k}\,\mathrm{d}X^{I_{k+1}}\wedge\cdots\wedge\mathrm{d}X^{I_n}.
\end{equation}
Note that this convention leads to an extra factor in the exterior derivative, e.g., in
\begin{equation}
\verified{}
\mathrm{d}_X\form{WX}{X}{n-2}=(n-1)\form{W}{X}{n-1}.
\end{equation}

It is natural to generalize this notation to the subspace defined in \eqref{app:subspaceS}. There the indices run in a subset $S$, and we choose the convention that the corresponding Levi-Civita symbol is positive for ascending indices; for example, if $S=\{1,3,4,7\}$, then $\epsilon_{1347}=+1$. This fixes the sign of localizing a lower form to a subspace, e.g.,
\begin{equation}
\fixed
\form{WX}{X}{n-2}\Big|_{x_i=0}=(-1)^{i-1}\,W_i\,\measure{X_{(i)}}{n-2}.
\end{equation}

In most of our discussions the integration contour is some generic simplex, denoted as $\Delta$ (or $\Delta_{n-1}$ if the dimension is to be emphasized). This is specified either by a set of $n$ vectors $V_i^I$ for its vertices or by a set of $n$ covectors $H_{iI}$ for its codim-1 boundaries. The vertices and codim-1 boundaries satisfy the relation (note the homogeneous coordinates are defined up to an overall scale)
\begin{equation}
V_i^I=\langle H_1\ldots\widehat{H_i}\ldots H_n\rangle^I,\quad
H_{iI}=\langle V_1\ldots\widehat{V_i}\ldots V_n\rangle_I.
\end{equation}

We encode the orientation of the simplex contour in the ordered sequence of its vertex labels and we assume the ascending sequence $[12\ldots n]$ to be the canonical one. We denote the codim-$k$ boundary lying on the intersection of $k$ hyperplanes $\{H_{i_1},\ldots,H_{i_k}\}$ by $\Delta_{(i_1\ldots i_k)}$, and we always assume the canonical orientation on it (and so we treat the subscripts to be symmetric). With this convention the bordism operator acts, e.g.,
\begin{equation}
\fixed
\partial\Delta=\sum_{i=1}^n(-1)^{i-1}\,\Delta_{(i)},
\end{equation}
and $\partial^2\equiv0$.

A projective integral in $\mathbb{CP}^{n-1}$ is by definition invariant under an arbitrary $\mathrm{PGL(n)}$ action. For an integral with a simplex contour we can thus always transform it into a frame where the contour is canonical, in the sense that its $i^{\rm th}$ boundary is defined by $x_i=0$. We call this frame the \textit{canonical frame}.

To make some computations more transparent, we also utilized a graphical notation for the Levi-Civita symbol, e.g.,
\begin{equation}
\fixed
\langle A_1\ldots A_n\rangle\equiv\epsilon_{I_1\ldots I_n}A_1^{I_1}\cdots A_n^{I_n}\equiv\parbox{2.5cm}{
\tikz{
\node [anchor=center] at (0,0) {$A_1$};
\node [anchor=center] at (.6,0) {$A_2$};
\node [anchor=center] at (2*.6,0) {$\cdots$};
\node [anchor=center] at (3*.6,0) {$A_n$};
\draw [black,thick] (0,.4) -- (3*.6,.4);
\draw [black,thick] (0,.4) -- +(0,-.15);
\draw [black,thick] (.6,.4) -- +(0,-.15);
\draw [black,thick] (3*.6,.4) -- +(0,-.15);
}},
\end{equation}
where each leg denotes a contraction between a covariant index and a contravariant index. While we keep in mind that the legs attached to each half-ladder are antisymmetric under exchange, we will leave the resulting possible overall sign implicit, which can be relatively easily fixed at the end. With this notation, for example, the usual determinant and inverse of a non-degenerate matrix $Q$ is represented as
\begin{equation}
\fixed
q\equiv\det Q\equiv\frac{1}{n!}\parbox{2.4cm}{
\tikz{
\node [anchor=center] at (0,0) {$Q$};
\node [anchor=center] at (.6,0) {$Q$};
\node [anchor=center] at (2*.6,0) {$\cdots$};
\node [anchor=center] at (3*.6,0) {$Q$};
\draw [black,thick] (0,.4) -- +(3*.6,0);
\draw [black,thick] (0,.4) -- +(0,-.15);
\draw [black,thick] (.6,.4) -- +(0,-.15);
\draw [black,thick] (3*.6,.4) -- +(0,-.15);
\draw [black,thick] (0,-.4) -- +(3*.6,0);
\draw [black,thick] (0,-.4) -- +(0,.15);
\draw [black,thick] (.6,-.4) -- +(0,.15);
\draw [black,thick] (3*.6,-.4) -- +(0,.15);
}},\qquad
Q^{-1}=\frac{1}{(n-1)!}\frac{\parbox{2cm}{
\tikz{
\node [anchor=center] at (.6,0) {$Q$};
\node [anchor=center] at (2*.6,0) {$\cdots$};
\node [anchor=center] at (3*.6,0) {$Q$};
\draw [black,thick] (0,.4) -- +(3*.6,0);
\draw [black,thick] (0,.4) -- +(0,-.15);
\draw [black,thick] (.6,.4) -- +(0,-.15);
\draw [black,thick] (3*.6,.4) -- +(0,-.15);
\draw [black,thick] (0,-.4) -- +(3*.6,0);
\draw [black,thick] (0,-.4) -- +(0,.15);
\draw [black,thick] (.6,-.4) -- +(0,.15);
\draw [black,thick] (3*.6,-.4) -- +(0,.15);
}}}{\det Q}.
\end{equation}
Also, the Schouten identity reads
\begin{equation}\label{app:eq:schouten}
\parbox{3.1cm}{
\tikz{
\node [anchor=center] at (-.6,0) {$A_0$};
\node [anchor=center] at (0,0) {$A_1$};
\node [anchor=center] at (.6,0) {$A_2$};
\node [anchor=center] at (2*.6,0) {$\cdots$};
\node [anchor=center] at (3*.6,0) {$A_n$};
\draw [red,thick] (-.6,.4) -- +(0,-.15);
\draw [black,thick] (0,.4) -- +(3*.6,0);
\draw [red,thick] (0,.4) -- +(0,-.15);
\draw [red,thick] (.6,.4) -- +(0,-.15);
\draw [red,thick] (3*.6,.4) -- +(0,-.15);
}}=\sum_{i=1}^n
\parbox{5.5cm}{
\tikz{
\node [anchor=center] at (-.6,0) {$A_i$};
\node [anchor=center] at (0,0) {$A_1$};
\node [anchor=center] at (.6,0) {$\cdots$};
\node [anchor=center] at (2*.6+.15,0) {$A_{i-1}$};
\node [anchor=center] at (3.5*.6,0) {$A_0$};
\node [anchor=center] at (4.5*.6+.15,0) {$A_{i+1}$};
\node [anchor=center] at (6*.6,0) {$\cdots$};
\node [anchor=center] at (7*.6,0) {$A_n$};
\draw [black,thick] (-.6,.4) -- +(0,-.15);
\draw [black,thick] (0,.4) -- +(7*.6,0);
\draw [black,thick] (0,.4) -- +(0,-.15);
\draw [black,thick] (2*.6,.4) -- +(0,-.15);
\draw [black,thick] (3.5*.6,.4) -- +(0,-.15);
\draw [black,thick] (4.5*.6,.4) -- +(0,-.15);
\draw [black,thick] (7*.6,.4) -- +(0,-.15);
}}.
\end{equation}
Equivalently we also say we apply a Schouten identity on the red legs on LHS above to obtain the RHS. Here again we are not keeping track of the signs.

\section{Aomoto Polylogarithms}\label{app:aomoto}

This appendix is devoted to the study of a generalization of polylogarithms, the \textit{Aomoto polylogarithms} \cite{aomoto1982,Goncharov.A.B.:2009tja}. The purpose is to should how a generic Aomoto polylogarithm comes about as a special case of the quadric integrals that we analyzed in the main body of the note.

First let us quickly review the definition and basic properties of these objects. 

Consider in $\mathbb{CP}^{n-1}$ a pair of simplices $\overline{\Delta}$ and $\underline{\Delta}$, $\overline{\Delta}$ specified by its vertices $\{V_i\}$ or boundaries $\{H_i\}$, and $\underline{\Delta}$ by its vertices $\{Z_i\}$ or boundaries $\{G_i\}$. The pair $(\overline{\Delta},\underline{\Delta})$ is called \textit{admissible} if they do not share any common faces of the same dimension.

Now for an admissible pair $(\overline{\Delta},\underline{\Delta})$ we take $\overline{\Delta}$ to be the integration contour as we did for the quadric integrals, and assign a top form $\Omega_{\underline{\Delta}}$ to $\underline{\Delta}$ that has logarithmic singularities on its codim-1 boundaries. $\Omega_{\underline{\Delta}}$ admits several equivalent definitions
\begin{equation}
\begin{split}
\Omega_{\underline{\Delta}}
&\equiv\mathrm{d}\log\left(\frac{G_1X}{G_nX}\right)\wedge\mathrm{d}\log\left(\frac{G_2X}{G_nX}\right)\wedge\cdots\wedge\mathrm{d}\log\left(\frac{G_{n-1}X}{G_nX}\right)\\
&\equiv\frac{\langle G_1G_2\ldots G_n\rangle\,\langle X\mathrm{d}X^{n-1}\rangle}{(G_1X)(G_2X)\cdots(G_nX)}
\equiv\frac{\langle Z_1Z_2\cdots Z_n\rangle^{n-1}\,\langle X\mathrm{d}^nX\rangle}{\langle XZ_2\cdots Z_n\rangle\cdots\langle XZ_1\cdots Z_{n-1}\rangle}.
\end{split}
\end{equation}
The Aomoto polylogarithm associated to this pair $(\overline{\Delta},\underline{\Delta})$ is defined as
\begin{equation}
\Lambda(\overline{\Delta},\underline{\Delta})=\int_{\overline{\Delta}}\Omega_{\underline{\Delta}}.
\end{equation}
This is known to be a pure function of transcendental weight $n-1$. The requirement of admissibility is to avoid divergence so as to guarantee that $\Lambda(\overline{\Delta},\underline{\Delta})$ is well-defined.

$\Lambda(\overline{\Delta},\underline{\Delta})$ is invariant under $(\overline{\Delta},\underline{\Delta})\mapsto\Lambda(g\overline{\Delta},g\underline{\Delta})$, $g\in\mathrm{PGL}(n)$. It is skew-symmetric in the orientation of both $\overline{\Delta}$ and $\underline{\Delta}$, and vanishes as long as any of them is degenerate. Also, due to the additive nature in both the contour simplex and the integrand simplex, the definition can naturally be extended to any pair of polytopes $(\overline{P},\underline{P})$ in $\mathbb{CP}^{n-1}$. For the time being we focus on a pair of simplices.

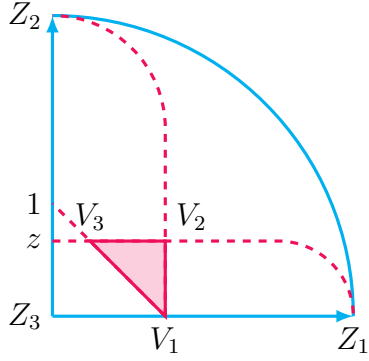
\begin{figure}[ht]
\begin{center}
\begin{tikzpicture}
\draw [ProcessBlue,very thick,-latex] (0,0) -- (4,0);
\draw [ProcessBlue,very thick,-latex] (0,0) -- (0,4);
\draw [ProcessBlue,very thick] (4,0) arc [start angle=0,end angle=90,radius=4];
\fill [OrangeRed,opacity=.2] (1.5,0) -- +(0,1) -- +(-1,1) -- cycle;
\draw [OrangeRed,very thick,dashed] (0,1) -- +(3,0) arc [start angle=90,end angle=0,radius=1];
\draw [OrangeRed,very thick,dashed] (1.5,0) -- +(0,2.5) arc [start angle=0,end angle=90,radius=1.5];
\draw [OrangeRed,very thick,dashed] (1.5,0) -- (0,1.5);
\draw [OrangeRed,very thick] (1.5,0) -- +(0,1) -- +(-1,1) -- cycle;
\node [anchor=north] at (4,0) {$Z_1$};
\node [anchor=east] at (0,4) {$Z_2$};
\node [anchor=east] at (0,0) {$Z_3$};
\node [anchor=east] at (0,1) {$z$};
\node [anchor=east] at (0,1.5) {$1$};
\node [anchor=north] at (1.5,0) {$V_1$};
\node [anchor=south west] at (1.5,1) {$V_2$};
\node [anchor=south] at (.5,1) {$V_3$};
\end{tikzpicture}
\caption{$(\overline\Delta,\underline\Delta)$ for $\mathrm{Li}_2(z)$.}
\label{app:fig:li2}
\end{center}
\end{figure}

In particular, all the classical polylogarithms are special Aomoto polylogarithms. As a simple example, consider the dilog
\begin{equation}
\text{Li}_2(z)=\int_{0\leq1-x_1\leq x_2\leq z}\frac{\mathrm{d}x_1}{x_1}\wedge\frac{\mathrm{d}x_2}{x_2}.
\end{equation}
This can be represented by an Aomoto polylogarithm where
\begin{align}
\overline{\Delta}:&\quad V_1=[1:0:1],\quad V_2=[1:z:1]\quad V_3=[1-z:z:1],\\
\underline{\Delta}:&\quad Z_i^I=\delta_i^I.
\end{align}
These are depicted in Figure \ref{app:fig:li2}, where the red shaded region indicates the contour $\overline{\Delta}$, while the blue lines indicates the boundaries of $\underline{\Delta}$.

\subsection{Differentiations}

To analyze the structure, without loss of generality let us vary the vertex $Z_1$
\begin{equation}\begin{split}
\mathrm{d}_{Z_1}\Lambda(\overline{\Delta},\underline{\Delta})
&=\int_{\overline{\Delta}}\frac{\measure{X}{n-1}\,\langle Z_1Z_2\cdots Z_n\rangle^{n-1}}{\langle XZ_2\cdots Z_n\rangle\cdots\langle XZ_1\cdots Z_{n-1}\rangle}\times\\
&\times\Big((n-1)\frac{\langle\delta Z_1Z_2\cdots Z_n\rangle}{\langle Z_1\cdots Z_n\rangle}-\frac{\langle X\delta Z_1Z_3\cdots Z_n\rangle}{\langle XZ_1Z_3\cdots Z_n\rangle}-\cdots-\frac{\langle X\delta Z_1 Z_2\cdots Z_{n-1}\rangle}{\langle X Z_1Z_2\cdots Z_{n-1}\rangle}\Big).
\end{split}\end{equation}
We split the first term into $n-1$ parts and combine with each of the latter terms. Such combination can be simplified, e.g.,
\begin{equation}
\frac{\langle\delta Z_1Z_2\cdots Z_n\rangle}{\langle Z_1\cdots Z_n\rangle}-\frac{\langle X\delta Z_1Z_3\cdots Z_n\rangle}{\langle XZ_1Z_3\cdots Z_n\rangle}
=\frac{\langle\delta Z_1Z_1Z_3\cdots Z_n\rangle\,\langle XZ_2\cdots Z_n\rangle}{\langle XZ_1Z_3\cdots Z_n\rangle\,\langle Z_1\cdots Z_n\rangle},
\end{equation}
and similarly for other combinations. Note there is a common factor $\frac{\langle XZ_2\cdots Z_n\rangle}{\langle Z_1\cdots Z_n\rangle}$, and so the expression reduces to
\begin{equation}\begin{split}
\mathrm{d}_{Z_1}\Lambda(\overline{\Delta},\underline{\Delta})
=\int_{\overline{\Delta}}&\frac{\measure{X}{n-1}\,\langle Z_1Z_2\cdots Z_n\rangle^{n-2}}{\langle XZ_3\cdots Z_1\rangle\cdots\langle XZ_1\cdots Z_{n-1}\rangle}\sum_{i=2}^{n}\frac{\langle \delta Z_1Z_1Z_2\cdots\widehat{Z_i}\cdots Z_n\rangle}{\langle XZ_1Z_2\cdots\widehat{Z_i}\cdots Z_n\rangle}.
\end{split}\end{equation}
In particular, the integrand now explicitly has no $\langle XZ_2\cdots Z_n\rangle$ pole. By Schouten identity we further have
\begin{equation}
\frac{\measure{X}{n-1}\,\langle \delta Z_1Z_1Z_2\cdots\widehat{Z_i}\cdots Z_n\rangle}{\langle XZ_1Z_2\cdots\widehat{Z_i}\cdots Z_n\rangle}
=\langle\delta Z_1\mathrm{d}X^{n-1}\rangle-(n-1)\frac{\langle\delta Z_1X\mathrm{d}^{n-2}X\rangle\,\langle \mathrm{d}XZ_1Z_2\cdots\widehat{Z_i}\cdots Z_n\rangle}{\langle XZ_1Z_2\cdots\widehat{Z_i}\cdots Z_n\rangle}.
\end{equation}
Substituting this into the previous expression, we observe that the integrand is a total derivative
\begin{equation}
\mathrm{d}_{Z_1}\Lambda(\overline{\Delta},\underline{\Delta})
=(n-1)\int_{\overline{\Delta}}\mathrm{d}_X\frac{\langle \delta Z_1X\mathrm{d}^{n-2}X\rangle\,\langle Z_1Z_2\cdots Z_n\rangle^{n-2}}{\langle XZ_3\cdots Z_1\rangle\cdots\langle XZ_1\cdots Z_{n-1}\rangle}.
\end{equation}

When we localize the integral onto the boundaries, for each boundary $H_i$ each invariant reduces as
\begin{equation}
\langle Z_1\cdots\rangle=(Z_1H_i)\langle\cdots\rangle,
\end{equation}
and similarly when $Z_1$ is substituted by $\delta Z_1$, where ``$\cdots$'' denotes the insertions of other points (on RHS the inserted points are projected points on $H_i$). And so by a simple counting of the powers we see that upon each boundary we have a dlog times a lower-dim integral of the same type
\begin{equation}
\mathrm{d}_{Z_1}\Lambda(\overline{\Delta},\underline{\Delta})
=(n-1)\sum_{i=1}^n\mathrm{d}_{Z_1}\big[\log(Z_1H_i)\big]\underbrace{\int_{\overline{\Delta}_{(i)}}\frac{\measure{Y}{n-2}\,\langle W_2\cdots W_n\rangle^{n-2}}{\langle Y W_3\cdots W_n\rangle\cdots\langle YW_2\cdots W_{n-1}\rangle}}_{\Lambda(\overline{\Delta}_{(i)},\underline{\Delta}_{(i)})},
\end{equation}
where the new integrand simplex $\underline{\Delta}_{(i)}$ is specified by the vertices $\{W_i\}$, where each $W_i$ the projection of $Z_i$ through $Z_1$ onto the chosen hyperplane of $H_i$.

Iterating this procedure for other vertices of $\underline{\Delta}$ as well as the lower weight Aomoto polylog induced on each boundary (and boundaries of boundaries, etc) result in the following structure of the symbols
\begin{equation}
\begin{split}
\mathcal{S}\Lambda(\overline{\Delta},\underline{\Delta})=&\sum_{\rho,\sigma\in S_n}\text{sgn}(\rho)\text{sgn}(\sigma)\times\\
&\times\langle V_{\rho(2)}Z_{\sigma(2)}\ldots Z_{\sigma(n)}\rangle\otimes\langle V_{\rho(2)}V_{\rho_3}Z_{\sigma(3)}\ldots Z_{\sigma(n)}\rangle\otimes\langle V_{\rho(2)}\ldots V_{\rho(n)}Z_{\sigma(n)}\rangle,
\end{split}
\end{equation}
where sgn denotes the sign of the permutation.

\subsection{Discontinuities}\label{app:sec:aomotodisc}

We now discuss how to understand the above structure of the symbols from the geometry setup. Here we try to make the presentation pedagogical.

Whenever a function is expressed in terms of a contour integral the singularity of the function is always encoded in the relation between the shape of the integral contour and the singularities of the integrand. Both of these may depend on the complex variables entering the function. As we vary the variables, the singularity of the integrand may hit the contour such that a potential singularity of the function is produces, which can in general be avoided by deforming the contour away from the integrand singularity. When such deformation does not exist, an actually singularity of the function is produced.

To illustrate this, consider the log function
\begin{equation}
\log(z)=\int_1^z\frac{\mathrm{d}x}{x}.
\end{equation}
As usual the integral on RHS makes sense when $z$ is some positive real number. In fact the contour can be chosen arbitrarily as long as it is homologous to the interval $[1,z]$ on the real axis. The integrand contains a pole at $x=0$ (and also one at $x=\infty$). As we change $z$, when $z\to0$ or $z\to\infty$ we see that the poles necessarily hit (the end of) the contour, indicating that the function contains singularities at $z=0$ and at $z=\infty$, which is indeed the case for $\log(z)$. To see why these are branch points, we bring $z$ from its original value to a negative value from one side of $0$ and then return to the original value from the other side. As a result the contour is only homologous to the original one up to an addition of a circle wrapping around $x=0$, which computes the corresponding discontinuity across the cut. On the other hand, this $S^1$ contour exactly computes a residue of the integrand at $x=0$.

View this toy example from a slightly different angle. Let us restrict the data to the real slice of $\mathbb{CP}^1$. The integrand simplex has its two boundaries located as $x=0$ and $x=\infty$, and the contour simplex is the segment $[1,z]$. It is easy to see that the branch cut is detected if and only if the boundaries of $\underline{\Delta}$ has a non-trivial intersection with $\overline{\Delta}$, as follows
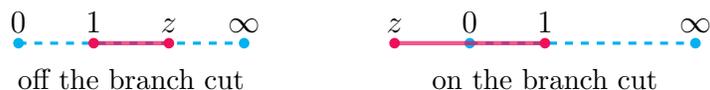
\begin{center}
\begin{tikzpicture}
\draw [ProcessBlue,very thick,dashed] (0,0) -- (3,0);
\fill [ProcessBlue] (0,0) circle(2pt);
\fill [ProcessBlue] (3,0) circle(2pt);
\draw [OrangeRed,ultra thick,opacity=.7] (1,0) -- (2,0);
\fill [OrangeRed] (1,0) circle(2pt);
\fill [OrangeRed] (2,0) circle(2pt);
\node [anchor=south] at (0,0) {$0$};
\node [anchor=south] at (1,0) {$1$};
\node [anchor=south] at (2,0) {$z$};
\node [anchor=south] at (3,0) {$\infty$};
\node [anchor=north] at (1.5,-.2) {\small off the branch cut};
\begin{scope}[xshift=6cm]
\draw [ProcessBlue,very thick,dashed] (0,0) -- (3,0);
\fill [ProcessBlue] (0,0) circle(2pt);
\fill [ProcessBlue] (3,0) circle(2pt);
\draw [OrangeRed,ultra thick,opacity=.7] (1,0) -- (-1,0);
\fill [OrangeRed] (1,0) circle(2pt);
\fill [OrangeRed] (-1,0) circle(2pt);
\node [anchor=south] at (0,0) {$0$};
\node [anchor=south] at (1,0) {$1$};
\node [anchor=south] at (-1,0) {$z$};
\node [anchor=south] at (3,0) {$\infty$};
\node [anchor=north] at (1,-.2) {\small on the branch cut};
\end{scope}
\end{tikzpicture}
\end{center}
Furthermore, once the branch cut is encountered, its corresponding discontinuity is identical to the residue of the integrand at the intersection.

The above picture immediately generalizes to arbitrary Aomoto polylogarithms. Let us temporarily return to the integral representation of $\mathrm{Li}_2(z)$ given previously. It is well-known this has a branch cut at $[1,\infty)$. Applying the above logic, we see this again detected by the intersection of the boundaries of $\underline{\Delta}$ with the contour, depending on the value of $z$.
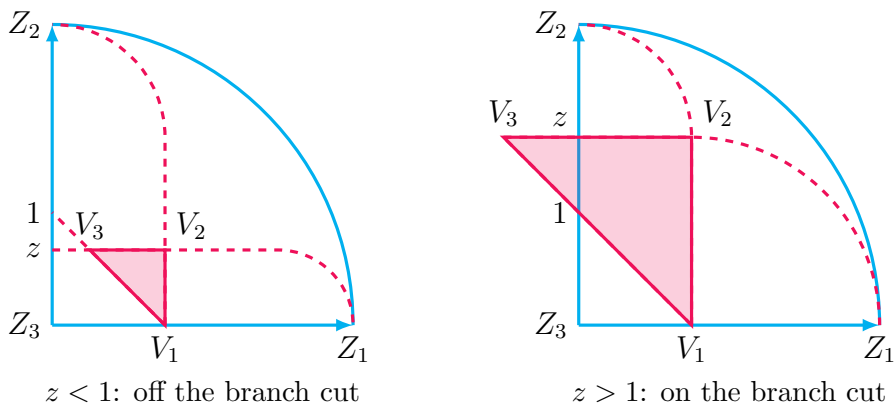
\begin{center}
\begin{tikzpicture}
\draw [ProcessBlue,very thick,-latex] (0,0) -- (4,0);
\draw [ProcessBlue,very thick,-latex] (0,0) -- (0,4);
\draw [ProcessBlue,very thick] (4,0) arc [start angle=0,end angle=90,radius=4];
\fill [OrangeRed,opacity=.2] (1.5,0) -- +(0,1) -- +(-1,1) -- cycle;
\draw [OrangeRed,very thick,dashed] (0,1) -- +(3,0) arc [start angle=90,end angle=0,radius=1];
\draw [OrangeRed,very thick,dashed] (1.5,0) -- +(0,2.5) arc [start angle=0,end angle=90,radius=1.5];
\draw [OrangeRed,very thick,dashed] (1.5,0) -- (0,1.5);
\draw [OrangeRed,very thick] (1.5,0) -- +(0,1) -- +(-1,1) -- cycle;
\node [anchor=north] at (4,0) {$Z_1$};
\node [anchor=east] at (0,4) {$Z_2$};
\node [anchor=east] at (0,0) {$Z_3$};
\node [anchor=east] at (0,1) {$z$};
\node [anchor=east] at (0,1.5) {$1$};
\node [anchor=north] at (1.5,0) {$V_1$};
\node [anchor=south west] at (1.5,1) {$V_2$};
\node [anchor=south] at (.5,1) {$V_3$};
\node [anchor=north] at (2,-.6) {\small $z<1$: off the branch cut};
\begin{scope}[xshift=7cm]
\draw [ProcessBlue,very thick,-latex] (0,0) -- (4,0);
\draw [ProcessBlue,very thick,-latex] (0,0) -- (0,4);
\draw [ProcessBlue,very thick] (4,0) arc [start angle=0,end angle=90,radius=4];
\fill [OrangeRed,opacity=.2] (1.5,0) -- +(0,2.5) -- (-1,2.5) -- cycle;
\draw [OrangeRed,very thick,dashed] (-1,2.5) -- +(2.5,0) arc [start angle=90,end angle=0,radius=2.5];
\draw [OrangeRed,very thick,dashed] (1.5,0) -- +(0,2.5) arc [start angle=0,end angle=90,radius=1.5];
\draw [OrangeRed,very thick,dashed] (1.5,0) -- (-1,2.5);
\draw [OrangeRed,very thick] (1.5,0) -- +(0,2.5) -- (-1,2.5) -- cycle;
\node [anchor=north] at (4,0) {$Z_1$};
\node [anchor=east] at (0,4) {$Z_2$};
\node [anchor=east] at (0,0) {$Z_3$};
\node [anchor=south east] at (0,2.5) {$z$};
\node [anchor=east] at (0,1.5) {$1$};
\node [anchor=north] at (1.5,0) {$V_1$};
\node [anchor=south west] at (1.5,2.5) {$V_2$};
\node [anchor=south] at (-1,2.5) {$V_3$};
\node [anchor=north] at (2,-.6) {\small $z>1$: on the branch cut};
\end{scope}
\end{tikzpicture}
\end{center}
To compute the discontinuity, we correspondingly calculate the residue of the integrand at the boundary $H_1$. This generates a one form and there is still one integration to be down, and the corresponding contour is defined exactly by the intersection. Hence
\begin{equation}
\underset{z>1}{\text{Disc}}\,\mathrm{Li}_2(z)=\int_1^z\underset{x_1=0}{\text{Res}}\,\omega_{\underline{\Delta}}=2\pi i\,\int_1^z\frac{\mathrm{d}x_2}{x_2}=2\pi i\,\log(z).
\end{equation}

In general, when the two simplices are at generic positions, the intersection is a union of several components, each of which corresponds to the intersection of $\overline{\Delta}$ and one specific codim-1 boundary of $\underline{\Delta}$, e.g.,
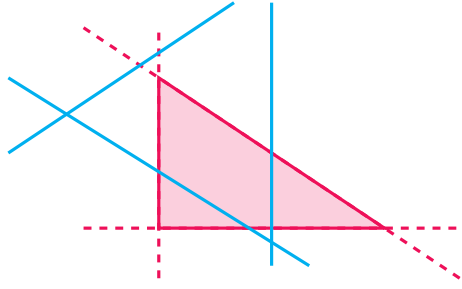
\begin{center}
\begin{tikzpicture}
\fill [OrangeRed,opacity=.2] (0,0) -- (3,0) -- (0,2) -- cycle;
\draw [OrangeRed,very thick,dashed] (-1,0) -- (4,0);
\draw [OrangeRed,very thick,dashed] (4,-2/3) -- (-1,8/3);
\draw [OrangeRed,very thick,dashed] (0,-2/3) -- (0,8/3);
\draw [OrangeRed,very thick] (0,0) -- (3,0) -- (0,2) -- cycle;
\draw [ProcessBlue,very thick] (-2,1) -- (1,3);
\draw [ProcessBlue,very thick] (-2,2) -- (2,-.5);
\draw [ProcessBlue,very thick] (1.5,3) -- (1.5,-.5);
\end{tikzpicture}
\end{center}
Following the above prescription, the discontinuity can always be decomposed into
\begin{equation}
\text{Disc}\,\Lambda(\overline{\Delta},\underline{\Delta})=
\sum_{i\in S}\int_{\overline{\Delta}\cap G_{i}}{\text{Res}}_{G_i}\,\omega_{\underline{\Delta}},
\end{equation}
where $S$ is the set of all labels for which the corresponding boundary hyperplane $G$ of $\underline{\Delta}$ has non-trivial intersection with $\overline{\Delta}$. Note that generically each intersection $\underline{\Delta}\cap G_i$ is some polytope in $\mathbb{CP}^{n-2}$.

In fact there is a refined decomposition of the discontinuity that identifies the discontinuity as projections. For this purpose it is convenient to got to the canonical frame (though not necessary). Let us begin by consider one particular contribution to the discontinuity from some $\overline{\Delta}\cap G_i$ such that the hyperplane $G_i$ separates one vertex $V_j$ of $\overline{\Delta}$ from the other vertices. When we compute the residue, we can explicitly specify the $S^1$ contour by integrating $x_j$ around the pole $G_jX=0$. The contribution is then obtained by integrating the resulting form over $\overline{\Delta}_{(j)}$, since the integration region of the remaining variables obviously remain the same. Here the lower form from the residue computation is manifestly identical to the image of projecting the face $\underline{\Delta}_{(i)}$ through the vertex $V_j$. To understand why this, note that in the canonical frame we can write
\begin{equation}
\underset{G_iX=0}{\text{Disc}}\,\omega_{\underline{\Delta}}=\frac{\measure{X_{(j)}}{n-2}}{\prod_{k\neq i}(G_iG_kV_jX)},
\end{equation}
where
\begin{equation}
(G_iG_kV_jX)\equiv (G_iX)(G_kV_j)-(G_iV_j)(G_kX)\propto(G_{iI_1}G_{kI_2}\epsilon^{I_1\ldots I_n})V_j^{J_1}X^{J_2}\epsilon_{J_1J_2I_3\ldots I_n}.
\end{equation}
Each such factor $(G_iG_kV_jX)$ thus defines a codim-1 hyperplane inside $H_j$ that is coplanar with both $V_j$ and $\underline{\Delta}_{(ik)}$ (which is the codim-1 boundary of $\Delta_{(i)}$).

As an example to illustrate, consider the following example
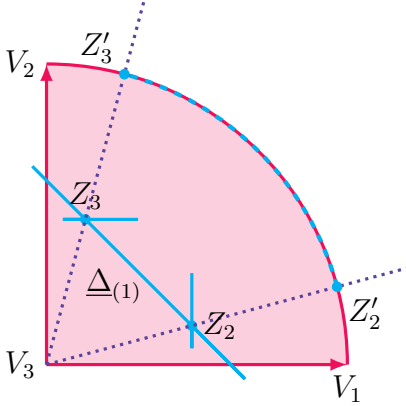
\begin{center}
\begin{tikzpicture}
\fill [OrangeRed,opacity=.2] (0,0) -- (4,0) arc [start angle=0,end angle=90,radius=4] -- cycle;
\draw [OrangeRed,very thick,-latex] (0,0) -- (4,0);
\draw [OrangeRed,very thick,-latex] (0,0) -- (0,4);
\draw [OrangeRed,very thick] (4,0) arc [start angle=0,end angle=90,radius=4];
\node [anchor=north] at (4,0) {$V_1$};
\node [anchor=east] at (0,4) {$V_2$};
\node [anchor=east] at (0,0) {$V_3$};
\path (15:2) -- +(-45:1) [draw=ProcessBlue,very thick] -- +(135:3);
\path (15:2) -- +(0,-.3) [draw=ProcessBlue,very thick] -- +(0,.7);
\path (75:2) -- +(-.3,0) [draw=ProcessBlue,very thick] -- +(.7,0);
\fill [ProcessBlue] (15:2) circle(2pt);
\fill [ProcessBlue] (75:2) circle(2pt);
\draw [Violet,very thick,dotted] (0,0) -- (15:5);
\draw [Violet,very thick,dotted] (0,0) -- (75:5);
\draw [ProcessBlue,ultra thick,dashed] (15:4) arc [start angle=15,end angle=75,radius=4];
\fill [ProcessBlue] (15:4) circle(2pt);
\fill [ProcessBlue] (75:4) circle(2pt);
\node [anchor=north east] at (45:2) {$\underline{\Delta}_{(1)}$};
\node [anchor=west] at (15:2) {$Z_2$};
\node [anchor=south] at (75:2) {$Z_3$};
\node [anchor=north west] at (15:4) {$Z'_2$};
\node [anchor=south east] at (75:4) {$Z'_3$};
\end{tikzpicture}
\end{center}
which illustrates a boundary hyperplane $G_1$ of $\underline{\Delta}$ isolating the vertex $V_3$ from the rest vertices of $\overline{\Delta}$. To compute the contribution from the intersection $\overline{\Delta}\cap G_1$, we project the face $\underline{\Delta}_{(1)}$ through $V_3$, whose image can be treated as the segment $(Z'_2Z'_3)$ on line at infinity $H_3X=0$. The residue is identical to the integrand associated to this image simplex, and we integrate it over $\overline{\Delta}_{(3)}$, which is also the projection of $\overline{\Delta}$ through $V_3$.

As pointed out before, in general we expect the intersection $\overline{\Delta}\cap G_i$ to be some complex instead of a single simplex. However, there is a combinatorically canonical way to decompose it into projections as described above. Here instead of a rigorous proof we just provide an example in $\mathbb{CP}^3$ to illustrated this.

\begin{figure}[ht]
\begin{center}
\begin{tikzpicture}
\coordinate [label=90:$V_1$] (v1) at (45:4);
\coordinate [label=135:$V_2$] (v2) at (45:1);
\coordinate [label=-90:$V_3$] (v3) at (-45:3);
\coordinate [label=0:$V_4$] (v4) at (-5:6);
\draw [OrangeRed,very thick,dashed] (v2) -- (v4);
\fill [OrangeRed,opacity=.2] (v1) -- (v2) -- (v3) -- (v4) -- cycle;
\draw [OrangeRed,very thick] (v1) -- (v2) -- (v3) -- (v4) -- cycle;
\draw [OrangeRed,very thick] (v1) -- (v3);
\node [anchor=south] at (v1) {$V_1$};
\node [anchor=south east] at (v2) {$V_2$};
\node [anchor=north] at (v3) {$V_3$};
\node [anchor=west] at (v4) {$V_4$};
\coordinate [label=180:$V'_2(V''_1)$] (p1) at (0,0);
\coordinate [label=-10:$V'_3$] (p2) at ($(v1)!.75!(v3)$);
\coordinate [label=20:$V'_4$] (p3) at ($(v1)!.5!(v4)$);
\coordinate [label=90:$V''_4$] (p4) at (intersection of v2--v4 and p1--p3);
\coordinate (p5) at (intersection of v2--v3 and p1--p3);
\coordinate [label=-100:$V''_3$] (p6) at (intersection of v2--v3 and p1--p2);
\fill [Violet,opacity=.1] (p1)-- (p2) -- (p3) -- cycle;
\fill [Violet,opacity=.2] (p6)-- (p2) -- (p3) -- (p4) -- cycle;
\draw [Violet,thick] (p5) -- (p1) -- (p2) -- (p3);
\draw [Violet,thick,dashed] (p3) -- (p5);
\draw [Violet,thick,dashed] (p4) -- (p6);
\draw [Violet,thick,dotted] (v1) -- (p1);
\draw [Violet,thick,dotted] (v1) -- (p2);
\draw [Violet,thick,dotted] (v1) -- (p3);
\draw [Violet,thick,dotted] (v2) -- (p4);
\draw [Violet,thick,dotted] (v2) -- (p6);
\draw [ProcessBlue,thick,dotted] (p2) -- ($(p1)!2.5!(p2)$);
\draw [ProcessBlue,thick,dotted] (p3) -- ($(p1)!1.5!(p3)$);
\node [anchor=north] at ($(p1)!1.4!(p3)$) {$G_i$};
\end{tikzpicture}
\end{center}
\vspace{-1.5em}\caption{Decomposition of a generic intersection into projections.}
\label{app:fig:decomposeintoprojections}
\end{figure}
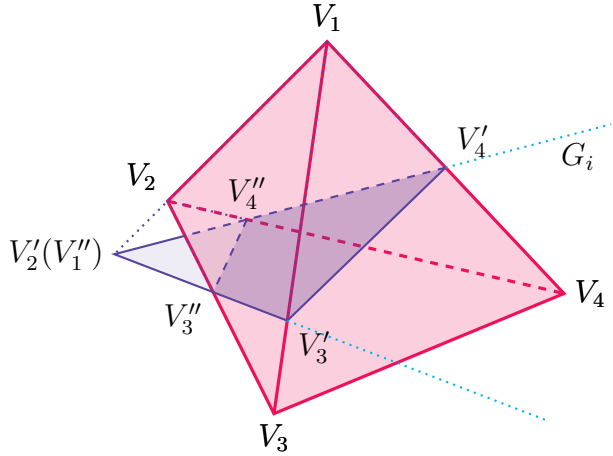

Consider, for example $G_i$ separates the vertices $\{V_1,V_2\}$ from $\{V_3,V_4\}$, then the intersection $\overline{\Delta}\cap G_i$ is some $4$-gon on $G_i$. Of course one can straightforwardly compute the residue at the singularity $G_iX=0$ and integrate over this $4$-gon to obtain the corresponding discontinuity. Despite of this, a more systematic treatment is to observe that this is identical to a summation of two projections of $\{\overline{\Delta},\underline{\Delta}_{(i)}\}$, one through $V_1$ and the other through $V_2$ (one can also do it for $V_3$ and $V_4$ instead). This is easy to see as illustrated in Figure \ref{app:fig:decomposeintoprojections}.
Here we visualize both projections using the hyperplane $G_i$, on which the face $\underline{\Delta}_{(i)}$ already leaves, hence we did not explicitly draw it. However, the contour $\overline{\Delta}$ has a non-trivial projection in this setup: it maps to the triangle $(V'_2V'_3V'_4)$ when projected through $V_1$, and to $(V''_3V''_4V''_1)$ through $V_2$. Note that $V'_2$ and $V''_1$ are the same point. It is then explicit that when the orientations are properly taken care of, the two image contours add up to the intersection $\overline{\Delta}\cap G_i$, which in this case is the $4$-gon $(V'_3V'_4V''_4V''_3)$. Since the integrand on $G_i$ from both projections is the same, we confirm that the contribution to the discontinuity from $\overline{\Delta}\cap G_i$ is decomposed into a projection through $V_1$ and another from $V_2$.

It is now obvious that in the most general situation, every time when we focus on a boundary hyperplane $G_i$ of $\underline{\Delta}$, by taking the set of $\overline{\Delta}$ vertices that fall on one side of $G_i$, the corresponding discontinuity is then the sum of projections through each of the vertices in the set (taking orientations into consideration).

\subsection{Symbols from the branch cuts and discontinuities}

From the elementary projection through a single vertex $V_j$, we see that whether such projection contributes to the discontinuity depends on whether the vertex $V_j$ is deformed to pass the hyperplane $G_i$ so that $G_i$ intersects $\overline{\Delta}$ non-trivially. This indicates that  every pair $(V_j,G_i)$ indicates a possible branch point at $G_iV_j=0$, or equivalently, at $\langle V_jZ_1\ldots\widehat{Z_i}\ldots Z_n\rangle=0$. As a result the first entries of $\mathcal{S}\Lambda(\overline{\Delta},\underline{\Delta})$ can be obtained by enumerating all these pairs.

As we move on to the second symbol entries, since the discontinuity is always a summation of elementary projections, it suffices to focus on a specific projection, say associated to the pair $(V_{j_1},G_{i_1})$. After projection the new $\overline{\Delta}'$ is defined by images of $\{V_1,\ldots,\widehat{V_{j_1}},\ldots,V_n\}$ and the new $\underline{\Delta}'$ by images of $\{Z_1,\ldots,\widehat{Z_{i_1}},\ldots,Z_n\}$. Again we enumerate all possible pairs extracted from $(\overline{\Delta}',\underline{\Delta}')$. Since these are projected through $V_{j_1}$, the corresponding second symbol entries thus consist of invariants of the form $\langle V_{j_1}V_{j_2}Z_1\ldots\widehat{Z_{i_1}}\ldots\widehat{Z_{i_2}}\ldots Z_n\rangle$, for some $i_2\neq i_1$ and $j_2\neq j_1$.

Iterating this procedure for the third entries and so on, we see that every time we should include the $V$ vertex we project through in the previous step into the contractions, and correspondingly exclude the $Z$ vertex (whose image is) dual to the hyperplane under consideration. When orientations are properties taken care of, this recovers the symbols of a generic Aomoto polylogarithm
\begin{equation}
\begin{split}
\mathcal{S}\Lambda(\overline{\Delta},\underline{\Delta})=&\sum_{\rho,\sigma\in S_n}\text{sgn}(\rho)\text{sgn}(\sigma)\times\\
&\times\langle V_{\rho(1)}Z_{\sigma(1)}\ldots Z_{\sigma(n-1)}\rangle\otimes\langle V_{\rho(1)}V_{\rho_2}Z_{\sigma(2)}\ldots Z_{\sigma(n-1)}\rangle\otimes\langle V_{\rho(1)}\ldots V_{\rho(n-1)}Z_{\sigma(n-1)}\rangle,
\end{split}
\end{equation}
which up to an overall constant, is the same as that obtained from the differentiation method.

Note that in the discussion of differentations we observed that the variation of each integrand vertex $Z_i$ corresponds to a projection through $Z_i$, while in the above we conclude that the discontinuities are always treated as projections through some contour vertex $V_j$. This is analogy is made more explicit by the detailed structure of the symbols we worked out in both methods. This can be quoted as a duality between Aomoto polylogarithms under the exchange of the contour and the integrand simplices
\begin{equation}
S\Lambda(\underline{\Delta},\overline{\Delta})=\left(\mathcal{S}\Lambda(\overline{\Delta},\underline{\Delta})\right)^{\rm R},
\end{equation}
where R is the same revertion of symbols that we defined in the discussion of duality among $e_n$ integrals (with even $n$).

\subsection{Generalizing Aomoto to polytopes}

The notion of Aomoto polylogarithms can be generalized by taking both $\overline\Delta$ and $\underline\Delta$ to be some polytopes. We again specify their data by vertices $V$'s and $Z$'s respectively. To obtain the corresponding symbols we can of course always start by triangulating $\overline\Delta$ and $\underline\Delta$ and sum up the symbols for each resulting pair of simplices. In fact, for these ``polytope polylogarithms'' we can apply the projection picture to obtain a more unified description.

\begin{figure}[ht]
\begin{center}
\begin{tikzpicture}
\coordinate (v1) at (0,0);
\coordinate (v2) at (160:2);
\coordinate (v3) at (20:2);
\coordinate (v4) at (0,1);
\coordinate (v5) at (105:3);
\coordinate (v6) at (50:2.2);
\coordinate (v7) at ($(v5)!2!(v6)$);
\coordinate (v8) at ($(v5)!1.6!(v2)$);
\coordinate (v9) at ($(v5)!2.2!(v4)$);
\coordinate (v10) at ($(v7)!1.3!(v8)$);
\coordinate (v11) at ($(v7)!1.25!(v9)$);
\coordinate (v12) at ($(v10)!1.4!(v11)$);
\draw [red,thick,dotted,-latex] (v5) -- (v7);
\draw [red,thick,dotted,-latex] (v5) -- (v8);
\draw [red,thick,dotted,-latex] (v5) -- (v9);
\draw [red,thick,dotted,-latex] (v7) -- (v10);
\draw [red,thick,dotted,-latex] (v7) -- (v11);
\draw [red,thick,dotted,-latex] (v10) -- (v12);
\draw [OrangeRed,very thick,dashed] (v2) -- (v3);
\fill [OrangeRed,opacity=.15] (v1) -- (v2) -- (v5) -- (v6) -- (v3) -- cycle;
\draw [OrangeRed,thick] (v1) -- (v2) -- (v5) -- (v4) -- cycle;
\draw [OrangeRed,thick] (v1) -- (v3) -- (v6) -- (v5);
\draw [OrangeRed,thick] (v4) -- (v6);
\fill [OrangeRed,opacity=.3] (v7) -- (v8) -- (v9) -- cycle;
\draw [OrangeRed,thick] (v7) -- (v8) -- (v9) -- cycle;
\draw [OrangeRed,thick] (v10) -- (v11);
\draw [WildStrawberry,fill=white] (v5) circle [radius=1.5pt]; 
\fill [WildStrawberry] (v2) circle [radius=1.5pt]; 
\fill [WildStrawberry] (v3) circle [radius=1.5pt]; 
\fill [WildStrawberry] (v4) circle [radius=1.5pt]; 
\fill [WildStrawberry] (v1) circle [radius=1.5pt]; 
\fill [WildStrawberry] (v6) circle [radius=1.5pt]; 
\draw [WildStrawberry,fill=white] (v7) circle [radius=1.5pt]; 
\fill [WildStrawberry] (v8) circle [radius=1.5pt]; 
\fill [WildStrawberry] (v9) circle [radius=1.5pt]; 
\draw [WildStrawberry,fill=white] (v10) circle [radius=1.5pt]; 
\fill [WildStrawberry] (v11) circle [radius=1.5pt]; 
\fill [WildStrawberry] (v12) circle [radius=1.5pt]; 
\node [anchor=south] at (v5) {$V_1$};
\node [anchor=east] at (v4) {$V_2$};
\node [anchor=south] at (v6) {$V_3$};
\node [anchor=west] at (v3) {$V_4$};
\node [anchor=east] at (v2) {$V_5$};
\node [anchor=north] at (v1) {$V_6$};
\node [anchor=west] at (v7) {$V'_3$};
\node [anchor=south east] at (v8) {$V'_5$};
\node [anchor=west] at (v9) {$V'_2$};
\node [anchor=east] at (v10) {$V''_5$};
\node [anchor=north] at (v11) {$V''_2$};
\node [anchor=west] at (v12) {$V'''_2$};
\end{tikzpicture}
\end{center}
\vspace{-1.5em}\caption{Projecting a polytope.}
\label{app:fig:projectingpolytope}
\end{figure}
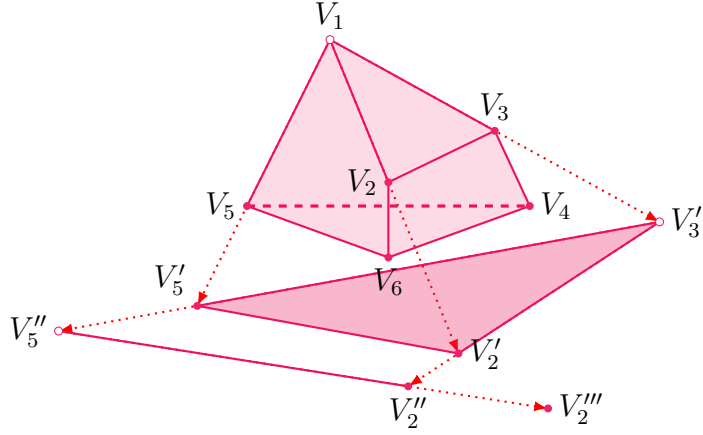

Since as discussed before the projection can be defined individually for $\overline\Delta$ and for $\underline\Delta$ and both are on the same footing in the symbols, it suffices to just focus on, e.g., the contour $\overline\Delta$. In order to illustrate, let us study an example in three dimensions, as depicted by the polytope in Figure \ref{app:fig:projectingpolytope} with vertices $\{V_1,\ldots,V_6\}$.
Here we first choose to project through $V_1$, which results in a triangle in 2d, with vertices $\{V'_2,V'_3,V'_5\}$, which are the images of $\{V_2,V_3,V_5\}$ respectively. This is in correspondence to the fact that in the original polytope $V_1$ is incident to three codim-1 faces. The image of $V_4$ (and $V_6$) lies within the face $(V'_3V'_5)$ (and $(V'_2V'_5)$), and thus do not characterize the new polytope from the projection. We then further choose to project through $V'_3$, obtaining a 1d polytope, which is just a segment with boundary points $V''_2$ and $V''_5$. These two points are images of $V'_2$ and $V'_5$ respectively. Then a third projection, e.g., through $V''_5$, further project it to a single point.

In the above example we chose the projecting point sequentially as $\{V_1,V'_3,V''_5\}$, which are tracked back to the vertices $\{V_1,V_3,V_5\}$ in the original contour polytope. This sequence of projections then maps to a particular set of symbol terms of the common form (recall the symbols of such case has length three)
\begin{equation}\label{app:eq:Vpattern}
\pm\langle V_1\cdots\rangle\otimes\langle V_1V_3\cdots\rangle\otimes\langle V_1V_3V_5\cdots\rangle.
\end{equation}
The relative sign in front is determined by the ordering $(V_1V_3V_5)$ relative to the one induced by $\overline\Delta$ on its boundary, or more explicitly, $\text{sgn}\langle V_1V_3V_5V_2\rangle$, which is the sign of the invariant made from the preimages of the sequence of vertices we collected from the projections. In the above the three ellipses refer to some combination of $Z$ vertices for the integrand $\underline\Delta$ that follow from projections similarly, but their ordering in the symbol entries are reversed. For example, if we first project $\underline\Delta$ through some $Z_1$, and then some $Z'_2$ and further some $Z''_3$. Then it leads to symbol terms of the form
\begin{equation}\label{app:eq:Zpattern}
\pm\langle\cdots Z_3Z_2Z_1\rangle\otimes\langle\cdots Z_2Z_1\rangle\otimes\langle\cdots Z_1\rangle.
\end{equation}
Each pair of legal patterns \eqref{app:eq:Vpattern} and \eqref{app:eq:Zpattern} allowed by the projections of $\overline\Delta$ and $\underline\Delta$ determine a particular symbol term that appears (where the sign is the product of those in \eqref{app:eq:Vpattern} and \eqref{app:eq:Zpattern}). $\mathcal{S}\Lambda(\overline\Delta,\underline\Delta)$ is then the summation of all such pairs. This holds for any pair of generic polytopes. By taking the dual of \eqref{app:eq:Vpattern}, it is easy to observe that the resulting expression is the same as \eqref{eq:polytopesymbols} presented in the introduction.

Note that in \eqref{app:eq:Vpattern} the combinations in each entry, i.e., $(V_1)$, $(V_1V_3)$ and $(V_1V_3V_5)$, corresponds to the actual dim-0, dim-1 and dim-2 faces of the original polytope. This is expected, as each particular symbol entry should admit of the interpretation is a pairing between some dim-$k$ face of $\overline\Delta$ and some codim-$k$ face of $\underline\Delta$ in order to capture the information about possible branch point of $\Lambda(\overline\Delta,\underline\Delta)$. The novelty in polytopes is that the same face can in general be represented by different subset of vertices incident to it, e.g., in Figure \ref{app:fig:projectingpolytope} $(V_1V_3V_5)$ and $(V_3V_4V_5)$ represents the same face. The projection picture exactly determines which subset of vertices have to enter each particular symbol term.

\subsection{Aomoto as a quadric}

In fact, the Aomoto polylogarithms can be treated just as special cases of the quadric integrals, because its integrand is explicitly a volume form, in the sense that we can switch it to an integral on another simplex $\widetilde{\underline{\Delta}}$ dual to $\underline{\Delta}$ (hence the vertices of $\widetilde{\underline{\Delta}}$ are specified by $\{G_i\}$)
\begin{equation}
\Lambda(\overline{\Delta},\underline{\Delta})=\int_{\overline{\Delta}}\measure{X}{n-1}\int_{\widetilde{\underline{\Delta}}}\measure{Y}{n-1}\,\frac{1}{(XY)^{n}}.
\end{equation}
By construction $Y$ is a covector and $X$ and $Y$ live in different $\mathbb{CP}^{n-1}$, but we can identify both $X$ and $Y$ as components in an ambient $\mathbb{CP}^{2n-2}$, for which we denote the coordinates as $\mathcal{X}=[\chi_1:\ldots:\chi_{2n-1}]$. Naively the integration contour is a certain complex (as was already obvious in the case of $n=1$). However, we can play a trick so as to identify the entire integral as an $E_{2n-1,1}$ integral (i.e., even with a simplex contour).

Since $\Lambda(\overline{\Delta},\underline{\Delta})$ enjoys invariance under arbitrary $\mathrm{PGL}(n)$ actions, we can first apply such an action to put the $Y$ integral into the canonical frame. Note that the $\mathrm{PGL}(n)$ action acts inversely on $X$, and so we have
\begin{equation}
\Lambda(\overline{\Delta},\underline{\Delta})=\int_{\mathbf{\Delta}}\measure{X}{n-1}\int_0^\infty\measure{Y}{n-1}\,\frac{1}{(XY)^{n}},
\end{equation}
where the new simplex contour $\mathbf{\Delta}$ is defined by the vertices
\begin{equation}
[V_iG_1:V_iG_2:\ldots:V_iG_n],\quad i=1,2,\ldots,n.
\end{equation}
Now let us fix $x_n=y_n=1$ and identify
\begin{equation}
x_i\equiv \chi_i,\quad y_i\equiv \chi_{i+n-1},\qquad i=1,2,\ldots,n-1.
\end{equation}
In the ambient space this corresponds to fixing $\chi_{2n-1}=1$, and we have
\begin{equation}
\Lambda(\overline{\Delta},\underline{\Delta})=\int_{\mathbf{\Delta}_{\rm ext}}\frac{\measure{\mathcal{X}}{2n-2}\,(L\mathcal{X})}{(\mathcal{X}Q\mathcal{X})^n},
\end{equation}
where the contour $\mathbf{\Delta}_{\rm ext}$ is defined by its vertices $\{\mathcal{V}_i\}$ specified as
\begin{equation}
\mathcal{V}_i^I=\begin{cases}[V_iG_1:\ldots:V_iG_{n-1}:0:\ldots:0:V_iG_n],&i=1,\ldots,n,\\\delta_{i-1}^I,&i=n+1,\ldots,2n-1.\end{cases}
\end{equation}
In the integrand we have $L=[0:\ldots:0:1]$, and
\begin{equation}
Q=\left(\begin{matrix}0&\frac{1}{2}\mathbf{1}_{n-1}&0\\\frac{1}{2}\mathbf{1}_{n-1}&0&0\\0&0&1\end{matrix}\right),
\end{equation}
with $\mathbf{1}_{n-1}$ an identity matrix of size $n-1$.

We can make this nicer by transform to the canonical frame in the ambient $\mathbb{CP}^{2n-2}$. In the canonical frame
\begin{equation}
L=[V_1G_n:\ldots:V_nG_n:0:\ldots:0],
\end{equation}
while the new quadric is
\begin{equation}
Q=\left(\begin{matrix}\mathbf{1}_{n}&\mathbf{R}\\\mathbf{R}^{\rm T}&\mathbf{0}_{n-1}\end{matrix}\right),\quad
\mathbf{R}_{IJ}=\frac{1}{2}\frac{V_IG_J}{V_IG_n}.
\end{equation}

As we apply the localization procedure to the $E_{2n-1,1}$ integral above, from the structure of $L$ we see it only receives contributions from boundaries $\mathbf{\Delta}_{(i)}$ for $i=1,2,\ldots,n$. Furthermore, only each $\mathbf{\Delta}_{(i)}$ the coefficients is trivial and we are left we purely an $e_{2n-2}$ integral, where the quadric is obtained from $Q$ by deleting the $i^{\rm th}$ row and column (hence completely independent of $V_i$)
\begin{equation}
\Lambda(\overline{\Delta},\underline{\Delta})=\sum_{i=1}^n\int_{\mathbf{\Delta}_{(i)}}\frac{\sqrt{\det Q_{(i)}}\,\measure{\mathcal{X}_{(i)}}{2n-3}}{(\mathcal{X}_{(i)}Q_{(i)}\mathcal{X}_{(i)})^{n-1}}.
\end{equation}

\section{Detailed Derivation of the dlog Coefficients}\label{app:derivation}

\flag{No further revisions.}

In this appendix we provide a detailed derivation of how the coefficients we obtained from localizing the $e_n$ integrals turn out to have dlog form.

\subsection{Coefficients for generic boundaries}\label{app:genericCijderivation}

In the case when the boundaries we localize to are at generic positions, i.e., not tangent to the quadric, we concluded in \eqref{Cijgeneric} that the coefficient associated to the contribution from $\Delta_{(ij)}$ is
\begin{equation}\label{appCijgeneric}
\fixed
C_{(ij)}=-\sqrt{\frac{q}{q_{(ij)}}}\left(H_iQ^{-1}\mathrm{d}QP(i)Q_{(i)}^{-1}P^{\rm T}(i)H_j+(i\leftrightarrow j)\right).
\end{equation}
Instead of doing a pure derivation, here we just perform an analytic check that the above expression matches
\begin{equation}\label{appdQHresult}
\fixed
\mathrm{d}_Q\log\,r(\mathbf{H}_{\{i,j\}})\equiv\mathrm{d}_Q\log\frac{\mathbf{H}_{ij}-\sqrt{\mathbf{H}_{ij}^2-\mathbf{H}_{ii}\mathbf{H}_{jj}}}{\mathbf{H}_{ij}+\sqrt{\mathbf{H}_{ij}^2-\mathbf{H}_{ii}\mathbf{H}_{jj}}},\qquad \mathbf{H}_{ij}\equiv H_iQ^{-1}H_j.
\end{equation}

First of all note that
\begin{equation}\label{appqqij}
\fixed
\frac{q_{(ij)}}{q}=\mathbf{H}_{ii}\mathbf{H}_{jj}-\mathbf{H}_{ij}^2,
\end{equation}
which can be easily verified using Schouten identities. We thus can rewrite \eqref{appdQHresult} as
\begin{equation}\label{appdrHij}
\fixed
\mathrm{d}_Q\log\,r(\mathbf{H}_{\{i,j\}})=\sqrt{-1}\sqrt{\frac{q}{q_{(ij)}}}\left(2\,\mathrm{d}_Q\mathbf{H}_{ij}-\frac{\mathbf{H}_{ij}}{\mathbf{H}_{ii}}\mathrm{d}_Q\mathbf{H}_{ii}-\frac{\mathbf{H}_{ij}}{\mathbf{H}_{jj}}\mathrm{d}_Q\mathbf{H}_{jj}\right).
\end{equation}
Hence we only need to check that the terms in the big parentheses of both \eqref{appCijgeneric} and \eqref{appdrHij} are identical. Let us focus on the first term in \eqref{appCijgeneric}. We have
\begin{equation}\label{appabigcontraction}
\begin{split}
\fixed
(n-1)!&(n-2)!\,q\,q_{(i)}\,(H_iQ^{-1}\mathrm{d}QP(i)Q_{(i)}^{-1}P^{\rm T}(i)H_j)\\
&=\parbox{6cm}{\tikz{
\node [anchor=center] at (0*.6,0) {$Q$};
\node [anchor=center] at (1*.6,0) {$\cdots$};
\node [anchor=center] at (2*.6,0) {$Q$};
\node [anchor=center] at (3*.6,.2) {$H_i$};
\node [anchor=center] at (4*.6,0) {$\mathrm{d}Q$};
\node [anchor=center] at (5*.6,0) {$Q$};
\node [anchor=center] at (6*.6,0) {$\cdots$};
\node [anchor=center] at (7*.6,0) {$Q$};
\node [anchor=center] at (8*.6,.2) {$H_i$};
\node [anchor=center] at (8*.6,-.2) {$H_i$};
\node [anchor=center] at (9*.6,-.2) {$H_j$};
\draw [black,thick] (0,.6) -- +(3*.6,0);
\draw [red,thick] (0*.6,.6) -- +(0,-.35);
\draw [red,thick] (2*.6,.6) -- +(0,-.35);
\draw [red,thick] (3*.6,.6) -- +(0,-.15);
\draw [black,thick] (4*.6,.6) -- +(4*.6,0);
\draw [red,thick] (4*.6,.6) -- +(0,-.35);
\draw [black,thick] (5*.6,.6) -- +(0,-.35);
\draw [black,thick] (7*.6,.6) -- +(0,-.35);
\draw [black,thick] (8*.6,.6) -- +(0,-.15);
\draw [black,thick] (0,-.6) -- +(4*.6,0);
\draw [black,thick] (0*.6,-.6) -- +(0,.35);
\draw [black,thick] (2*.6,-.6) -- +(0,.35);
\draw [black,thick] (4*.6,-.6) -- +(0,.35);
\draw [black,thick] (5*.6,-.6) -- +(4*.6,0);
\draw [black,thick] (5*.6,-.6) -- +(0,.35);
\draw [black,thick] (7*.6,-.6) -- +(0,.35);
\draw [black,thick] (8*.6,-.6) -- +(0,.15);
\draw [black,thick] (9*.6,-.6) -- +(0,.15);
}}\\
&=(n-1)\parbox{7.3cm}{\tikz{
\node [anchor=center] at (-1*.6,0) {$Q$};
\node [anchor=center] at (0*.6,0) {$\cdots$};
\node [anchor=center] at (1*.6,0) {$Q$};
\node [anchor=center] at (2*.6,0) {$\mathrm{d}Q$};
\node [anchor=center] at (3*.6,.2) {$H_i$};
\node [anchor=center] at (4*.6,0) {$Q$};
\node [anchor=center] at (5*.6,0) {$Q$};
\node [anchor=center] at (6*.6,0) {$\cdots$};
\node [anchor=center] at (7*.6,0) {$Q$};
\node [anchor=center] at (8*.6,.2) {$H_i$};
\node [anchor=center] at (8*.6,-.2) {$H_i$};
\node [anchor=center] at (9*.6,-.2) {$H_j$};
\draw [black,thick] (-1*.6,.6) -- +(4*.6,0);
\draw [black,thick] (-1*.6,.6) -- +(0,-.35);
\draw [black,thick] (1*.6,.6) -- +(0,-.35);
\draw [black,thick] (2*.6,.6) -- +(0,-.35);
\draw [black,thick] (3*.6,.6) -- +(0,-.15);
\draw [black,thick] (4*.6,.6) -- +(4*.6,0);
\draw [black,thick] (4*.6,.6) -- +(0,-.35);
\draw [black,thick] (5*.6,.6) -- +(0,-.35);
\draw [black,thick] (7*.6,.6) -- +(0,-.35);
\draw [black,thick] (8*.6,.6) -- +(0,-.15);
\draw [black,thick] (-1*.6,-.6) -- +(5*.6,0);
\draw [black,thick] (-1*.6,-.6) -- +(0,.35);
\draw [black,thick] (1*.6,-.6) -- +(0,.35);
\draw [black,thick] (2*.6,-.6) -- +(0,.35);
\draw [red,thick] (4*.6,-.6) -- +(0,.35);
\draw [black,thick] (5*.6,-.6) -- +(4*.6,0);
\draw [red,thick] (5*.6,-.6) -- +(0,.35);
\draw [red,thick] (7*.6,-.6) -- +(0,.35);
\draw [red,thick] (8*.6,-.6) -- +(0,.15);
\draw [red,thick] (9*.6,-.6) -- +(0,.15);
}}\\
&=
\parbox{3cm}{\tikz{
\node [anchor=center] at (0*.6,0) {$Q$};
\node [anchor=center] at (1*.6,0) {$\cdots$};
\node [anchor=center] at (2*.6,0) {$Q$};
\node [anchor=center] at (3*.6,0) {$\mathrm{d}Q$};
\node [anchor=center] at (4*.6,.2) {$H_i$};
\node [anchor=center] at (4*.6,-.2) {$H_j$};
\draw [black,thick] (0,.6) -- +(4*.6,0);
\draw [black,thick] (0*.6,.6) -- +(0,-.35);
\draw [black,thick] (2*.6,.6) -- +(0,-.35);
\draw [black,thick] (3*.6,.6) -- +(0,-.35);
\draw [black,thick] (4*.6,.6) -- +(0,-.15);
\draw [black,thick] (0,-.6) -- +(4*.6,0);
\draw [black,thick] (0*.6,-.6) -- +(0,.35);
\draw [black,thick] (2*.6,-.6) -- +(0,.35);
\draw [black,thick] (3*.6,-.6) -- +(0,.35);
\draw [black,thick] (4*.6,-.6) -- +(0,.15);
}}\,(n-1)!\,q_{(i)}-\parbox{5.5cm}{\tikz{
\node [anchor=center] at (0*.6,0) {$Q$};
\node [anchor=center] at (1*.6,0) {$\cdots$};
\node [anchor=center] at (2*.6,0) {$Q$};
\node [anchor=center] at (3*.6,0) {$\mathrm{d}Q$};
\node [anchor=center] at (4*.6,.2) {$H_i$};
\node [anchor=center] at (4*.6,-.2) {$H_i$};
\draw [black,thick] (0,.6) -- +(4*.6,0);
\draw [black,thick] (0*.6,.6) -- +(0,-.35);
\draw [black,thick] (2*.6,.6) -- +(0,-.35);
\draw [black,thick] (3*.6,.6) -- +(0,-.35);
\draw [black,thick] (4*.6,.6) -- +(0,-.15);
\draw [black,thick] (0,-.6) -- +(4*.6,0);
\draw [black,thick] (0*.6,-.6) -- +(0,.35);
\draw [black,thick] (2*.6,-.6) -- +(0,.35);
\draw [black,thick] (3*.6,-.6) -- +(0,.35);
\draw [black,thick] (4*.6,-.6) -- +(0,.15);
\node [anchor=center] at (5*.6,0) {$Q$};
\node [anchor=center] at (6*.6,0) {$\cdots$};
\node [anchor=center] at (7*.6,0) {$Q$};
\node [anchor=center] at (8*.6,.2) {$H_i$};
\node [anchor=center] at (8*.6,-.2) {$H_j$};
\draw [black,thick] (5*.6,.6) -- +(3*.6,0);
\draw [black,thick] (5*.6,.6) -- +(0,-.35);
\draw [black,thick] (7*.6,.6) -- +(0,-.35);
\draw [black,thick] (8*.6,.6) -- +(0,-.15);
\draw [black,thick] (5*.6,-.6) -- +(3*.6,0);
\draw [black,thick] (5*.6,-.6) -- +(0,.35);
\draw [black,thick] (7*.6,-.6) -- +(0,.35);
\draw [black,thick] (8*.6,-.6) -- +(0,.15);
}},
\end{split}
\end{equation}
where the third line is obtained from the second line by a Schouten identity on the red legs therein, and similarly from the third line to the last line. This indicates that
\begin{equation}
\begin{split}
\fixed
H_iQ^{-1}\mathrm{d}QP(i)Q_{(i)}^{-1}P^{\rm T}(i)H_j)
&=\frac{1}{q}\,\mathrm{d}_Q(q\mathbf{H}_{ij})-\frac{\mathbf{H}_{ij}}{q_{(i)}}\,\mathrm{d}_Q(q\mathbf{H}_{ii})\\
&=\mathrm{d}_Q\mathbf{H}_{ij}-\frac{\mathbf{H}_{ij}}{\mathbf{H}_{ii}}\mathrm{d}_Q\mathbf{H}_{ii},
\end{split}
\end{equation}
where in the second line we have used the fact that $q_{(i)}=q\mathbf{H}_{ii}$. The second term in \eqref{appCijgeneric} is related to the above just by switching the labels $i$ and $j$, which together with \eqref{appdrHij} shows that
\begin{equation}
C_{(ij)}=\sqrt{-1}\,\mathrm{d}_Q\log\,r(\mathbf{H}_{\{i,j\}}).
\end{equation}

\subsection{Coefficients for boundaries tangent to $Q$}
\label{app:tangentCijderivation}

When $H_iQ^{-1}H_i=0$, we first focus on the contribution \eqref{coefH1H2dQ}
\begin{equation}
\fixed
C_{ij}=-\sqrt{\frac{q}{q_{(ij)}}}
H_iQ^{-1}\mathrm{d}QP(i)\mathring{Q}_{(i)}^{-1}P(i)^{\rm T}H_j.
\end{equation}
In this case we have
\begin{align}\fixed
\mathring{q}_{(i)}&\equiv\frac{1}{(n-2)!}\parbox{3cm}{\tikz{
\node [anchor=center] at (0*.7+.15,0) {$Q_{(i)}$};
\node [anchor=center] at (1*.7+.15,0) {$\cdots$};
\node [anchor=center] at (2*.7+.15,0) {$Q_{(i)}$};
\node [anchor=center] at (3*.7+.15,.2) {$R_{(i)}$};
\node [anchor=center] at (3*.7+.15,-.2) {$R_{(i)}$};
\node [anchor=south] at (1.5*.7,.5) {\scriptsize $(n-1)$};
\draw [black,thick] (0,.6) -- +(3*.7,0);
\draw [black,thick] (0*.7,.6) -- +(0,-.35);
\draw [black,thick] (2*.7,.6) -- +(0,-.35);
\draw [black,thick] (3*.7,.6) -- +(0,-.15);
\node [anchor=north] at (1.5*.7,-.5) {\scriptsize $(n-1)$};
\draw [black,thick] (0,-.6) -- +(3*.7,0);
\draw [black,thick] (0*.7,-.6) -- +(0,.35);
\draw [black,thick] (2*.7,-.6) -- +(0,.35);
\draw [black,thick] (3*.7,-.6) -- +(0,.15);
}}\equiv\frac{1}{(n-2)!}\parbox{3cm}{\tikz{
\node [anchor=center] at (0*.6,0) {$Q$};
\node [anchor=center] at (1*.6,0) {$\cdots$};
\node [anchor=center] at (2*.6,0) {$Q$};
\node [anchor=center] at (3*.6,.2) {$R$};
\node [anchor=center] at (3*.6,-.2) {$R$};
\node [anchor=center] at (4*.6,.2) {$H_i$};
\node [anchor=center] at (4*.6,-.2) {$H_i$};
\node [anchor=south] at (2*.6,.5) {\scriptsize $(n)$};
\draw [black,thick] (0*.6,.6) -- +(4*.6,0);
\draw [black,thick] (0*.6,.6) -- +(0,-.35);
\draw [black,thick] (2*.6,.6) -- +(0,-.35);
\draw [black,thick] (3*.6,.6) -- +(0,-.15);
\draw [black,thick] (4*.6,.6) -- +(0,-.15);
\node [anchor=north] at (2*.6,-.5) {\scriptsize $(n)$};
\draw [black,thick] (0*.6,-.6) -- +(4*.6,0);
\draw [black,thick] (0*.6,-.6) -- +(0,.35);
\draw [black,thick] (2*.6,-.6) -- +(0,.35);
\draw [black,thick] (3*.6,-.6) -- +(0,.15);
\draw [black,thick] (4*.6,-.6) -- +(0,.15);
}}.
\end{align}

First from the generic identity \eqref{appqqij} we have $q_{(ij)}/q=-\mathbf{H}_{ij}^2$.
Expressed in terms of contractions in the original space we have
\begin{equation}
\begin{split}
\fixed
(n-1)!(n-3)!\,q\,\mathring{q}_{(i)}\,&(H_iQ^{-1}\delta QP(i)\mathring{Q}_{(i)}^{-1}P(i)^{\rm T}H_j)\\
&=\parbox{6cm}{\tikz{
\node [anchor=center] at (0*.6,0) {$Q$};
\node [anchor=center] at (1*.6,0) {$\cdots$};
\node [anchor=center] at (2*.6,0) {$Q$};
\node [anchor=center] at (3*.6,.2) {$H_i$};
\node [anchor=center] at (4*.6,0) {$\delta Q$};
\node [anchor=center] at (5*.6,0) {$Q$};
\node [anchor=center] at (6*.6,0) {$\cdots$};
\node [anchor=center] at (7*.6,0) {$Q$};
\node [anchor=center] at (8*.6,.2) {$R$};
\node [anchor=center] at (8*.6,-.2) {$R$};
\node [anchor=center] at (9*.6,.2) {$H_i$};
\node [anchor=center] at (9*.6,-.2) {$H_i$};
\node [anchor=center] at (10*.6,-.2) {$H_j$};
\draw [black,thick] (0,.6) -- +(3*.6,0);
\draw [red,thick] (0*.6,.6) -- +(0,-.35);
\draw [red,thick] (2*.6,.6) -- +(0,-.35);
\draw [red,thick] (3*.6,.6) -- +(0,-.15);
\draw [black,thick] (4*.6,.6) -- +(5*.6,0);
\draw [red,thick] (4*.6,.6) -- +(0,-.35);
\draw [black,thick] (5*.6,.6) -- +(0,-.35);
\draw [black,thick] (7*.6,.6) -- +(0,-.35);
\draw [black,thick] (8*.6,.6) -- +(0,-.15);
\draw [black,thick] (9*.6,.6) -- +(0,-.15);
\draw [black,thick] (0,-.6) -- +(4*.6,0);
\draw [black,thick] (0*.6,-.6) -- +(0,.35);
\draw [black,thick] (2*.6,-.6) -- +(0,.35);
\draw [black,thick] (4*.6,-.6) -- +(0,.35);
\draw [black,thick] (5*.6,-.6) -- +(5*.6,0);
\draw [black,thick] (5*.6,-.6) -- +(0,.35);
\draw [black,thick] (7*.6,-.6) -- +(0,.35);
\draw [black,thick] (8*.6,-.6) -- +(0,.15);
\draw [black,thick] (9*.6,-.6) -- +(0,.15);
\draw [black,thick] (10*.6,-.6) -- +(0,.15);
}}\\
&=(n-1)\parbox{7.3cm}{\tikz{
\node [anchor=center] at (-1*.6,0) {$Q$};
\node [anchor=center] at (0*.6,0) {$\cdots$};
\node [anchor=center] at (1*.6,0) {$Q$};
\node [anchor=center] at (2*.6,0) {$\delta Q$};
\node [anchor=center] at (3*.6,.2) {$H_i$};
\node [anchor=center] at (4*.6,0) {$Q$};
\node [anchor=center] at (5*.6,0) {$Q$};
\node [anchor=center] at (6*.6,0) {$\cdots$};
\node [anchor=center] at (7*.6,0) {$Q$};
\node [anchor=center] at (8*.6,.2) {$R$};
\node [anchor=center] at (8*.6,-.2) {$R$};
\node [anchor=center] at (9*.6,.2) {$H_i$};
\node [anchor=center] at (9*.6,-.2) {$H_i$};
\node [anchor=center] at (10*.6,-.2) {$H_j$};
\draw [black,thick] (-1*.6,.6) -- +(4*.6,0);
\draw [black,thick] (-1*.6,.6) -- +(0,-.35);
\draw [black,thick] (1*.6,.6) -- +(0,-.35);
\draw [black,thick] (2*.6,.6) -- +(0,-.35);
\draw [black,thick] (3*.6,.6) -- +(0,-.15);
\draw [black,thick] (4*.6,.6) -- +(5*.6,0);
\draw [black,thick] (4*.6,.6) -- +(0,-.35);
\draw [black,thick] (5*.6,.6) -- +(0,-.35);
\draw [black,thick] (7*.6,.6) -- +(0,-.35);
\draw [black,thick] (8*.6,.6) -- +(0,-.15);
\draw [black,thick] (9*.6,.6) -- +(0,-.15);
\draw [black,thick] (-1*.6,-.6) -- +(5*.6,0);
\draw [black,thick] (-1*.6,-.6) -- +(0,.35);
\draw [black,thick] (1*.6,-.6) -- +(0,.35);
\draw [black,thick] (2*.6,-.6) -- +(0,.35);
\draw [red,thick] (4*.6,-.6) -- +(0,.35);
\draw [black,thick] (5*.6,-.6) -- +(5*.6,0);
\draw [red,thick] (5*.6,-.6) -- +(0,.35);
\draw [red,thick] (7*.6,-.6) -- +(0,.35);
\draw [red,thick] (8*.6,-.6) -- +(0,.15);
\draw [red,thick] (9*.6,-.6) -- +(0,.15);
\draw [red,thick] (10*.6,-.6) -- +(0,.15);
}},
\end{split}
\end{equation}
where the third line is obtained by applying Schouten identity among the red legs in the second line. Then a further Schouten identity among the red legs in the third line, together with the tangent condition, leads to two pieces
\begin{equation}\label{A9}
\fixed
\frac{n-1}{n-2}\left(
\parbox{3cm}{\tikz{
\node [anchor=center] at (0*.6,0) {$Q$};
\node [anchor=center] at (1*.6,0) {$\cdots$};
\node [anchor=center] at (2*.6,0) {$Q$};
\node [anchor=center] at (3*.6,0) {$\delta Q$};
\node [anchor=center] at (4*.6,.2) {$H_i$};
\node [anchor=center] at (4*.6,-.2) {$H_j$};
\draw [black,thick] (0,.6) -- +(4*.6,0);
\draw [black,thick] (0*.6,.6) -- +(0,-.35);
\draw [black,thick] (2*.6,.6) -- +(0,-.35);
\draw [black,thick] (3*.6,.6) -- +(0,-.35);
\draw [black,thick] (4*.6,.6) -- +(0,-.15);
\draw [black,thick] (0,-.6) -- +(4*.6,0);
\draw [black,thick] (0*.6,-.6) -- +(0,.35);
\draw [black,thick] (2*.6,-.6) -- +(0,.35);
\draw [black,thick] (3*.6,-.6) -- +(0,.35);
\draw [black,thick] (4*.6,-.6) -- +(0,.15);
}}\,(n-2)!\,\mathring{q}_{(i)}-\parbox{6cm}{\tikz{
\node [anchor=center] at (0*.6,0) {$Q$};
\node [anchor=center] at (1*.6,0) {$\cdots$};
\node [anchor=center] at (2*.6,0) {$Q$};
\node [anchor=center] at (3*.6,0) {$\delta Q$};
\node [anchor=center] at (4*.6,.2) {$H_i$};
\node [anchor=center] at (4*.6,-.2) {$R$};
\draw [black,thick] (0,.6) -- +(4*.6,0);
\draw [black,thick] (0*.6,.6) -- +(0,-.35);
\draw [black,thick] (2*.6,.6) -- +(0,-.35);
\draw [black,thick] (3*.6,.6) -- +(0,-.35);
\draw [black,thick] (4*.6,.6) -- +(0,-.15);
\draw [black,thick] (0,-.6) -- +(4*.6,0);
\draw [black,thick] (0*.6,-.6) -- +(0,.35);
\draw [black,thick] (2*.6,-.6) -- +(0,.35);
\draw [black,thick] (3*.6,-.6) -- +(0,.35);
\draw [black,thick] (4*.6,-.6) -- +(0,.15);
\node [anchor=center] at (5*.6,0) {$Q$};
\node [anchor=center] at (6*.6,0) {$\cdots$};
\node [anchor=center] at (7*.6,0) {$Q$};
\node [anchor=center] at (8*.6,.2) {$R$};
\node [anchor=center] at (8*.6,-.2) {$H_j$};
\node [anchor=center] at (9*.6,.2) {$H_i$};
\node [anchor=center] at (9*.6,-.2) {$H_i$};
\draw [black,thick] (5*.6,.6) -- +(4*.6,0);
\draw [black,thick] (5*.6,.6) -- +(0,-.35);
\draw [black,thick] (7*.6,.6) -- +(0,-.35);
\draw [black,thick] (8*.6,.6) -- +(0,-.15);
\draw [black,thick] (9*.6,.6) -- +(0,-.15);
\draw [black,thick] (5*.6,-.6) -- +(4*.6,0);
\draw [black,thick] (5*.6,-.6) -- +(0,.35);
\draw [black,thick] (7*.6,-.6) -- +(0,.35);
\draw [black,thick] (8*.6,-.6) -- +(0,.15);
\draw [black,thick] (9*.6,-.6) -- +(0,.15);
}}
\right).
\end{equation}
Note that taking the extra factors into consideration the first term above already has the $\mathrm{d}_Q\log$ form. For the second term, note that
\begin{equation}
\fixed
\parbox{2.5cm}{\tikz{
\node [anchor=center] at (0*.6,0) {$Q$};
\node [anchor=center] at (1*.6,0) {$\ldots$};
\node [anchor=center] at (2*.6,0) {$Q$};
\node [anchor=center] at (3*.6,.2) {$H_i$};
\node [anchor=center] at (3*.6,-.2) {$R$};
\draw [black,thick] (0,.6) -- (3*.6,.6);
\draw [black,thick] (0,.6) -- +(0,-.35);
\draw [black,thick] (2*.6,.6) -- +(0,-.35);
\draw [black,thick] (3*.6,.6) -- +(0,-.15);
\draw [black,thick] (0,-.6) -- (3*.6,-.6);
\draw [black,thick] (0,-.6) -- +(0,.35);
\draw [black,thick] (2*.6,-.6) -- +(0,.35);
\draw [red,thick] (3*.6,-.6) -- +(0,.15);
}}
\frac{
\parbox{3cm}{\tikz{
\node [anchor=center] at (5*.6,0) {$Q$};
\node [anchor=center] at (6*.6,0) {$\cdots$};
\node [anchor=center] at (7*.6,0) {$Q$};
\node [anchor=center] at (8*.6,.2) {$R$};
\node [anchor=center] at (8*.6,-.2) {$H_j$};
\node [anchor=center] at (9*.6,.2) {$H_i$};
\node [anchor=center] at (9*.6,-.2) {$H_i$};
\draw [black,thick] (5*.6,.6) -- +(4*.6,0);
\draw [black,thick] (5*.6,.6) -- +(0,-.35);
\draw [black,thick] (7*.6,.6) -- +(0,-.35);
\draw [black,thick] (8*.6,.6) -- +(0,-.15);
\draw [black,thick] (9*.6,.6) -- +(0,-.15);
\draw [black,thick] (5*.6,-.6) -- +(4*.6,0);
\draw [red,thick] (5*.6,-.6) -- +(0,.35);
\draw [red,thick] (7*.6,-.6) -- +(0,.35);
\draw [red,thick] (8*.6,-.6) -- +(0,.15);
\draw [red,thick] (9*.6,-.6) -- +(0,.15);
}}}{\quad\mathring{q}\;\parbox{3cm}{\tikz{
\node [anchor=center] at (0*.6,0) {$Q$};
\node [anchor=center] at (1*.6,0) {$\ldots$};
\node [anchor=center] at (2*.6,0) {$Q$};
\node [anchor=center] at (3*.6,.2) {$H_i$};
\node [anchor=center] at (3*.6,-.2) {$H_j$};
\draw [black,thick] (0,.6) -- (3*.6,.6);
\draw [black,thick] (0,.6) -- +(0,-.35);
\draw [black,thick] (2*.6,.6) -- +(0,-.35);
\draw [black,thick] (3*.6,.6) -- +(0,-.15);
\draw [black,thick] (0,-.6) -- (3*.6,-.6);
\draw [black,thick] (0,-.6) -- +(0,.35);
\draw [black,thick] (2*.6,-.6) -- +(0,.35);
\draw [black,thick] (3*.6,-.6) -- +(0,.15);
}}}=(n-2)!,
\end{equation}
as can be verified by a schouten identity on the red legs. This turns the second term into a dlog as well. Altogether we thus obtain
\begin{equation}\label{tangentCijapp}
\fixed
C_{ij}=\sqrt{-1}\,\text{sign}(\mathbf{H}_{ij})\left(
\mathrm{d}_Q\log(q\,\mathbf{H}_{ij})
-\mathrm{d}_Q\log\left(\parbox{2.5cm}{\tikz{
\node [anchor=center] at (0*.6,0) {$Q$};
\node [anchor=center] at (1*.6,0) {$\ldots$};
\node [anchor=center] at (2*.6,0) {$Q$};
\node [anchor=center] at (3*.6,.2) {$H_i$};
\node [anchor=center] at (3*.6,-.2) {$R$};
\draw [black,thick] (0,.6) -- (3*.6,.6);
\draw [black,thick] (0,.6) -- +(0,-.35);
\draw [black,thick] (2*.6,.6) -- +(0,-.35);
\draw [black,thick] (3*.6,.6) -- +(0,-.15);
\draw [black,thick] (0,-.6) -- (3*.6,-.6);
\draw [black,thick] (0,-.6) -- +(0,.35);
\draw [black,thick] (2*.6,-.6) -- +(0,.35);
\draw [black,thick] (3*.6,-.6) -- +(0,.15);
}}\right)\right),
\end{equation}
which is \eqref{tangentCij}. \flag{I also numerically verified the relative sign.}

Now let us switch to the other contribution $C_{ji}$ to the same boundary $\Delta_{(ij)}$ but obtained by localization via $\Delta_{(j)}$. In the case when $H_jQ^{-1}H_j=0$ also holds, then the result is the same as \eqref{tangentCijapp} by with the labels $i$ and $j$ switched
\begin{equation}
\fixed
C_{ji}=\sqrt{-1}\,\text{sign}(\mathbf{H}_{ij})\left(
\mathrm{d}_Q\log(q\,\mathbf{H}_{ij})
-\mathrm{d}_Q\log\left(\parbox{2.5cm}{\tikz{
\node [anchor=center] at (0*.6,0) {$Q$};
\node [anchor=center] at (1*.6,0) {$\ldots$};
\node [anchor=center] at (2*.6,0) {$Q$};
\node [anchor=center] at (3*.6,.2) {$H_j$};
\node [anchor=center] at (3*.6,-.2) {$R$};
\draw [black,thick] (0,.6) -- (3*.6,.6);
\draw [black,thick] (0,.6) -- +(0,-.35);
\draw [black,thick] (2*.6,.6) -- +(0,-.35);
\draw [black,thick] (3*.6,.6) -- +(0,-.15);
\draw [black,thick] (0,-.6) -- (3*.6,-.6);
\draw [black,thick] (0,-.6) -- +(0,.35);
\draw [black,thick] (2*.6,-.6) -- +(0,.35);
\draw [black,thick] (3*.6,-.6) -- +(0,.15);
}}\right)\right).
\end{equation}
As we are going to argue in Appendix \ref{app:cancelation}, when a hyperplane is tangent to $Q$ the second dlog above can always be dropped, because this coefficient is universal to all codim-2 boundaries in this hyperplane, the induced integrals on which however cancel away. Thus we conclude that when both $H_i$ and $H_j$ are tangent to $Q$, we have
\begin{equation}
\fixed
C_{(ij)}\equiv C_{ij}+C_{ji}=-\sqrt{-1}\,\text{sign}(\mathbf{H}_{ij})\,\mathrm{d}\log((q\,\mathbf{H}_{ij})^2).
\end{equation}

If instead $H_j$ is at generic positions, then we have
\begin{equation}
\fixed
C_{ji}=-\sqrt{\frac{q}{q_{(ij)}}}H_jQ^{-1}\mathrm{d}QP(j)Q_{(j)}^{-1}P^{\rm T}(j)H_i
\end{equation}
The last contraction above is
\begin{equation}
\begin{split}
\fixed
(n-1)!&(n-2)!\,q\,q_{(j)}\,(H_jQ^{-1}\mathrm{d}QP(j)Q_{(j)}^{-1}P^{\rm T}(j)H_i)\\
&=
\parbox{3cm}{\tikz{
\node [anchor=center] at (0*.6,0) {$Q$};
\node [anchor=center] at (1*.6,0) {$\cdots$};
\node [anchor=center] at (2*.6,0) {$Q$};
\node [anchor=center] at (3*.6,0) {$\mathrm{d}Q$};
\node [anchor=center] at (4*.6,.2) {$H_i$};
\node [anchor=center] at (4*.6,-.2) {$H_j$};
\draw [black,thick] (0,.6) -- +(4*.6,0);
\draw [black,thick] (0*.6,.6) -- +(0,-.35);
\draw [black,thick] (2*.6,.6) -- +(0,-.35);
\draw [black,thick] (3*.6,.6) -- +(0,-.35);
\draw [black,thick] (4*.6,.6) -- +(0,-.15);
\draw [black,thick] (0,-.6) -- +(4*.6,0);
\draw [black,thick] (0*.6,-.6) -- +(0,.35);
\draw [black,thick] (2*.6,-.6) -- +(0,.35);
\draw [black,thick] (3*.6,-.6) -- +(0,.35);
\draw [black,thick] (4*.6,-.6) -- +(0,.15);
}}\,(n-1)!\,q_{(j)}-\parbox{5.5cm}{\tikz{
\node [anchor=center] at (0*.6,0) {$Q$};
\node [anchor=center] at (1*.6,0) {$\cdots$};
\node [anchor=center] at (2*.6,0) {$Q$};
\node [anchor=center] at (3*.6,0) {$\mathrm{d}Q$};
\node [anchor=center] at (4*.6,.2) {$H_j$};
\node [anchor=center] at (4*.6,-.2) {$H_j$};
\draw [black,thick] (0,.6) -- +(4*.6,0);
\draw [black,thick] (0*.6,.6) -- +(0,-.35);
\draw [black,thick] (2*.6,.6) -- +(0,-.35);
\draw [black,thick] (3*.6,.6) -- +(0,-.35);
\draw [black,thick] (4*.6,.6) -- +(0,-.15);
\draw [black,thick] (0,-.6) -- +(4*.6,0);
\draw [black,thick] (0*.6,-.6) -- +(0,.35);
\draw [black,thick] (2*.6,-.6) -- +(0,.35);
\draw [black,thick] (3*.6,-.6) -- +(0,.35);
\draw [black,thick] (4*.6,-.6) -- +(0,.15);
\node [anchor=center] at (5*.6,0) {$Q$};
\node [anchor=center] at (6*.6,0) {$\cdots$};
\node [anchor=center] at (7*.6,0) {$Q$};
\node [anchor=center] at (8*.6,.2) {$H_i$};
\node [anchor=center] at (8*.6,-.2) {$H_j$};
\draw [black,thick] (5*.6,.6) -- +(3*.6,0);
\draw [black,thick] (5*.6,.6) -- +(0,-.35);
\draw [black,thick] (7*.6,.6) -- +(0,-.35);
\draw [black,thick] (8*.6,.6) -- +(0,-.15);
\draw [black,thick] (5*.6,-.6) -- +(3*.6,0);
\draw [black,thick] (5*.6,-.6) -- +(0,.35);
\draw [black,thick] (7*.6,-.6) -- +(0,.35);
\draw [black,thick] (8*.6,-.6) -- +(0,.15);
}},
\end{split}
\end{equation}
which directly follows from \eqref{appabigcontraction}. This shows that
\begin{equation}
\fixed
C_{ji}=\sqrt{-1}\,\text{sign}(\mathbf{H}_{ij})\,(\mathrm{d}_Q\log(q\,\mathbf{H}_{ij})-\mathrm{d}_Q\log q_{(j)}).
\end{equation}
Hence in the case when $H_i$ is tangent to $Q$ but $H_j$ is not, we have
\begin{equation}
\fixed
C_{(ij)}=\sqrt{-1}\,\text{sign}(\mathbf{H}_{ij})\,\mathrm{d}_Q\log\frac{(q\,\mathbf{H}_{ij})^2}{q_{(j)}}.
\end{equation}

\section{Proof of Cancellations}\label{app:cancelation}

\flag{No further revisions.}

In this appendix we verify the cancellation \eqref{spuriousterms}
\begin{equation}\label{app:eq:spuriousterms}
\fixed
\sum_{\Delta_{(ik)}\subset\Delta_{(i)}}\text{sign}(\mathbf{H}_{ik})\int_{\Delta_{(ik)}}\frac{\sqrt{-q_{(ik)}}\,\measure{X_{(ik)}}{n-3}}{(X_{(ik)}Q_{(ik)}X_{(ik)})^{\frac{n-2}{2}}}=0.
\end{equation}
First recall the form of the expression in the normalization factor
\begin{equation}
\fixed
q_{(ik)}=
\frac{1}{(n-2)!}\parbox{3cm}{
\tikz{
\node [anchor=center] at (0*.6,0) {$Q$};
\node [anchor=center] at (1*.6,0) {$\ldots$};
\node [anchor=center] at (2*.6,0) {$Q$};
\node [anchor=center] at (3*.6,.2) {$H_i$};
\node [anchor=center] at (3*.6,-.2) {$H_i$};
\node [anchor=center] at (4*.6,.2) {$H_k$};
\node [anchor=center] at (4*.6,-.2) {$H_k$};
\draw [black,thick] (0,.6) -- (4*.6,.6);
\draw [black,thick] (0,.6) -- +(0,-.35);
\draw [black,thick] (2*.6,.6) -- +(0,-.35);
\draw [black,thick] (3*.6,.6) -- +(0,-.15);
\draw [black,thick] (4*.6,.6) -- +(0,-.15);
\draw [black,thick] (0,-.6) -- (4*.6,-.6);
\draw [black,thick] (0,-.6) -- +(0,.35);
\draw [black,thick] (2*.6,-.6) -- +(0,.35);
\draw [black,thick] (3*.6,-.6) -- +(0,.15);
\draw [black,thick] (4*.6,-.6) -- +(0,.15);
}}.
\end{equation}
Under the condition $H_iQ^{-1}H_i=0$, it is easy to verify that
\begin{equation}
\fixed
\sqrt{q_{(ik)}q}=
\frac{1}{(n-1)!}\parbox{2.5cm}{
\tikz{
\node [anchor=center] at (0*.6,0) {$Q$};
\node [anchor=center] at (1*.6,0) {$\ldots$};
\node [anchor=center] at (2*.6,0) {$Q$};
\node [anchor=center] at (3*.6,.2) {$H_i$};
\node [anchor=center] at (3*.6,-.2) {$H_k$};
\draw [black,thick] (0,.6) -- (3*.6,.6);
\draw [black,thick] (0,.6) -- +(0,-.35);
\draw [black,thick] (2*.6,.6) -- +(0,-.35);
\draw [black,thick] (3*.6,.6) -- +(0,-.15);
\draw [black,thick] (0,-.6) -- (3*.6,-.6);
\draw [black,thick] (0,-.6) -- +(0,.35);
\draw [black,thick] (2*.6,-.6) -- +(0,.35);
\draw [black,thick] (3*.6,-.6) -- +(0,.15);
}},
\end{equation}
by applying Schouten identity twice. Similarly we also have the following identity
\begin{equation}
\begin{split}
\fixed
\frac{1}{((n-1)!)^2}\parbox{2.5cm}{
\tikz{
\node [anchor=center] at (0*.6,0) {$Q$};
\node [anchor=center] at (1*.6,0) {$\ldots$};
\node [anchor=center] at (2*.6,0) {$Q$};
\node [anchor=center] at (3*.6,.2) {$H_i$};
\node [anchor=center] at (3*.6,-.2) {$H_k$};
\draw [black,thick] (0,.6) -- (3*.6,.6);
\draw [black,thick] (0,.6) -- +(0,-.35);
\draw [black,thick] (2*.6,.6) -- +(0,-.35);
\draw [black,thick] (3*.6,.6) -- +(0,-.15);
\draw [black,thick] (0,-.6) -- (3*.6,-.6);
\draw [black,thick] (0,-.6) -- +(0,.35);
\draw [black,thick] (2*.6,-.6) -- +(0,.35);
\draw [black,thick] (3*.6,-.6) -- +(0,.15);
}}
\parbox{2.5cm}{
\tikz{
\node [anchor=center] at (0*.6,0) {$Q$};
\node [anchor=center] at (1*.6,0) {$\ldots$};
\node [anchor=center] at (2*.6,0) {$Q$};
\node [anchor=center] at (3*.6,.2) {$H_i$};
\node [anchor=center] at (3*.6,-.2) {$R$};
\draw [black,thick] (0,.6) -- (3*.6,.6);
\draw [black,thick] (0,.6) -- +(0,-.35);
\draw [black,thick] (2*.6,.6) -- +(0,-.35);
\draw [black,thick] (3*.6,.6) -- +(0,-.15);
\draw [black,thick] (0,-.6) -- (3*.6,-.6);
\draw [black,thick] (0,-.6) -- +(0,.35);
\draw [black,thick] (2*.6,-.6) -- +(0,.35);
\draw [black,thick] (3*.6,-.6) -- +(0,.15);
}}&=
\frac{q}{(n-2)!}\parbox{3cm}{
\tikz{
\node [anchor=center] at (0*.6,0) {$Q$};
\node [anchor=center] at (1*.6,0) {$\ldots$};
\node [anchor=center] at (2*.6,0) {$Q$};
\node [anchor=center] at (3*.6,.2) {$H_i$};
\node [anchor=center] at (3*.6,-.2) {$H_i$};
\node [anchor=center] at (4*.6,.2) {$H_k$};
\node [anchor=center] at (4*.6,-.2) {$R$};
\draw [black,thick] (0,.6) -- (4*.6,.6);
\draw [black,thick] (0,.6) -- +(0,-.35);
\draw [black,thick] (2*.6,.6) -- +(0,-.35);
\draw [black,thick] (3*.6,.6) -- +(0,-.15);
\draw [black,thick] (4*.6,.6) -- +(0,-.15);
\draw [black,thick] (0,-.6) -- (4*.6,-.6);
\draw [black,thick] (0,-.6) -- +(0,.35);
\draw [black,thick] (2*.6,-.6) -- +(0,.35);
\draw [black,thick] (3*.6,-.6) -- +(0,.15);
\draw [black,thick] (4*.6,-.6) -- +(0,.15);
}}\\
&\equiv\frac{q}{(n-2)!}
\parbox{3.5cm}{
\tikz{
\node [anchor=center] at (0*.8+.15,0) {$Q_{(i)}$};
\node [anchor=center] at (1*.8+.15,0) {$\ldots$};
\node [anchor=center] at (2*.8+.15,0) {$Q_{(i)}$};
\node [anchor=center] at (3*.8+.2,.2) {$(H_k)_{(i)}$};
\node [anchor=center] at (3*.8+.2,-.2) {$R_{(i)}$};
\draw [black,thick] (0,.6) -- (3*.8,.6);
\draw [black,thick] (0,.6) -- +(0,-.35);
\draw [black,thick] (2*.8,.6) -- +(0,-.35);
\draw [black,thick] (3*.8,.6) -- +(0,-.15);
\draw [black,thick] (0,-.6) -- (3*.8,-.6);
\draw [black,thick] (0,-.6) -- +(0,.35);
\draw [black,thick] (2*.8,-.6) -- +(0,.35);
\draw [black,thick] (3*.8,-.6) -- +(0,.15);
}},
\end{split}
\end{equation}
for an arbitrary covector $R$. The expression in the second line can be treated exactly as being pulled out of the measure in localizing to the codim-2 boundaries, and so we are now able to lift the LHS of \eqref{app:eq:spuriousterms} back into a single integral on the codim-1 boundary $H_i$, which is
\begin{equation}
\fixed
\Big(
\parbox{2.5cm}{
\tikz{
\node [anchor=center] at (0*.6,0) {$Q$};
\node [anchor=center] at (1*.6,0) {$\ldots$};
\node [anchor=center] at (2*.6,0) {$Q$};
\node [anchor=center] at (3*.6,.2) {$H_i$};
\node [anchor=center] at (3*.6,-.2) {$R$};
\draw [black,thick] (0,.6) -- (3*.6,.6);
\draw [black,thick] (0,.6) -- +(0,-.35);
\draw [black,thick] (2*.6,.6) -- +(0,-.35);
\draw [black,thick] (3*.6,.6) -- +(0,-.15);
\draw [black,thick] (0,-.6) -- (3*.6,-.6);
\draw [black,thick] (0,-.6) -- +(0,.35);
\draw [black,thick] (2*.6,-.6) -- +(0,.35);
\draw [black,thick] (3*.6,-.6) -- +(0,.15);
}}
\Big)^{-1}\sqrt{-q}\int_{\Delta_{(i)}}
\mathrm{d}_X\!\Bigg[
\frac{\Big\langle\Big(
\parbox{3.4cm}{
\tikz{
\node [anchor=center] at (0*.8+.15,0) {$Q_{(i)}$};
\node [anchor=center] at (1*.8+.15,0) {$\ldots$};
\node [anchor=center] at (2*.8+.15,0) {$Q_{(i)}$};
\node [anchor=center] at (3*.8+.2,-.2) {$R_{(i)}$};
\draw [black,thick] (0,.6) -- (3*.8,.6);
\draw [black,thick] (0,.6) -- +(0,-.35);
\draw [black,thick] (2*.8,.6) -- +(0,-.35);
\draw [black,thick] (3*.8,.6) -- +(0,-.15);
\draw [black,thick] (0,-.6) -- (3*.8,-.6);
\draw [black,thick] (0,-.6) -- +(0,.35);
\draw [black,thick] (2*.8,-.6) -- +(0,.35);
\draw [black,thick] (3*.8,-.6) -- +(0,.15);
}}\Big)X_{(i)}\mathrm{d}^{n-3}X_{(i)}\Big\rangle
}{(X_{(i)}Q_{(i)}X_{(i)})^{\frac{n-2}{2}}}
\Bigg].
\end{equation}
Calculating the total derivative leads to the standard integrand on the codim-1 boundary times the coefficient
\begin{equation}
\fixed
\parbox{4cm}{
\tikz{
\node [anchor=center] at (0*.8+.15,0) {$Q_{(i)}$};
\node [anchor=center] at (1*.8+.15,0) {$\ldots$};
\node [anchor=center] at (2*.8+.15,0) {$Q_{(i)}$};
\node [anchor=center] at (3*.8+.15,-.2) {$R_{(i)}$};
\node [anchor=center] at (3*.8+.5,.2) {$Q_{(i)}X_{(i)}$};
\draw [black,thick] (0,.6) -- (3*.8,.6);
\draw [black,thick] (0,.6) -- +(0,-.35);
\draw [black,thick] (2*.8,.6) -- +(0,-.35);
\draw [black,thick] (3*.8,.6) -- +(0,-.15);
\draw [black,thick] (0,-.6) -- (3*.8,-.6);
\draw [black,thick] (0,-.6) -- +(0,.35);
\draw [black,thick] (2*.8,-.6) -- +(0,.35);
\draw [black,thick] (3*.8,-.6) -- +(0,.15);
}}=0,
\end{equation}
which obviously vanishes since $\det Q_{(i)}=0$.

\section{Details of the Symbol Terms for Scalar Hexagon}\label{app:hexagon}

\flag{No further revisions.}

In this appendix we list out all the non-vanishing symbol terms for the scalar hexagon in 6d discussed in Section \ref{sec:hexagon6d}.

Terms starting with $\underline{24}$ are
\begin{align}
\underline{24}\otimes\underline{13}\otimes\underline{56}&=u_1^2\otimes u_1^4\otimes\frac{x_+(1-u_2x_-)}{x_-(1-u_2x_+)},\\
\underline{24}\otimes\underline{15}\otimes\underline{36}&=u_1^2\otimes \frac{1}{u_1^2(1-u_1)^2}\otimes\frac{x_+}{x_-},\\
\underline{24}\otimes\underline{16}\otimes\underline{35}&=u_1^2\otimes \frac{u_3}{u_1}\otimes\frac{1-u_2x_-}{1-u_2x_+},\\
\underline{24}\otimes\underline{35}\otimes\underline{16}&=u_1^2\otimes u_1^4u_2^2\otimes\frac{x_+(1-u_3x_-)}{x_-(1-u_3x_+)},\\
\underline{24}\otimes\underline{36}\otimes\underline{15}&=u_1^2\otimes \frac{u_1^3}{u_3}\otimes\frac{1-u_1x_-}{1-u_1x_+},\\
\underline{24}\otimes\underline{56}\otimes\underline{13}&=u_1^2\otimes \frac{1}{u_1u_3}\otimes\frac{1-u_3x_-}{1-u_3x_+}.
\end{align}

Terms starting with $\underline{35}$ are
\begin{align}
\underline{35}\otimes\underline{12}\otimes\underline{46}&=u_2^2\otimes \frac{1}{u_2}\otimes\frac{1-u_3x_-}{1-u_3x_+},\\
\underline{35}\otimes\underline{14}\otimes\underline{26}&=u_2^2\otimes u_2^3\otimes\frac{1-u_2x_-}{1-u_2x_+},\\
\underline{35}\otimes\underline{16}\otimes\underline{24}&=u_2^2\otimes \frac{1}{u_2}\otimes\frac{1-u_1x_-}{1-u_1x_+},\\
\underline{35}\otimes\underline{24}\otimes\underline{16}&=u_2^2\otimes u_1^2u_2^4\otimes\frac{x_+(1-u_3x_-)}{x_-(1-u_3x_+)},\\
\underline{35}\otimes\underline{26}\otimes\underline{14}&=u_2^2\otimes \frac{1}{u_2^2(1-u_2)^2}\otimes\frac{x_+}{x_-},\\
\underline{35}\otimes\underline{46}\otimes\underline{12}&=u_2^2\otimes u_2^4u_3^2\otimes\frac{x_+(1-u_1x_-)}{x_-(1-u_1x_+)}.
\end{align}

Terms starting with $\underline{46}$ are
\begin{align}
\underline{46}\otimes\underline{12}\otimes\underline{35}&=u_3^2\otimes \frac{u_1}{u_3}\otimes\frac{1-u_2x_-}{1-u_2x_+},\\
\underline{46}\otimes\underline{13}\otimes\underline{25}&=u_3^2\otimes \frac{1}{u_3^2(1-u_3)^2}\otimes\frac{x_+}{x_-},\\
\underline{46}\otimes\underline{15}\otimes\underline{23}&=u_3^2\otimes u_3^4\otimes\frac{x_+(1-u_2x_-)}{x_-(1-u_2x_+)},\\
\underline{46}\otimes\underline{23}\otimes\underline{15}&=u_3^2\otimes \frac{1}{u_1u_3}\otimes\frac{1-u_1x_-}{1-u_1x_+},\\
\underline{46}\otimes\underline{25}\otimes\underline{13}&=u_3^2\otimes \frac{u_3^3}{u_1}\otimes\frac{1-u_3x_-}{1-u_3x_+},\\
\underline{46}\otimes\underline{35}\otimes\underline{12}&=u_3^2\otimes u_2^2u_3^4\otimes\frac{x_+(1-u_1x_-)}{x_-(1-u_1x_+)}.
\end{align}

\flag{I have checked consistency of these expressions, confirming that they sum to the correct result.}

\section{Integrals with a Degenerate Quadric}\label{app:degenerateQ}

In this appendix we provide a detailed account on integrals where the quadric is degenerate. We have already encountered in Section \ref{sec:tangentconfig} a simple case in the discussion of $e$ integrals where the boundaries are tangent to the quadric, such that a degenerate quadric of corank-1 is induced on the boundaries. Here we discuss the most general situations.

\subsection{The null space of $Q$}\label{app:Qnullspace}

When $Q$ is degenerate $q\equiv\det Q=0$; the first non-trivial invariant we can possibly construct is to do the same contraction with Levi-Civita but with $r$ $Q$'s replaced by $r$ reference covectors $R_a$
\begin{equation}
\mathring{q}=\parbox{4.2cm}{
\tikz{
\node [anchor=center] at (0,0) {$Q$};
\node [anchor=center] at (.6,0) {$Q$};
\node [anchor=center] at (1.2,0) {$\cdots$};
\node [anchor=center] at (1.8,0) {$Q$};
\node [anchor=center] at (2.4,.2) {$R_1$};
\node [anchor=center] at (3,.2) {$\cdots$};
\node [anchor=center] at (3.6,.2) {$R_r$};
\node [anchor=center] at (2.4,-.2) {$R_1$};
\node [anchor=center] at (3,-.2) {$\cdots$};
\node [anchor=center] at (3.6,-.2) {$R_r$};
\draw [black,thick] (0,.6) -- (3.6,.6);
\draw [black,thick] (0,.6) -- (0,.25);
\draw [black,thick] (.6,.6) -- (.6,.25);
\draw [black,thick] (1.8,.6) -- (1.8,.25);
\draw [black,thick] (2.4,.6) -- (2.4,.45);
\draw [black,thick] (3.6,.6) -- (3.6,.45);
\draw [black,thick] (0,-.6) -- (3.6,-.6);
\draw [black,thick] (0,-.6) -- (0,-.25);
\draw [black,thick] (.6,-.6) -- (.6,-.25);
\draw [black,thick] (1.8,-.6) -- (1.8,-.25);
\draw [black,thick] (2.4,-.6) -- (2.4,-.45);
\draw [black,thick] (3.6,-.6) -- (3.6,-.45);
}}.
\end{equation}
Associated with it is a modified inverse
\begin{equation}
\mathring{Q}^{-1}=\parbox{4cm}{
\tikz{
\node [anchor=center] at (.6,0) {$Q$};
\node [anchor=center] at (1.2,0) {$\cdots$};
\node [anchor=center] at (1.8,0) {$Q$};
\node [anchor=center] at (2.4,.2) {$R_1$};
\node [anchor=center] at (3,.2) {$\cdots$};
\node [anchor=center] at (3.6,.2) {$R_r$};
\node [anchor=center] at (2.4,-.2) {$R_1$};
\node [anchor=center] at (3,-.2) {$\cdots$};
\node [anchor=center] at (3.6,-.2) {$R_r$};
\draw [black,thick] (0,.6) -- (3.6,.6);
\draw [black,thick] (0,.6) -- (0,.25);
\draw [black,thick] (.6,.6) -- (.6,.25);
\draw [black,thick] (1.8,.6) -- (1.8,.25);
\draw [black,thick] (2.4,.6) -- (2.4,.45);
\draw [black,thick] (3.6,.6) -- (3.6,.45);
\draw [black,thick] (0,-.6) -- (3.6,-.6);
\draw [black,thick] (0,-.6) -- (0,-.25);
\draw [black,thick] (.6,-.6) -- (.6,-.25);
\draw [black,thick] (1.8,-.6) -- (1.8,-.25);
\draw [black,thick] (2.4,-.6) -- (2.4,-.45);
\draw [black,thick] (3.6,-.6) -- (3.6,-.45);
}}\Big/\mathring{q}.
\end{equation}

The reference covectors $\{R_1,\ldots,R_r\}$ also induce a set of null vectors of $Q$
\begin{equation}\label{nullspacebasis}
N_a:=
\parbox{5.2cm}{
\tikz{
\node [anchor=center] at (0,.2) {$R_a$};
\node [anchor=center] at (.6,0) {$Q$};
\node [anchor=center] at (2*.6,0) {$\ldots$};
\node [anchor=center] at (3*.6,0) {$Q$};
\node [anchor=center] at (4*.6,.2) {$R_1$};
\node [anchor=center] at (5*.6,.2) {$\ldots$};
\node [anchor=center] at (6*.6,.2) {$\widehat{a}$};
\node [anchor=center] at (7*.6,.2) {$\ldots$};
\node [anchor=center] at (8*.6,.2) {$R_{r}$};
\node [anchor=center] at (4*.6,-.2) {$R_1$};
\node [anchor=center] at (5*.6,-.2) {$\ldots$};
\node [anchor=center] at (6*.6,-.2) {$\widehat{a}$};
\node [anchor=center] at (7*.6,-.2) {$\ldots$};
\node [anchor=center] at (8*.6,-.2) {$R_{r}$};
\draw [black,thick] (0,.6) -- (8*.6,.6);
\draw [black,thick] (0,.6) -- (0,.45);
\draw [black,thick] (.6,.6) -- (.6,.25);
\draw [black,thick] (3*.6,.6) -- (3*.6,.25);
\draw [black,thick] (4*.6,.6) -- (4*.6,.45);
\draw [black,thick] (8*.6,.6) -- (8*.6,.45);
\draw [black,thick] (0,-.6) -- (8*.6,-.6);
\draw [black,thick] (0,-.6) -- (0,-.45);
\draw [black,thick] (.6,-.6) -- (.6,-.25);
\draw [black,thick] (3*.6,-.6) -- (3*.6,-.25);
\draw [black,thick] (4*.6,-.6) -- (4*.6,-.45);
\draw [black,thick] (8*.6,-.6) -- (8*.6,-.45);
}},
\end{equation}
where $\widehat{a}$ indicates that we omit $R_a$ in the corresponding contraction. To see why this is true, we contract the remaining index above with $Q$, then we have (by applying Schouten identity on the red indices below)
\begin{equation}\label{nullcondition}
(QN_a)_I=\parbox{5.5cm}{
\tikz{
\node [anchor=center] at (0,.2) {$R_a$};
\node [anchor=center] (q) at (0,-.2) {$Q$};
\draw [thick,red] (q.west) --+(-.2,0);
\node [anchor=center] at (.6,0) {$Q$};
\node [anchor=center] at (2*.6,0) {$\ldots$};
\node [anchor=center] at (3*.6,0) {$Q$};
\node [anchor=center] at (4*.6,.2) {$R_1$};
\node [anchor=center] at (5*.6,.2) {$\ldots$};
\node [anchor=center] at (6*.6,.2) {$\widehat{a}$};
\node [anchor=center] at (7*.6,.2) {$\ldots$};
\node [anchor=center] at (8*.6,.2) {$R_{r}$};
\node [anchor=center] at (4*.6,-.2) {$R_1$};
\node [anchor=center] at (5*.6,-.2) {$\ldots$};
\node [anchor=center] at (6*.6,-.2) {$\widehat{a}$};
\node [anchor=center] at (7*.6,-.2) {$\ldots$};
\node [anchor=center] at (8*.6,-.2) {$R_{r}$};
\draw [black,thick] (0,.6) -- (8*.6,.6);
\draw [red,thick] (0,.6) -- (0,.45);
\draw [red,thick] (.6,.6) -- (.6,.25);
\draw [red,thick] (3*.6,.6) -- (3*.6,.25);
\draw [red,thick] (4*.6,.6) -- (4*.6,.45);
\draw [red,thick] (8*.6,.6) -- (8*.6,.45);
\draw [black,thick] (0,-.6) -- (8*.6,-.6);
\draw [black,thick] (0,-.6) -- (0,-.45);
\draw [black,thick] (.6,-.6) -- (.6,-.25);
\draw [black,thick] (3*.6,-.6) -- (3*.6,-.25);
\draw [black,thick] (4*.6,-.6) -- (4*.6,-.45);
\draw [black,thick] (8*.6,-.6) -- (8*.6,-.45);
}}
=\sum_{b=1}^{r}\frac{R_{bI}}{n-r}
\parbox{5cm}{
\tikz{
\node [anchor=center] at (.6,0) {$Q$};
\node [anchor=center] at (2*.6,0) {$\ldots$};
\node [anchor=center] at (3*.6,0) {$Q$};
\node [anchor=center] at (4*.6,.2) {$R_1$};
\node [anchor=center] at (5*.6,.2) {$\ldots$};
\node [anchor=center] at (6*.6,.2) {$\widehat{b}$};
\node [anchor=center] at (7*.6,.2) {$\ldots$};
\node [anchor=center] at (8*.6,.2) {$R_{r}$};
\node [anchor=center] at (4*.6,-.2) {$R_1$};
\node [anchor=center] at (5*.6,-.2) {$\ldots$};
\node [anchor=center] at (6*.6,-.2) {$\widehat{a}$};
\node [anchor=center] at (7*.6,-.2) {$\ldots$};
\node [anchor=center] at (8*.6,-.2) {$R_{r}$};
\draw [black,thick] (.6,.6) -- (8*.6,.6);
\draw [black,thick] (.6,.6) -- (.6,.25);
\draw [black,thick] (3*.6,.6) -- (3*.6,.25);
\draw [black,thick] (4*.6,.6) -- (4*.6,.45);
\draw [black,thick] (8*.6,.6) -- (8*.6,.45);
\draw [black,thick] (.6,-.6) -- (8*.6,-.6);
\draw [black,thick] (.6,-.6) -- (.6,-.25);
\draw [black,thick] (3*.6,-.6) -- (3*.6,-.25);
\draw [black,thick] (4*.6,-.6) -- (4*.6,-.45);
\draw [black,thick] (8*.6,-.6) -- (8*.6,-.45);
}},
\end{equation}
where each contraction on RHS contains $(n-r+1)$ $Q$'s and thus vanishes. 

It is easy to observe the orthogonality
\begin{equation}
R_{a_I}N_b^I=\mathring{q}\delta_{ab}.
\end{equation}
Hence we see that a necessary condition for $\mathring{q}$ to be nonzero is that none of the $R$'s induces a hyperplane that fully contains the null space of $Q$ as a subspace. When this is the case the null space $\mathfrak{N}_Q$ of $Q$ is the set of all linear combinations of $\{N_1,\ldots,N_r\}$ quotienting an overall scaling
\begin{equation}
\mathfrak{N}_Q=\{\sum_{a=1}^r\alpha_aN_a|[\alpha_1:\ldots:\alpha_r]\in\mathbb{CP}^{r-1}\}.
\end{equation}
Correspondingly this naturally leads to a projection onto the null space
\begin{equation}
\mathring{P}^I_J=\mathring{q}^{-1}\sum_{a=1}^rN_a^IR_{aJ},\qquad
\mathring{P}^2=\mathring{P}.
\end{equation}
Furthermore, we have
\begin{equation}
\mathring{Q}^{-1}Q=\frac{1}{n-r}(\mathbf{1}-\mathring{P}).
\end{equation}

\subsection{Tensor integrals with a degenerate $Q$, special case}\label{app:degenerateQintegralspecial}

We first study the integrals $E_{n,k}$ with nontrivial tensor numerators, $k>0$, under the assumption that $\mathfrak{N}_Q$ is null to $T$, i.e.,
\begin{equation}\label{nullcondition}
\quad (TN)_{I_1I_2\ldots I_{k-1}}=0,\quad\forall N\in\mathfrak{N}_Q.
\end{equation}
Equivalently this means that $T\mathring{P}W=0$ for arbitrary vector $W$. This null condition is sufficient for all our physical applications, as it turns out to be satisfied by any one-loop Feynman integrals.

As with the case of non-degenerate $Q$, here it is natural to use the modified inverse $\mathring{Q}^{-1}$, with which we have the following identity
\begin{equation}
\begin{split}
\mathrm{d}\left[\frac{\langle(\mathring{Q}^{-1}T)[X^{k-1}]X\mathrm{d}^{n-2}X\rangle}{(XQX)^{\frac{n+k-2}{2}}}\right]=&-(k-1)\frac{\measure{X}{n-1}\,(\text{tr}_{\mathring{Q}^{-1}}T)[X^{k-2}]}{(XQX)^{\frac{n+k-2}{2}}}\\
&+(n+k-2)\frac{\measure{X}{n-1}\,(XQ\mathring{Q}^{-1}T)[X^{k-1}]}{(XQX)^{\frac{n+k}{2}}}.
\end{split}
\end{equation}
Using Shouten identity we can verify that
\begin{equation}\label{tensorreduce}
(n-r)(XQ\mathring{Q}^{-1}T)[X^{k-1}]=T[X^k]+(T\mathring{P}X)[X^{k-1}],
\end{equation}
where the second term on RHS vanishes due to the null condition. Hence the integrand again decomposes into a total derivative plus a term of the same form but with a new tensor numerator with its rank lowered by two. This second term vanishes for $E_{n,1}$ integrals (as is obvious by its coefficient).

We can now localized the total derivative part to the codim-1 boundaries, and the original integral decomposes into
\begin{equation}\label{degenerateQtensorderivative}
\begin{split}
E_{n,k}=&\frac{(n-r)}{n+k-2}\sum_{i=1}^n\int_{\Delta_{(i)}}\frac{\measure{X_{(i)}}{n-2}\,(H_i\mathring{Q}^{-1}T)[(P(i)X_{(i)})^{k-1}]}{(X_{(i)}Q_{(i)}X_{(i)})^{\frac{n+k-2}{2}}}\\
&+\frac{(k-1)(n-r)}{n+k-2}\int_\Delta\frac{\measure{X}{n-1}\,(\text{tr}_{\mathring{Q}^{-1}}T)[X^{k-2}]}{(XQX)^{\frac{n+k-2}{2}}}.
\end{split}
\end{equation}

In principle we should be able to repeat this analysis to fully decompose the integral $I$ into $e$ integrals. For example, if both $n$ and $k$ are even, obviously the decomposition contains the highest weight term (up to a constant factor)
\begin{equation}
((\mathring{Q}^{-1})^{I_1I_2}\ldots(\mathring{Q}^{-1})^{I_{k-1}I_k}T_{I_1I_2\ldots I_k})\,e_{n}.
\end{equation}
In practice, however, we need to make sure that the null condition \eqref{nullcondition} is preserved for each of the new integrals on the codim-1 boundaries.

In general we expect the corank of $Q_{(i)}$ reduces by one as compared to $\mathrm{corank}Q$. Without loss of generality we can choose the reference covectors in the new space to be $\{P(i)R_1,P(i)R_2,\ldots,P(i)R_{r-1}\}$ (these references can be different from the ones in the first decomposition). They induce a basis of $\mathfrak{N}_{Q_{(i)}}$. We can easily see that for any $a=1,\ldots,r-1$
\begin{equation}
T_{(i)}N_{a(i)}=\parbox{5.5cm}{
\tikz{
\node [anchor=center] at (0,.2) {$R_a$};
\node [anchor=center] at (0,-.2) {$T$};
\node [anchor=center] at (.6,0) {$Q$};
\node [anchor=center] at (1.2,0) {$\cdots$};
\node [anchor=center] at (1.8,0) {$Q$};
\node [anchor=center] at (2.4,.2) {$H_i$};
\node [anchor=center] at (5*.6,.2) {$R_1$};
\node [anchor=center] at (6*.6,.2) {$\cdots$};
\node [anchor=center] at (7*.6,.2) {$\widehat{a}$};
\node [anchor=center] at (8*.6,.2) {$\cdots$};
\node [anchor=center] at (9*.6+.2,.2) {$R_{r-1}$};
\node [anchor=center] at (2.4,-.2) {$H_i$};
\node [anchor=center] at (5*.6,-.2) {$R_1$};
\node [anchor=center] at (6*.6,-.2) {$\cdots$};
\node [anchor=center] at (7*.6,-.2) {$\widehat{a}$};
\node [anchor=center] at (8*.6,-.2) {$\cdots$};
\node [anchor=center] at (9*.6+.2,-.2) {$R_{r-1}$};
\draw [black,thick] (0,.6) -- (5.4,.6);
\draw [black,thick] (0,.6) -- (0,.45);
\draw [black,thick] (.6,.6) -- (.6,.25);
\draw [black,thick] (1.8,.6) -- (1.8,.25);
\draw [black,thick] (2.4,.6) -- (2.4,.45);
\draw [black,thick] (3,.6) -- (3,.45);
\draw [black,thick] (5.4,.6) -- (5.4,.45);
\draw [black,thick] (0,-.6) -- (5.4,-.6);
\draw [black,thick] (0,-.6) -- (0,-.45);
\draw [black,thick] (.6,-.6) -- (.6,-.25);
\draw [black,thick] (1.8,-.6) -- (1.8,-.25);
\draw [black,thick] (2.4,-.6) -- (2.4,-.45);
\draw [black,thick] (3,-.6) -- (3,-.45);
\draw [black,thick] (5.4,-.6) -- (5.4,-.45);
}}
\end{equation}
where the RHS is equivalent to $T$ contracted with a null vector in the original space, thus equaling to zero. This guarantees that we can continue to drop off the second term on RHS of \eqref{tensorreduce} as we repeatedly apply the decomposition.

As a result, we see that under the null condition on the numerator, the structure of the decomposition of an $E_{n,k}$ integral onto the $E_{n}$ integrals follows similarly with the cases with non-degenerate $Q$. The actual behavior of the transcendental weights of course has to depend on whether the induced quadric in each $E_n$ in the decomposition is still degenerate, which we will discuss later.

\subsection{Tensor integrals with a degenerate $Q$, generic case}

Now let us consider the situation when the tensor numerate is generic. Although these are not directly needed for the study of Feynman integrals, we keep the general discussion here, having in mind that they might as well find applications in other physical objects.

The aim is again to localize a given tensor integral to the boundaries of the contour, but here we cannot directly follow the procedure discussed in the previous subsection, due to the absence of the null condition.

To acquire some intuition, let us start with the simplest case $E_{m,1}$, with corank$Q=1$. As was argued in Section \ref{sec:ecaseofdegenerateQ}, if we were to use the ansatz of the form $\langle WX\mathrm{d}^{n-2}X\rangle$ for the lower form inside the exterior derivative, then it necessarily implies that the numerator $L$ under study have to satisfy the null condition. This forces us to go to the next simplest ansatz. Consider the identity
\begin{equation}
\mathrm{d}\left[\frac{\langle S[X^2]X\mathrm{d}^{n-2}X\rangle}{(XQX)^{\frac{n+1}{2}}}\right]=-2\frac{\measure{X}{n-1}\,(\text{tr}S)[X]}{(XQX)^{\frac{n+1}{2}}}+(n+1)\frac{\measure{X}{n-1}\,(XQS)[X^2]}{(XQX)^{\frac{n+3}{2}}},
\end{equation}
where $S$ is a rank-$(1,2)$ tensor symmetric in its two lower indices, and $\text{tr}S$ a covector obtained from $S$ by contracting its upper index and a lower one. (Naively there is a simpler ansatz, where $S$ is instead a rank-$(1,1)$ tensor, but that cannot work simply due to the mismatch in the exponent of the denominator when comparing with $E_{n,1}$.)

Observe that if we manage to find an $S$ such that
\begin{equation}
(\text{tr}S)_I=L_I,\text{ and }(XQS)_{IJ}=0,\quad\forall I,J,
\end{equation}
then it allows to localize the $E_{n,1}$ integral. We are indeed able to construct such a solution, even without the need of any reference vector, which is
\begin{equation}
S=\frac{\parbox{4cm}{
\tikz{
\node [anchor=center] at (0,0) {$Q$};
\node [anchor=center] at (.6,0) {$\cdots$};
\node [anchor=center] at (2*.6,0) {$Q$};
\node [anchor=center] at (3*.6,0) {$L$};
\node [anchor=center] at (4*.6,0) {$L$};
\node [anchor=center] at (5*.6,0) {$L$};
\draw [black,thick] (-.6,.4) -- +(3*.6,0);
\draw [black,thick] (-.6,.4) -- +(0,-.15);
\draw [black,thick] (0,.4) -- +(0,-.15);
\draw [black,thick] (2*.6,.4) -- +(0,-.15);
\draw [black,thick] (0,-.4) -- +(3*.6,0);
\draw [black,thick] (0,-.4) -- +(0,.15);
\draw [black,thick] (2*.6,-.4) -- +(0,.15);
\draw [black,thick] (3*.6,-.4) -- +(0,.15);
\draw [black,thick] (4*.6,-.4) -- +(0,.15);
\draw [black,thick] (5*.6,-.4) -- +(0,.15);
}}}{\parbox{3cm}{
\tikz{
\node [anchor=center] at (-.6,0) {$L$};
\node [anchor=center] at (0,0) {$Q$};
\node [anchor=center] at (.6,0) {$\cdots$};
\node [anchor=center] at (2*.6,0) {$Q$};
\node [anchor=center] at (3*.6,0) {$L$};
\draw [black,thick] (-.6,.4) -- +(3*.6,0);
\draw [black,thick] (-.6,.4) -- +(0,-.15);
\draw [black,thick] (0,.4) -- +(0,-.15);
\draw [black,thick] (2*.6,.4) -- +(0,-.15);
\draw [black,thick] (0,-.4) -- +(3*.6,0);
\draw [black,thick] (0,-.4) -- +(0,.15);
\draw [black,thick] (2*.6,-.4) -- +(0,.15);
\draw [black,thick] (3*.6,-.4) -- +(0,.15);
}}}.
\end{equation}
We thus obtain the decomposition
\begin{equation}
E_{n,1}=\sum_{i=1}^n\frac{\parbox{3cm}{
\tikz{
\node [anchor=center] at (-.6,0) {$H_i$};
\node [anchor=center] at (0,0) {$Q$};
\node [anchor=center] at (.6,0) {$\cdots$};
\node [anchor=center] at (2*.6,0) {$Q$};
\node [anchor=center] at (3*.6,0) {$L$};
\draw [black,thick] (-.6,.4) -- +(3*.6,0);
\draw [black,thick] (-.6,.4) -- +(0,-.15);
\draw [black,thick] (0,.4) -- +(0,-.15);
\draw [black,thick] (2*.6,.4) -- +(0,-.15);
\draw [black,thick] (0,-.4) -- +(3*.6,0);
\draw [black,thick] (0,-.4) -- +(0,.15);
\draw [black,thick] (2*.6,-.4) -- +(0,.15);
\draw [black,thick] (3*.6,-.4) -- +(0,.15);
}}}{\parbox{3cm}{
\tikz{
\node [anchor=center] at (-.6,0) {$L$};
\node [anchor=center] at (0,0) {$Q$};
\node [anchor=center] at (.6,0) {$\cdots$};
\node [anchor=center] at (2*.6,0) {$Q$};
\node [anchor=center] at (3*.6,0) {$L$};
\draw [black,thick] (-.6,.4) -- +(3*.6,0);
\draw [black,thick] (-.6,.4) -- +(0,-.15);
\draw [black,thick] (0,.4) -- +(0,-.15);
\draw [black,thick] (2*.6,.4) -- +(0,-.15);
\draw [black,thick] (0,-.4) -- +(3*.6,0);
\draw [black,thick] (0,-.4) -- +(0,.15);
\draw [black,thick] (2*.6,-.4) -- +(0,.15);
\draw [black,thick] (3*.6,-.4) -- +(0,.15);
}}}
\int_{\Delta_{(i)}}\frac{\measure{X_{(i)}}{n-2}\,(LP(i)X_{(i)})^2}{(X_{(i)}Q_{(i)}X_{(i)})^{\frac{n+1}{2}}}.
\end{equation}
Note that the decomposition discussed here for integrals with a generic tensor and a degenerate quadric is qualitatively different from the cases we analyzed before, where either the quadric is non-degenerate or the numerator meets the null condition. In this decomposition the rank of the numerator increases instead of decreases as we localize to the boundaries.


Now let us generalize the above result to higher-rank tensors as well as high-corank $Q$. In fact we can make the above observation more systematic. Let us first still focus on $T_{n,1}$ but assume $Q$ to have arbitrary corank $r$. Then we can apply the following identity
\begin{equation}
\begin{split}
\mathrm{d}\left[\frac{(LX)\,\langle (\mathring{Q}^{-1}QX)X\mathrm{d}^{n-2}X\rangle}{(XQX)^{\frac{n+1}{2}}}\right]&=-\frac{\text{tr}(\mathring{Q}^{-1}Q)\,(LX)\,\measure{X}{n-1}}{(XQX)^{\frac{n+1}{2}}}+\frac{(L\mathring{Q}^{-1}QX)\,\measure{X}{n-1}}{(XQX)^{\frac{n+1}{2}}}\\
&\quad+(n+1)\frac{(XQ\mathring{Q}^{-1}QX)(LX)\,\measure{X}{n-1}}{(XQX)^{\frac{n+3}{2}}}.
\end{split}
\end{equation}
Given the fact that
\begin{equation}
\text{tr}\mathring{Q}^{-1}Q=1,\qquad
(XQ\mathring{Q}^{-1}QX)=\frac{1}{n-r}(XQX),
\end{equation}
we see the first and the third term are both proportional to the integrand of $E_{n,1}$ up to a constant. While the second term still leads to some other $E_{n,1}$, its linear numerator $(L\mathring{Q}^{-1}QX)$ obviously satisfies the null condition, and so can be further analyzed with the method described before.  Hence we have
\begin{equation}
\begin{split}
\frac{(LX)\,\measure{X}{n-1}}{(XQX)^{\frac{n+1}{2}}}&=\frac{n-r}{r+1}\left(\mathrm{d}\left[\frac{(LX)\,\langle (\mathring{Q}^{-1}QX)X\mathrm{d}^{n-2}X\rangle}{(XQX)^{\frac{n+1}{2}}}\right]-\frac{(L\mathring{Q}^{-1}QX)\,\measure{X}{n-1}}{(XQX)^{\frac{n+1}{2}}}\right)\\
&=\frac{n-r}{r+1}\,\mathrm{d}\left[\frac{(LX)\,\langle (\mathring{Q}^{-1}QX)X\mathrm{d}^{n-2}X\rangle}{(XQX)^{\frac{n+1}{2}}}-\frac{\langle (\mathring{Q}^{-1}Q\mathring{Q}^{-1}L)X\mathrm{d}^{n-2}X\rangle}{(XQX)^{\frac{n-1}{2}}}\right].
\end{split}
\end{equation}
In particular, when corank$Q=1$ the expression reduces to the solution we previously found.

It is straightforward to generalize the above construction to higher-rank tensors. Here it is helpful to temporarily treat all the indices of $T$ to be distinct, some of which might satisfy the null condition while some other not. Then most generally we have
\begin{equation}
\begin{split}
\frac{T[X^k]\,\measure{X}{n-1}}{(XQX)^{\frac{n+k}{2}}}&=\frac{n-r}{r+1}\;\mathrm{d}\!\left[\frac{T[X^k]\,\langle (\mathring{Q}^{-1}QX)X\mathrm{d}^{n-2}X\rangle}{(XQX)^{\frac{n+k}{2}}}\right]\\
&\quad-\frac{n-r}{r+1}\sum_{a=1}^k\frac{(XQ\mathring{Q}^{-1}T)_{\hat{a}}[X^{k-1}]\,\measure{X}{n-1}}{(XQX)^{\frac{n+k}{2}}},
\end{split}
\end{equation}
where $(XQ\mathring{Q}^{-1}T)_{\hat{a}}$ denotes the tensor obtained by contracting the $a^{\rm th}$ index of $T$ with $XQ\mathring{Q}$. This means that in each of the remaining term we are projecting one leg of $T$ to a subspace that satisfies the null condition. Furthermore, note that
\begin{equation}
Q\mathring{Q}^{-1}Q\mathring{Q}^{-1}=\frac{1}{n-r} Q\mathring{Q}^{-1},
\end{equation}
which indicates that the null condition on any leg of $T$ is preserved by the above operation. As a result, applying the same operation at most $k$ times will yield, apart from a total derivative, a remaining $E_{n,k}$ integral that completely satisfies the null condition, which then can be treated by the method in the previous subsection.

\subsection{$E_n$ with a degenerate $Q$}

With the above tools we are able to further simplify the $E_{n}$ integrals with corank$Q=r$ by localizing into lower dimensional space. This is achieved by picking up any pair $(R_a,N_a)$ with a specific label $a\in\{1,2,\ldots,r\}$, and applying the following identity
\begin{equation}
\frac{\langle X\mathrm{d}^{n-1}X\rangle}{(XQX)^{\frac{n}{2}}}=-\frac{n-1}{\mathring{q}}\,\mathrm{d}\Big[\frac{(R_aX)\,\langle N_aX\mathrm{d}^{n-2}X\rangle}{(XQX)^{\frac{n}{2}}}\Big],
\end{equation}
which is a consequence of $N_aQX=0$, following \eqref{nullcondition}. Hence when $Q$ is degenerate the integral potentially has lower weight than the case of a non-degenerate $Q$.

\subsection{Projection of the degenerate quadric}\label{app:projectdegenerateQ}

Previously when we described the discontinuity prescription for the $e_n$ integrals (with even $n$) we implicitly assumed that the procedure of carrying out the $S^2$ contour integrations can be performed $\frac{n}{2}$ times until it leads to $1$, so that correspondingly we obtain a symbol of length $\frac{n}{2}$. Although we did not verify this directly, it is indirectly guaranteed by its connection to the differentiation method as was discussed in \ref{subsec:eqdis2dif}.

However, more generally when the quadric is degenerate this no longer holds and some new features arise. We assume corank$Q=r$. Since $\det Q=0$ we do not normalize the $E_n$ integrals to $e_n$.

Let us first check the behavior of $Q$ under the projection $\mathfrak{p}_{(S)}:Q\mapsto Q^{(S)}$. Obviously this map is well-defined only when the cardinality $|S|\leq n-r$, since otherwise we necessarily have $\det Q_{S,S}=0$. 

There are three special situations that we need to take care.

\subsubsection{Extreme projection}

In the extreme case the projection is trivial
\begin{equation}
\mathfrak{p}_{(S)}Q=\mathbf{0},\qquad\text{iff }|S|=n-r.
\end{equation}
This is a direct consequence of the identity $Q\mathring{Q}^{-1}Q=Q$ (for $\mathring{Q}^{-1}$ associated to any set of $r$ independent covectors). Correspondingly the affine transformation brings the original quadric into merely a product $w_iw_j$, and so the spherical contour is not well-defined.

This indicates that once the rank of the quadric is no bigger than 2, we are not able to use the spherical contours to detect the properties of $E_n$, and they have to be treated separately.

\subsubsection{Next-to-extreme projection}

When we have instead $|S|=n-r-1$, the projection leads to a matrix of rank 1, i.e., the direct product of a covector
\begin{equation}
Q^{(S)}\equiv Q_{\widehat{S},\widehat{S}}-Q_{\widehat{S},S}Q_{S,S}^{-1}Q_{S,\widehat{S}}=L^{(S)}L^{(S)}.
\end{equation}
To see this, note that for any null vector $N_a$ of $Q$, in components we have
\begin{equation}
Q_{i,\widehat{S}}N_{\widehat{(S)}}+Q_{i,S}N_{(S)}=0,\quad\forall i,
\end{equation}
and so
\begin{equation}
Q^{(S)}N_{\widehat{(S)}}=Q_{\widehat{S},\widehat{S}}N_{\widehat{(S)}}+Q_{\widehat{S},S}Q_{S,S}^{-1}Q_{S,S}N_{(S)}=0,\quad\forall N\in\mathfrak{N}_Q.
\end{equation}
This indicates that $\mathfrak{N}_Q\cap\Delta^{(S)}$ forms the null space of $Q^{(S)}$, which carves out a $\mathbb{CP}^{r-1}$ in $\mathbb{CP}^r$.

In this case the original quadric projects to a linears factor squared. Hence the remaining integral only produces a rational function and our discontinuity analysis can properly terminate. Explicitly
\begin{equation}\label{degeneratelastintegral}
\mathfrak{p}_{(S)}E_n\propto\int_{\Delta^{(S)}}\frac{\langle X_{(\widehat{S})}\mathrm{d}X_{(\widehat{S})}^{r}\rangle}{(L^{(S)}X_{(\widehat{S})})^{r+1}}=\frac{1}{r!\,\prod_{I\in \widehat{S}}L_I^{(S)}}.
\end{equation}
The product in the final answer can be easily written in terms of $Q$, although there is not a unique way. Note that $Q^{(S)}_{ij}=L^{(S)}_iL^{(S)}_j$. Specifically, if $r$ is odd, we can pick up any partition $\sigma$ of $\widehat{S}$ into $\frac{r+1}{2}$ pairs, and then
\begin{equation}
\prod_{I\in\widehat{S}}L^{(S)}_I=Q^{(S)}_{\sigma_1\sigma_2}\cdots Q^{(S)}_{\sigma_r\sigma_{r+1}},\quad\text{odd }r,\;\;\forall \sigma.
\end{equation}
If $r$ is even instead, we can pick out any label $I\in\widehat{S}$ and then
\begin{equation}
\prod_{I\in\widehat{S}}L^{(S)}_I=\frac{\prod_{J\in\widehat{S},J\neq I}Q^{(S)}_{IJ}}{(Q^{(S)}_{II})^{\frac{r}{2}}},\quad\text{even }r,\;\;\forall I\in\widehat{S}.
\end{equation}

\subsubsection{Next-to-next-to-extreme projection}

When we have $|S|=n-r-2$, the projected quadric degenerates into two linear factors. This is true because we can further compute the discriminant of the projected quadric wrst any remaining variables, say some $x_i$, which associates to an additional projection $\mathfrak{p}_{(i)}Q^{(S)}=Q^{(S\cup\{i\})}$, but from the above discussion $X_{\widehat{(S\cup\{i\})}}Q^{(S\cup\{i\})}X_{\widehat{(S\cup\{i\})}}$ is just a perfect square. Hence we can identify in this case
\begin{equation}
Q_{IJ}^{(S)}=\frac{1}{2}\left(L_I^{(S),1}L_J^{(S),2}+L_I^{(S),2}L_J^{(S),1}\right).
\end{equation}

As commented before the spherical contours are not well-defined for $Q^{(S)}$. Fortunately the resulting integrals
\begin{equation}
\mathfrak{p}_{(S)}E_n\propto\int_{\Delta^{(S)}}\frac{\langle X_{\widehat{(S)}}\mathrm{d}X_{\widehat{(S)}}^{r+1}\rangle}{(L^{(S),1}X_{\widehat{(S)}})^{\frac{r+2}{2}}(L^{(S),2}X_{\widehat{(S)}})^{\frac{r+2}{2}}}
\end{equation}
can be easily computed. 

To easily see this, we introduce one more Feynman parameter to turn the above expression into
\begin{equation}
\int_0^\infty\mathrm{d}\alpha\,\alpha^{\frac{r}{2}}\int_{\Delta^{(S)}}\frac{\langle X_{\widehat{(S)}}\mathrm{d}X_{\widehat{(S)}}^{r+1}\rangle}{((L^{(S),1}+\alpha L^{(S),2})X_{\widehat{(S)}})^{r+2}}.
\end{equation}
The integration over $\Delta^{(S)}$ is performed in the same way as in the case of next-to-extreme projection, resulting in
\begin{equation}
\int_0^\infty\mathrm{d}\alpha\,\frac{\alpha^{\frac{r}{2}}}{r!\,\prod_{I\in\widehat{S}}(L^{(S),1}+\alpha L^{(S),2})}.
\end{equation}
The situation now divides into two cases. When $r$ is odd, the $\alpha$ integration leads to a rational function, which is
\begin{equation}
(-1)^{\frac{r+1}{2}}\pi\sum_{i=1}^{r+2}\frac{(L^{(S),1}_iL^{(S),2}_i)^{\frac{r}{2}}}{\prod_{j\neq i}(L^{(S),1}_iL^{(S),2}_j-L^{(S),1}_jL^{(S),2}_i)},\qquad \text{odd }r.
\end{equation}
When $r$ is even, the $\alpha$ integration leads to a weight-1 function, which is
\begin{equation}
(-1)^{\frac{r}{2}}\sum_{i=1}^{r+2}\frac{(L^{(S),1}_iL^{(S),2}_i)^{\frac{r}{2}}\,\log\left(\frac{L^{(S),1}_i}{L^{(S),2}_i}\right)}{\prod_{j\neq i}(L^{(S),1}_iL^{(S),2}_j-L^{(S),1}_jL^{(S),2}_i)},\qquad \text{even }r.
\end{equation}

\section{Proof of the Duality}\label{app:duality}

To understand why this is true, it is sufficient to focus on an arbitrary symbol entry, say, some $(\rho_{2k-1}\rho_{2k})$ in a partition $\rho$. From previous discussions in the differentiation method we know that in the original space the relevant $2\times2$ matrix for this symbol entry has the form
\begin{equation}\label{relevantminordifferentiation}
\parbox{6cm}{
\tikz{
\node [anchor=center] at (.5,0) {$Q$};
\node [anchor=center] at (2*.5,0) {$\ldots$};
\node [anchor=center] at (3*.5,0) {$Q$};
\node [anchor=center] at (4.5*.5,.2) {$H_{2k+1}$};
\node [anchor=center] at (6.5*.5,.2) {$H_{2k+2}$};
\node [anchor=center] at (8*.5+.2,.2) {$\ldots$};
\node [anchor=center] at (9.5*.5,.2) {$H_{n}$};
\node [anchor=center] at (11*.5+.2,.2) {$H_{\alpha}$};
\node [anchor=center] at (4.5*.5,-.2) {$H_{2k+1}$};
\node [anchor=center] at (6.5*.5,-.2) {$H_{2k+2}$};
\node [anchor=center] at (8*.5+.2,-.2) {$\ldots$};
\node [anchor=center] at (9.5*.5,-.2) {$H_{n}$};
\node [anchor=center] at (11*.5+.2,-.2) {$H_{\beta}$};
\draw [black,thick] (.5,.6) -- (11.5*.5,.6);
\draw [black,thick] (.5,.6) -- (.5,.25);
\draw [black,thick] (3*.5,.6) -- (3*.5,.25);
\draw [black,thick] (4*.5,.6) -- (4*.5,.45);
\draw [black,thick] (6*.5,.6) -- (6*.5,.45);
\draw [black,thick] (9.5*.5,.6) -- (9.5*.5,.45);
\draw [black,thick] (11.5*.5,.6) -- (11.5*.5,.45);
\draw [black,thick] (.5,-.6) -- (11.5*.5,-.6);
\draw [black,thick] (.5,-.6) -- (.5,-.25);
\draw [black,thick] (3*.5,-.6) -- (3*.5,-.25);
\draw [black,thick] (4*.5,-.6) -- (4*.5,-.45);
\draw [black,thick] (6*.5,-.6) -- (6*.5,-.45);
\draw [black,thick] (9.5*.5,-.6) -- (9.5*.5,-.45);
\draw [black,thick] (11.5*.5,-.6) -- (11.5*.5,-.45);
}},\quad\alpha,\beta\in\{2k-1,2k\}.
\end{equation}
As we go to the dual space, and again use the differentiation method, since each symbol term expected to be reversed, the relevant matrix is now
\begin{equation}
\parbox{6cm}{
\tikz{
\node [anchor=center] at (0,0) {$Q^{-1}$};
\node [anchor=center] at (1.25*.6,0) {$\ldots$};
\node [anchor=center] at (2.5*.6,0) {$Q^{-1}$};
\node [anchor=center] at (3.5*.6,.2) {$V_1$};
\node [anchor=center] at (4.5*.6,.2) {$V_2$};
\node [anchor=center] at (5.5*.6,.2) {$\ldots$};
\node [anchor=center] at (7*.6,.2) {$V_{2k-2}$};
\node [anchor=center] at (8.5*.6,.2) {$V_{\alpha}$};
\node [anchor=center] at (3.5*.6,-.2) {$V_1$};
\node [anchor=center] at (4.5*.6,-.2) {$V_2$};
\node [anchor=center] at (5.5*.6,-.2) {$\ldots$};
\node [anchor=center] at (7*.6,-.2) {$V_{2k-2}$};
\node [anchor=center] at (8.5*.6,-.2) {$V_{\beta}$};
\draw [black,thick] (-.2,.6) -- +(8.5*.6+.2,0);
\draw [black,thick] (-.2,.6) -- +(0,-.35);
\draw [black,thick] (2.25*.6,.6) -- +(0,-.35);
\draw [black,thick] (3.5*.6,.6) -- +(0,-.15);
\draw [black,thick] (4.5*.6,.6) -- +(0,-.15);
\draw [black,thick] (6.5*.6,.6) -- +(0,-.15);
\draw [black,thick] (8.5*.6,.6) -- +(0,-.15);
\draw [black,thick] (-.2,-.6) -- +(8.5*.6+.2,0);
\draw [black,thick] (-.2,-.6) -- +(0,.35);
\draw [black,thick] (2.25*.6,-.6) -- +(0,.35);
\draw [black,thick] (3.5*.6,-.6) -- +(0,.15);
\draw [black,thick] (4.5*.6,-.6) -- +(0,.15);
\draw [black,thick] (6.5*.6,-.6) -- +(0,.15);
\draw [black,thick] (8.5*.6,-.6) -- +(0,.15);
}},\quad\alpha,\beta\in\{2k-1,2k\}.
\end{equation}
In the first expression we feed in $2k-1$ $Q$'s and in the second we feed in $n-2k+1$ $Q^{-1}$'s. 

In order that the statement \eqref{reversionid} holds, we need these two matrices to be identical up to an overall factor and a possible simultaneous exchange between rows and columns. To prove this, start from the second expression and we apply the relation
\begin{equation}
V_1^{I_1}\cdots V_{2k-2}^{I_{2k-2}}V_{\alpha}^{I_{\alpha}}\epsilon_{I_1\ldots I_{n}}=H_{\alpha^{\rm c},J_{\alpha^{\rm c}}}H_{2k+1,J_{2k+1}}\cdots H_{n,J_{n}}\epsilon^{I_1\ldots I_{2k-2}I_\alpha J_{\alpha^{\rm c}}J_{2k+1}\ldots J_{n}}\epsilon_{I_1\ldots I_{n}},
\end{equation}
where $\alpha^{\rm c}$ denotes the label other than $\alpha$ in the set $\{2k-2,2k\}$. This turn the expression into
\begin{equation}
\parbox{2.4cm}{
\tikz{
\node [anchor=center] at (0,0) {$Q$};
\node [anchor=center] at (.6,0) {$Q$};
\node [anchor=center] at (2*.6,0) {$\cdots$};
\node [anchor=center] at (3*.6,0) {$Q$};
\draw [black,thick] (0,.4) -- +(3*.6,0);
\draw [black,thick] (0,.4) -- +(0,-.15);
\draw [black,thick] (.6,.4) -- +(0,-.15);
\draw [black,thick] (3*.6,.4) -- +(0,-.15);
\draw [black,thick] (0,-.4) -- +(3*.6,0);
\draw [black,thick] (0,-.4) -- +(0,.15);
\draw [black,thick] (.6,-.4) -- +(0,.15);
\draw [black,thick] (3*.6,-.4) -- +(0,.15);
}}\times
\parbox{6cm}{
\tikz{
\node [anchor=center] at (0,0) {$Q^{-1}$};
\node [anchor=center] at (1.25*.6,0) {$\ldots$};
\node [anchor=center] at (2.5*.6,0) {$Q^{-1}$};
\node [anchor=center] at (7*.6,.8) {$H_{2k+1}$};
\node [anchor=center] at (8.5*.6,.8) {$\cdots$};
\node [anchor=center] at (10*.6,.8) {$H_{n}$};
\node [anchor=center] at (11.5*.6,.8) {$H_{\alpha^{\rm c}}$};
\node [anchor=center] at (7*.6,-.8) {$H_{2k+1}$};
\node [anchor=center] at (8.5*.6,-.8) {$\cdots$};
\node [anchor=center] at (10*.6,-.8) {$H_{n}$};
\node [anchor=center] at (11.5*.6,-.8) {$H_{\beta^{\rm c}}$};
\draw [black,thick] (-.2,.6) -- +(5.5*.6+.2,0);
\draw [black,thick] (-.2,.6) -- +(0,-.35);
\draw [black,thick] (2.25*.6,.6) -- +(0,-.35);
\draw [black,thick] (3.5*.6,.6) -- +(0,-.2);
\draw [black,thick] (4*.6,.6) -- +(0,-.2);
\node [anchor=center] at (4.75*.6,.45) {$\cdots$};
\draw [black,thick] (5.5*.6,.6) -- +(0,-.2);
\draw [black,thick] (3.5*.6,.4) -- +(8*.6-.1,0);
\draw [black,thick] (6.5*.6,.4) -- +(0,.15);
\draw [black,thick] (9.7*.6,.4) -- +(0,.15);
\draw [black,thick] (11.5*.6-.1,.4) -- +(0,.15);
\draw [black,thick] (-.2,-.6) -- +(5.5*.6+.2,0);
\draw [black,thick] (-.2,-.6) -- +(0,.35);
\draw [black,thick] (2.25*.6,-.6) -- +(0,.35);
\draw [black,thick] (3.5*.6,-.6) -- +(0,.2);
\draw [black,thick] (4*.6,-.6) -- +(0,.2);
\node [anchor=center] at (4.75*.6,-.5) {$\cdots$};
\draw [black,thick] (5.5*.6,-.6) -- +(0,.2);
\draw [black,thick] (3.5*.6,-.4) -- +(8*.6-.1,0);
\draw [black,thick] (6.5*.6,-.4) -- +(0,-.15);
\draw [black,thick] (9.7*.6,-.4) -- +(0,-.15);
\draw [black,thick] (11.5*.6-.1,-.4) -- +(0,-.15);
}}
\end{equation}
In the above we have inserted a factor $\det{Q}$, which of course does no harm. We can then apply a Schouten identity to merge the two factors into
\begin{equation}
\parbox{10cm}{
\tikz{
\node [anchor=center] at (-4.5*.6,0) {$Q$};
\node [anchor=center] at (-3.5*.6,0) {$\cdots$};
\node [anchor=center] at (-2.5*.6,0) {$Q$};
\node [anchor=center] at (0,0) {$Q^{-1}$};
\node [anchor=center] at (1.25*.6,0) {$\ldots$};
\node [anchor=center] at (2.5*.6,0) {$Q^{-1}$};
\node [anchor=center] at (5.5*.6,.8) {$Q$};
\node [anchor=center] at (7*.6,.8) {$H_{2k+1}$};
\node [anchor=center] at (8.5*.6,.8) {$\cdots$};
\node [anchor=center] at (10*.6,.8) {$H_{n}$};
\node [anchor=center] at (11.5*.6,.8) {$H_{\alpha^{\rm c}}$};
\node [anchor=center] at (7*.6,-.8) {$H_{2k+1}$};
\node [anchor=center] at (8.5*.6,-.8) {$\cdots$};
\node [anchor=center] at (10*.6,-.8) {$H_{n}$};
\node [anchor=center] at (11.5*.6,-.8) {$H_{\beta^{\rm c}}$};
\draw [black,thick] (-.9,.6) -- +(5*.6+.9,0);
\draw [black,thick] (-.2,.6) -- +(0,-.35);
\draw [black,thick] (2.25*.6,.6) -- +(0,-.35);
\draw [black,thick] (3.5*.6,.6) -- +(0,-.2);
\node [anchor=center] at (4.25*.6,.45) {$\cdots$};
\draw [black,thick] (5*.6,.6) -- +(0,-.2);
\draw [black,thick] (5.5*.6,.6) -- +(0,-.2);
\draw [black,thick] (3.5*.6,.4) -- +(8*.6-.1,0);
\draw [black,thick] (6.5*.6,.4) -- +(0,.15);
\draw [black,thick] (9.7*.6,.4) -- +(0,.15);
\draw [black,thick] (11.5*.6-.1,.4) -- +(0,.15);
\draw [black,thick] (-4.5*.6,1.2) -- (5.5*.6,1.2);
\draw [black,thick] (-4.5*.6,1.2) -- +(0,-.95);
\draw [black,thick] (-2.5*.6,1.2) -- +(0,-.95);
\draw [black,thick] (5.5*.6,1.2) -- +(0,-.15);
\draw [black,thick] (-4.5*.6,-.4) -- (-.9,-.4);
\draw [black,thick] (-4.5*.6,-.4) -- +(0,.15);
\draw [black,thick] (-2.5*.6,-.4) -- +(0,.15);
\draw [black,thick] (-.9,-.4) -- +(0,1);
\draw [black,thick] (-.2,-.6) -- +(5.5*.6+.2,0);
\draw [black,thick] (-.2,-.6) -- +(0,.35);
\draw [black,thick] (2.25*.6,-.6) -- +(0,.35);
\draw [black,thick] (3.5*.6,-.6) -- +(0,.2);
\draw [black,thick] (4*.6,-.6) -- +(0,.2);
\node [anchor=center] at (4.75*.6,-.5) {$\cdots$};
\draw [black,thick] (5.5*.6,-.6) -- +(0,.2);
\draw [black,thick] (3.5*.6,-.4) -- +(8*.6-.1,0);
\draw [black,thick] (6.5*.6,-.4) -- +(0,-.15);
\draw [black,thick] (9.7*.6,-.4) -- +(0,-.15);
\draw [black,thick] (11.5*.6-.1,-.4) -- +(0,-.15);
}}
\end{equation}
By further applying Schouten identities twice, one can show that this is actually proportional to
\begin{equation}
\parbox{6cm}{
\tikz{
\node [anchor=center] at (0,0) {$Q^{-1}$};
\node [anchor=center] at (1.25*.6,0) {$\ldots$};
\node [anchor=center] at (2.5*.6,0) {$Q^{-1}$};
\node [anchor=center] at (5.5*.6,0) {$Q$};
\node [anchor=center] at (7*.6,.8) {$H_{2k+1}$};
\node [anchor=center] at (8.5*.6,.8) {$\cdots$};
\node [anchor=center] at (10*.6,.8) {$H_{n}$};
\node [anchor=center] at (11.5*.6,.8) {$H_{\alpha^{\rm c}}$};
\node [anchor=center] at (7*.6,-.8) {$H_{2k+1}$};
\node [anchor=center] at (8.5*.6,-.8) {$\cdots$};
\node [anchor=center] at (10*.6,-.8) {$H_{n}$};
\node [anchor=center] at (11.5*.6,-.8) {$H_{\beta^{\rm c}}$};
\draw [black,thick] (-.2,.6) -- +(5*.6+.2,0);
\draw [black,thick] (-.2,.6) -- +(0,-.35);
\draw [black,thick] (2.25*.6,.6) -- +(0,-.35);
\draw [black,thick] (3.5*.6,.6) -- +(0,-.2);
\node [anchor=center] at (4.25*.6,.45) {$\cdots$};
\draw [black,thick] (5*.6,.6) -- +(0,-.2);
\draw [black,thick] (5.5*.6,.4) -- +(0,-.15);
\draw [black,thick] (3.5*.6,.4) -- +(8*.6-.1,0);
\draw [black,thick] (6.5*.6,.4) -- +(0,.15);
\draw [black,thick] (9.7*.6,.4) -- +(0,.15);
\draw [black,thick] (11.5*.6-.1,.4) -- +(0,.15);
\draw [black,thick] (-.2,-.6) -- +(5*.6+.2,0);
\draw [black,thick] (-.2,-.6) -- +(0,.35);
\draw [black,thick] (2.25*.6,-.6) -- +(0,.35);
\draw [black,thick] (3.5*.6,-.6) -- +(0,.2);
\node [anchor=center] at (4.25*.6,-.5) {$\cdots$};
\draw [black,thick] (5*.6,-.6) -- +(0,.2);
\draw [black,thick] (5.5*.6,-.4) -- +(0,.15);
\draw [black,thick] (3.5*.6,-.4) -- +(8*.6-.1,0);
\draw [black,thick] (6.5*.6,-.4) -- +(0,-.15);
\draw [black,thick] (9.7*.6,-.4) -- +(0,-.15);
\draw [black,thick] (11.5*.6-.1,-.4) -- +(0,-.15);
}}
\end{equation}
So we have one more $Q^{-1}$ and one more $Q$ contracting with $H$'s via the Levi-Civita, but one less contractions between two Levi-Civita symbols. Now the idea is clear: we can keep multiplying the expression but more $\det Q$ and apply the same operations. After iterating this procedure $2k$ times in total, we mange to recover the relevant submatrix \eqref{relevantminordifferentiation} in the original space, but with the label $\alpha,\beta$ substituted by $\alpha^{\rm c},\beta^{\rm c}$, and so this resulting matrix is related to \eqref{relevantminordifferentiation} exactly by a simultaneous exchange of rows and columns, thus yielding the same ratio of roots.

\section{Feynman Parametrization of Generic Number of Cuts}\label{app:sec:morecuts}

In this appendix we generalize the discussion in Section \ref{sec:FPofcuts} to arbitrary cuts and derive their corresponding Feynman parametrization. Again the essential object under study is the $d$-gon in $d$ spacetime dimensions, but here we consider cutting some $c$ propagators labeled by $\{i_1,\ldots,i_c\}$. We use the same notation $\mathcal{Y}$ for the associated region variables to keep in mind that they can be massive propagators. Generically the cut integral reads
\begin{equation}
\dot{I}_d^{(i_1\ldots i_c)}=\int_0^\infty\langle X_{(i_1\ldots i_c)}\mathrm{d}^{d-c-1}X_{(i_1\ldots i_c)}\rangle\int\frac{\mathrm{d}^{d+2}Y\,\delta(Y\cdot Y)}{\text{vol.}\mathrm{GL}(1)}\frac{\delta(Y\cdot \mathcal{Y}_{i_1})\cdots\delta(Y\cdot\mathcal{Y}_{i_c})}{(Y\cdot W_{(i_1\ldots i_c)})^{d-c}},
\end{equation}
where
\begin{equation}
W_{(i_!\ldots i_c)}=\sum_{j\notin\{i_1,\ldots,i_c\}}x_j\mathcal{Y}_j.
\end{equation}

We first integrate against the delta constraints $\prod\delta(Y\cdot\mathcal{Y})$. Let us reparametrize the loop momentum as follows
\begin{equation}
Y^I=\sum_{j=1}^{d+2-c}\alpha_jK^I_{\;J}Z_j^I+\sum_{j=1}^c\beta_j\mathcal{Y}_{i_j}.
\end{equation}
Here $K$ is an operator projecting any vector to the hyperplane constrained by $\{\mathcal{Y}_{i_1I},\ldots,\mathcal{Y}_{i_cI}\}$, constructed as
\begin{equation}
K_{IJ}=\frac{
\parbox{3.6cm}{\tikz{
\node [anchor=center] at (0*.6,0) {$\eta$};
\node [anchor=center] at (1*.6,0) {$\cdots$};
\node [anchor=center] at (2*.6,0) {$\eta$};
\node [anchor=center] at (3*.6+.1,.2) {$\mathcal{Y}_{i_1}$};
\node [anchor=center] at (3*.6+.1,-.2) {$\mathcal{Y}_{i_1}$};
\node [anchor=center] at (4*.6+.1,.2) {$\cdots$};
\node [anchor=center] at (4*.6+.1,-.2) {$\cdots$};
\node [anchor=center] at (5*.6+.1,.2) {$\mathcal{Y}_{i_c}$};
\node [anchor=center] at (5*.6+.1,-.2) {$\mathcal{Y}_{i_c}$};
\draw [black,thick] (-1*.6,.6) -- +(6*.6,0);
\draw [black,thick] (-1*.6,.6) -- +(0,-.15);
\draw [black,thick] (0*.6,.6) -- +(0,-.35);
\draw [black,thick] (2*.6,.6) -- +(0,-.35);
\draw [black,thick] (3*.6,.6) -- +(0,-.15);
\draw [black,thick] (5*.6,.6) -- +(0,-.15);
\draw [black,thick] (-1*.6,-.6) -- +(6*.6,0);
\draw [black,thick] (-1*.6,-.6) -- +(0,.15);
\draw [black,thick] (0*.6,-.6) -- +(0,.35);
\draw [black,thick] (2*.6,-.6) -- +(0,.35);
\draw [black,thick] (3*.6,-.6) -- +(0,.15);
\draw [black,thick] (5*.6,-.6) -- +(0,.15);
}}
}{(d-c-1)!\,G(\{\mathcal{Y}_{i_1},\ldots,\mathcal{Y}_{i_c})\}},
\end{equation}
where $G$ denotes the Gram determinant. The Levi-Civita symbols above takes Lorentz indices, and $\eta$ denotes the inverse metric of the ambient space. Obviously $K\mathcal{Y}_j=0$ $(\forall j\in\{i+1,\ldots,i_c\})$, and satisfies
\begin{equation}
KK=K,
\end{equation}
and one can check this reduces to the expressions for the projector in the previous subsection in the case of unitarity cuts. $\{Z_j\}$ are some reference vectors, and for convenience they can be chosen to be orthonormal wrst the projector $K$, i.e.,
\begin{equation}
Z_iKZ_j=\delta_{ij}.
\end{equation}

With the above setup the integral turns into
\begin{equation}
\begin{split}
\dot{I}_d^{(i_1\ldots i_c)}=&\int_0^\infty\langle X_{(i_1\ldots i_c)}\mathrm{d}^{d-c-1}X_{(i_1\ldots i_c)}\rangle\int\frac{\mathrm{d}^{d+2-c}\alpha\,\mathrm{d}^c\beta\,\delta(\sum_j\alpha_j^2+(\sum_j\beta_j\mathcal{Y}_{i_j})^2)}{\text{vol.}\mathrm{GL}(1)}\times\\
&\times\sqrt{G(\{\mathcal{Y}_{i_1},\ldots,\mathcal{Y}_{i_c})\}}\frac{\delta((\sum_j\beta_j\mathcal{Y}_{i_j})\cdot \mathcal{Y}_{i_1})\cdots\delta((\sum_j\beta_j\mathcal{Y}_{i_j})\cdot\mathcal{Y}_{i_c})}{((\sum_j\alpha_jZ_j)KW_{(i_1\ldots i_c)}+(\sum_j\beta_j\mathcal{Y}_{i_j})\cdot W_{(i_1\ldots i_c)})^{d-c}}\\
=&\int_0^\infty\frac{\langle X_{(i_1\ldots i_c)}\mathrm{d}^{d-c-1}X_{(i_1\ldots i_c)}\rangle}{\sqrt{G(\{\mathcal{Y}_{i_1},\ldots,\mathcal{Y}_{i_c}\})}}\int\frac{\mathrm{d}^{d+2-c}\alpha\,\delta(\sum_j\alpha_j^2)}{\text{vol.}\mathrm{GL}(1)}\frac{1}{((\sum_j\alpha_jZ_j)KW_{(i_1\ldots i_c)})^{d-c}}.
\end{split}
\end{equation}
The remaining $\alpha$ integrals are the same as the one for the Feynman parametrization of an $(d+2-c)$-gon in $d+2-c$ dimensions, and so we immediately obtain
\begin{equation}
\dot{I}_d^{(i_1\ldots i_c)}=
\int_0^\infty\frac{\langle X_{(i_1\ldots i_c)}\mathrm{d}^{d-c-1}X_{(i_1\ldots i_c)}\rangle}{\sqrt{G(\{\mathcal{Y}_{i_1},\ldots,\mathcal{Y}_{i_c}\})}}
\frac{1}{\left(\sum_{j=1}^{d+2-c}(W_{(i_1\ldots i_c)}KZ_j)(Z_jKW_{(i_1\ldots i_c)})\right)^{\frac{d-c}{2}}}.
\end{equation}
Note that the operator $K'\equiv\sum_j(KZ_j)(Z_jK)$ satisfies
\begin{equation}
K'K'=K',
\end{equation}
and for arbitrary vector $Y$
\begin{equation}
K'Y=\sum_{j=1}^{d+2-c}\alpha_jKZ_j=KY,
\end{equation}
and so it is the same projector $K'=K$. Hence we conclude
\begin{equation}
\dot{I}_d^{(i_1\ldots i_c)}=
\int_0^\infty\frac{\langle X_{(i_1\ldots i_c)}\mathrm{d}^{d-c-1}X_{(i_1\ldots i_c)}\rangle}{\sqrt{G(\{\mathcal{Y}_{i_1},\ldots,\mathcal{Y}_{i_c}\})}}
\frac{1}{(W_{(i_1\ldots i_c)}KW_{(i_1\ldots i_c)})^{\frac{d-c}{2}}}.
\end{equation}

\newpage

\bibliographystyle{JHEP}
\bibliography{feynmanparameters}

\end{document}